\documentstyle[11pt]{article}
\newfont{\blackb}{msbm10 scaled\magstep1}
\def\Bbb#1{\hbox{\blackb #1}}

\newtheorem{bbb}{Proposition}[section]
\newtheorem{ccc}{Lemma}[section]
\newtheorem{ddd}{Theorem}[section]
\newtheorem{eee}{Remark}[section]
\newtheorem{fff}{Corollary}[section]
\newtheorem{ggg}{Definition}[section]

\newtheorem{bbbb}{Proposition}[subsection]
\newtheorem{cccc}{Lemma}[subsection]

\newtheorem{ffff}{Corollary}[subsection]

\begin{document}
\baselineskip=12pt
\frenchspacing
\abovedisplayshortskip=4.25pt
\belowdisplayshortskip=4.25pt
\abovedisplayskip=4.25pt
\belowdisplayskip=4.25pt
\title{Higher Order Asymptotics of the Modified 
Non-Linear Schr\"{o}dinger Equation}
\author{A. H. Vartanian\footnote{{\sf E-mail: 
arthur@pdmi.ras.ru}} \\
Steklov Mathematical Institute \\
Fontanka 27 \\
St.~Petersburg 191011 \\
Russia}
\date{11 April 1998}
\maketitle 
\begin{abstract}
\noindent
Using the matrix Riemann-Hilbert factorisation approach for 
non-linear evolution systems which take the form of Lax-pair 
isospectral deformations, the higher order asymptotics as $t 
\! \to \! \pm \infty$ $(x/t \! \sim \! {\cal O}(1))$ of the 
solution to the Cauchy problem for the modified non-linear 
Schr\"{o}dinger equation, $i \partial_{t} u \! + \! \frac{
1}{2} \partial_{x}^{2} u \! + \! \vert u \vert^{2} u \! + \! 
i s \partial_{x} (\vert u \vert^{2} u) \! = \! 0$, $s \! \in 
\! \Bbb R_{> 0}$, which is a model for non-linear pulse 
propagation in optical fibres in the subpicosecond time scale, 
are obtained: also derived are analogous results for two 
gauge-equivalent non-linear evolution equations; in particular, 
the derivative non-linear Schr\"{o}dinger equation, $i 
\partial_{t} q \! + \! \partial_{x}^{2} q \! - \! i 
\partial_{x}(\vert q \vert^{2} q) \! = \! 0$.

\vspace{1.35cm}
{\bf AMS subject classifications.} 35Q15, 35Q55, 58F07, 78A60
 
\vspace{0.50cm}
{\bf PACS.} 02.30.Jr, 42.81.Dp, 42.65.Tg, 02.30.Mv

\vspace{0.50cm}
{\bf Abbreviated title.} Higher Order Asymptotics of the MNLSE

\vspace{0.50cm}
{\bf Key words.} asymptotics, Riemann-Hilbert problem, solitons, 
optical fibres
\end{abstract}
\clearpage
\section{Introduction}
A model for the space-time evolution of the slowly-varying 
amplitude of the complex field (pulse) envelope in optical 
fibres in the subpicosecond time scale is the modified 
non-linear Schr\"{o}dinger equation (MNLSE) \cite{a1},
\begin{eqnarray}
&i \partial_{t} u + \frac{1}{2} \partial^{2}_{x} u + \vert u 
\vert^{2} u + i s \partial_{x} (\vert u \vert^{2} u) = 0,&
\end{eqnarray}
$s \! \in \! \Bbb R_{>0}$. In order to study its asymptotics 
in the solitonless sector as $t \! \rightarrow \! \pm \infty$ 
$(x/t \! \sim \! {\cal O}(1))$, it was shown in \cite{a2} that, 
since its associated matrix Riemann-Hilbert (RH) factorisation 
problem does not possess the canonical normalisation, the MNLSE 
can be reduced to a gauge-equivalent non-linear evolution 
equation (NLEE) whose associated (solvable) matrix RH factorisation 
problem possesses the canonical normalisation: in fact, it was 
shown in \cite{a2} that, for $Q(x,t)$ satisfying the following 
NLEE,
\begin{eqnarray}
&i {\partial}_{t} Q + {\partial}_{x}^{2} Q + i Q^{2} 
{\partial}_{x} \overline{Q} + \frac{1}{2} Q \vert Q 
\vert^{4} = 0,&
\end{eqnarray}
with initial condition $Q(x,\! 0) \! \in \! {\cal S}(\Bbb R;\! 
\Bbb C)$ (see the beginning of Sec.~2, Notational Conventions, 
for the definition of the class ${\cal S}(D;\! \Bbb C)$ and 
$\overline{(\bullet)})$, the function $q(x,t)$, which is defined 
by the following gauge transformation (see Proposition~2.3)
\begin{eqnarray}
&q(x,t) := Q(x,t)((\Psi^{-1}(x,t;0))_{11})^{2},& 
\end{eqnarray}
with initial condition $q(x,\! 0) \! \in \! {\cal S}(\Bbb R;\! 
\Bbb C)$, satisfies the derivative non-linear Schr\"{o}dinger 
equation (DNLSE),
\begin{eqnarray}
&i {\partial}_{t} q + {\partial}_{x}^{2} q - i {\partial}_{x} 
(\vert q \vert^{2} q) = 0,&
\end{eqnarray}
and $u(x,t)$, under the following (scaling and Galilean) 
transformation,
\begin{eqnarray} 
&u(x,t) := \frac{1}{\sqrt{2 s}} \exp \{\frac{i}{s} 
(x-\frac{t}{2 s})\} q(\frac{t}{s} - x,\frac{t}{2}),&
\end{eqnarray}
with initial condition $u(x,\! 0) \! \in \! {\cal S}(\Bbb R;\! 
\Bbb C)$, satisfies the MNLSE. A systematic method for the 
leading order asymptotic analysis of soliton-bearing equations 
(integrable in the sense of the inverse scattering method (ISM) 
\cite{a3,a4,a5}) solvable through a canonical, oscillatory matrix 
RH factorisation problem was developed by Deift and Zhou 
\cite{a6} (see, also, \cite{a7,a8,a9}). Using the Deift-Zhou 
\cite{a6} procedure (with amendments \cite{a2,a10}), the following 
leading-order results were proven (in Theorems~1.1 and 1.2 below, 
one should keep the upper signs as $t \! \to \! +\infty$ and the 
lower signs as $t \! \to \! -\infty$ everywhere):
\begin{ddd}[{\rm \cite{a2,a10}}]
For $\vert \vert r \vert \vert_{{\cal L}^{\infty}(\Bbb R;
\Bbb C)} \! := \! \sup_{\lambda \in \Bbb R} \vert r(\lambda) 
\vert \! < \! 1$, where $r(\lambda)$ is the reflection 
coefficient associated with the direct scattering problem 
for the operator $\partial_{x} \! - \! U (x,t;\! \lambda)$ 
(see Proposition~2.1), and $Q(x,\!0)$ and $q(x,\! 0)$, 
respectively, $\! \in \! {\cal S}(\Bbb R;\! \Bbb C)$, as $t 
\! \to \! \pm \infty$ and $x \! \to \! \mp \infty$ such that 
$\lambda_{0} \! := \! \frac{1}{2} \sqrt{-\frac{x}{t}} \! > 
\! M_{1} \, (> \! 0)$ and $(x,t) \! \in \! \Bbb R^{2} \! 
\setminus \! \Omega_{n}$,
\begin{eqnarray*}
&Q(x,t) = \sqrt{\pm \frac{\nu(\lambda_{0})}{2 \lambda_{0}^{2} 
t}} e^{i \{4 \lambda_{0}^{4} t \mp \nu(\lambda_{0}) \ln \! 
\vert t \vert + \phi^{\pm}(\lambda_{0}) + \widehat{\Phi}^{\pm}
(\lambda_{0}) + \frac{\pi}{2}\}} + {\cal O} \! \left(
\frac{C_{1}(\lambda_{0}) \ln \vert t \vert}{\lambda_{0}^{2} t} 
\right) \!,& \\
&q(x,t) = \sqrt{\pm \frac{\nu(\lambda_{0})}{2 \lambda_{0}^{2} 
t}} e^{i \{4 \lambda_{0}^{4} t \mp \nu(\lambda_{0}) \ln \! 
\vert t \vert + \phi^{\pm}(\lambda_{0}) + \widehat{\Phi}^{\pm}
(\lambda_{0}) + \epsilon^{\pm}(\lambda_{0}) + \frac{\pi}{2}\}} 
+ {\cal O} \! \left(\frac{C_{2}(\lambda_{0}) \ln \vert t \vert}{
\lambda_{0}^{2} t} \right) \!,&
\end{eqnarray*}
where $\Omega_{n}$, $\nu(\lambda_{0})$, $\phi^{\pm}(\lambda_{0})$, 
and $\widehat{\Phi}^{\pm}(\lambda_{0})$ are given in Theorem~2.2, 
$\epsilon^{\pm}(\lambda_{0})$ are given in Theorem~2.3, with $C_{1,
2}(\lambda_{0}) \! \in \! {\cal S}(\Bbb R_{> M_{1}};\! \Bbb C)$, 
and, as $t \! \to \! \pm \infty$ and $x \! \to \! \pm \infty$ such 
that $\mu_{0} \! := \! \frac{1}{2} \sqrt{\frac{x}{t}} \! > \! 
M_{2} \, (> \! 0)$ and $(x,t) \! \in \! \Bbb R^{2} \! \setminus \! 
\mho_{n}$,
\begin{eqnarray*}
&Q(x,t) = \sqrt{\mp \frac{\nu(i \mu_{0})}{2 \mu_{0}^{2} t}} 
e^{i \{4 \mu_{0}^{4} t \mp \nu(i \mu_{0}) \ln \! \vert 
t \vert + \phi^{\pm \prime}(\mu_{0}) + \widehat{\Phi}^{\pm 
\prime}(\mu_{0}) + \pi \}} + {\cal O} \! \left(\frac{
C_{3}(\mu_{0}) \ln \vert t \vert}{\mu_{0}^{2} t} \right) \!,& \\
&q(x,t) = \sqrt{\mp \frac{\nu(i \mu_{0})}{2 \mu_{0}^{2} t}} 
e^{i \{4 \mu_{0}^{4} t \mp \nu(i \mu_{0}) \ln \! \vert 
t \vert + \phi^{\pm \prime}(\mu_{0}) + \widehat{\Phi}^{\pm 
\prime}(\mu_{0}) + \epsilon^{\pm \prime}(\mu_{0}) + \pi\}} + 
{\cal O} \! \left(\frac{C_{4}(\mu_{0}) \ln \vert t \vert}{
\mu_{0}^{2} t} \right) \!,&
\end{eqnarray*}
where $\mho_{n}$, $\nu(i\mu_{0})$, $\phi^{\pm \prime}(\mu_{0})$, 
and $\widehat{\Phi}^{\pm \prime}(\mu_{0})$ are given in 
Theorem~2.2, $\epsilon^{\pm \prime}(\mu_{0})$ are given in 
Theorem~2.3, and $C_{3,4}(\mu_{0}) \! \in \! {\cal S}(\Bbb R_{
>M_{2}};\! \Bbb C)$.
\end{ddd}
\begin{ddd}[{\rm \cite{a2,a10}}]
For $\vert \vert r \vert \vert_{{\cal L}^{\infty}(\Bbb R;\Bbb 
C)} \! := \! \sup_{\lambda \in \Bbb R} \vert r(\lambda) 
\vert \! < \! 1$ and $u(x,\!0) \! \in \! {\cal S}(\Bbb R;\! \Bbb 
C)$, as $t \! \to \! \pm \infty$ and $x \! \to \! \pm \infty$ 
such that $\widehat{\lambda}_{0} \! := \! \sqrt{\frac{1}{2}
(\frac{x}{t} \! - \! \frac{1}{s})} \! > \! M_{3} \, (> \! 0)$, 
$\frac{x}{t} \! > \! \frac{1}{s}$, $s \! \in \! \Bbb R_{>0}$, and 
$(x,t) \! \in \! \Bbb R^{2} \setminus \widetilde{\Omega}_{n}$,
\begin{eqnarray*}
&u(x,t) = \sqrt{\pm \frac{\nu(\widehat{\lambda}_{0})}{2 
\widehat{\lambda}_{0}^{2}st}} e^{i \{2(\widehat{\lambda}
_{0}^{2} + \frac{1}{2s})^{2} t \mp \nu(\widehat{\lambda}
_{0}) \ln \! \vert t \vert + \phi^{\pm}(\widehat{\lambda}
_{0}) + \widetilde{\Phi}^{\pm}(\widehat{\lambda}_{0}) + 
\frac{\pi}{2}\}} + {\cal O} \! \left(\frac{C_{5}(\widehat{
\lambda}_{0}) \ln \vert t \vert}{\widehat{\lambda}_{0}^{2} 
t} \right) \!,&
\end{eqnarray*}
where $\widetilde{\Omega}_{n}$ and $\widetilde{\Phi}^{\pm}(
\widehat{\lambda}_{0})$ are given in Theorem~2.4, with $C_{5}
(\widehat{\lambda}_{0}) \! \in \! {\cal S}(\Bbb R_{> M_{3}};\! 
\Bbb C)$, and, as $t \! \to \! \pm \infty$ and $x \! \to \! \mp 
\infty$ or $\pm \infty$ such that $\widehat{\mu}_{0} \! := \! 
\sqrt{\frac{1}{2}(\frac{1}{s} \! - \! \frac{x}{t})} \! > \! M_{
4} \, (> \! 0)$, $\frac{x}{t} \! < \! \frac{1}{s}$, $s \! \in 
\! \Bbb R_{>0}$, and $(x,t) \! \in \! \Bbb R^{2} \setminus 
\widetilde{\mho}_{n}$,
\begin{eqnarray*}
&u(x,t) = \sqrt{\mp \frac{\nu(i\widehat{\mu}_{0})}{2 \widehat{
\mu}_{0}^{2}st}} e^{i \{2(\widehat{\mu}_{0}^{2} - \frac{
1}{2s})^{2} t \mp \nu(i\widehat{\mu}_{0}) \ln \! \vert t \vert 
+ \phi^{\pm \prime}(\widehat{\mu}_{0}) + \widetilde{\Phi}^{\pm 
\prime}(\widehat{\mu}_{0}) + \pi\}} + {\cal O} \! 
\left(\frac{C_{6}(\widehat{\mu}_{0}) \ln \vert t \vert}{
\widehat{\mu}_{0}^{2} t} \right) \!,&
\end{eqnarray*}
where $\widetilde{\mho}_{n}$ and $\widetilde{\Phi}^{\pm \prime}
(\widehat{\mu}_{0})$ are given in Theorem~2.4, and $C_{6}
(\widehat{\mu}_{0}) \! \in \! {\cal S}(\Bbb R_{> M_{4}};\! 
\Bbb C)$.
\end{ddd}

Recently, Deift and Zhou \cite{a11} introduced a higher-order 
generalisation of their previous procedure \cite{a6}, and 
showed, using the defocusing non-linear Schr\"{o}dinger (NLS) 
and modified Korteweg-de Vries (MKdV) equations, respectively, 
as their illustrative examples, how, in principle, to obtain 
the full asymptotic expansion to all orders as $t \! \to \! 
+\infty$ $(x/t \! \sim \! {\cal O}(1))$ for integrable systems 
solvable through a canonical, oscillatory matrix RH factorisation 
problem: in order to decipher the structure of the higher-order 
interaction of modes for the solution of the MNLSE in the 
solitonless sector as $t \! \to \! \pm \infty$ $(x/t \! \sim \! 
{\cal O}(1))$, the higher order generalisation of the Deift-Zhou 
\cite{a11} procedure is applied (see, also, \cite{a12} for some 
recent results pertaining to non-soliton solutions of the NLS 
equation, and \cite{a13} for the higher order asymptotics of 
the KdV equation in the solitonless sector); moreover, even 
though specific design-related issues and/or 
feasibility studies for the MNLSE are not considered in 
this article, it is also instructive to consider its asymptotics
in the solitonless sector as $t \! \to \! \pm \infty$ $(x/t \! 
\sim \! {\cal O}(1))$ as it may find potential application(s) 
as a model for (propagation in) the optical fibre element used, 
say, in a non-linear fibre-loop mirror (whose transmission 
characteristics are intensity dependent), since such devices, 
when designed specifically to transmit the central part of a 
pulse and block its low-intensity wings, can compress any 
non-soliton pulse(s) \cite{a1}. 

Even though the application of the higher-order generalisation 
of the Deift-Zhou \cite{a11} procedure gives a systematic 
justification for the asymptotic form of the solutions to the 
Cauchy problems for Eq.~(2), the DNLSE, and MNLSE, unlike the 
NLS and MKdV equations, whose associated recurrence relations 
for the coefficients of the asymptotic expansions are explicitly 
(algebraically) solvable in closed form, the following two novel 
and substantial complications, associated with the 
non-analyticity of the reflection coefficient, $r(\lambda)$ 
(for the direct scattering problem of the operator 
$\partial_{x} \! - \! U(x,t;\! \lambda)$: see Proposition~2.1), 
at the origin of the complex plane of the auxiliary 
spectral parameter $(\lambda)$, are encountered: (1) 
$(D^{m}(r(\lambda) \vert_{\lambda \in \Bbb R})) \vert_{\lambda=0} 
\! \not= \! (D^{m}(r(\lambda) \vert_{\lambda \in i \Bbb R})) \vert
_{\lambda=0}$, where $D \! := \! \frac{d}{d\lambda}$, and $m \! = 
\! 1,3,5,\ldots$, in which case, there are two independent sets 
of constant coefficients associated with the Taylor series 
expansion of the reflection coefficient about $\lambda \! = \! 
0$; and (2) $(r(\lambda) \vert_{\lambda \in \Bbb R}) \vert_{
\lambda=0} \! = \! (r(\lambda) \vert_{\lambda \in i \Bbb R}) 
\vert_{\lambda=0} \! = \! 0$ \cite{a2}. Despite the fact 
that a set of recurrence relations for the coefficients of 
the asymptotic expansions of the solutions to Eq.~(2), the 
DNLSE, and MNLSE as $t \! \to \! \pm \infty$ $(x/t \! \sim \! 
{\cal O}(1))$ can be derived, their closed form solutions are 
not possible to obtain due to the fact that a subset of the 
coefficients appearing in these recurrence relations, associated 
with the two independent sets of (odd order) derivatives of the 
reflection coefficient at the origin, {\bf must} be determined 
independently. It is shown, in principle, how to overcome this 
complication, and the next three non-trivial terms beyond the 
leading-order ones given in Theorems~1.1 and 1.2 are calculated 
for the solutions of Eq.~(2), the DNLSE, and MNLSE.

This paper is organised as follows. In Sec.~2: (1) the matrix 
RH factorisation problem for the solution of the gauge-equivalent 
NLEE, Eq.~(2), is stated; (2) the Beals-Coifman \cite{a14} 
formulation for the solution of a (matrix) RH problem on an 
oriented contour is succinctly recapitulated; and (3) the results 
of this paper are summarised as Theorems~2.2--2.4. In Sec.~3, the 
higher-order generalisation of the Deift-Zhou \cite{a11} procedure 
is carried out in order to determine the (form of the) asymptotic 
expansion to all orders as $t \! \to \! +\infty$ $(x/t \! \sim \! 
{\cal O}(1))$ of the solution to the Cauchy (initial-value) problem 
for $Q(x,t)$, and recurrence relations for a few of its leading 
expansion coefficients are explicitly derived. In Sec.~4, a certain 
set of $\lambda_{0}$-dependent numbers, which are expressed in 
terms of the two independent sets of the Taylor series expansion 
coefficients for the reflection coefficient about the origin of 
the complex plane of the auxiliary spectral parameter, are 
calculated, and are shown, in conjunction with the (first few) 
leading recurrence formulae derived in Sec.~3, to lead to 
linear integro-differential equations for the asymptotic series 
expansion coefficients for $Q(x,t)$, which can---in principle---be 
rewritten as integral equations of the first kind, with Laplace 
transform-type kernels, whose inversion, concomitant with a 
$\sigma_{3}$-induced symmetry reduction, leads to their 
solution. In Sec.~5, the phase integral (Eq.~(3)), $((
\Psi^{-1}(x,t;0))_{11})^{2}$, is evaluated asymptotically as $t 
\! \to \! +\infty$ $(x/t \! \sim \! {\cal O}(1))$. In Sec.~6, 
the above asymptotic paradigm is briefly presented for the case 
when $t \! \to \! -\infty$ $(x/t \! \sim \! {\cal O}(1))$; 
furthermore, the results obtained in this paper constitute a 
formal proof of Lemmae~5.2 and 6.3.1 which were stated, but not 
proved, in \cite{a10}.
\begin{eee}
{\sf Strictly speaking, this paper is a hybrid continuation of 
\cite{a2} and \cite{a10}, and is not completely self-contained: 
theorems, concepts, definitions, and formulae {}from these references 
are used with minimal explanations.\/} {\rm Even though $\Bbb C \Bbb 
P^{1} \! := \! \Bbb C \cup \{\infty\}$ is the standard definition 
for the Riemann sphere (the complex plane with the point at 
infinity), throughout this paper, the simplified notation $\Bbb 
C$ is used to denote the Riemann sphere.\/}
\end{eee}
\section{The Riemann-Hilbert (RH) Problem, The Beals-Coifman 
(BC) Formulation, and Summary of Results}
In this section, the (general) matrix RH factorisation problem 
is stated, the BC \cite{a14} formulation for the solution of a 
(matrix) RH problem on an oriented contour is briefly reviewed, 
and the results of this paper are summarised as Theorems~2.2--2.4. 
Before doing so, however, notations and definitions 
which are used throughout the paper are introduced:
\begin{flushleft}
{\bf Notational Conventions}
\end{flushleft}
\begin{enumerate}
{\em \item[(1)] $e_{\alpha \beta}$, $\alpha,\beta \! 
\in \! \{1,2\}$, denote $2 \! \times \! 2$ matrices with 
entry $1$ in $(\alpha \, \beta)$, $(e_{\alpha \beta})_{ij} 
\! := \! \delta_{\alpha i} \delta_{\beta j}$, $i,j \! 
\in \! \{1,2\}$, where $\delta_{ij}$ is the Kronecker 
delta;
\item[(2)] ${\rm I} \! := \! e_{11} \! + \! e_{22}
\! = \! {\rm diag}(1,1)$ denotes the $2 \! \times \! 
2$ identity matrix;
\item[(3)] $\sigma_{3} \! := \! e_{11} 
\! - \! e_{22} \! = \! {\rm diag}(1,-1)$, $\sigma_{-} 
\! := \! e_{21}$, $\sigma_{+} \! := \! e_{12}$, 
and $\sigma_{1} \! := \! \sigma_{-} \! + \! 
\sigma_{+};$
\item[(4)] for a scalar $\varpi$ and a $2 \! 
\times \! 2$ matrix $\Upsilon$, $\varpi^{{\rm ad} 
(\sigma_{3})} \Upsilon \! := \! \varpi^{\sigma_{3}}
\Upsilon \varpi^{-\sigma_{3}};$
\item[(5)] $\overline{(\bullet)}$ 
denotes complex conjugation of $(\bullet);$
\item[(6)] $M_{2} 
(\Bbb C)$ denotes the $2 \! \times \! 2$ complex matrix 
algebra with the following inner product $((\cdot,\! \cdot) 
\colon M_{2} (\Bbb C) \! \times \! M_{2} (\Bbb C) \! 
\rightarrow \! \Bbb C)$, $\forall \, a,b \! \in \! M_{2} 
(\Bbb C)$, $(a,\! b) \! := \! {\rm tr} (\overline{b} 
a)$, and (for $a \! \in \! M_{2}(\Bbb C))$ the norm on 
$M_{2}(\Bbb C)$ is defined as $\vert a \vert \! := \! 
\sqrt{(a,\! a)};$
\item[(7)] ${\cal L}^{p}(D;\! M_{2}(\Bbb C)) 
\! := \! \{\mathstrut f; \, f \colon D \! \rightarrow 
\! M_{2} (\Bbb C), \, \vert \vert f \vert \vert_{{\cal 
L}^{p}(D;M_{2}(\Bbb C))} \! := \! (\int_{D} \vert 
f(\xi) \vert^{p} \vert d \xi \vert)^{1/p} \! < \! 
\infty, p \! \in \! \{1,2\}\};$
\item[(8)] ${\cal L}^{\infty}(D;\! M_{2}(\Bbb C)) \! := \! 
\{\mathstrut g; g \colon D \! \rightarrow \! M_{2}(\Bbb C), 
\vert \vert g \vert \vert_{{\cal L}^{\infty}(D;M_{2}(\Bbb 
C))} \! := \! \max\limits_{1 \leq i,j \leq 2} \sup\limits_{
\xi \in D} \vert g_{ij}(\xi) \vert \! < \! \infty\};$
\item[(9)] for $D$ an unbounded domain of $\Bbb 
R \cup i \Bbb R$, let ${\cal S}(D;\! \Bbb C)$ 
(resp.~${\cal S}(D;\! M_{2}(\Bbb C)))$ denote the Schwartz 
class on $D$, i.e., the class of smooth $\Bbb C$-valued 
(resp.~$M_{2}(\Bbb C)$-valued) functions $f(x) \colon D \! 
\rightarrow \! \Bbb C$ (resp.~$F(x) \colon D \! \rightarrow 
\! M_{2}(\Bbb C))$ which together with all derivatives tend 
to zero faster than any positive power of $\vert x \vert^{
-1}$ as $\vert x \vert \! \rightarrow \! \infty$.}
\end{enumerate}

As is clear {}from the discussion at the beginning of the 
Introduction, the strategy here is as follows: (1) solve 
Eq.~(2) for $Q(x,t)$; (2) use the gauge transformation 
(Eq.~(3)) to solve the DNLSE; and (3) via Eq.~(5), solve 
the MNLSE. To recall the relations between the solutions 
of these NLEEs, the following propositions are formulated:
\begin{bbb}[{\rm \cite{a15}}]
The necessary and sufficient condition for the compatibility of 
the following system of linear ODEs (the Lax pair) for arbitrary 
$\lambda \! \in \! \Bbb C$,
\begin{eqnarray}
&\partial_{x} \Psi(x,t;\! \lambda)=U(x,t;\! \lambda) \Psi(x,t;
\! \lambda), \, \, \, \, \, \, \, \, \, \partial_{t} \Psi(x,t;
\! \lambda)=V(x,t;\! \lambda) \Psi(x,t;\! \lambda),&
\end{eqnarray}
where
\begin{eqnarray*}
&U(x,t;\! \lambda) \! = \! - i \lambda^{2} \sigma_{3} \! + 
\! \lambda (\overline{Q} \sigma_{-} \! + \! Q \sigma_{+}) 
\! - \! \frac{i}{2} \vert Q \vert^{2} \sigma_{3},& \\ 
&V(x,t;\! \lambda)=2 \lambda^{2} U(x,t;\! \lambda) \! - \! 
i \lambda ((\partial_{x} \overline{Q}) \sigma_{-} \! - \! 
(\partial_{x} Q) \sigma_{+}) \! + \! (\frac{i}{4} \vert 
Q \vert^{4} \! + \! \frac{1}{2}(\overline{Q} \partial_{x} 
Q \! - \! Q \partial_{x} \overline{Q})) \sigma_{3},& 
\end{eqnarray*}
is that $Q(x,t)$ satisfies Eq.~(2).
\end{bbb}
\begin{bbb}[{\rm \cite{a2,a10}}]
Let $Q(x,t)$ be a solution of Eq.~(2). Then there exists a 
corresponding solution of system~(6) such that $\Psi(x,t;0)$ 
is a diagonal matrix.
\end{bbb}
\begin{bbb}[{\rm \cite{a2,a10}}]
Let $Q(x,t)$ be a solution of Eq.~(2) and $\Psi(x,t;\! \lambda)$ 
the corresponding solution of system~(6) given in Proposition~2.2. 
Set $\Psi_{q}(x,t;\! \lambda) \! := \! \Psi^{-1}(x,t;0) \Psi
(x,t;\! \lambda)$. Then
\begin{eqnarray*}
\partial_{x} \Psi_{q}(x,t;\! \lambda)={\cal U}_{q}(x,t;\! \lambda) 
\Psi_{q}(x,t;\! \lambda), \, \, \, \, \, \, \, \, \, \, \partial_{
t} \Psi_{q}(x,t;\! \lambda)={\cal V}_{q}(x,t;\! \lambda) \Psi_{q}
(x,t;\! \lambda),&
\end{eqnarray*}
where 
\begin{eqnarray*}
&{\cal U}_{q}(x,t;\! \lambda) = - i \lambda^{2} \sigma_{3} 
+ \lambda (\overline{q} \sigma_{-} + q \sigma_{+}),& 
\end{eqnarray*}
\begin{eqnarray*}
&{\cal V}_{q}(x,t;\! \lambda) = \left(\begin{array}{cc}
- 2 i \lambda^{4} - i \lambda^{2} \vert q \vert^{2} & 2 
\lambda^{3} q + i \lambda \partial_{x} q + \lambda 
\vert q \vert^{2} q \\
2 \lambda^{3} \overline{q} - i \lambda \partial_{x} 
\overline{q} + \lambda \vert q \vert^{2} \overline{q} & 
2 i \lambda^{4} + i \lambda^{2} \vert q \vert^{2} 
\end{array} \right) \!,&
\end{eqnarray*}
with $q(x,t)$ defined by Eq.~(3), is the ``Kaup-Newell'' 
{\rm \cite{a16}} Lax pair for the DNLSE.
\end{bbb}
\begin{bbb}[{\rm \cite{a2,a10}}]
If $q(x,t)$ is a solution of the DNLSE such that $q(x,\! 0) \! 
\in \! {\cal S}(\Bbb R;\! \Bbb C)$, then $u(x,t)$, defined by 
Eq.~(5), satisfies the MNLSE with initial condition $u(x,\! 0) 
\! \in \! {\cal S}(\Bbb R;\! \Bbb C)$.
\end{bbb}
\begin{eee}
{\rm A convention is now adopted which is adhered to {\em 
sensus strictu\/} throughout the paper: for each segment 
of an oriented contour, according to the given orientation, 
the ``$+$'' side is to the left and the ``$-$'' side is to 
the right as one traverses the contour in the direction of 
the orientation; hence, $(\bullet)_{+}$ and $(\bullet)_{-}$ 
denote, respectively, the non-tangential limits (boundary 
values) of $(\bullet)$ on an oriented contour {}from the 
``$+$'' (left) and ``$-$'' (right) sides.\/}
\end{eee}
\vspace{-0.30cm}
\begin{figure}[bht]
\begin{center}
\unitlength=1cm
\vspace{-0.825cm}
\begin{picture}(6,6)(0,0)
\thicklines
\put(2.5,0){\vector(0,1){1.25}}
\put(2.25,1.25){\makebox(0,0){$\scriptstyle{}+$}}
\put(2.75,1.25){\makebox(0,0){$\scriptstyle{}-$}}
\put(2.75,3.75){\makebox(0,0){$\scriptstyle{}+$}}
\put(2.25,3.75){\makebox(0,0){$\scriptstyle{}-$}}
\put(1.25,2.25){\makebox(0,0){$\scriptstyle{}+$}}
\put(1.25,2.75){\makebox(0,0){$\scriptstyle{}-$}}
\put(3.75,2.25){\makebox(0,0){$\scriptstyle{}-$}}
\put(3.75,2.75){\makebox(0,0){$\scriptstyle{}+$}}
\put(2.5,1.25){\line(0,1){1.25}}
\put(2.5,5){\vector(0,-1){1.25}}
\put(2.5,2.5){\line(0,1){1.25}}
\put(2.5,2.5){\vector(-1,0){1.25}}
\put(1.25,2.5){\line(-1,0){1.25}}
\put(2.5,2.5){\vector(1,0){1.25}}
\put(3.75,2.5){\line(1,0){1.25}}
\put(3.85,3.95){\makebox(0,0){$\scriptstyle{}\Im 
(\lambda^{2}) \, > \, 0$}}
\put(3.85,1.05){\makebox(0,0){$\scriptstyle{}\Im 
(\lambda^{2}) \, < \, 0$}}
\put(1.15,1.05){\makebox(0,0){$\scriptstyle{}\Im 
(\lambda^{2}) \, > \, 0$}}
\put(1.15,3.95){\makebox(0,0){$\scriptstyle{}\Im 
(\lambda^{2}) \, < \, 0$}}
\put(4.75,2.75){\makebox(0,0){$\scriptstyle{}\Re 
(\lambda)$}}
\put(3.0,4.85){\makebox(0,0){$\scriptstyle{}\Im 
(\lambda)$}}
\end{picture}
\vspace{-0.5cm}
\end{center}
\caption{Continuous spectrum ${\widehat \Gamma}$.}
\end{figure}
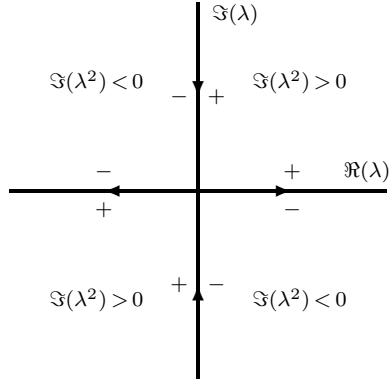

Before stating the (general) matrix RH problem, it is 
convenient to introduce the following preamble: {\em let ${
\cal Z}_{d} \! := \! \cup_{i=1}^{N} (\{\pm \lambda_{i}\} 
\! \cup \! \{\pm \overline{\lambda_{i}}\})$ and $\widehat{
\Gamma} \! := \! \{\mathstrut \lambda; \, \Im(\lambda^{2}
) \! = \! 0\}$ (oriented as in Fig.~1) denote, respectively, 
the discrete and continuous spectra of the operator $\partial
_{x} \! - \! U(x,t;\! \lambda)$, and $\sigma_{\pounds} \! 
:= \! {\rm Spec}(\partial_{x} \! - \! U) \! = \! {\cal Z}_{d} 
\! \cup \! \widehat{\Gamma}$ $({\cal Z}_{d} \cap \widehat{
\Gamma} \! = \! \emptyset)$.\/}
\begin{ccc}[{\rm \cite{a10}}]
Let $Q(x,t)$, as a function of $x$, $\! \in \! {\cal S}(\Bbb 
R;\! \Bbb C)$. Set $m(x,t;\! \lambda) \! := \! \Psi(x,t;\! 
\lambda) \exp\{\linebreak[4] i (\lambda^{2} x \! + \! 
2 \lambda^{4} t) \sigma_{3}\}$. Then: (1) the bounded discrete 
set ${\cal Z}_{d}$ is finite $({\rm card} ({\cal Z}_{d}) \! < 
\! \infty)$; (2) the poles of $m(x,t;\! \lambda)$ are simple; 
(3) the first (resp.~second) column of $m(x,t;\! \lambda)$ has 
poles at $\{\pm \lambda_{i}\}_{i=1}^{N}$ (resp.~$\{\pm 
\overline{\lambda_{i}}\}_{i=1}^{N})$; and (4) $\forall 
\, t \! \in \! \Bbb R$ the function $m(x,t;\! \lambda) 
\colon \Bbb C \! \setminus \! ({\cal Z}_{d} \! \cup \! 
\widehat{\Gamma}) \! \rightarrow \! {\rm SL\/}(2,\! \Bbb C)$ 
solves the following RH problem:
\begin{enumerate}
\item[a.] $m(x,t;\! \lambda)$ is meromorphic $\forall \, 
\lambda \! \in \! \Bbb C \! \setminus \! \widehat{\Gamma};$
\item[b.] $m(x,t;\! \lambda)$ satisfies the following 
jump conditions,
\begin{eqnarray*}
&m_{+}(x,t;\! \lambda)=m_{-}(x,t;\! \lambda) \exp\{-i(\lambda^{
2} x \! + \! 2\lambda^{4}t) {\rm ad} (\sigma_{3})\} G(\lambda), 
\, \, \, \, \, \, \, \, \lambda \! \in \! \widehat{\Gamma},&
\end{eqnarray*}
where
\begin{eqnarray*}
&G(\lambda) \! := \! \left(\! \begin{array}{cc}
1 - r(\lambda) \overline{r(\overline{\lambda})} & r(\lambda) \\
- \overline{r(\overline{\lambda})} & 1 
\end{array} \! \right) := ({\rm I} \! - \! w^{-}(\lambda))^{-1}
({\rm I} \! + \! w^{+}(\lambda)),&
\end{eqnarray*}
$r(\lambda)$, the reflection coefficient associated with the 
direct scattering problem for the operator $\partial_{x} \! 
- \! U(x,t;\! \lambda)$, satisfies $r(-\lambda) \! = \! - 
r(\lambda)$, and $r(\lambda) \! \in \! {\cal S}(\widehat{
\Gamma};\! \Bbb C);$
\item[c.] for the simple poles of $m(x,t;\! \lambda)$ at $\{
\pm \lambda_{j}\}_{j=1}^{N}$ and $\{\pm \overline{\lambda_{j}}
\}_{j=1}^{N}$, there exist nilpotent matrices $\{v_{j}(x,t) 
\sigma_{-}\}_{j=1}^{N}$ and $\{\overline{v_{j}(x,t)} \sigma_{
+}\}_{j=1}^{N}$, respectively, such that the residues, for $1 
\! \leq \! j \! \leq \! N$, satisfy the {\rm (BC \cite{a14})} 
polar conditions,
\begin{eqnarray*}
&{\rm res} (m(x,t;\! \lambda);\! \lambda_{j}) = \lim\limits_{
\lambda \rightarrow \lambda_{j}} \! m(x,t;\! \lambda) v_{j}
(x,t) \sigma_{-},& \\
&{\rm res} (m(x,t;\! \lambda);\! -\lambda_{j}) = - 
\sigma_{3} {\rm res} (m(x,t;\! \lambda);\! \lambda_{j}) 
\sigma_{3},& \\
&{\rm res} (m(x,t;\! \lambda);\! \overline{\lambda_{j}} \,) 
= \lim\limits_{\lambda \rightarrow \overline{\lambda_{
j}}} \! m(x,t;\! \lambda) \overline{v_{j}(x,t)} \sigma_{+},& \\
&{\rm res} (m(x,t;\! \lambda);\! -\overline{\lambda_{j}} \,)=- 
\sigma_{3} {\rm res} (m(x,t;\! \lambda);\! \overline{\lambda_{
j}} \,) \sigma_{3},& 
\end{eqnarray*}
where $\lambda_{j} \! := \! \Delta_{j} \exp \{\frac{i}{2} 
(\pi \! - \! \gamma_{j})\}$, $\Delta_{j} \! > \! 0$, $\gamma_{
j} \! \in \! (0,\! \pi)$, $v_{j}(x,t) \! := \! C_{j} \exp 
\{2 i(\lambda_{j}^{2} x \! + \! 2 \lambda_{j}^{4} t)\}$, and 
$C_{j}$ are complex constants associated with the direct 
scattering problem for the operator $\partial_{x} \! - \! 
U(x,t;\! \lambda);$ 
\item[d.] as $\lambda \! \rightarrow \! \infty$, $\lambda 
\! \in \! \Bbb C \! \setminus \! ({\cal Z}_{d} \! \cup \! 
\widehat{\Gamma})$,
\begin{eqnarray*}
&m(x,t;\! \lambda) = {\rm I} + {\cal O}(\lambda^{-1}).& 
\end{eqnarray*}
\end{enumerate}
\end{ccc}
\begin{ccc}
Let $\vert \vert r \vert \vert_{{\cal L}^{\infty}(\Bbb R;\Bbb 
C)} \! := \! \sup_{\lambda \in \Bbb R} \vert r(\lambda) \vert 
\! < \! 1$. Then: (1) the RH problem formulated in Lemma~2.1 
is uniquely solvable; (2) $\Psi(x,t;\! \lambda) \! = \! m(x,t;
\! \lambda) \exp \{-i(\lambda^{2} x \! + \! 2 \lambda^{4} t) 
\sigma_{3}\}$ is the solution of system~(6) with
\begin{eqnarray}
&Q(x,t) := 2 i \! \lim\limits_{\lambda \rightarrow \infty \atop 
\lambda \, \in \, \Bbb C \setminus ({\cal Z}_{d} \cup \widehat{
\Gamma})}(\lambda m(x,t;\! \lambda))_{12};&
\end{eqnarray}
(3) the function $Q(x,t)$ defined by Eq.~(7) satisfies Eq.~(2), 
and 
\begin{eqnarray}
&q(x,t) := Q(x,t)((m^{-1}(x,t;0))_{11})^{2}&
\end{eqnarray}
satisfies the DNLSE; and (4) $m(x,t;\! \lambda)$ possesses 
the following symmetry reductions, $m(x,t;\linebreak[4] 
\lambda) \! = \! \sigma_{3} m(x,t;\! -\lambda) \sigma_{3}$ 
and $m(x,t;\! \lambda) \! = \! \sigma_{1} \overline{m(x,t;\! 
\overline{\lambda})} \sigma_{1}$.

If $r(\lambda) \! \in \! {\cal S}(\widehat{\Gamma};\! \Bbb C)$, 
then, for any $t \! \in \! \Bbb R$, $Q(x,t)$ (resp.~$q(x,t))$, 
as a function of $x$, $\! \in \! {\cal S}(\Bbb R;\! \Bbb C)$.
\end{ccc}

{\em Proof.\/} The solvability of the RH problem (formulated 
in Lemma~2.1) is a consequence of Theorem~9.3 in \cite{a17}
(Zhou's skew Schwarz reflection invariant symmetry arguments: 
see, also, \cite{a18}) and the vanishing winding number of $1 
\! - \! r(\lambda) \overline{r(\overline{\lambda})}$, $\int_{
\widehat{\Gamma}} d(\arg(1 \! - \! r(\lambda) \overline{r(
\overline{\lambda})})) \! = \! \sum_{l \in \{{> \atop <}\}} \! 
{\rm s}(l) {\rm n}(l) \! = \! 0$, where ${\rm s}(>) \! = \! 
-{\rm s}(<) \! = \! 1$, and ${\rm n}({> \atop <}) \! := \! {\rm 
card}(\{\mathstrut \lambda_{j}; \, \Im(\lambda_{j}^{2}) {> \atop 
<} 0\})$, items {\em (2)} and {\em (4)} can be verified via 
straightforward calculations, and the fact that $q(x,t)$ 
(Eq.~(8)) satisfies the DNLSE follows {}from Eq.~(3) and the 
definition of $m(x,t;\! \lambda)$. 
\hfill \rule{6.5pt}{6.5pt}
\begin{eee}
{\rm The solvability of the RH problem for $\vert \vert r 
\vert \vert_{{\cal L}^{\infty}(\widehat{\Gamma})} \! < \! 1$ 
(resp.~$\vert \vert r \vert \vert_{{\cal L}^{\infty}(\Bbb R;
\Bbb C)} \! < \! 1)$ and ${\cal Z}_{d} \! \equiv \! \emptyset$ 
(resp.~${\cal Z}_{d} \! \not\equiv \! \emptyset)$ for 
sufficiently large $\vert t \vert$ was proved in {\rm \cite{a2}} 
(resp.~{\rm \cite{a10}}). \underline{{\bf Note}}: the condition 
$\vert \vert r \vert \vert_{{\cal L}^{\infty}(\widehat{\Gamma})} 
\! < \! 1$ which appears in \cite{a2} is restrictive, and can be 
replaced by the weaker condition $\vert \vert r \vert \vert_{{
\cal L}^{\infty}(\Bbb R;\Bbb C)} \! < \! 1$.\/}
\end{eee}
\begin{eee}
{\rm The RH problem formulated in Lemma~2.1 subsumes the ``full''
spectrum, $\sigma_{\pounds}$, of the operator $\partial_{x} \! - 
\! U(x,t;\! \lambda)$, namely, discrete, ${\cal Z}_{d}$, and 
continuous, $\widehat{\Gamma}$, with $\sigma_{\pounds} \! = \!
{\cal Z}_{d} \cup \widehat{\Gamma}$ $({\cal Z}_{d} \cap \widehat{
\Gamma} \! = \! \emptyset)$: the leading-order asymptotics as $t 
\! \to \! \pm \infty$ $(x/t \! \sim \! {\cal O}(1))$ for the 
solutions of Eq.~(2), the DNLSE, and MNLSE corresponding to the 
``full'' spectrum, $\sigma_{\pounds}$, are given in \cite{a10}; 
in this paper, however, a somewhat different set of asymptotics 
is considered, namely, the higher order asymptotics as $t \! \to 
\! \pm \infty$ $(x/t \! \sim \! {\cal O}(1))$ of the ``continuum 
component'' (corresponding to $\widehat{\Gamma})$ of the solutions 
to these NLEEs (the so-called ``non-soliton'' asymptotics), and 
the ``effect'' of the discrete spectrum, ${\cal Z}_{d}$, is 
``felt'' (or manifests itself) as Blaschke-Potapov-type factors 
\cite{a3,a5} which appear in the form of finite products 
multiplying the reflection coefficient, $r(\lambda)$, in the 
respective jump (conjugation) matrices of the ``asymptotic'' RH 
problems (see Sec.~4, Lemma~4.1 and Sec.~6, Lemma~6.1.1): in 
essence, oscillatory matrix RH problems will be considered 
\cite{a19}. Loosely speaking, and abusing nomenclature, the 
higher order asymptotics as $t \! \to \! \pm \infty$ $(x/t \! 
\sim \! {\cal O}(1))$ in the real plane, $\Bbb R^{2}$, but ``off 
the soliton trajectories'' \cite{a10} are considered, namely, for 
arbitrarily chosen but fixed $n$ such that $1 \! \leq \! n \! \leq 
\! N$, $N \! \in \! \Bbb Z_{\geq 1}$, and for those $\gamma_{n} \! 
\in \! (\frac{\pi}{2},\! \pi)$ (resp.~$\gamma_{n} \! \in \! (0,
\! \frac{\pi}{2}))$, the following asymptotics are investigated: 
(1) for $Q(x,t)$ and $q(x,t)$, with $Q(x,\! 0)$ and $q(x,\! 0)$, 
respectively, $\in \! {\cal S}(\Bbb R;\! \Bbb C)$, $(x,t) \! \to 
\! (\mp \infty,\pm \infty)$ (resp.~$(x,t) \! \to \! (\pm \infty,
\pm \infty))$ such that $\lambda_{0} \! := \! \frac{1}{2} 
\sqrt{-\frac{x}{t}} \! > \! M_{1} \, (> \! 0)$ (resp.~$\mu_{0} 
\! := \! \frac{1}{2} \sqrt{\frac{x}{t}} \! > \! M_{2} \, (> 
\! 0))$ and $(x,t) \! \in \! \Bbb R^{2} \! \setminus \! \Omega_{
n}$ (resp.~$(x,t) \! \in \! \Bbb R^{2} \! \setminus \! \mho_{n})$ 
(see Theorem~2.2 for the definitions of $\Omega_{n}$ and $\mho_{
n}$); and (2) for $u(x,t)$, with $u(x,\! 0) \! \in \! {\cal S}(
\Bbb R;\! \Bbb C)$, $(x,t) \! \to \! (\pm \infty,\pm \infty)$ 
(resp.~$(x,t) \! \to \! (\mp \infty,\pm \infty)$ or $(\pm \infty,
\pm \infty))$ such that $\widehat{\lambda}_{0} \! := \! \sqrt{
\frac{1}{2}(\frac{x}{t} \! - \! \frac{1}{s})} \! > \! M_{3} \, 
(> \! 0)$, $\frac{x}{t} \! > \! \frac{1}{s}$, $s \! \in \! \Bbb 
R_{>0}$ (resp.~$\widehat{\mu}_{0} \! := \! \sqrt{\frac{1}{2}(
\frac{1}{s} \! - \! \frac{x}{t})} \! > \! M_{4} \, (> \! 0)$, 
$\frac{x}{t} \! < \! \frac{1}{s}$, $s \! \in \! \Bbb R_{>0})$, 
and $(x,t) \! \in \! \Bbb R^{2} \! \setminus \! \widetilde{
\Omega}_{n}$ (resp.~$(x,t) \! \in \! \Bbb R^{2} \! \setminus 
\! \widetilde{\mho}_{n})$ (see Theorem~2.4 for the definitions 
of $\widetilde{\Omega}_{n}$ and $\widetilde{\mho}_{n}$).\/}
\end{eee}

Before proceeding any further, a self-contained synopsis of the 
BC \cite{a14} formulation (see, also, \cite{a17,a18,a20}) for 
the solution of an oscillatory matrix RH problem on an oriented 
contour is presented. The matrix RH problem on an oriented contour 
$L$ consists of finding a $M_{2}(\Bbb C)$-valued function ${\cal 
X}(\lambda)$ such that: (1) ${\cal X}(\lambda)$ is piecewise 
holomorphic $\forall \, \lambda \! \in \! \Bbb C \! \setminus \! 
L$; (2) ${\cal X}_{+}(\lambda) \! = \! {\cal X}_{-}(\lambda) {\cal 
G}(\lambda)$, $\lambda \! \in \! L$, for jump matrix ${\cal G}(
\lambda) \colon L \! \to \! M_{2}(\Bbb C)$; and (3) as $\lambda 
\! \to \! \infty$, $\lambda \! \in \! \Bbb C \! \setminus \! L$, 
${\cal X}(\lambda) \! = \! {\rm I} \! + \! {\cal O}(\lambda^{-1})$. 
Writing the jump matrix in the following factorised form, ${\cal 
G}(\lambda) \! := \! ({\rm I} \! - \! w^{-}_{x,t}(\lambda))^{-1}
({\rm I} \! + \! w^{+}_{x,t}(\lambda))$, $\lambda \! \in \! L$, 
where $w^{\pm}_{x,t}(\lambda) \! \in \! \cap_{p \in \{2,\infty\}} 
{\cal L}^{p}(L;\! M_{2}(\Bbb C))$ (with $\vert \vert w^{\pm}_{x,t}
(\cdot) \vert \vert_{\cap_{k \in \{2,\infty\}} {\cal L}^{k}(L;M_{
2}(\Bbb C))} \! := \! \sum_{k \in \{2,\infty\}} \vert \vert w^{
\pm}_{x,t}(\cdot) \vert \vert_{{\cal L}^{k}(L;M_{2}(\Bbb C))})$, 
respectively, are nilpotent off-diagonal upper/lower 
triangular $2 \! \times \! 2$ matrices, define $w_{x,t}(\lambda) 
\! := \! w^{-}_{x,t}(\lambda) \! + \! w^{+}_{x,t}(\lambda)$, 
and introduce the operator $C_{w_{x,t}}$ on ${\cal L}^{2}(L;\! 
M_{2}(\Bbb C))$ as $C_{w_{x,t}} f \! := \! C_{+}(f \, w^{-}
_{x,t}) \! + \! C_{-}(f \, w^{+}_{x,t})$, where $C_{\pm} \colon 
{\cal L}^{2}(L;\! M_{2}(\Bbb C)) \! \to \! {\cal L}^{2}(L;\! M_{
2}(\Bbb C))$ denote the Cauchy operators, $(C_{\pm}f)(\lambda) 
\! := \! \! \lim\limits_{\lambda^{\prime} \to \lambda \atop 
\lambda^{\prime} \in {\pm} \, {\rm side} \, {\rm of} \, L} \! 
\int_{\raise-0.45ex\hbox{$\scriptstyle{}L$}} \frac{f(\xi)}{(
\xi-\lambda^{\prime})} \frac{d \xi}{2 \pi i}$.
\begin{ddd}[{\rm \cite{a14}}]
If $\mu(\lambda) \! \in \! {\rm I} \! \oplus \! {\cal L}^{2}(L;\! 
M_{2}(\Bbb C))$ solves the following linear singular integral 
equation,
\begin{eqnarray*}
&(\underline{{\bf Id}} - C_{w_{x,t}}) \mu = {\rm I},&
\end{eqnarray*}
where $\underline{{\bf Id}}$ is the identity operator on ${\rm I} 
\! \oplus \! {\cal L}^{2}(L;\! M_{2}(\Bbb C))$, then the solution 
of the RH problem for ${\cal X}(\lambda)$ is 
\begin{eqnarray*}
{\cal X}(\lambda) = {\rm I} + \int_{L} \frac{\mu(\xi) w_{x,t}
(\xi)}{(\xi - \lambda)} \frac{d \xi}{2 \pi i}, \, \, \, \, \, 
\, \, \, \, \lambda \in \Bbb C \! \setminus \! L,
\end{eqnarray*}
where $\mu(\lambda) \! = \! {\cal X}_{+}(\lambda)({\rm I} \! + 
\! w^{+}_{x,t}(\lambda))^{-1} \! = \! {\cal X}_{-}(\lambda)({\rm 
I} \! - \! w^{-}_{x,t}(\lambda))^{-1}$.
\end{ddd}

{}From Lemma~2.2, Remark~2.3, and Theorem~2.1, the solution of 
the Cauchy problem for Eq.~(2) is
\begin{eqnarray}
Q(x,t)=-i \left([\sigma_{3},
\int_{\raise-0.39ex\hbox{$\scriptstyle{}\widehat{\Gamma}$}}
((\underline{{\bf Id}} - C_{w_{x,t}})^{-1} {\rm I})(\xi) 
w_{x,t}(\xi) \frac{d\xi}{2 \pi i}] \right)_{12},
\end{eqnarray}
where $w_{x,t}(\lambda) \! := \! \sum_{l \in \{\pm\}} \! w^{
l}_{x,t}(\lambda)$, and $w^{\pm}_{x,t}(\lambda) \! := \! \exp 
\{-i(\lambda^{2} x \! + \! 2 \lambda^{4} t){\rm ad}(\sigma_{
3})\} w^{\pm}(\lambda)$.
\begin{eee}
{\rm To facilitate the reading of the results presented in 
Theorems~2.2--2.4 (see below), as well as those appearing 
throughout the paper, the following preamble is necessary: 
(1) $M \! \in \! \Bbb R_{>0}$ is a fixed constant; (2) the 
``symbols'' $\underline{c}(z)$ and $c^{{\cal S}}(z)$, 
respectively, which appear in the various error estimations 
are to be understood as follows, $\underline{c}(z) \! \in \! 
{\cal L}^{\infty}(\Bbb R_{>M};\! \Bbb C)$ and $c^{{\cal S}}
(z) \! \in \! {\cal S}(\Bbb R_{>M};\! \Bbb C)$; and (3) even 
though the symbols $\underline{c}(z)$ and $c^{{\cal S}}(z)$ 
are \underline{{\bf not}} equal and should properly be denoted 
as $c_{1}(z)$, $c_{2}(z)$, etc., the simplified ``notations'' 
$\underline{c}(z)$ and $c^{{\cal S}}(z)$ are retained throughout 
since the principal concern here is not their explicit functional 
$z$-dependence, but rather, the functional class(es) to which 
they belong.\/}
\end{eee}
\begin{eee}
{\rm In Theorems~2.2--2.4 (see below), one should keep the 
upper signs as $t \! \to \! +\infty$ and the lower signs as 
$t \! \to \! -\infty$ everywhere, and, concerning the constants 
${\cal K}_{1}^{\pm}$ which appear in these theorems, the reader 
should refer to Sec.~4, Remark~4.2.\/}
\end{eee}

In this paper, the following results are proven:
\begin{ddd}
Let $m(x,t;\! \lambda)$ be the solution of the RH problem 
formulated in Lemma~2.1 with the condition $\vert \vert r 
\vert \vert_{{\cal L}^{\infty}(\Bbb R;\Bbb C)} \! < \! 1$ and 
$Q(x,t)$, the solution of Eq.~(2), be defined by Eq.~(7). Then, 
for $Q(x,0) \! \in \! {\cal S}(\Bbb R;\! \Bbb C)$, as $t \! \to 
\! \pm \infty$ and $x \! \to \! \mp \infty$ such that $\lambda
_{0} \! := \! \frac{1}{2} \sqrt{-\frac{x}{t}} \! > \! M$ and 
$(x,t) \! \in \! \Bbb R^{2} \setminus \Omega_{n}$, $\Omega_{n} 
\! := \! \{\mathstrut (x,t); \, x \! - \! 4 t \Delta_{n}^{2} 
\cos \gamma_{n} \! = \! {\cal O}(1)\}$, for those $\gamma_{n} 
\! \in \! (\frac{\pi}{2},\! \pi)$,
\begin{eqnarray}
&Q(x,t) \! = \! Q_{\pm}(x,t) \! + \! {\cal O} \! \left(\frac{
\underline{c}(\lambda_{0})(\ln \vert t \vert)^{\varepsilon(\pm)
}}{(\pm \lambda_{0}^{2} t)^{2}} \right) \!,&
\end{eqnarray}
where
\begin{eqnarray}
&Q_{\pm}(x,t) \! := \! \frac{e^{i \tau_{\pm}(\lambda_{0};t)} u^{
\pm}_{1,1,0}(\lambda_{0})}{\sqrt{\pm t}} \! + \! \frac{u^{\pm}_{
-1,2,0}(\lambda_{0})}{(\pm t)} \! + \! \frac{e^{i \tau_{\pm}(
\lambda_{0};t)} \sum_{q=0}^{2} u^{\pm}_{1,3,q}(\lambda_{0})(\ln 
\vert t \vert)^{q}}{(\pm t)^{3/2}} \! + \! \frac{u^{\pm}_{-1,3,
0}(\lambda_{0})}{(\pm t)^{3/2}},& \\
&\tau_{\pm}(z_{1};z_{2}) \! := \! 4 z_{1}^{4} z_{2} \! \mp \! 
\nu(z_{1}) \ln \! \vert z_{2} \vert,& \\
&\nu(z) \! := \! - \frac{1}{2 \pi} \ln (1 \! - \! \vert r(z) 
\vert^{2}),& \\
&u^{\pm}_{1,1,0}(z) \! = \! \sqrt{\frac{\nu(z)}{2 z^{2}}} \exp\{
i(\phi^{\pm}(z)+\widehat{\Phi}^{\pm}(z) + \frac{\pi}{2})\},& \\
&\phi^{+}(z) \! = \! \frac{1}{\pi} \int_{0}^{z} \ln \! \vert \xi
^{2} \! - \! z^{2} \vert d \ln (1 \! - \! \vert r(\xi) \vert^{2}) 
\! - \! \frac{1}{\pi} \int_{0}^{\infty} \ln \! \vert \xi^{2} \! + 
\! z^{2} \vert d \ln (1 \! + \! \vert r (i \xi) \vert^{2}),& \\
&\phi^{-}(z) \! = \! \frac{1}{\pi} \int_{z}^{\infty} \ln \! \vert 
\xi^{2} \! - \! z^{2} \vert d \ln (1 \! - \! \vert r (\xi) \vert
^{2}),& \\
&\widehat{\Phi}^{\pm}(z) \! = \! \pm \arg \Gamma (i \nu(z)) \! + 
\! \arg r(z) \! \mp \! 3 \nu(z) \ln 2 \! + \! \frac{(2 \pm 1) 
\pi}{4} \! + \! 2 \! \sum\limits_{l \in L_{\pm}} \! \! \arg \! 
\left(\! \frac{(z-\overline{\lambda_{l}})(z + \overline{\lambda_{
l}})}{(z - \lambda_{l})(z + \lambda_{l})} \! \right) \!,& \\
&u^{\pm}_{-1,2,0}(z) \! = \! \mp \frac{i(r^{\prime}(0)-r^{\prime}
(i0)) \exp \{2i(\vartheta_{\pm}(z)+2 \sum_{l \in L_{\pm}} \! 
\gamma_{l})\}}{\pi z^{2} 2^{3}},& \\
&r^{\prime}(0) \! := \! (\frac{d r(\lambda)}{d \lambda} \vert_{
\lambda \in \Bbb R}) \vert_{\lambda=0}, \, \, \, \, \, \, \, \, 
r^{\prime}(i0) \! := \! (\frac{d r(\lambda)}{d \lambda} \vert_{
\lambda \in i \Bbb R}) \vert_{\lambda=0},& \\
&\vartheta_{+}(z) \! = \! - \int_{0}^{z} \! \frac{\ln (1 - \vert 
r(\xi) \vert^{2})}{\xi} \frac{d \xi}{\pi} \! + \! \int_{0}^{
\infty} \! \frac{\ln (1 + \vert r(i \xi) \vert^{2})}{\xi} \frac{
d \xi}{\pi}, \, \, \, \, \, \, \vartheta_{-}(z) \! = \! -\int_{
z}^{\infty} \! \frac{\ln (1-\vert r(\xi) \vert^{2})}{\xi} 
\frac{d \xi}{\pi},& \\
&u^{\pm}_{-1,3,0}(z) \! = \! \mp \frac{i(r^{\prime}(0)-r^{\prime}
(i0)) \exp \{2i(\vartheta_{\pm}(z)+2 \sum_{l \in L_{\pm}} \! 
\gamma_{l})\} {\cal K}_{1}^{\pm}}{\pi z^{3} 2^{3}},& \\
&u^{\pm}_{1,3,2}(z) \! = \! \frac{\pm i(\frac{\partial \nu(z)}{
\partial z})^{2} u^{\pm}_{1,1,0}(z) + \nu(z)(\frac{\partial \nu
(z)}{\partial z})^{2} \{\overline{u^{\pm}_{1,1,0}(z)} \exp\{2 i 
\varphi_{\pm}(z)\}-u^{\pm}_{1,1,0}(z)\}}{64 z^{2}},& \\
&u^{\pm}_{1,3,1}(z) \! = \! \pm (i \! \mp \! \nu(z)){\cal R}^{
\pm}_{o}(z) \! + \! \nu(z) \exp\{2 i \varphi_{\pm}(z)\} \overline{
{\cal R}^{\pm}_{o}(z)},& \\
&{\cal R}^{\pm}_{o}(z) \! := \! \mp 2 i u^{\pm}_{1,3,2}(z) \! \pm 
\! \frac{i(\frac{\partial u^{\pm}_{1,1,0}(z)}{\partial z})(\frac{
\partial \nu(z)}{\partial z})}{32 z^{2}} \! \pm \! \frac{i(\frac{
\partial^{2} \nu(z)}{\partial z^{2}}) u^{\pm}_{1,1,0}(z)}{64 z^{
2}} \! \mp \! \frac{i(\frac{\partial \nu(z)}{\partial z}) u^{\pm}
_{1,1,0}(z)}{64 z^{3}}& \nonumber \\
&\! \! \! \! \! \! \! \! \! \! \! \! \! \! \! \! \! \! \! \! \! 
\! \! \! \! \! \! \! \! \! \! \! \! \! \! \! \! \! \! \! \! \! 
\! \! \! \! \! \! \! \! \! \! \! \! \! \! \! \! \! \! \! \! \! 
\! \! \! \! \! \! \! \! \! \! \! \! \! \! \! \! \! \! \! \! \! 
\! \! \! \! \! \! \! \! \! \! \! \! \! \! - \, \frac{\nu(z)(
\frac{\partial \nu(z)}{\partial z}) u^{\pm}_{1,1,0}(z)}{16 
z^{3}},& \\
&u^{\pm}_{1,3,0}(z) \! = \! \pm (i \! \mp \! \nu(z)){\cal S}^{
\pm}_{o}(z) \! + \! \nu(z) \exp\{2 i \varphi_{\pm}(z)\} 
\overline{{\cal S}^{\pm}_{o}(z)},& \\
&{\cal S}^{\pm}_{o}(z) \! := \! \mp iu^{\pm}_{1,3,1}(z) \! - \! 
\frac{(\frac{\partial^{2} u^{\pm}_{1,1,0}(z)}{\partial z^{2}})
}{64 z^{2}} \! + \! \frac{(\frac{\partial u^{\pm}_{1,1,0}(z)}{
\partial z})}{64 z^{3}} \! \pm \! \frac{i \overline{(\frac{
\partial u^{\pm}_{1,1,0}(z)}{\partial z})} \nu(z) \exp\{2i 
\varphi_{\pm}(z)\}}{16 z^{3}}& \nonumber \\
&\! \! \! \! \! \! \! \! \! \! \! \! \! \! \! \! \! \! \! \! \! 
\! \! \! \! \! \! \! \! \! \! \! \! \! \! \! \! \! \! \! \! \! 
\! \! \! \! \! \! \! \! \! \! \! \! \! \! \! \! \! \! \! \! \! 
\! \! \! \! \! \! \! \! \! \! \! \! \! \! \! \! \! \! \! \! \! 
\! \! \! \! \! \! \! \! \! \! \! \! \! \! \! \! \! \! \! \! \! 
\! \! - \, \frac{(\nu(z))^{2} u^{\pm}_{1,1,0}(z)}{8 z^{4}},& \\
&\varphi_{\pm}(z) \! := \! \phi^{\pm}(z) \! + \! \widehat{\Phi}
^{\pm}(z) \! + \! \frac{\pi}{2},&
\end{eqnarray}
$\Gamma(\cdot)$ is the gamma function {\rm \cite{a21}}, $\sum_{l 
\in L_{+}} \! := \! \sum_{l=n+1}^{N}$, $\sum_{l \in L_{-}} \! := 
\! \sum_{l=1}^{n-1}$, ${\cal K}_{1}^{\pm} \! \in \! \Bbb C$, and 
$\varepsilon(\pm) \! := \! \frac{1}{2}(1 \! \pm \! 1)$, and, as 
$t \! \to \! \pm \infty$ and $x \! \to \! \pm \infty$ such that 
$\mu_{0} \! := \! \frac{1}{2} \sqrt{\frac{x}{t}} \! > \! M$ and 
$(x,t) \! \in \! \Bbb R^{2} \setminus \mho_{n}$, $\mho_{n} \! := 
\! \{\mathstrut (x,t); \, x \! - \! 4 t \Delta_{n}^{2} \cos \gamma
_{n} \! = \! {\cal O}(1)\}$, for those $\gamma_{n} \! \in \! (0,\! 
\frac{\pi}{2})$,
\begin{eqnarray}
&Q(x,t) \! = \! Q_{\pm}^{\prime}(x,t) \! + \! {\cal O} \! \left(
\frac{\underline{c}(\mu_{0})(\ln \vert t \vert)^{\varepsilon(\pm)
}}{(\pm \mu_{0}^{2} t)^{2}} \right) \!,&
\end{eqnarray}
where
\begin{eqnarray}
&Q_{\pm}^{\prime}(x,t) \! := \! \frac{e^{i \tau_{\pm}^{\prime}(
\mu_{0};t)} u^{\pm \prime}_{1,1,0}(\mu_{0})}{\sqrt{\pm t}} \! + 
\! \frac{u^{\pm \prime}_{-1,2,0}(\mu_{0})}{(\pm t)} \! + \! 
\frac{e^{i \tau_{\pm}^{\prime}(\mu_{0};t)} \sum_{q=0}^{2}u^{\pm 
\prime}_{1,3,q}(\mu_{0})(\ln \vert t \vert)^{q}}{(\pm t)^{3/2}} 
\! + \! \frac{u^{\pm \prime}_{-1,3,0}(\mu_{0})}{(\pm t)^{3/2}},
& \\
&\tau_{\pm}^{\prime}(z_{1};z_{2}) \! := \! 4 z_{1}^{4} z_{2} \! 
\mp \! \nu(i z_{1}) \ln \! \vert z_{2} \vert,& \\
&\nu(iz) \! := \! - \frac{1}{2 \pi} \ln (1 \! + \! \vert r(iz) 
\vert^{2}),& \\
&u^{\pm \prime}_{1,1,0}(z) \! = \! \sqrt{-\frac{\nu(i z)}{2 z^{
2}}} \exp\{i(\phi^{\pm \prime}(z)+\widehat{\Phi}^{\pm \prime}(z) 
+ \pi)\},& \\
&\phi^{+ \prime}(z) \! = \! \frac{1}{\pi} \int_{0}^{z} \ln \! \vert 
\xi^{2} \! - \! z^{2} \vert d \ln (1 \! + \! \vert r(i \xi) \vert^{
2}) \! - \! \frac{1}{\pi} \int_{0}^{\infty} \ln \! \vert \xi^{2} \! 
+ \! z^{2} \vert d \ln (1 \! - \! \vert r (\xi) \vert^{2}),& \\
&\phi^{- \prime}(z) \! = \! \frac{1}{\pi} \int_{z}^{\infty} \ln \! 
\vert \xi^{2} \! - \! z^{2} \vert d \ln (1 \! + \! \vert r (i \xi) 
\vert^{2}),& \\
&\widehat{\Phi}^{\pm \prime}(z) \! = \! \pm \arg \Gamma (i \nu(iz)) 
\! + \! \arg r(iz) \! \mp \! 3 \nu(iz) \ln 2 \! - \! \frac{(2 \mp 
1) \pi}{4} \! - \! 2 \! \sum\limits_{l \in L_{\pm}} \! \! \arg \! 
\left(\! \frac{(z-\overline{\lambda_{l}})(z+\overline{\lambda_{l}}
)}{(z-\lambda_{l})(z+\lambda_{l})} \! \right) \!,& \\
&u^{\pm \prime}_{-1,2,0}(z) \! = \! \mp \frac{(r^{\prime}(0) 
- r^{\prime}(i0)) \exp\{2i(\vartheta_{\pm}^{\prime}(z) - 2 
\sum_{l \in L_{\pm}} \! \gamma_{l})\}}{\pi z^{2} 2^{3}},& 
\\
&\vartheta_{+}^{\prime}(z) \! = \! - \int_{0}^{z} \! \frac{\ln (1 
+ \vert r(i \xi) \vert^{2})}{\xi} \frac{d \xi}{\pi} \! + \! \int_{
0}^{\infty} \! \frac{\ln (1 - \vert r(\xi) \vert^{2})}{\xi} \frac{
d \xi}{\pi}, \, \, \, \, \, \, \vartheta_{-}^{\prime}(z) \! = \! - 
\int_{z}^{\infty} \! \frac{\ln(1+\vert r(i\xi) \vert^{2})}{\xi} 
\frac{d \xi}{\pi},& \\
&u^{\pm \prime}_{-1,3,0}(z) \! = \! \mp \frac{i(r^{\prime}(0) 
- r^{\prime}(i0)) \exp \{2i(\vartheta_{\pm}^{\prime}(z) - 2 
\sum_{l \in L_{\pm}} \! \gamma_{l})\} {\cal K}_{1}^{\pm}}{\pi 
z^{3} 2^{3}},& \\
&u^{\pm \prime}_{1,3,2}(z) \! = \! \frac{\pm i(\frac{\partial 
\nu(iz)}{\partial z})^{2} u^{\pm \prime}_{1,1,0}(z) - \nu(iz)
(\frac{\partial \nu(i z)}{\partial z})^{2} \{\overline{u^{\pm 
\prime}_{1,1,0}(z)} \exp\{2 i \varphi_{\pm}^{\prime}(z)\}+u^{
\pm \prime}_{1,1,0}(z)\}}{64 z^{2}},& \\
&u^{\pm \prime}_{1,3,1}(z) \! = \! \pm (i \! \mp \! \nu(iz)){
\cal R}^{\pm \prime}_{o}(z) \! - \! \nu(i z) \exp\{2i\varphi_{
\pm}^{\prime}(z)\} \overline{{\cal R}^{\pm \prime}_{o}(z)},& \\
&{\cal R}^{\pm \prime}_{o}(z) \! := \! \mp 2iu^{\pm \prime}_{1,
3,2}(z) \! \pm \! \frac{i(\frac{\partial u^{\pm \prime}_{1,1,0}
(z)}{\partial z})(\frac{\partial \nu(i z)}{\partial z})}{32 z^{
2}} \! \pm \! \frac{i(\frac{\partial^{2} \nu(i z)}{\partial z^{
2}}) u^{\pm \prime}_{1,1,0}(z)}{64 z^{2}} \! \mp \! \frac{i(
\frac{\partial \nu(i z)}{\partial z}) u^{\pm \prime}_{1,1,0}(
z)}{64 z^{3}}& \nonumber \\
&\! \! \! \! \! \! \! \! \! \! \! \! \! \! \! \! \! \! \! \! \! 
\! \! \! \! \! \! \! \! \! \! \! \! \! \! \! \! \! \! \! \! \! 
\! \! \! \! \! \! \! \! \! \! \! \! \! \! \! \! \! \! \! \! \! 
\! \! \! \! \! \! \! \! \! \! \! \! \! \! \! \! \! \! \! \! \! 
\! \! \! \! \! \! \! \! \! \! \! \! \! \! - \, \frac{\nu(iz)(
\frac{\partial \nu(i z)}{\partial z}) u^{\pm \prime}_{1,1,0}(
z)}{16 z^{3}},& \\
&u^{\pm \prime}_{1,3,0}(z) \! = \! \pm (i \! \mp \! \nu(iz)){
\cal S}^{\pm \prime}_{o}(z) \! - \! \nu(iz) \exp\{2i\varphi_{
\pm}^{\prime}(z)\} \overline{{\cal S}^{\pm \prime}_{o}(z)},& 
\\
&{\cal S}^{\pm \prime}_{o}(z) \! := \! \mp i u^{\pm \prime}_{1,
3,1}(z) \! - \! \frac{(\frac{\partial^{2} u^{\pm \prime}_{1,1,
0}(z)}{\partial z^{2}})}{64 z^{2}} \! + \! \frac{(\frac{\partial 
u^{\pm \prime}_{1,1,0}(z)}{\partial z})}{64 z^{3}} \! \pm \! 
\frac{i \overline{(\frac{\partial u^{\pm \prime}_{1,1,0}(z)}{
\partial z})} \nu(i z) \exp\{2 i \varphi_{\pm}^{\prime}(z)\}}{
16 z^{3}}& \nonumber \\
&\! \! \! \! \! \! \! \! \! \! \! \! \! \! \! \! \! \! \! \! \! 
\! \! \! \! \! \! \! \! \! \! \! \! \! \! \! \! \! \! \! \! \! 
\! \! \! \! \! \! \! \! \! \! \! \! \! \! \! \! \! \! \! \! \! 
\! \! \! \! \! \! \! \! \! \! \! \! \! \! \! \! \! \! \! \! \! 
\! \! \! \! \! \! \! \! \! \! \! \! \! \! \! \! \! \! \! \! \! 
\! - \, \frac{(\nu(iz))^{2} u^{\pm \prime}_{1,1,0}(z)}{8 z^{4}
},& \\
&\varphi_{\pm}^{\prime}(z) \! := \! \phi^{\pm \prime}(z) \! + \! 
\widehat{\Phi}^{\pm \prime}(z) \! + \! \frac{\pi}{2}.&
\end{eqnarray}
\end{ddd}
\begin{ddd}
Let $m(x,t;\! \lambda)$ be the solution of the RH problem 
formulated in Lemma~2.1 with the condition $\vert \vert r 
\vert \vert_{{\cal L}^{\infty}(\Bbb R;\Bbb C)} \! < \! 1$ and 
$q(x,t)$, the solution of the DNLSE (Eq.~(4)), be defined by 
Eq.~(8). Then, for $q(x,0) \! \in \! {\cal S}(\Bbb R;\! \Bbb 
C)$, as $t \! \to \! \pm \infty$ and $x \! \to \! \mp \infty$ 
such that $\lambda_{0} \! := \! \frac{1}{2} \sqrt{-\frac{x}{t}} 
\! > \! M$ and $(x,t) \! \in \! \Bbb R^{2} \setminus \Omega_{
n}$, $\Omega_{n} \! := \! \{\mathstrut (x,t); \, x \! - \! 4 t 
\Delta_{n}^{2} \cos \gamma_{n} \! = \! {\cal O}(1)\}$, for 
those $\gamma_{n} \! \in \! (\frac{\pi}{2},\! \pi)$,
\begin{eqnarray}
&q(x,t) \! = \! Q_{\pm}(x,t) \exp\{i \arg q_{\pm}(x,t)\} \! + 
\! {\cal O} \! \left(\frac{c^{{\cal S}}(\lambda_{0})(\ln \vert 
t \vert)^{2}}{(\pm \lambda_{0}^{2} t)^{2}} \right) \!,&
\end{eqnarray}
where $Q_{\pm}(x,t)$ are given in Theorem~2.2, Eqs.~(11)--(27), 
and
\begin{eqnarray}
&\arg q_{\pm}(x,t) \! = \! \epsilon^{\pm}(\lambda_{0}) \! - \! 
\sqrt{\frac{2}{\pm t}} 
\int_{\raise-0.45ex\hbox{$\scriptstyle{}\lambda_{0}$}}^{\infty} 
\! \frac{\sqrt{\nu(\xi)}}{\xi^{2}}(R_{r} \sin(\kappa_{\pm}(\xi;
t)) \! - \! R_{i} \cos(\kappa_{\pm}(\xi;t))) \frac{d \xi}{\pi}& 
\nonumber \\
&\pm \frac{\vert R(0) \vert^{2}}{\pi^{2} 2^{4} \lambda_{0}^{2}
(\pm t)} \! - \! \frac{\sqrt{2}}{(\pm t)} 
\int_{\raise-0.45ex\hbox{$\scriptstyle{}\lambda_{0}$}}^{\infty} 
\! \frac{\sqrt{\nu(\xi)}}{\xi^{3}}(a^{\pm} \sin(\kappa
_{\pm}(\xi;t)) \! - \! b^{\pm} \cos(\kappa_{\pm}(\xi;t))) 
\frac{d \xi}{\pi}& \nonumber \\
&\pm \frac{1}{2^{3}(\pm t)} 
\int_{\raise-0.45ex\hbox{$\scriptstyle{}\lambda_{0}$}}^{\infty} 
(\frac{4(\nu(\xi))^{2}}{\xi^{5}} \! + \! \frac{\nu(\xi)(
\frac{\partial \nu(\xi)}{\partial \xi})}{\xi^{4}} \! - \! \frac{
\nu(\xi)(\frac{\partial^{2} \nu(\xi)}{\partial \xi^{2}})}{\xi^{
3}} \! - \! \frac{(\frac{\partial \nu(\xi)}{\partial \xi})^{2}}
{\xi^{3}} \! \pm \! \frac{(\frac{\partial \nu(\xi)}{\partial 
\xi})(\frac{\partial \varphi_{\pm}(\xi)}{\partial \xi})}{\xi
^{3}}& \nonumber \\
&\pm \frac{\nu(\xi)(\frac{\partial^{2} \varphi_{\pm}(\xi)}{
\partial \xi^{2}})}{\xi^{3}} \! \mp \! \frac{3 \nu(\xi)(\frac{
\partial \varphi_{\pm}(\xi)}{\partial \xi})}{\xi^{4}}) d \xi 
\! \mp \! \frac{\ln \vert t \vert}{2^{3}(\pm t)} \int_{
\raise-0.45ex\hbox{$\scriptstyle{}\lambda_{0}$}}^{\infty} 
(\frac{(\frac{\partial \nu(\xi)}{\partial \xi})^{2}}{\xi^{
3}} \! - \! \frac{3 \nu(\xi)(\frac{\partial \nu(\xi)}{\partial 
\xi})}{\xi^{4}} \! + \! \frac{\nu(\xi)(\frac{\partial^{2} \nu(
\xi)}{\partial \xi^{2}})}{\xi^{3}}) d \xi,& \\
&\epsilon^{\pm}(z) \! := \! - 2 \vartheta_{\pm}(z),& \\
&\kappa_{\pm}(z;t) \! := \! \tau_{\pm}(z;t) \! + \! \phi^{\pm}
(z) \! + \! \widehat{\Phi}^{\pm}(z) \! + \! \frac{\pi}{2} \! + 
\! \epsilon^{\pm}(z) \! - \! 4 \sum\limits_{l \in L_{\pm}} \! 
\gamma_{l},& \\
&R(0) \! := \! r^{\prime}(0) \! - \! r^{\prime}(i0), \, \, \, 
\, \, \, \, R_{r} \! := \! \Re\{R(0)\}, \, \, \, \, \, \, \, 
R_{i} \! := \! \Im\{R(0)\},& \\
&a^{\pm} \! := \! R_{r}{\cal K}_{1,r}^{\pm} \! - \! R_{i}{\cal 
K}_{1,i}^{\pm}, \, \, \, \, \, \, \, b^{\pm} \! := \! R_{r}{
\cal K}_{1,i}^{\pm} \! + \! R_{i}{\cal K}_{1,r}^{\pm},& \\
&{\cal K}_{1,r}^{\pm} \! := \! \Re\{{\cal K}_{1}^{\pm}\}, \, \, 
\, \, \, \, \, {\cal K}_{1,i}^{\pm} \! := \! \Im\{{\cal K}_{1}
^{\pm}\},&
\end{eqnarray}
and, as $t \! \to \! \pm \infty$ and $x \! \to \! \pm \infty$ such 
that $\mu_{0} \! := \! \frac{1}{2} \sqrt{\frac{x}{t}} \! > \! M$ 
and $(x,t) \! \in \! \Bbb R^{2} \setminus \mho_{n}$, $\mho_{n} \! 
:= \! \{\mathstrut (x,t); \, x \! - \! 4 t \Delta_{n}^{2} \cos 
\gamma_{n} \! = \! {\cal O}(1)\}$, for those $\gamma_{n} \! \in \! 
(0,\! \frac{\pi}{2})$,
\begin{eqnarray}
&q(x,t) \! = \! Q_{\pm}^{\prime}(x,t) \exp\{i \arg q_{\pm}^{
\prime}(x,t)\} \! + \! {\cal O} \! \left(\frac{c^{{\cal S}}(
\mu_{0})(\ln \vert t \vert)^{2}}{(\pm \mu_{0}^{2} t)^{2}} 
\right) \!,&
\end{eqnarray}
where $Q_{\pm}^{\prime}(x,t)$ are given in Theorem~2.2, 
Eqs.~(29)--(44), and
\begin{eqnarray}
&\arg q_{\pm}^{\prime}(x,t) \! = \! \epsilon^{\pm \prime}(\mu_{
0}) \! + \! \sqrt{\frac{2}{\pm t}} \int_{\mu_{0}}^{\infty} \! 
\frac{\sqrt{-\nu(i\xi)}}{\xi^{2}}(R_{r} \cos(\kappa_{\pm}^{
\prime}(\xi;t)) \! + \! R_{i} \sin(\kappa_{\pm}^{\prime}(\xi;t)
)) \frac{d \xi}{\pi}& \nonumber \\
&\mp \frac{\vert R(0) \vert^{2}}{\pi^{2} 2^{4} \mu_{0}^{2}(\pm 
t)} \! + \! \frac{\sqrt{2}}{(\pm t)} \int_{\mu_{0}}^{\infty} \! 
\frac{\sqrt{-\nu(i \xi)}}{\xi^{3}}(a^{\pm} \sin(\kappa_{\pm}^{
\prime}(\xi;t)) \! - \! b^{\pm} \cos(\kappa_{\pm}^{\prime}(\xi;
t))) \frac{d \xi}{\pi}& \nonumber \\
&\pm \frac{1}{2^{3}(\pm t)} \int_{\mu_{0}}^{\infty} \! (\frac{4
(\nu(i\xi))^{2}}{\xi^{5}} \! + \! \frac{\nu(i\xi)(\frac{\partial 
\nu(i\xi)}{\partial \xi})}{\xi^{4}} \! - \! \frac{\nu(i\xi)(
\frac{\partial^{2} \nu(i\xi)}{\partial \xi^{2}})}{\xi^{3}} \! - 
\! \frac{(\frac{\partial \nu(i\xi)}{\partial \xi})^{2}}{\xi^{3}} 
\! \pm \! \frac{(\frac{\partial \nu(i\xi)}{\partial \xi})(\frac{
\partial \varphi_{\pm}^{\prime}(\xi)}{\partial \xi})}{\xi^{3}}& 
\nonumber \\
&\pm \frac{\nu(i\xi)(\frac{\partial^{2} \varphi_{\pm}^{\prime}
(\xi)}{\partial \xi^{2}})}{\xi^{3}} \! \mp \! \frac{3 \nu(i\xi)
(\frac{\partial \varphi_{\pm}^{\prime}(\xi)}{\partial \xi})}{
\xi^{4}}) d \xi \! \mp \! \frac{\ln \vert t \vert}{2^{3}(\pm 
t)} \int_{\mu_{0}}^{\infty} \! (\frac{(\frac{\partial \nu(i 
\xi)}{\partial \xi})^{2}}{\xi^{3}} \! - \! \frac{3 \nu(i\xi)(
\frac{\partial \nu(i\xi)}{\partial \xi})}{\xi^{4}} \! + \! 
\frac{\nu(i\xi)(\frac{\partial^{2} \nu(i\xi)}{\partial \xi^{
2}})}{\xi^{3}}) d \xi,& \\
&\epsilon^{\pm \prime}(z) \! := \! - 2 \vartheta_{\pm}^{\prime}
(z),& \\
&\kappa_{\pm}^{\prime}(z;t) \! := \! \tau_{\pm}^{\prime}(z;t) 
\! + \! \phi^{\pm \prime}(z) \! + \! \widehat{\Phi}^{\pm \prime}
(z) \! + \! \pi \! + \! \epsilon^{\pm \prime}(z) \! + \! 4 
\sum\limits_{l \in L_{\pm}} \! \gamma_{l}.&
\end{eqnarray}
\end{ddd}
\begin{ddd}
Let $m(x,t;\! \lambda)$ be the solution of the RH problem 
formulated in Lemma~2.1 with the condition $\vert \vert r \vert 
\vert_{{\cal L}^{\infty}(\Bbb R;\Bbb C)} \! < \! 1$ and $u(x,t)$, 
the solution of the MNLSE (Eq.~(1)), be defined by Eq.~(5) in 
terms of the function $q(x,t)$ given in Theorem~2.3. Then, for 
$u(x,0) \! \in \! {\cal S}(\Bbb R;\! \Bbb C)$, as $t \! \to \! 
\pm \infty$ and $x \! \to \! \pm \infty$ such that $\widehat{
\lambda}_{0} \! := \! \sqrt{\frac{1}{2}(\frac{x}{t} \! - \! 
\frac{1}{s})} \! > \! M$, $\frac{x}{t} \! > \! \frac{1}{s}$, $s 
\! \in \! \Bbb R_{>0}$, and $(x,t) \! \in \! \Bbb R^{2} \setminus 
\widetilde{\Omega}_{n}$, $\widetilde{\Omega}_{n} \! := \! \{
\mathstrut (x,t); \, - x \! + \! t(\frac{1}{s} \! - \! 2 \Delta
_{n}^{2} \cos \gamma_{n}) \! = \! {\cal O}(1)\}$, for those 
$\gamma_{n} \! \in \! (\frac{\pi}{2},\! \pi)$,
\begin{eqnarray}
&u(x,t) \! = \! \widehat{v}_{\pm}(x,t) \widehat{w}_{\pm}(x,t) 
\! + \! {\cal O} \! \left(\frac{c^{{\cal S}}(\widehat{\lambda}
_{0})(\ln \vert t \vert)^{2}}{(\pm \widehat{\lambda}_{0}^{2} 
t)^{2}} \right) \!,&
\end{eqnarray}
where
\begin{eqnarray}
&\widehat{v}_{\pm}(x,t) \! = \! \frac{e^{i \widehat{\tau}_{
\pm}(\widehat{\lambda}_{0};t)} v^{\pm}_{1,1,0}(\widehat{
\lambda}_{0})}{\sqrt{\pm t}} \! + \! \frac{v^{\pm}_{-1,2,0}(
\widehat{\lambda}_{0})}{(\pm t)} \! + \! \frac{e^{i \widehat{
\tau}_{\pm}(\widehat{\lambda}_{0};t)} \sum_{q=0}^{2}v^{\pm}_{
1,3,q}(\widehat{\lambda}_{0})(\ln \vert t \vert)^{q}}{(\pm t)
^{3/2}} \! + \! \frac{v^{\pm}_{-1,3,0}(\widehat{\lambda}_{0})
}{(\pm t)^{3/2}},& \\
&\widehat{\tau}_{\pm}(\widehat{\lambda}_{0};t) \! := \! \tau
_{\pm}(\widehat{\lambda}_{0};t/2),& \\
&v^{\pm}_{1,1,0}(\widehat{\lambda}_{0}) \! := \! \frac{1}{
\sqrt{s}} u^{\pm}_{1,1,0}(\widehat{\lambda}_{0}),& \\
&v^{\pm}_{-1,2,0}(\widehat{\lambda}_{0}) \! := \! \sqrt{\frac{
2}{s}} u^{\pm}_{-1,2,0}(\widehat{\lambda}_{0}),& \\
&v^{\pm}_{1,3,0}(\widehat{\lambda}_{0}) \! := \! \frac{2}{
\sqrt{s}} (u^{\pm}_{1,3,0}(\widehat{\lambda}_{0}) \! - \! 
u^{\pm}_{1,3,1}(\widehat{\lambda}_{0}) \ln 2 \! + \! u^{
\pm}_{1,3,2}(\widehat{\lambda}_{0})(\ln 2)^{2}),& \\
&v^{\pm}_{1,3,1}(\widehat{\lambda}_{0}) \! := \! \frac{2}{
\sqrt{s}} (u^{\pm}_{1,3,1}(\widehat{\lambda}_{0}) \! - \! 
2 u^{\pm}_{1,3,2}(\widehat{\lambda}_{0}) \ln 2),& \\
&v^{\pm}_{1,3,2}(\widehat{\lambda}_{0}) \! := \! \frac{2}{
\sqrt{s}} u^{\pm}_{1,3,2}(\widehat{\lambda}_{0}),& \\
&v^{\pm}_{-1,3,0}(\widehat{\lambda}_{0}) \! := \! \frac{2}{
\sqrt{s}} u^{\pm}_{-1,3,0}(\widehat{\lambda}_{0}),& \\
&\widehat{w}_{\pm}(x,t) \! = \! \exp \! \left\{ \! i \! \left(
\! \epsilon^{\pm}(\widehat{\lambda}_{0}) \! - \! \frac{2}{
\sqrt{\pm t}} 
\int_{\raise-0.65ex\hbox{$\scriptstyle{}\widehat{\lambda}_{
0}$}}^{\infty} \! \frac{\sqrt{\nu(\xi)}}{\xi^{2}}(R_{r} \sin 
(\widehat{\kappa}_{\pm}(\xi;t)) \! - \! R_{i} \cos(\widehat{
\kappa}_{\pm}(\xi;t))) \frac{d \xi}{\pi} \right. \right.& 
\nonumber \\
&\pm \frac{\vert R(0) \vert^{2}}{\pi^{2}2^{3} \widehat{\lambda}
_{0}^{2}(\pm t)} \! - \! \frac{2 \sqrt{2}}{(\pm t)} \int_{
\raise-0.65ex\hbox{$\scriptstyle{}\widehat{\lambda}_{0}$}}^{
\infty} \! \frac{\sqrt{\nu(\xi)}}{\xi^{3}}(a^{\pm} \sin(
\widehat{\kappa}_{\pm}(\xi;t)) \! - \! b^{\pm} \cos(\widehat{
\kappa}_{\pm}(\xi;t))) \frac{d \xi}{\pi}& \nonumber \\
&\pm \frac{1}{2^{2}(\pm t)} 
\int_{\raise-0.65ex\hbox{$\scriptstyle{}\widehat{\lambda}_{
0}$}}^{\infty}(\frac{4(\nu(\xi))^{2}}{\xi^{5}} \! + \! \frac{
\nu(\xi)(\frac{\partial \nu(\xi)}{\partial \xi}) \ln(e/8)}{
\xi^{4}} \! + \! \frac{\nu(\xi)(\frac{\partial^{2} \nu(\xi)}{
\partial \xi^{2}}) \ln(2/e)}{\xi^{3}} \! + \! \frac{(\frac{
\partial \nu(\xi)}{\partial \xi})^{2} \ln(2/e)}{\xi^{3}}& 
\nonumber \\
&\pm \frac{(\frac{\partial \nu(\xi)}{\partial \xi})(\frac{
\partial \varphi_{\pm}(\xi)}{\partial \xi})}{\xi^{3}} \! \pm 
\! \frac{\nu(\xi)(\frac{\partial^{2} \varphi_{\pm}(\xi)}{
\partial \xi^{2}})}{\xi^{3}} \! \mp \! \frac{3 \nu(\xi)(\frac{
\partial \varphi_{\pm}(\xi)}{\partial \xi})}{\xi^{4}}) d \xi 
\! \mp \! \frac{\ln \vert t \vert}{2^{2}(\pm t)} \int_{
\raise-0.65ex\hbox{$\scriptstyle{}\widehat{\lambda}_{0}$}}^{
\infty}(\frac{(\frac{\partial \nu(\xi)}{\partial \xi})^{2}}{
\xi^{3}}& \nonumber \\
&\left. \left. - \frac{3 \nu(\xi)(\frac{\partial \nu(\xi)}{
\partial \xi})}{\xi^{4}} \! + \! \frac{\nu(\xi)(\frac{\partial
^{2} \nu(\xi)}{\partial \xi^{2}})}{\xi^{3}}) d \xi \! + \! 
\frac{t}{2 s^{2}}(4 \widehat{\lambda}_{0}^{2} s \! + \! 1) \! 
\right) \! \right\} \!,& \\
&\widehat{\kappa}_{\pm}(z;t) \! := \! \widehat{\tau}_{\pm}(z;t) 
\! + \! \phi^{\pm}(z) \! + \! \widehat{\Phi}^{\pm}(z) \! + \! 
\frac{\pi}{2} \! + \! \epsilon^{\pm}(z) \! - \! 4 \sum\limits_{
l \in L_{\pm}} \! \gamma_{l},&
\end{eqnarray}
and, as $t \! \to \! \pm \infty$ and $x \! \to \! \mp \infty$ 
or $\pm \infty$ such that $\widehat{\mu}_{0} \! := \! \sqrt{
\frac{1}{2}(\frac{1}{s} \! - \! \frac{x}{t})} \! > \! M$, 
$\frac{x}{t} \! < \! \frac{1}{s}$, $s \! \in \! \Bbb R_{>0}$, 
and $(x,t) \! \in \! \Bbb R^{2} \! \setminus \! \widetilde{
\mho}_{n}$, $\widetilde{\mho}_{n} \! := \! \{\mathstrut (x,t); 
\, - x \! + \! t(\frac{1}{s} \! - \! 2 \Delta_{n}^{2} \cos 
\gamma_{n}) \! = \! {\cal O}(1)\}$, for those $\gamma_{n} \! 
\in \! (0,\! \frac{\pi}{2})$,
\begin{eqnarray}
&u(x,t) \! = \! \widetilde{v}_{\pm}(x,t) \widetilde{w}_{\pm}
(x,t) \! + \! {\cal O} \! \left(\frac{c^{{\cal S}}(\widehat{
\mu}_{0})(\ln \vert t \vert)^{2}}{(\pm \widehat{\mu}_{0}^{2} 
t)^{2}} \right) \!,&
\end{eqnarray}
where
\begin{eqnarray}
&\widetilde{v}_{\pm}(x,t) \! = \! \frac{e^{i \widetilde{\tau}
_{\pm}^{\prime}(\widehat{\mu}_{0};t)} v^{\pm \prime}_{1,1,0}(
\widehat{\mu}_{0})}{\sqrt{\pm t}} \! + \! \frac{v^{\pm \prime}
_{-1,2,0}(\widehat{\mu}_{0})}{(\pm t)} \! + \! \frac{e^{i 
\widetilde{\tau}_{\pm}^{\prime}(\widehat{\mu}_{0};t)} \sum_{
q=0}^{2} v^{\pm \prime}_{1,3,q}(\widehat{\mu}_{0})(\ln \vert 
t \vert)^{q}}{(\pm t)^{3/2}} \! + \! \frac{v^{\pm \prime}_{-1,
3,0}(\widehat{\mu}_{0})}{(\pm t)^{3/2}},& \\
&\widetilde{\tau}_{\pm}^{\prime}(\widehat{\mu}_{0};t) \! := \! 
\tau_{\pm}^{\prime}(\widehat{\mu}_{0};t/2),& \\
&v^{\pm \prime}_{1,1,0}(\widehat{\mu}_{0}) \! := \! \frac{1}{
\sqrt{s}} u^{\pm \prime}_{1,1,0}(\widehat{\mu}_{0}),& \\
&v^{\pm \prime}_{-1,2,0}(\widehat{\mu}_{0}) \! := \! \sqrt{
\frac{2}{s}} u^{\pm \prime}_{-1,2,0}(\widehat{\mu}_{0}),& \\
&v^{\pm \prime}_{1,3,0}(\widehat{\mu}_{0}) \! := \! \frac{2}{
\sqrt{s}} (u^{\pm \prime}_{1,3,0}(\widehat{\mu}_{0}) \! - \! 
u^{\pm \prime}_{1,3,1}(\widehat{\mu}_{0}) \ln 2 \! + \! u^{
\pm \prime}_{1,3,2}(\widehat{\mu}_{0})(\ln 2)^{2}),& \\
&v^{\pm \prime}_{1,3,1}(\widehat{\mu}_{0}) \! := \! \frac{2}{
\sqrt{s}} (u^{\pm \prime}_{1,3,1}(\widehat{\mu}_{0}) \! - \! 
2 u^{\pm \prime}_{1,3,2}(\widehat{\mu}_{0}) \ln 2),& \\
&v^{\pm \prime}_{1,3,2}(\widehat{\mu}_{0}) \! := \! \frac{2}{
\sqrt{s}} u^{\pm \prime}_{1,3,2}(\widehat{\mu}_{0}),& \\
&v^{\pm \prime}_{-1,3,0}(\widehat{\mu}_{0}) \! := \! \frac{2}
{\sqrt{s}} u^{\pm \prime}_{-1,3,0}(\widehat{\mu}_{0}),& \\
&\widetilde{w}_{\pm}(x,t) \! = \! \exp \! \left\{\! i \! \left(
\! \epsilon^{\pm \prime}(\widehat{\mu}_{0}) \! + \! \frac{2}{
\sqrt{\pm t}} \int_{\raise-0.55ex\hbox{$\scriptstyle{}\widehat{
\mu}_{0}$}}^{\infty} \! \frac{\sqrt{-\nu(i\xi)}}{\xi^{2}}(R_{r} 
\cos(\widetilde{\kappa}_{\pm}^{\prime}(\xi;t)) \! + \! R_{i} 
\sin(\widetilde{\kappa}_{\pm}^{\prime}(\xi;t))) \frac{d \xi}{
\pi} \right. \right.& \nonumber \\
&\mp \frac{\vert R(0) \vert^{2}}{\pi^{2} 2^{3} \widehat{\mu}
_{0}^{2}(\pm t)} \! + \! \frac{2 \sqrt{2}}{(\pm t)} \int_{
\raise-0.55ex\hbox{$\scriptstyle{}\widehat{\mu}_{0}$}}^{\infty} 
\! \frac{\sqrt{-\nu(i \xi)}}{\xi^{3}}(a^{\pm} \sin(\widetilde{
\kappa}_{\pm}^{\prime}(\xi;t)) \! - \! b^{\pm} \cos(\widetilde{
\kappa}_{\pm}^{\prime}(\xi;t))) \frac{d \xi}{\pi}& \nonumber \\
&\pm \frac{1}{2^{2}(\pm t)} 
\int_{\raise-0.55ex\hbox{$\scriptstyle{}\widehat{\mu}_{0}$}}^{
\infty} (\frac{4(\nu(i\xi))^{2}}{\xi^{5}} \! + \! \frac{
\nu(i\xi)(\frac{\partial \nu(i\xi)}{\partial \xi}) \ln(e/8)}{
\xi^{4}} \! + \! \frac{\nu(i\xi)(\frac{\partial^{2} \nu(i\xi)
}{\partial \xi^{2}}) \ln(2/e)}{\xi^{3}} \! + \! \frac{(\frac{
\partial \nu(i\xi)}{\partial \xi})^{2} \ln(2/e)}{\xi^{3}}& 
\nonumber \\
&\pm \frac{(\frac{\partial \nu(i\xi)}{\partial \xi})(\frac{
\partial \varphi_{\pm}^{\prime}(\xi)}{\partial \xi})}{\xi^{3}} 
\! \pm \! \frac{\nu(i\xi)(\frac{\partial^{2} \varphi_{\pm}^{
\prime}(\xi)}{\partial \xi^{2}})}{\xi^{3}} \! \mp \! \frac{3 
\nu(i\xi)(\frac{\partial \varphi_{\pm}^{\prime}(\xi)}{\partial 
\xi})}{\xi^{4}}) d \xi \! \mp \! \frac{\ln \vert t \vert}{2^{
2}(\pm t)} \int_{\raise-0.55ex\hbox{$\scriptstyle{}\widehat{
\mu}_{0}$}}^{\infty}(\frac{(\frac{\partial \nu(i\xi)}{\partial 
\xi})^{2}}{\xi^{3}}& \nonumber \\
&\left. \left. - \frac{3 \nu(i\xi)(\frac{\partial \nu(i\xi)}{
\partial \xi})}{\xi^{4}} \! + \! \frac{\nu(i\xi)(\frac{\partial
^{2} \nu(i\xi)}{\partial \xi^{2}})}{\xi^{3}}) d \xi \! - \! 
\frac{t}{2s^{2}}(4 \widehat{\mu}_{0}^{2} s \! - \! 1) \! 
\right) \! \right\} \!,& \\
&\widetilde{\kappa}_{\pm}^{\prime}(z;t) \! := \! \widetilde{
\tau}_{\pm}^{\prime}(z;t) \! + \! \phi^{\pm \prime}(z) \! + 
\! \widehat{\Phi}^{\pm \prime}(z) \! + \! \pi \! + \! 
\epsilon^{\pm \prime}(z) \! + \! 4 \sum\limits_{l \in L_{\pm}} 
\! \gamma_{l}.&
\end{eqnarray}
\end{ddd}
\begin{eee}
{\rm The term-by-term differentiation of the asymptotic 
expansions given in Theorems~2.2--2.4 will not be proven 
here since the proof is identical to that given in Sec.~4 
of \cite{a11}. In this paper, the proof of the asymptotic 
expansions for $Q(x,t)$ and $q(x,t)$ (resp.~$u(x,t))$ is 
presented for the cases $(x,t) \! \to \! (\mp \infty,\pm 
\infty)$ (resp.~$(x,t) \! \to \! (\pm \infty,\pm \infty))$ 
such that $\lambda_{0} \! > \! M$ (resp.~$\widehat{\lambda}
_{0} \! > \! M)$ and $(x,t) \! \in \! \Bbb R^{2} \! \setminus 
\! \Omega_{n}$ (resp.~$(x,t) \! \in \! \Bbb R^{2} \! \setminus 
\! \widetilde{\Omega}_{n})$ for those $\gamma_{n} \! \in \! 
(\frac{\pi}{2},\! \pi)$: the results for the remaining domains 
of the $(x,t)$-plane are obtained analogously. For a recent 
account of the asymptotic behaviour of non-soliton solutions 
of the defocusing NLS equation (resp.~DNLSE) as $t \! \to \! 
+\infty$ such that $x/t \! \sim \! {\cal O}(1)$ (resp.~$t \! 
\to \! \pm \infty$ such that $x/t \! \sim \! {\cal O}(1))$ in 
weighted Sobolev spaces, see \cite{a22} (resp.~\cite{a23}); 
furthermore, for some recent results pertaining to the 
solvability of RH problems in weighted Sobolev spaces, see 
\cite{a24}.\/}
\end{eee}
\section{Higher Order Deift-Zhou Theory}
In this section, employing the higher-order generalisation of the 
Deift-Zhou \cite{a11} procedure, the asymptotic expansion to all 
orders as $t \! \to \! +\infty$ and $x \! \to \! -\infty$ such 
that $\lambda_{0} \! > \! M$ for $Q(x,t)$ given in Lemma~3.4 (see 
below) is systematically derived: hereafter, all explicit $x,t$ 
dependences are suppressed, except where absolutely necessary.
\begin{figure}[bht]
\begin{center}
\unitlength=1cm
\begin{picture}(12,4)(0,2.5)
\thicklines
\put(10,5){\makebox(0,0){$\scriptstyle{}\bullet$}}
\put(2,5){\makebox(0,0){$\scriptstyle{}\bullet$}}
\put(6,5){\makebox(0,0){$\scriptstyle{}\bullet$}}
\put(10.05,4.6){\makebox(0,0){$\scriptstyle{}+\lambda_{0}$}}
\put(2.05,4.6){\makebox(0,0){$\scriptstyle{}-\lambda_{0}$}}
\put(6,4.6){\makebox(0,0){$\scriptstyle{}0$}}
\put(10,5){\vector(1,1){0.7}}
\put(10.5,5.5){\line(1,1){0.5}}
\put(10,5){\line(1,-1){0.5}}
\put(11,4){\vector(-1,1){0.5}}
\put(10,5){\line(-1,1){0.5}}
\put(9,6){\vector(1,-1){0.5}}
\put(10,5){\vector(-1,-1){0.7}}
\put(9.5,4.5){\line(-1,-1){0.5}}
\put(6,5){\vector(1,1){0.7}}
\put(6.5,5.5){\line(1,1){0.5}}
\put(6,5){\line(1,-1){0.5}}
\put(7,4){\vector(-1,1){0.5}}
\put(6,5){\line(-1,1){0.5}}
\put(5,6){\vector(1,-1){0.5}}
\put(6,5){\vector(-1,-1){0.7}}
\put(5,4){\line(1,1){0.4}}
\put(2,5){\vector(1,1){0.7}}
\put(2.5,5.5){\line(1,1){0.5}}
\put(2,5){\line(1,-1){0.5}}
\put(3,4){\vector(-1,1){0.5}}
\put(2,5){\line(-1,1){0.5}}
\put(1,6){\vector(1,-1){0.5}}
\put(2,5){\vector(-1,-1){0.7}}
\put(1,4){\line(1,1){0.4}} 
\put(11.5,6.25){\makebox(0,0)[r]{$\scriptstyle{}\varsigma_{{\cal 
B}}^{(1)}$}}
\put(11.5,3.75){\makebox(0,0)[r]{$\scriptstyle{}\varsigma_{{\cal 
B}}^{(4)}$}}
\put(8.5,6.25){\makebox(0,0)[l]{$\scriptstyle{}\varsigma_{{\cal 
B}}^{(3)}$}}
\put(8.5,3.75){\makebox(0,0)[l]{$\scriptstyle{}\varsigma_{{\cal 
B}}^{(2)}$}}
\put(7.5,6.25){\makebox(0,0)[r]{$\scriptstyle{}\varsigma_{{\cal 
C}}^{(1)}$}}
\put(7.5,3.75){\makebox(0,0)[r]{$\scriptstyle{}\varsigma_{{\cal 
C}}^{(4)}$}}
\put(4.5,6.25){\makebox(0,0)[l]{$\scriptstyle{}\varsigma_{{\cal 
C}}^{(3)}$}}
\put(4.5,3.75){\makebox(0,0)[l]{$\scriptstyle{}\varsigma_{{\cal 
C}}^{(2)}$}}
\put(3.5,6.25){\makebox(0,0)[r]{$\scriptstyle{}\varsigma_{{\cal 
A}}^{(1)}$}}
\put(3.5,3.75){\makebox(0,0)[r]{$\scriptstyle{}\varsigma_{{\cal 
A}}^{(4)}$}}
\put(0.5,6.25){\makebox(0,0)[l]{$\scriptstyle{}\varsigma_{{\cal 
A}}^{(3)}$}}
\put(0.5,3.75){\makebox(0,0)[l]{$\scriptstyle{}\varsigma_{{\cal 
A}}^{(2)}$}}
\put(10,2.75){\makebox(0,0){$\scriptstyle{}\varsigma_{{\cal B}} 
:= \bigcup\limits_{k=1}^{4} \varsigma_{{\cal B}}^{(k)}$}}
\put(6,2.75){\makebox(0,0){$\scriptstyle{}\varsigma_{{\cal C}} 
:= \bigcup\limits_{k=1}^{4}\varsigma_{{\cal C}}^{(k)}$}}
\put(2,2.75){\makebox(0,0){$\scriptstyle{}\varsigma_{{\cal A}} 
:= \bigcup\limits_{k=1}^{4}\varsigma_{{\cal A}}^{(k)}$}}
\end{picture}
\end{center}
\vspace{-0.5cm}
\caption{}
\end{figure}
\begin{ccc}[{\rm \cite{a2}}]
Set $\varsigma \! := \! \cup_{l \in \{{\cal B},{\cal A},{\cal 
C}\}} \varsigma_{l}$ (Fig.~2). As $t \! \to \! +\infty$ and 
$x \! \to \! - \infty$ such that $\lambda_{0} \! > \! M$, there 
exists a unique function $m^{\varsigma}(\lambda) \colon \Bbb C 
\setminus \varsigma \! \to \! {\rm SL}(2,\! \Bbb C)$ which solves 
the following RH problem:
\begin{enumerate}
\item[(1)] $m^{\varsigma}(\lambda)$ is piecewise holomorphic 
$\forall \, \lambda \! \in \! \Bbb C \! \setminus \! \varsigma;$
\item[(2)] $m^{\varsigma}(\lambda)$ satisfies the following
jump conditions,
\begin{eqnarray*}
&m^{\varsigma}_{+}(\lambda)=m^{\varsigma}_{-}(\lambda)v^{\varsigma}
(\lambda), \, \, \, \, \, \, \, \, \lambda \in \varsigma,&
\end{eqnarray*}
where
\begin{eqnarray*}
&v^{\varsigma}(\lambda) \vert_{\varsigma_{j}^{(1)} \cup 
\varsigma_{j}^{(2)}}={\rm I} + {\cal R}_{j}^{(1,2)}(\lambda;\! 
\lambda_{0})(\delta^{+}(\lambda;\! \lambda_{0}))^{-2} \exp\{2i
t \rho(\lambda;\! \lambda_{0})\} \sigma_{-},& \\
&v^{\varsigma}(\lambda) \vert_{\varsigma_{j}^{(3)} \cup 
\varsigma_{j}^{(4)}}={\rm I} + {\cal R}_{j}^{(3,4)}(\lambda;\! 
\lambda_{0})(\delta^{+}(\lambda;\! \lambda_{0}))^{2} \exp\{-2i 
t \rho(\lambda;\! \lambda_{0})\} \sigma_{+},& \\
&v^{\varsigma}(\lambda) \vert_{\varsigma_{{\cal C}}^{(1)} \cup 
\varsigma_{{\cal C}}^{(2)}}={\rm I} + {\cal R}_{{\cal C}}^{(1,
2)}(\lambda;\! \lambda_{0})(\delta^{+}(\lambda;\! \lambda_{0}))
^{2} \exp\{-2it\rho(\lambda;\! \lambda_{0})\} \sigma_{+},& \\
&v^{\varsigma}(\lambda) \vert_{\varsigma_{{\cal C}}^{(3)} \cup 
\varsigma_{{\cal C}}^{(4)}}={\rm I} + {\cal R}_{{\cal C}}^{(3,
4)}(\lambda;\! \lambda_{0})(\delta^{+}(\lambda;\! \lambda_{0}))
^{-2} \exp\{2it\rho(\lambda;\! \lambda_{0})\} \sigma_{-},&
\end{eqnarray*}
$j \! \in \! \{{\cal B},{\cal A}\}$, $\rho(\lambda;\! \lambda_{
0}) \! = \! 2 \lambda^{2}(\lambda^{2} \! - \! 2 \lambda_{0}^{2}
)$, $\{0,\pm \lambda_{0}\} \! = \! \{\mathstrut \lambda^{\prime}; 
\, \partial_{\lambda} \rho(\lambda;\! \lambda_{0}) \vert_{\lambda
=\lambda^{\prime}} \! = \! 0\}$ are the first-order stationary 
phase points,
\begin{eqnarray*}
&\delta^{+}(\lambda;\! \lambda_{0})=\exp \! \left\{ \sum\limits
_{l \in \{\pm\}} \! \left(\! \int_{0}^{l\lambda_{0}} \frac{\ln(1
-\vert r(\xi) \vert^{2})}{(\xi - \lambda)} \frac{d\xi}{2 \pi i}+ 
\int_{li\infty}^{i0} \frac{\ln(1-r(\xi)\overline{r(\overline{\xi}
)})}{(\xi-\lambda)} \frac{d\xi}{2\pi i} \right) \! \right\} \!,&
\end{eqnarray*}
and $\{{\cal R}_{{\cal B}}^{(l,l+1)}(\lambda;\! \lambda_{0}),
{\cal R}_{{\cal A}}^{(l,l+1)}(\lambda;\! \lambda_{0}),{\cal 
R}_{{\cal C}}^{(l,l+1)}(\lambda;\! \lambda_{0})\}_{l \in \{1,
3\}}$ are some rational functions which decay to zero as 
$\lambda \! \to \! \infty$, $\lambda \! \in \! \varsigma 
\setminus \cup_{k \in \{{\cal B},{\cal A},{\cal C}\}} 
\{{\rm sgn}(k) \lambda_{0}\}$, where ${\rm sgn}({\cal B}) \! 
= \! - {\rm sgn}({\cal A}) \! = \! 1$ and ${\rm sgn}({\cal C}
) \! = \! 0$, and have, respectively, Taylor series expansions 
in $\{\mathstrut \lambda^{\prime}; \, \vert \lambda^{\prime} 
\! - \! {\rm sgn}(k) \lambda_{0} \vert \linebreak[4] < \! 
\varepsilon\} \! \cap \! \varsigma_{k}$, $k \in \{{\cal B},{
\cal A},{\cal C}\}$, where $\varepsilon$ is an arbitrarily 
fixed, sufficiently small positive real number;
\item[(3)] as $\lambda \! \to \! \infty$, $\lambda \! \in \! 
\Bbb C \! \setminus \! \varsigma$,
\begin{eqnarray*}
&m^{\varsigma}(\lambda)={\rm I}+{\cal O}(\lambda^{-1}).&
\end{eqnarray*}
\end{enumerate}
Moreover, for arbitrary $l^{\prime} \! \in \! \Bbb Z_{\geq 1}$,
\begin{eqnarray*}
&Q(x,t)=2i\lim\limits_{\lambda \to \infty \atop \lambda \, \in 
\, \Bbb C \setminus \varsigma}(\lambda m^{\varsigma}(x,t;\! 
\lambda))_{12} + {\cal O} \! \left(\frac{\underline{c}(\lambda
_{0})}{(\lambda_{0}^{2} t)^{l^{\prime}}} \right) \!,&
\end{eqnarray*}
with $Q(x,\! 0) \! \in \! {\cal S}(\Bbb R;\! \Bbb C)$, satisfies 
Eq.~(2), and $m^{\varsigma}(\lambda)$ satisfies the following 
symmetry reductions, $m^{\varsigma}(\lambda) \! = \! \sigma_{3} 
m^{\varsigma}(-\lambda) \sigma_{3}$ and $m^{\varsigma}(\lambda) 
\! = \! \sigma_{1} \overline{m^{\varsigma}(\overline{\lambda})} 
\sigma_{1}$.
\end{ccc}

Analysing the signature graph of $\Re(it\rho(\lambda;\! 
\lambda_{0}))$, it was shown in \cite{a2} that, as $t \! \to 
\! +\infty$ and $x \! \to \! -\infty$ such that $\lambda_{0} 
\! > \! M$, $\exists \, \varepsilon_{o} \! \in \! \Bbb R_{>0}$ 
such that $-{\rm I} \! + \! v^{\varsigma}(\lambda) \vert_{
\varsigma_{k}^{(j)} \cup \varsigma_{k}^{(j+1)}} \! \sim \! 
{\cal O}(e^{-\varepsilon_{o} t})$, $k \! \in \! \{{\cal B},{
\cal A},{\cal C}\}$, $j \! \in \! \{1,3\}$.
\begin{bbb}
Set ${\rm sgn}({\cal B}) \! = \! - {\rm sgn}({\cal A}) \! = 1$, 
${\rm sgn}({\cal C}) \! = \! 0$, and $z_{{\cal B}}(\lambda_{0})
\! = \! z_{{\cal A}}(\lambda_{0}) \! = \! z_{{\cal C}}(\lambda_{
0}) / \sqrt{2} \linebreak[4] := \! z(16 \lambda_{0}^{2})^{-1/2}$. 
For $j \! \in \! \{{\cal B},{\cal A},{\cal C}\}$ and $\lambda \! 
\in \! \varsigma_{j} \! \setminus \! \{{\rm sgn}(j) \lambda_{0}\}$, 
change variables according to the rule $\lambda \! - \! {\rm sgn}
(j) \lambda_{0} \! \to \! z_{j}(\lambda_{0}) / \sqrt{t} \, \, 
(\not=\! 0)$. Then as $t \! \to \! +\infty$ and $x \! \to \! 
-\infty$ such that $\lambda_{0} \! > \! M$, for arbitrary $N_{o} 
\! \in \! \Bbb Z_{\geq 1}$,
\begin{eqnarray*}
&v^{\varsigma}({\rm sgn}(k) \lambda_{0} \! + \! \frac{z_{k}(
\lambda_{0})}{\sqrt{t}}) \! = \! \exp\{i(2 \lambda_{0}^{4} t 
\! - \! \frac{\nu(\lambda_{0})}{2} \ln t){\rm ad}(\sigma_{3})\} 
V^{k}(z;\! \lambda_{0}),& \\
&V^{k}(z;\! \lambda_{0}) \! := \! {\rm I} + e^{i\{(-8 \lambda_{
0}^{2} z^{2}+\nu(\lambda_{0}) \ln z){\rm ad}(\sigma_{3})\}} 
\sum\limits_{p=0}^{N_{o}} \sum\limits_{q=0}^{p} \frac{V^{k}_{p,
q}(z;\lambda_{0})(\ln t)^{q}}{t^{p/2}} + {\cal E}^{k}(z;\! 
\lambda_{0}),& \\
&v^{\varsigma}({\rm sgn}({\cal C}) \lambda_{0} \! + \! \frac{z_{
{\cal C}}(\lambda_{0})}{\sqrt{t}}) \! := \! V^{{\cal C}}(z;\! 
\lambda_{0}) \! = \! {\rm I} \! + \! e^{i\{4 \lambda_{0}^{2} z^{
2} {\rm ad}(\sigma_{3})\}} \sum\limits_{p=0}^{N_{o}} \sum\limits
_{q=0}^{p} \frac{V^{{\cal C}}_{p,q}(z;\lambda_{0})(\ln t)^{q}}{t
^{p + 1/2}} \! + \! {\cal E}^{{\cal C}}(z;\! \lambda_{0}),&
\end{eqnarray*}
$k \! \in \! \{{\cal B},{\cal A}\}$, where, for $0 \! \leq \! p 
\! \leq \! N_{o}$ and $0 \! \leq \! q \! \leq \! p$,
\begin{eqnarray*}
&\vert \vert e^{i\{(-8 \lambda_{0}^{2} (\cdot)^{2}+\nu(\lambda
_{0}) \ln(\cdot)){\rm ad}(\sigma_{3})\}} V^{k}_{p,q}(\cdot;\! 
\lambda_{0}) \vert \vert_{\cap_{l \in \{1,2,\infty\}}{\cal L}^{
l}(\varsigma^{\prime}_{k} \setminus \{0\};M_{2}(\Bbb C))} 
< \infty,& \\
&\vert \vert e^{i\{4 \lambda_{0}^{2}(\cdot)^{2} {\rm ad}(
\sigma_{3})\}} V^{{\cal C}}_{p,q}(\cdot;\! \lambda_{0}) \vert 
\vert_{\cap_{l \in \{1,2,\infty\}}{\cal L}^{l}(\varsigma^{
\prime}_{{\cal C}} \setminus \{0\};M_{2}(\Bbb C))} < \infty,& 
\\
&\vert \vert {\cal E}^{k}(\cdot;\! \lambda_{0}) \vert \vert
_{\cap_{l \in \{1,2,\infty\}}{\cal L}^{l}(\varsigma^{\prime}
_{k} \setminus \{0\};M_{2}(\Bbb C))} = {\cal O} \! \left(
\frac{c^{{\cal S}}(\lambda_{0})(\ln t)^{N_{o}+1}}{(\lambda_{
0}^{2} t)^{(N_{o}+1)/2}} \right) \!,& \\
&\vert \vert {\cal E}^{{\cal C}}(\cdot;\! \lambda_{0}) \vert 
\vert_{\cap_{l \in \{1,2,\infty\}}{\cal L}^{l}(\varsigma^{
\prime}_{{\cal C}} \setminus \{0\};M_{2}(\Bbb C))} = {\cal O} 
\! \left(\frac{\underline{c}(\lambda_{0})(\ln t)^{N_{o}+1}}{
(\lambda_{0}^{2} t)^{N_{o}+3/2}} \right) \!,&
\end{eqnarray*}
with $\varsigma_{k}^{\prime}$ denoting the shifted and scaled 
version of $\varsigma_{k}$ (centred at the origin), and 
$\varsigma_{{\cal C}}^{\prime}$ denoting the scaled version 
of $\varsigma_{{\cal C}}$.
\end{bbb}

{\em Proof.\/} The case $k \! = \! {\cal B} \, (\leftrightarrow 
\! \lambda_{0})$ is considered in detail: the cases $k \! = \! 
{\cal A} \, (\leftrightarrow \! -\lambda_{0})$ and ${\cal C} \, 
(\leftrightarrow \! 0)$ are treated analogously. Using the fact 
that $\vert \vert r \vert \vert_{{\cal L}^{\infty}(\Bbb R;\Bbb 
C)} \! < \! 1$ and $r(\lambda) \! \in \! {\cal S}(\widehat{\Gamma};
\! \Bbb C)$, and recalling that $\nu(\lambda_{0}) \! := \! -\frac{
1}{2\pi} \ln(1 \! - \! \vert r(\lambda_{0}) \vert^{2}) \, (> \! 
0)$, one expands, via an integration by parts argument, the 
integral ${\rm I}(\lambda_{0}) \! := \! \int_{0}^{\lambda_{0}} \! 
\frac{\ln(1-\vert r(\xi) \vert^{2})}{(\xi-\lambda)} \frac{d\xi}
{2 \pi i}$, which appears in the expression for $\delta^{+}(
\lambda;\! \lambda_{0})$ given in Lemma~3.1, and shows that, for 
arbitrary $N_{o} \! \in \! \Bbb Z_{\geq 1}$, ${\rm I}(\lambda_{0}
) \! = \! i \nu(\lambda_{0}) \ln \! \vert \lambda \! - \! \lambda
_{0} \vert \! + \! \sum_{m=1}^{N_{o}}(I^{(1)}_{m}(\lambda_{0}) 
\ln \! \vert \lambda \! - \! \lambda_{0} \vert \! + \! I^{(2)}_{
m}(\lambda_{0}))(\lambda \! - \! \lambda_{0})^{m} \! + \! {\cal 
O}(c^{{\cal S}}(\lambda_{0}) \ln \! \vert \lambda \! - \! \lambda
_{0} \vert (\lambda \! - \! \lambda_{0})^{N_{o}+1})$, where $I^{
(j)}_{m}(\lambda_{0}) \! \in \! {\cal S}(\Bbb R_{>M};\! \Bbb C)$, 
$j \! \in \! \{1,2\}$, $1 \! \leq \! m \! \leq \! N_{o}$; 
furthermore, one expands $\rho(\lambda;\! \lambda_{0})$ given 
in Lemma~3.1 in a Taylor series about $\lambda_{0}$ and shows 
that $\rho(\lambda;\! \lambda_{0}) \! = \! -2 \lambda_{0}^{4} 
\! + \! 8 \lambda_{0}^{2}(\lambda \! - \! \lambda_{0})^{2} \! 
+ \! 8 \lambda_{0}(\lambda \! - \! \lambda_{0})^{3} \! + \! 2
(\lambda \! - \! \lambda_{0})^{4}$. Now, using the expressions 
for $v^{\varsigma}(\lambda) \vert_{\varsigma_{{\cal B}}^{(j)} 
\cup \varsigma_{{\cal B}}^{(j+1)}}$, $j \! \in \! \{1,3\}$, 
given in Lemma~3.1, expanding ${\cal R}_{{\cal B}}^{(j,j+1)}
(\lambda;\! \lambda_{0})$, $j \! \in \! \{1,3\}$, in Taylor 
series' about $\lambda_{0}$, and recalling the above expansions 
for ${\rm I}(\lambda_{0})$ and $\rho(\lambda;\! \lambda_{0})$, 
one makes the transformation (shift and scaling) given in the 
Proposition, $\lambda \! - \! \lambda_{0} \! = \! z(16 \lambda
_{0}^{2}t)^{-1/2} \, (\not= \! 0)$, and obtains the result for 
$v^{\varsigma}(\lambda_{0} \! + \! z(16\lambda_{0}^{2}t)^{-1/2}
)$ stated in the Proposition by expanding in reciprocal powers 
of $(\lambda_{0}^{2}t)^{1/2}$: the estimates for $\vert \vert 
e^{i\{(-8\lambda_{0}^{2}(\cdot)^{2}+\nu(\lambda_{0})\ln(\cdot))
{\rm ad}(\sigma_{3})\}} V^{{\cal B}}_{p,q}(\cdot;\! \lambda_{0}) 
\vert \vert_{\cap_{l \in \{1,2,\infty\}} {\cal L}^{l}(\varsigma
_{{\cal B}}^{\prime} \setminus \{0\};M_{2}(\Bbb C))}$ and $\vert 
\vert {\cal E}^{{\cal B}}(\cdot;\! \lambda_{0}) \vert \vert_{
\cap_{l \in \{1,2,\infty\}} {\cal L}^{l}(\varsigma_{{\cal B}}
^{\prime} \setminus \{0\};M_{2}(\Bbb C))}$ are a straightforward, 
but onerous, generalisation of those given in the proof of 
Lemma~4.1 in \cite{a10}. Recalling the expressions for 
$v^{\varsigma}(\lambda) \vert_{\varsigma_{l}^{(j)} \cup 
\varsigma_{l}^{(j+1)}}$, $j \! \in \! \{1,3\}$, $l \! \in \! 
\{{\cal A},{\cal C}\}$, given in Lemma~3.1, and noting that, in 
the case of the first-order stationary phase point at the origin 
\cite{a2}, $(r(\lambda) \vert_{\lambda \in \Bbb R}) \vert_{
\lambda=0} \! = \! (r(\lambda)\vert_{\lambda \in i \Bbb R}) 
\vert_{\lambda=0} \! = \! 0$, proceeding analogously as indicated 
above, one obtains the results stated in the Proposition. 
\hfill \rule{6.5pt}{6.5pt}
\begin{ggg}
{\rm Let ${\cal N}(\bullet;\! M_{2}(\Bbb C))$ denote the linear 
vector space of bounded linear operators acting {}from $\cup_{p 
\in \{2,\infty\}} {\cal L}^{p}(\bullet;\! M_{2}(\Bbb C))$ into 
${\cal L}^{2}(\bullet;\! M_{2}(\Bbb C))$.\/}
\end{ggg}
\begin{bbb}
As $t \! \to \! +\infty$ and $x \! \to \! -\infty$ such that 
$\lambda_{0} \! > \! M$, for arbitrary $N_{o} \! \in \! \Bbb 
Z_{\geq 1}$,
\begin{eqnarray*}
&(C_{w^{k}} f)(\widetilde{w};\! \lambda_{0})=\sum\limits_{p=0}^{
N_{o}} \sum\limits_{q=0}^{p} \frac{(\ln t)^{q}}{t^{p/2}}(\widehat{
C}^{k}_{p,q}f)(\widetilde{w};\! \lambda_{0})+(\widehat{{\cal E}}^{
k}f)(\widetilde{w}; \! \lambda_{0}),& \\
&(C_{w^{{\cal C}}}f)(\widetilde{w};\! \lambda_{0})=\sum\limits_{
p=0}^{N_{o}} \sum\limits_{q=0}^{p} \frac{(\ln t)^{q}}{t^{p+1/2}}
(\widehat{C}^{{\cal C}}_{p,q}f)(\widetilde{w};\! \lambda_{0})+
(\widehat{{\cal E}}^{{\cal C}}f)(\widetilde{w}; \! \lambda_{0}),&
\end{eqnarray*}
$k \! \in \! \{{\cal B},{\cal A}\}$, where, for $0 \! \leq \! p 
\! \leq \! N_{o}$ and $0 \! \leq \! q \! \leq \! p$,
\begin{eqnarray*}
&\vert \vert \widehat{C}^{k}_{p,q} \vert \vert_{{\cal N}
(\varsigma_{k}^{\prime} \setminus \{0\};M_{2}(\Bbb C))} 
< \infty, \, \, \, \, \, \, \, \, \, \, \, \, \vert \vert 
\widehat{C}^{{\cal C}}_{p,q} \vert \vert_{{\cal N}(\varsigma
_{{\cal C}}^{\prime} \setminus \{0\};M_{2}(\Bbb C))} < 
\infty,& \\
&\vert \vert \widehat{{\cal E}}^{k} \vert \vert_{{\cal N}
(\varsigma_{k}^{\prime} \setminus \{0\};M_{2}(\Bbb C))} =
{\cal O} \! \left(\frac{c^{{\cal S}}(\lambda_{0})(\ln t)^{
N_{o}+1}}{(\lambda_{0}^{2} t)^{(N_{o}+1)/2}} \right) \!, \, 
\, \, \, \, \, \, \vert \vert \widehat{{\cal E}}^{{\cal C}} 
\vert \vert_{{\cal N}(\varsigma_{{\cal C}}^{\prime} \setminus 
\{0\};M_{2}(\Bbb C))} = {\cal O} \! \left(\frac{\underline{c}
(\lambda_{0})(\ln t)^{N_{o}+1}}{(\lambda_{0}^{2} t)^{N_{o}+
3/2}} \right) \!.&
\end{eqnarray*}
\end{bbb}

{\em Proof.\/} The case $k \! = \! {\cal B} \, (\leftrightarrow 
\! \lambda_{0})$ is considered in detail: the cases $k \! = \! 
{\cal A} \, (\leftrightarrow \! -\lambda_{0})$ and ${\cal C} \, 
(\leftrightarrow \! 0)$ are treated analogously. Recalling the 
BC \cite{a14} formulation (Theorem~2.1 and the paragraph 
superseding it), and setting $w^{+}_{x,t}(z) \! := \! w^{{\cal 
B},+}_{x,t}(z) \! = \! V^{{\cal B}}(z;\! \lambda_{0}) \! - \! 
{\rm I}$ and $w^{-}_{x,t}(z) \! := \! w^{{\cal B},-}_{x,t}(z) 
\! = \! 0$ (for the justification of this last step, see 
\cite{a20}, pp.~293--294, {\em Proof of Theorem~3.14 and 
Proposition~1.9\/}, and \cite{a24}), where the superscript ${
\cal B}$ refers to the first-order stationary phase point at 
$\lambda_{0}$, the operator $C_{w_{x,t}} \! := \! C_{w^{\cal 
B}}$ on $z \! \in \! \varsigma_{{\cal B}}^{\prime} \! \setminus 
\! \{0\}$, for arbitrary $f(z) \! \in \! \cup_{l \in \{2,\infty
\}}{\cal L}^{l}(\varsigma_{{\cal B}}^{\prime} \! \setminus \! 
\{0\};\! M_{2}(\Bbb C))$, is represented as
\begin{eqnarray*}
&(C_{w^{{\cal B}}}f)(\widetilde{w};\! \lambda_{0}) \! = \! C_{-}
(f(V^{{\cal B}}(z;\! \lambda_{0})-{\rm I})) \! = \! \lim\limits_{
w^{\prime} \to \widetilde{w} \atop w^{\prime} \in - {\rm side} \,
{\rm of} \varsigma_{{\cal B}}^{\prime} \setminus \{0\}} 
\int\limits_{\varsigma_{{\cal B}}^{\prime} \setminus \{0\}} \frac{
f(\xi)(V^{{\cal B}}(\xi;\lambda_{0})-{\rm I})}{(\xi-w^{\prime})} 
\frac{d\xi}{2\pi i};&
\end{eqnarray*}
hence, using the expression for $V^{{\cal B}}(z;\! \lambda_{0})$ 
given in Proposition~3.1, defining, for $0 \! \leq \! p \! \leq 
\! N_{o}$ and $0 \! \leq \! q \! \leq \! p$, with $N_{o} \! \in 
\! \Bbb Z_{\geq 1}$ and arbitrary,
\begin{eqnarray*}
&(\widehat{C}^{{\cal B}}_{p,q}f)(\widetilde{w};\! \lambda_{0}) 
\! := \! \lim\limits_{w^{\prime} \to \widetilde{w} \atop w^{
\prime} \in - {\rm side} \, {\rm of} \varsigma_{{\cal B}}^{\prime} 
\setminus \{0\}} \int\limits_{\varsigma_{{\cal B}}^{\prime} 
\setminus \{0\}} \frac{f(\xi) e^{i\{(-8 \lambda_{0}^{2} \xi^{2}+
\nu(\lambda_{0}) \ln \xi){\rm ad}(\sigma_{3})\}} V^{{\cal B}}_{
p,q}(\xi;\lambda_{0})}{(\xi-w^{\prime})} \frac{d\xi}{2\pi i},& \\
&(\widehat{{\cal E}}^{{\cal B}}f)(\widetilde{w};\! \lambda_{0}) 
\! := \! \lim\limits_{w^{\prime} \to \widetilde{w} \atop w^{
\prime} \in - {\rm side} \, {\rm of} \varsigma_{{\cal B}}^{\prime} 
\setminus \{0\}} \int\limits_{\varsigma_{{\cal B}}^{\prime} 
\setminus \{0\}} \frac{f(\xi){\cal E}^{{\cal B}}(\xi;\lambda_{0})}
{(\xi-w^{\prime})} \frac{d\xi}{2\pi i},&
\end{eqnarray*}
and using the estimates for $\vert \vert e^{i\{(-8\lambda_{0}^{2}
(\cdot)^{2}+\nu(\lambda_{0})\ln(\cdot)){\rm ad}(\sigma_{3})\}} 
V^{{\cal B}}_{p,q}(\cdot;\! \lambda_{0}) \vert \vert_{\cap_{l \in 
\{1,2,\infty\}}{\cal L}^{l}(\varsigma_{{\cal B}}^{\prime} 
\setminus \{0\};M_{2}(\Bbb C))}$ and $\vert \vert {\cal E}^{{\cal 
B}}(\cdot;\! \lambda_{0}) \vert \vert_{\cap_{l \in \{1,2,\infty\}
} {\cal L}^{l}(\varsigma_{{\cal B}}^{\prime} \setminus \{0\};M_{
2}(\Bbb C))}$ given in Proposition~3.1, one obtains the results 
stated in the Proposition. For the cases $k \! = \! {\cal A} \, 
(\leftrightarrow \! -\lambda_{0})$ and ${\cal C} \, 
(\leftrightarrow \! 0)$, one proceeds analogously as described 
above, except with the following definitions:
\begin{eqnarray*}
&(\widehat{C}^{{\cal A}}_{p,q}f)(\widetilde{w};\! \lambda_{0}) 
\! := \! \lim\limits_{w^{\prime} \to \widetilde{w} \atop 
w^{\prime} \in - {\rm side} \, {\rm of} \varsigma_{{\cal A}}^{
\prime} \setminus \{0\}} \int\limits_{\varsigma_{{\cal A}}^{
\prime} \setminus \{0\}} \frac{f(\xi) e^{i\{(-8 \lambda_{0}^{2} 
\xi^{2}+\nu(\lambda_{0})\ln \xi){\rm ad}(\sigma_{3})\}} V^{{\cal 
A}}_{p,q}(\xi;\lambda_{0})}{(\xi-w^{\prime})} \frac{d\xi}{2\pi 
i},& \\
&(\widehat{C}^{{\cal C}}_{p,q}f)(\widetilde{w};\! \lambda_{0}) 
\! := \! \lim\limits_{w^{\prime} \to \widetilde{w} \atop w
^{\prime} \in - {\rm side} \, {\rm of} \varsigma_{{\cal C}}^{
\prime} \setminus \{0\}} \int\limits_{\varsigma_{{\cal C}}^{
\prime} \setminus \{0\}} \frac{f(\xi) e^{i\{4 \lambda_{0}^{2} 
\xi^{2}{\rm ad}(\sigma_{3})\}} V^{{\cal C}}_{p,q}(\xi;\lambda_{
0})}{(\xi-w^{\prime})} \frac{d \xi}{2\pi i},& \\
&(\widehat{{\cal E}}^{{\cal A}}f)(\widetilde{w};\! \lambda_{0}) 
\! := \! \lim\limits_{w^{\prime} \to \widetilde{w} \atop w^{
\prime} \in - {\rm side} \, {\rm of} \varsigma_{{\cal A}}^{\prime} 
\setminus \{0\}} \int\limits_{\varsigma_{{\cal A}}^{\prime} 
\setminus \{0\}} \frac{f(\xi){\cal E}^{{\cal A}}(\xi;\lambda_{0})}
{(\xi-w^{\prime})} \frac{d\xi}{2\pi i},& \\
&(\widehat{{\cal E}}^{{\cal C}}f)(\widetilde{w};\! \lambda_{0}) \!
:= \! \lim\limits_{w^{\prime} \to \widetilde{w} \atop w^{
\prime} \in - {\rm side} \, {\rm of} \varsigma_{{\cal C}}^{\prime} 
\setminus \{0\}} \int\limits_{\varsigma_{{\cal C}}^{\prime} 
\setminus \{0\}} \frac{f(\xi){\cal E}^{{\cal C}}(\xi;\lambda_{0})}
{(\xi-w^{\prime})} \frac{d\xi}{2\pi i},&
\end{eqnarray*}
where, for $z \! \in \! \varsigma_{{\cal A}}^{\prime} \! \setminus 
\! \{0\}$, $f(z) \! \in \! \cup_{l \in \{2,\infty\}}{\cal L}^{l}(
\varsigma_{{\cal A}}^{\prime} \! \setminus \! \{0\};\! M_{2}(\Bbb 
C))$, and, for $z \! \in \! \varsigma_{{\cal C}}^{\prime} \! 
\setminus \! \{0\}$, $f(z) \! \in \! \cup_{l \in \{2,\infty\}}{
\cal L}^{l}(\varsigma_{{\cal C}}^{\prime} \! \setminus \! \{0\};\! 
M_{2}(\Bbb C))$. \hfill \rule{6.5pt}{6.5pt}
\begin{ccc}[{\rm \cite{a2}}]
As $t \! \to \! +\infty$ and $x \! \to \! -\infty$ such 
that $\lambda_{0} \! > \! M$, $\exists \, \epsilon_{0,0}^{k}
(\lambda_{0}) \! \in \! \Bbb R_{>0}$, with $\epsilon_{0,0}^{
k}(\lambda_{0}) \! < \! \infty$, $k \! \in \! \{{\cal B},{
\cal A},{\cal C}\}$, such that
\begin{eqnarray*}
&\vert \vert (\underline{{\bf Id}}^{k} - \widehat{C}^{k}_{
0,0}(\cdot;\! \lambda_{0}))^{-1} \vert \vert_{{\cal N}(
\varsigma_{k}^{\prime} \setminus \{0\};M_{2}(\Bbb C))} \leq
\epsilon_{0,0}^{k}(\lambda_{0}),&
\end{eqnarray*}
where $\underline{{\bf Id}}^{k}$ denotes the identity operator 
on $\cup_{p \in \{2,\infty\}}{\cal L}^{p}(\varsigma_{k}^{\prime} 
\! \setminus \! \{0\};\! M_{2}(\Bbb C))$.
\end{ccc}
\begin{eee}
{\rm More explicitly, it was shown in \cite{a2} that, as $t \! 
\to \! +\infty$ and $x \! \to \! -\infty$ such that $\lambda_{0} 
\! > \! M$, $\epsilon_{0,0}^{k}(\lambda_{0}) \! = \! \underline{
c}(\lambda_{0})(1 \! - \! \vert \vert r \vert \vert_{{\cal L}^{
\infty}(\Bbb R;\Bbb C)})^{-1}$, $k \! \in \! \{{\cal B},{\cal A},
{\cal C}\}$.\/}
\end{eee}
\begin{ggg}
{\rm For some unbounded domain (open simply-connected set) $D 
\! \subset \! \Bbb C$ and a $M_{2}(\Bbb C)$-valued function 
$F(\cdot)$, the ``notation'' $\vert \vert F(\cdot) \vert \vert
_{\cup_{p \in \{2,\infty\}} {\cal L}^{p}(D;M_{2}(\Bbb C)) \to 
{\cal L}^{2}(D;M_{2}(\Bbb C))}$ is to be understood as follows: 
(1) $F \colon \cup_{p \in \{2,\infty\}} {\cal L}^{p}(D;\! 
M_{2}(\Bbb C)) \! \to \! {\cal L}^{2}(D;\! M_{2}(\Bbb C))$; and 
(2) the norm is taken as $(\vert \vert F(\infty) \vert \vert^{2}
_{{\cal L}^{\infty}(D;M_{2}(\Bbb C))} \! + \! \vert \vert 
F(\cdot) \! - \! F(\infty) \vert \vert^{2}_{{\cal L}^{2}(D;M_{2}
(\Bbb C))})^{1/2}$.\/}
\end{ggg}
\begin{bbb}
As $t \! \to \! +\infty$ and $x \! \to \! -\infty$ such that 
$\lambda_{0} \! > \! M$, for arbitrary $N_{o} \! \in \! \Bbb 
Z_{\geq 1}$,
\begin{eqnarray*}
&\mu^{k}(\widetilde{w};\! \lambda_{0}) = {\rm I} + \sum\limits_{
p=0}^{N_{o}} \sum\limits_{q=0}^{p} \frac{\mu^{k}_{p,q}(\widetilde{
w};\lambda_{0})(\ln t)^{q}}{t^{p/2}}+\widetilde{{\cal E}}^{k}_{\mu}
(\widetilde{w};\! \lambda_{0}),& \\
&\mu^{{\cal C}}(\widetilde{w};\! \lambda_{0})={\rm I}+\sum\limits_{
p^{\prime}=1}^{2} \sum\limits_{p=0}^{N_{o}} \sum\limits_{q=0}^{p} 
\frac{\mu^{{\cal C}}_{p^{\prime},p,q}(\widetilde{w};\lambda_{0})(
\ln t)^{q}}{t^{p+p^{\prime}/2}}+\widetilde{{\cal E}}^{{
\cal C}}_{\mu}(\widetilde{w};\! \lambda_{0}),&
\end{eqnarray*}
$k \! \in \! \{{\cal B},{\cal A}\}$, where, for $1 \! \leq \! 
p^{\prime} \! \leq \! 2$, $0 \! \leq \! p \! \leq \! N_{o}$, 
and $0 \! \leq \! q \! \leq \! p$,
\begin{eqnarray*}
&\vert \vert \mu^{k}_{p,q}(\cdot;\! \lambda_{0}) \vert \vert_{
\cup_{l \in \{2,\infty\}} {\cal L}^{l}(\varsigma_{k}^{\prime} 
\setminus \{0\};M_{2}(\Bbb C)) \to {\cal L}^{2}(\varsigma_{k}^{
\prime} \setminus \{0\};M_{2}(\Bbb C))} < \infty,& \\
&\vert \vert \mu^{{\cal C}}_{p^{\prime},p,q}(\cdot;\! \lambda_{
0}) \vert \vert_{\cup_{l \in \{2,\infty\}} {\cal L}^{l}(\varsigma
_{{\cal C}}^{\prime} \setminus \{0\};M_{2}(\Bbb C)) \to {\cal 
L}^{2}(\varsigma_{{\cal C}}^{\prime} \setminus \{0\};M_{2}(\Bbb 
C))} < \infty,& \\
&\vert \vert \widetilde{{\cal E}}^{k}_{\mu}(\cdot;\! \lambda_{
0}) \vert \vert_{\cup_{l \in \{2,\infty\}} {\cal L}^{l}(\varsigma
_{k}^{\prime} \setminus \{0\};M_{2}(\Bbb C)) \to {\cal L}^{2}(
\varsigma_{k}^{\prime} \setminus \{0\};M_{2}(\Bbb C))} = {\cal O} 
\! \left(\frac{c^{{\cal S}}(\lambda_{0})(\ln t)^{N_{o}+1}}{
(\lambda_{0}^{2} t)^{(N_{o}+1)/2}} \right) \!,& \\
&\vert \vert \widetilde{{\cal E}}^{{\cal C}}_{\mu}(\cdot;\! 
\lambda_{0}) \vert \vert_{\cup_{l \in \{2,\infty\}}{\cal L}^{l}
(\varsigma_{{\cal C}}^{\prime} \setminus \{0\};M_{2}(\Bbb C)) \to 
{\cal L}^{2}(\varsigma_{{\cal C}}^{\prime} \setminus \{0\};M_{2}(
\Bbb C))} = {\cal O} \! \left(\frac{\underline{c}(\lambda_{0})
(\ln t)^{N_{o}+1}}{(\lambda_{0}^{2} t)^{N_{o}+3/2}} \right) 
\!.&
\end{eqnarray*}
\end{bbb}

{\em Proof.\/}
For $k \! \in \! \{{\cal B},{\cal A},{\cal C}\}$, define $\mu^{
k}(\widetilde{w};\! \lambda_{0}) \! := \! (\underline{{\bf Id}}
^{k} \! - \! C_{w^{k}}(\widetilde{w};\! \lambda_{0}))^{-1} {\rm 
I}$, where $\underline{{\bf Id}}^{k}$ denotes the identity operator 
on $\cup_{p \in \{2,\infty\}}{\cal L}^{p}(\varsigma_{k}^{\prime} \! 
\setminus \! \{0\};M_{2}(\Bbb C))$. Since, {}from 
Lemma~3.2, as $t \! \to \! +\infty$ and $x \! \to \! -\infty$ 
such that $\lambda_{0} \! > \! M$, $\exists \, \epsilon_{0,0}^{
k}(\lambda_{0}) \! \in \! \Bbb R_{>0}$ with $\epsilon_{0,0}^{k}
(\lambda_{0}) \! < \! \infty$ such that
\begin{eqnarray*}
&\vert \vert (\underline{{\bf Id}}^{k}-\widehat{C}^{k}_{0,0}
(\cdot;\! \lambda_{0}))^{-1} {\rm I} \vert \vert_{\cup_{l \in 
\{2,\infty\}} {\cal L}^{l}(\varsigma_{k}^{\prime} \setminus 
\{0\};M_{2}(\Bbb C)) \to {\cal L}^{2}(\varsigma_{k}^{\prime} 
\setminus \{0\};M_{2}(\Bbb C))} \! \leq \! \epsilon_{0,0}^{k}
(\lambda_{0}),&
\end{eqnarray*}
the results stated in the Proposition follow {}from those given 
in Proposition~3.2, via the method of successive approximations, 
by expanding $\{\mu^{k}(\widetilde{w};\! \lambda_{0})\}_{k \in 
\{{\cal B},{\cal A},{\cal C}\}}$ in von Neumann-type series' as 
$t \! \to \! +\infty$ $(x/t \! \sim \! {\cal O}(1))$ (see, also, 
Part II of \cite{a5}). \hfill \rule{6.5pt}{6.5pt}
\begin{ggg}
{\rm Let $\{m^{\varsigma}_{k}(\lambda)\}_{k \in \{{\cal B},{
\cal A},{\cal C}\}}$ denote the solutions of the corresponding 
RH problems on $\{v^{\varsigma}(\lambda) \vert_{\varsigma_{k}^{
(p)} \cup \varsigma_{k}^{(p+1)}},\varsigma_{k}\}_{p \in \{1,3\} 
\atop k \in \{{\cal B},{\cal A},{\cal C}\}}$, and $\{m^{
\varsigma^{\prime}}_{k}(\widetilde{w})\}_{k \in \{{\cal B},{
\cal A},{\cal C}\}}$ denote the solutions of the associated RH 
problems on $\{v^{\varsigma^{\prime}}(\widetilde{w}) \vert_{
\varsigma_{k}^{\prime (p)} \cup \varsigma_{k}^{\prime (p+1)}},
\varsigma_{k}^{\prime}\}_{p \in \{1,3\} \atop k \in \{{\cal B},
{\cal A},{\cal C}\}}$.\/}
\end{ggg}
\begin{bbb}
As $t \! \to \! +\infty$ and $x \! \to \! -\infty$ such that 
$\lambda_{0} \! > \! M$, for arbitrary $N_{o} \! \in \! \Bbb 
Z_{\geq 1}$,
\begin{eqnarray*}
&m^{\varsigma}_{k}(\lambda) = {\rm I} + e^{\frac{i}{2} \{(4 
\lambda_{0}^{4}t-\nu(\lambda_{0}) \ln t){\rm ad}(\sigma_{3})
\}} \sum\limits_{p=1}^{N_{o}} \sum\limits_{q=0}^{p-1} \frac{
m^{k}_{p,q}(\lambda;\lambda_{0})(\ln t)^{q}}{t^{p/2}}+E^{k}_{
m^{\varsigma}}(\lambda;\! \lambda_{0}),& \\
&m^{\varsigma}_{{\cal C}}(\lambda) = {\rm I} + \sum\limits_{
p^{\prime}=0}^{1} \sum\limits_{p=1}^{N_{o}} \sum\limits_{q=0}
^{p-1} \frac{m^{{\cal C}}_{p^{\prime},p,q}(\lambda;\lambda_{0}
)(\ln t)^{q}}{t^{p+p^{\prime}/2}} + E^{{\cal C}}_{m^{\varsigma
}}(\lambda;\! \lambda_{0}),&
\end{eqnarray*}
$k \! \in \! \{{\cal B},{\cal A}\}$, where, for $0 \! \leq \! 
p^{\prime} \! \leq \! 1$, $1 \! \leq \! p \! \leq \! N_{o}$, 
and $0 \! \leq \! q \! \leq \! p \! - \! 1$,
\begin{eqnarray*}
&\vert \vert m^{k}_{p,q}(\cdot;\! \lambda_{0}) \vert \vert_{{
\cal L}^{\infty}(\Bbb C \setminus \varsigma_{k};M_{2}(\Bbb C))} 
< \infty, \, \, \, \, \, \, \, \, \, \, \, \vert \vert m^{{\cal 
C}}_{p^{\prime},p,q}(\cdot;\! \lambda_{0}) \vert \vert_{{\cal L}
^{\infty}(\Bbb C \setminus \varsigma_{{\cal C}};M_{2}(\Bbb C))} 
< \infty,& \\
&\vert \vert E^{k}_{m^{\varsigma}}(\cdot;\! \lambda_{0}) \vert 
\vert_{{\cal L}^{\infty}(\Bbb C \setminus \varsigma_{k};M_{2}(
\Bbb C))} = {\cal O} \! \left(\frac{c^{{\cal S}}(\lambda_{0})(
\ln t)^{N_{o}}}{(\lambda_{0}^{2} t)^{(N_{o}+1)/2}} \right) 
\!,& \\
&\vert \vert E^{{\cal C}}_{m^{\varsigma}}(\cdot;\! \lambda_{0}
) \vert \vert_{{\cal L}^{\infty}(\Bbb C \setminus \varsigma_{{
\cal C}};M_{2}(\Bbb C))} \! = \! {\cal O} \! \left(\frac{
\underline{c}(\lambda_{0})(\ln t)^{N_{o}}}{(\lambda_{0}^{2} 
t)^{N_{o}+1}} \right) \!.&
\end{eqnarray*}
\end{bbb}

{\em Proof.\/} {}From Theorem~2.1, Lemma~3.1, the 
change-of-variable rule in Proposition~3.1, and the proof of 
Proposition~3.2, it follows that,
\begin{eqnarray*}
&m^{\varsigma}_{k}(\lambda)=e^{\frac{i}{2}\{(4 \lambda_{0}^{4}
t-\nu(\lambda_{0}) \ln t){\rm ad}(\sigma_{3})\}} m^{\varsigma
^{\prime}}_{k}((16\lambda_{0}^{2}t)^{1/2}(\lambda-{\rm sgn}(k)
\lambda_{0}))& \\
&={\rm I} + e^{\frac{i}{2}\{(4\lambda_{0}^{4}t-\nu(\lambda_{0}) 
\ln t){\rm ad}(\sigma_{3})\}} \int\limits_{(16 \lambda_{0}^{2} 
t)^{1/2}(\varsigma_{k} \setminus \{{\rm sgn}(k) \lambda_{0}\})} 
\frac{\mu^{k}(\xi;\lambda_{0})(V^{k}(\xi;\lambda_{0})-{\rm I})}
{(\xi-(16\lambda_{0}^{2}t)^{1/2}(\lambda-{\rm sgn}(k) \lambda_{
0}))} \frac{d \xi}{2 \pi i},& \\
&m^{\varsigma}_{{\cal C}}(\lambda) = m^{\varsigma^{\prime}}_{{
\cal C}}((8\lambda_{0}^{2}t)^{1/2} \lambda) = {\rm I} + 
\int\limits_{(8 \lambda_{0}^{2} t)^{1/2}(\varsigma_{{\cal C}} 
\setminus \{0\})} \frac{\mu^{{\cal C}}(\xi;\lambda_{0})(V^{{\cal 
C}}(\xi;\lambda_{0})-{\rm I})}{(\xi-(8 \lambda_{0}^{2} t)^{1/2} 
\lambda)} \frac{d \xi}{2 \pi i},&
\end{eqnarray*}
$k \! \in \! \{{\cal B},{\cal A}\}$, with ${\rm sgn}({\cal B}) 
\! = \! - {\rm sgn}({\cal A}) \! = \! 1$. The results stated 
in the Proposition now follow {}from the asymptotic expansions 
given in Propositions~3.1 and 3.3, and the following geometric 
progressions, $(\xi \! - \! (16 \lambda_{0}^{2} t)^{1/2}(\lambda 
\! - \! {\rm sgn}(k) \lambda_{0}))^{-1} \! = \! -\sum_{l \in \Bbb 
Z_{\geq 0}} \! \xi^{l}((16 \lambda_{0}^{2} t)^{1/2}(\lambda \! - 
\! {\rm sgn}(k) \lambda_{0}))^{-(l+1)}$ and $(\xi \! - \! (8 
\lambda_{0}^{2} t)^{1/2} \lambda)^{-1} \! = \! -\sum_{l \in \Bbb 
Z_{\geq 0}} \! \xi^{l}((8 \lambda_{0}^{2} t)^{1/2} \lambda)^{-(
l+1)}$. \hfill \rule{6.5pt}{6.5pt}

Let $\Xi_{k}$ be sufficiently small, counter-clockwise-oriented, 
mutually disjoint oblongs centred at ${\rm sgn}(k) \lambda_{0}$, 
$k \! \in \! \{{\cal B},{\cal A},{\cal C}\}$, where ${\rm sgn}(
{\cal B}) \! = \! -{\rm sgn}({\cal A}) \! = \! 1$ and ${\rm sgn}
({\cal C}) \! = \! 0$ (see Fig.~3).
\begin{figure}[bht]
\begin{center}
\unitlength=1cm
\begin{picture}(12,4)(0,2.5)
\thicklines
\put(10,5){\makebox(0,0){$\scriptstyle{}\bullet$}}
\put(10,6.85){\makebox(0,0){$\scriptstyle{}\Xi_{{\cal B}}$}}
\put(2,5){\makebox(0,0){$\scriptstyle{}\bullet$}}
\put(2,6.85){\makebox(0,0){$\scriptstyle{}\Xi_{{\cal A}}$}}
\put(6,5){\makebox(0,0){$\scriptstyle{}\bullet$}}
\put(6,6.85){\makebox(0,0){$\scriptstyle{}\Xi_{{\cal C}}$}}
\put(10.05,4.6){\makebox(0,0){$\scriptstyle{}+\lambda_{0}$}}
\put(10,4.8){\oval(3.5,3.5)}
\put(11.75,4){\vector(0,1){1}}
\put(2.05,4.6){\makebox(0,0){$\scriptstyle{}-\lambda_{0}$}}
\put(2,4.8){\oval(3.5,3.5)}
\put(3.75,4){\vector(0,1){1}}
\put(6,4.6){\makebox(0,0){$\scriptstyle{}0$}}
\put(6,4.8){\oval(3.5,3.5)}
\put(7.75,4){\vector(0,1){1}}
\put(10,5){\vector(1,1){0.7}}
\put(10.5,5.5){\line(1,1){0.5}}
\put(10,5){\line(1,-1){0.5}}
\put(11,4){\vector(-1,1){0.5}}
\put(10,5){\line(-1,1){0.5}}
\put(9,6){\vector(1,-1){0.5}}
\put(10,5){\vector(-1,-1){0.7}}
\put(9.5,4.5){\line(-1,-1){0.5}}
\put(6,5){\vector(1,1){0.7}}
\put(6.5,5.5){\line(1,1){0.5}}
\put(6,5){\line(1,-1){0.5}}
\put(7,4){\vector(-1,1){0.5}}
\put(6,5){\line(-1,1){0.5}}
\put(5,6){\vector(1,-1){0.5}}
\put(6,5){\vector(-1,-1){0.7}}
\put(5,4){\line(1,1){0.4}}
\put(2,5){\vector(1,1){0.7}}
\put(2.5,5.5){\line(1,1){0.5}}
\put(2,5){\line(1,-1){0.5}}
\put(3,4){\vector(-1,1){0.5}}
\put(2,5){\line(-1,1){0.5}}
\put(1,6){\vector(1,-1){0.5}}
\put(2,5){\vector(-1,-1){0.7}}
\put(1,4){\line(1,1){0.4}}
\put(11.25,5.5){\makebox(0,0)[r]{$\scriptstyle{}\varsigma_{{\cal 
B}}^{(1)}$}}
\put(11.25,4.5){\makebox(0,0)[r]{$\scriptstyle{}\varsigma_{{\cal 
B}}^{(4)}$}}
\put(8.75,5.5){\makebox(0,0)[l]{$\scriptstyle{}\varsigma_{{\cal 
B}}^{(3)}$}}
\put(8.75,4.5){\makebox(0,0)[l]{$\scriptstyle{}\varsigma_{{\cal 
B}}^{(2)}$}}
\put(7.25,5.5){\makebox(0,0)[r]{$\scriptstyle{}\varsigma_{{\cal 
C}}^{(1)}$}}
\put(7.25,4.5){\makebox(0,0)[r]{$\scriptstyle{}\varsigma_{{\cal 
C}}^{(4)}$}}
\put(4.75,5.5){\makebox(0,0)[l]{$\scriptstyle{}\varsigma_{{\cal 
C}}^{(3)}$}}
\put(4.75,4.5){\makebox(0,0)[l]{$\scriptstyle{}\varsigma_{{\cal 
C}}^{(2)}$}}
\put(3.25,5.5){\makebox(0,0)[r]{$\scriptstyle{}\varsigma_{{\cal 
A}}^{(1)}$}}
\put(3.25,4.5){\makebox(0,0)[r]{$\scriptstyle{}\varsigma_{{\cal 
A}}^{(4)}$}}
\put(0.75,5.5){\makebox(0,0)[l]{$\scriptstyle{}\varsigma_{{\cal 
A}}^{(3)}$}}
\put(0.75,4.5){\makebox(0,0)[l]{$\scriptstyle{}\varsigma_{{\cal 
A}}^{(2)}$}}
\put(10,3.65){\makebox(0,0){$\scriptstyle{}\varsigma_{{\cal B}} 
:= \bigcup\limits_{k=1}^{4} \varsigma_{{\cal B}}^{(k)}$}}
\put(6,3.65){\makebox(0,0){$\scriptstyle{}\varsigma_{{\cal C}} 
:= \bigcup\limits_{k=1}^{4}\varsigma_{{\cal C}}^{(k)}$}}
\put(2,3.65){\makebox(0,0){$\scriptstyle{}\varsigma_{{\cal A}} 
:= \bigcup\limits_{k=1}^{4}\varsigma_{{\cal A}}^{(k)}$}}
\end{picture}
\end{center}
\vspace{-1.15cm}
\caption{}
\end{figure}
\vspace{-0.20cm}
\begin{fff}
As $t \! \to \! +\infty$ and $x \! \to \! -\infty$ such that 
$\lambda_{0} \! > \! M$, for arbitrary $N_{o} \! \in \! \Bbb 
Z_{\geq 1}$,
\begin{eqnarray*}
&\vert \vert m^{k}_{p,q}(\cdot;\! \lambda_{0}) \vert \vert_{{\cal 
L}^{\infty}(\Xi_{k};M_{2}(\Bbb C))} < \infty, \, \, \, \, \, \, 
\, \, \, \, \, \, \vert \vert m^{{\cal C}}_{p^{\prime},p,q}(\cdot;
\! \lambda_{0}) \vert \vert_{{\cal L}^{\infty}(\Xi_{{\cal C}};M_{
2}(\Bbb C))} < \infty,& \\
&\vert \vert E^{k}_{m^{\varsigma}}(\cdot;\! \lambda_{0}) \vert 
\vert_{{\cal L}^{\infty}(\Xi_{k};M_{2}(\Bbb C))} = {\cal O} \! 
\left(\frac{c^{{\cal S}}(\lambda_{0})(\ln t)^{N_{o}}}{(\lambda
_{0}^{2} t)^{(N_{o}+1)/2}} \right) \!,& \\
&\vert \vert E^{{\cal C}}_{m^{\varsigma}}(\cdot;\! \lambda_{0}) 
\vert \vert_{{\cal L}^{\infty}(\Xi_{{\cal C}};M_{2}(\Bbb C))} = 
{\cal O} \! \left(\frac{\underline{c}(\lambda_{0})(\ln t)^{N_{
o}}}{(\lambda_{0}^{2} t)^{N_{o}+1}} \right) \!,&
\end{eqnarray*}
where $k \! \in \! \{{\cal B},{\cal A}\}$, $0 \! \leq \! p^{
\prime} \! \leq \! 1$, $1 \! \leq \! p \! \leq \! N_{o}$, and 
$0 \! \leq \! q \! \leq \! p \! - \! 1$.
\end{fff}

{\em Proof.\/}
Since, {}from Lemma~3.1, the solution of the RH problem $(m^{
\varsigma}(\lambda),v^{\varsigma}(\lambda),\varsigma)$ is 
bounded, i.e., $\vert \vert m^{\varsigma}(\cdot) \vert \vert_{
{\cal L}^{\infty}(\Bbb C \setminus \varsigma;M_{2}(\Bbb C))} 
\! < \! \infty$, the results stated in the Corollary are an 
immediate consequence of Proposition~3.4. 
\hfill \rule{6.5pt}{6.5pt}

Following Eqs.~(2.38) and {\em Remark~2.49\/} in \cite{a11}, 
make the following definition (see Fig.~4):
\begin{ggg}
{\rm Set}
\begin{eqnarray*}
&m^{\Xi}(\lambda) := \left\{\begin{array}{cc} 
m^{\varsigma}(\lambda), & \lambda \in \Bbb C \setminus 
\bigcup\limits_{k \in \{{\cal B},{\cal A},{\cal C}\}}(
{\rm int} \, \Xi_{k} \cup \Xi_{k}), \\
m^{\varsigma}(\lambda)(m_{k}^{\varsigma}(\lambda))^{-1}, 
& \lambda \in {\rm int} \, \Xi_{k}.
\end{array} \right.&
\end{eqnarray*}
\end{ggg}
\begin{ccc}
Set $\Xi \! := \! \cup_{l \in \{{\cal B},{\cal A},{\cal C}\}} 
\Xi_{l}$. As $t \! \to \! +\infty$ and $x \! \to \! -\infty$ such 
that $\lambda_{0} \! > \! M$, there exists a unique function $m^{
\Xi}(\lambda) \colon \Bbb C \! \setminus \! \Xi \! \to \! {\rm SL}
(2,\! \Bbb C)$ which solves the following RH problem:
\begin{enumerate}
\item[(1)] $m^{\Xi}(\lambda)$ is piecewise holomorphic $\forall 
\, \lambda \! \in \! \Bbb C \! \setminus \! \Xi;$
\item[(2)] $m^{\Xi}(\lambda)$ satisfies the following jump 
conditions,
\begin{eqnarray*}
&m^{\Xi}_{+}(\lambda) = m^{\Xi}_{-}(\lambda) v^{\Xi}(\lambda), 
\, \, \, \, \, \, \, \, \lambda \in \Xi,&
\end{eqnarray*}
where, for $l \! \in \! \{{\cal B},{\cal A},{\cal C}\}$,
\begin{eqnarray*}
&v^{\Xi}(\lambda) \vert_{\Xi_{l}} := (v^{\Xi}_{-}(\lambda))^{-1} 
v^{\Xi}_{+}(\lambda) \vert_{\Xi_{l}}=(m^{\varsigma}_{l}
(\lambda))^{-1};&
\end{eqnarray*}
\item[(3)] as $\lambda \! \to \! \infty$, $\lambda \! \in \! \Bbb 
C \! \setminus \! \Xi$,
\begin{eqnarray*}
&m^{\Xi}(\lambda) = {\rm I} + {\cal O}(\lambda^{-1}).&
\end{eqnarray*}
\end{enumerate}
Moreover, for arbitrary $l^{\prime} \! \in \! \Bbb Z_{\geq 1}$,
\begin{eqnarray*}
&Q(x,t)=2i\lim\limits_{\lambda \to \infty \atop \lambda \, \in \, 
\Bbb C \setminus \Xi}(\lambda m^{\Xi}(x,t;\! \lambda))_{12} + 
{\cal O} \! \left(\frac{\underline{c}(\lambda_{0})}{(\lambda_{0}
^{2} t)^{l^{\prime}}} \right) \!,&
\end{eqnarray*}
with $Q(x,\! 0) \! \in \! {\cal S}(\Bbb R;\! \Bbb C)$, satisfies 
Eq.~(2), and $m^{\Xi}(\lambda)$ satisfies the following symmetry 
reductions, $m^{\Xi}(\lambda) \! = \! \sigma_{3}m^{\Xi}(-\lambda) 
\sigma_{3}$ and $m^{\Xi}(\lambda) \! = \! \sigma_{1} \overline{m
^{\Xi}(\overline{\lambda})} \sigma_{1}$.
\end{ccc}

{\em Proof.\/} Follows {}from Lemma~3.1, the fact that 
$m^{\varsigma}_{+}(\lambda) \! = \! m^{\varsigma}_{-}(\lambda) 
\! = \! m^{\varsigma}(\lambda) \, \, \forall \, \, \lambda \! 
\in \! \Bbb C \! \setminus \! \varsigma$, and the definition 
of $m^{\Xi}(\lambda)$. \hfill \rule{6.5pt}{6.5pt}
\begin{figure}[bht]
\begin{center}
\unitlength=1cm
\begin{picture}(12,4)(0,2.5)
\thicklines
\put(10,5){\makebox(0,0){$\scriptstyle{}\bullet$}}
\put(10,5){\oval(2.5,2.5)}
\put(11.25,4.5){\vector(0,1){1}}
\put(2,5){\makebox(0,0){$\scriptstyle{}\bullet$}}
\put(2,5){\oval(2.5,2.5)}
\put(3.25,4.5){\vector(0,1){1}}
\put(6,5){\makebox(0,0){$\scriptstyle{}\bullet$}}
\put(6,5){\oval(2.5,2.5)}
\put(7.25,4.5){\vector(0,1){1}}
\put(10,4.65){\makebox(0,0){$\scriptstyle{}+\lambda_{0}$}}
\put(2,4.65){\makebox(0,0){$\scriptstyle{}-\lambda_{0}$}}
\put(6,4.65){\makebox(0,0){$\scriptstyle{}0$}}
\put(2,3){\makebox(0,0){$\scriptstyle{}v^{\Xi}(\lambda) \vert
_{\Xi_{{\cal A}}} \, = \, (m_{{\cal A}}^{\varsigma}(\lambda))
^{-1}$}}
\put(2,6.55){\makebox(0,0){$\scriptstyle{}\Xi_{{\cal A}}$}}
\put(6,3){\makebox(0,0){$\scriptstyle{}v^{\Xi}(\lambda) \vert
_{\Xi_{{\cal C}}} \, = \, (m_{{\cal C}}^{\varsigma}(\lambda))
^{-1}$}}
\put(6,6.55){\makebox(0,0){$\scriptstyle{}\Xi_{{\cal C}}$}}
\put(10,3){\makebox(0,0){$\scriptstyle{}v^{\Xi}(\lambda) \vert
_{\Xi_{{\cal B}}} \, = \, (m_{{\cal B}}^{\varsigma}(\lambda))
^{-1}$}}
\put(10,6.55){\makebox(0,0){$\scriptstyle{}\Xi_{{\cal B}}$}}
\end{picture}
\end{center}
\vspace{-1.0cm}
\caption{}
\end{figure}
\vspace{-0.15cm}
\begin{ggg}
Let $i,j,k \! \in \! \{{\cal B},{\cal A},{\cal C}\}$. For $i \! 
\not= \! j \! \not= \! k$, set
\begin{eqnarray*}
&m^{\sharp}_{i}(\lambda) := \left\{\begin{array}{cc} 
m^{\varsigma}_{i}(\lambda), & \lambda \in \Xi_{i}, \\
{\rm I}, & \lambda \in \Xi_{j} \cup \Xi_{k}.
\end{array} \right.&
\end{eqnarray*}
\end{ggg}
\begin{bbb}
The operator $C_{w^{\Xi}}$ for the inverse problem on $\Xi$ is 
given by
\begin{eqnarray*}
&C_{w^{\Xi}}(\lambda;\! \lambda_{0})=\sum\limits_{X \in \{{\cal 
B},{\cal A},{\cal C}\}} X^{\Xi}(\lambda;\! \lambda_{0}),&
\end{eqnarray*}
where, for any $f \! \in \! \cup_{p \in \{2,\infty\}}{\cal L}
^{p}(\Xi;\! M_{2}(\Bbb C))$ and $X \! \not= \! Y \! \not= \! 
Z \! \in \! \{{\cal B},{\cal A},{\cal C}\}$,
\begin{eqnarray*}
&(X^{\Xi} f)(\lambda;\! \lambda_{0}) = 
\left\{\begin{array}{cc}
C_{-}(f((m_{X}^{\sharp})^{-1} - {\rm I})), & \lambda \in \Xi_{X}, 
\\
0, & \lambda \in \Xi_{Y} \cup \Xi_{Z},
\end{array} \right.&
\end{eqnarray*}
and, for arbitrary $N_{o} \! \in \! \Bbb Z_{\geq 1}$,
\begin{eqnarray*}
&\mu^{\Xi}(\lambda;\! \lambda_{0})={\rm I}+\sum\limits_{j=1}^{
N_{o}} \! \left(\sum\limits_{X \in \{{\cal B},{\cal A},{\cal C}
\}} X^{\Xi}(\lambda;\! \lambda_{0}) \right)^{j} \! + E^{\Xi}(
\lambda;\! \lambda_{0}),&
\end{eqnarray*}
with
\begin{eqnarray*}
&\vert \vert E^{\Xi}(\cdot;\! \lambda_{0}) \vert \vert_{\cup_{
l \in \{2,\infty\}} {\cal L}^{l}(\Xi;M_{2}(\Bbb C)) \to {\cal 
L}^{2}(\Xi;M_{2}(\Bbb C))} = {\cal O} \! \left(\frac{\underline{
c}(\lambda_{0})(\ln t)^{N_{o}}}{(\lambda_{0}^{2} t)^{(N_{o}+1)/2}
} \right) \!.&
\end{eqnarray*}
\end{bbb}

{\em Proof.\/} {}From Theorem~2.1 (and the paragraph preceding it: 
the analogue of the BC \cite{a14} operator $C_{w_{x,t}}$ on 
$\Xi$, denoted here as $C_{w^{\Xi}})$, Lemma~3.3 (the expression 
for $v^{\Xi}(\lambda) \vert_{\Xi_{l}}$, $l \! \in \! \{{\cal B},
{\cal A},{\cal C}\})$, Definition~3.5, and setting $\{w_{\Xi_{l}
}^{-}(\lambda)\}_{l \in \{{\cal B},{\cal A},{\cal C}\}} \! = \! 
0$ and $w_{\Xi_{k}}^{+}(\lambda) \! = \! v^{\Xi}(\lambda) \vert_{
\Xi_{k}} \! - \! {\rm I} \! = \! (m^{\varsigma}_{k}(\lambda))^{-1} 
\! - \! {\rm I}$, $k \! \in \! \{{\cal B},{\cal A},{\cal C}\}$, 
one obtains, {}from the asymptotic expansions given 
in Corollary~3.1, the expressions for $C_{w^{\Xi}}(\lambda;\! 
\lambda_{0})$ and $(X^{\Xi} f)(\lambda;\! \lambda_{0})$, $X \! 
\in \! \{{\cal B},{\cal A},{\cal C}\}$, given in the Proposition: 
now, using the method of successive approximations, one expands, 
as $t \! \to \! +\infty$ and $x \! \to \! -\infty$ such that 
$\lambda_{0} \! > \! M$, the function $\mu^{\Xi}(\lambda;\! 
\lambda_{0}) \! := \! (\underline{{\bf Id}}^{\Xi} \! - \! C_{w^{
\Xi}}(\lambda;\! \lambda_{0}))^{-1} {\rm I}$, where $\underline{
{\bf Id}}^{\Xi}$ denotes the identity operator on $\cup_{p \in 
\{2,\infty\}}{\cal L}^{p}(\Xi;\! M_{2}(\Bbb C))$, in a von 
Neumann-type series (see, also, Part II of \cite{a5}), and 
obtains, for arbitrary $N_{o} \! \in \! \Bbb Z_{\geq 1}$, the 
result for $\mu^{\Xi}(\lambda;\! \lambda_{0})$ and the 
estimation for $\vert \vert E^{\Xi}(\cdot;\! \lambda_{0}) \vert 
\vert_{\cup_{l \in \{2,\infty\}} {\cal L}^{l}(\Xi;M_{2}(\Bbb C)) 
\to {\cal L}^{2}(\Xi;M_{2}(\Bbb C))}$ stated in the Proposition. 
\hfill \rule{6.5pt}{6.5pt}
\begin{bbb}
As $t \! \to \! +\infty$ and $x \! \to \! -\infty$ such that 
$\lambda_{0} \! > \! M$, for arbitrary $N_{o} \! \in \! \Bbb 
Z_{\geq 1}$,
\begin{eqnarray*}
&X^{\Xi}(\lambda;\! \lambda_{0}) = \sum\limits_{p=1}^{N_{o}} 
\sum\limits_{q=0}^{p-1} \frac{(\ln t)^{q} \widetilde{X}_{p,q}
(\lambda;\lambda_{0})}{t^{p/2}} + E_{X}(\lambda;\! \lambda_{
0}),& \\
&{\cal C}^{\Xi}(\lambda;\! \lambda_{0}) = \sum\limits_{p^{
\prime}=0}^{1} \sum\limits_{p=1}^{N_{o}} \sum\limits_{q=0}^{
p-1} \frac{(\ln t)^{q} {\cal C}_{p^{\prime},p,q}(\lambda;
\lambda_{0})}{t^{p+p^{\prime}/2}} + E_{{\cal C}}(\lambda;\! 
\lambda_{0}),&
\end{eqnarray*}
where $X \! \in \! \{{\cal B},{\cal A}\}$, $\widetilde{X}_{p,q}
(\lambda;\! \lambda_{0}) \! := \! \exp \{\frac{i}{2} \varpi
_{2}(\lambda_{0})\} X_{p,q}(\lambda;\! \lambda_{0}) \exp \{- 
\frac{i}{2} \varpi_{2}(\lambda_{0})\}$, with $\varpi_{2}(\lambda
_{0}) \! := \! (4 \lambda_{0}^{4} t \! - \! \nu(\lambda_{0}
) \ln t) {\rm ad} (\sigma_{3})$, and, for $0 \! \leq \! p^{
\prime} \! \leq \! 1$, $1 \! \leq \! p \! \leq \! N_{o}$, and 
$0 \! \leq \! q \! \leq \! p \! - \! 1$,
\begin{eqnarray*}
&\vert \vert X_{p,q}(\cdot;\! \lambda_{0}) \vert \vert_{\cup_{l 
\in \{2,\infty\}} {\cal L}^{l}(\Xi_{X};M_{2}(\Bbb C)) \to {\cal 
L}^{2}(\Xi_{X};M_{2}(\Bbb C))} < \infty,& \\
&\vert \vert {\cal C}_{p^{\prime},p,q}(\cdot;\! \lambda_{0}) 
\vert \vert_{\cup_{l \in \{2,\infty\}} {\cal L}^{l}(\Xi_{{\cal 
C}};M_{2}(\Bbb C)) \to {\cal L}^{2}(\Xi_{{\cal C}};M_{2}(\Bbb 
C))} < \infty,& \\
&\vert \vert E_{X}(\cdot;\! \lambda_{0}) \vert \vert_{\cup_{l 
\in \{2,\infty\}} {\cal L}^{l}(\Xi_{X};M_{2}(\Bbb C)) \to {\cal 
L}^{2}(\Xi_{X};M_{2}(\Bbb C))} = {\cal O} \! \left(\frac{c^{{
\cal S}}(\lambda_{0})(\ln t)^{N_{o}}}{(\lambda_{0}^{2} t)^{(N_{
o}+1)/2}} \right) \!,& \\
&\vert \vert E_{{\cal C}}(\cdot;\! \lambda_{0}) \vert \vert_{
\cup_{l \in \{2,\infty\}} {\cal L}^{l}(\Xi_{{\cal C}};M_{2}(
\Bbb C)) \to {\cal L}^{2}(\Xi_{{\cal C}};M_{2}(\Bbb C))}={\cal 
O} \! \left(\frac{\underline{c}(\lambda_{0})(\ln t)^{N_{o}}}{
(\lambda_{0}^{2} t)^{N_{o}+1}} \right) \!.&
\end{eqnarray*}
\end{bbb}

{\em Proof.\/} The results stated in the Proposition are a 
consequence of those given in Propositions~3.4 and 3.5, the 
definition of $m^{\sharp}_{i}(\lambda)$, and the analogues 
of Eqs.~(2.46) in \cite{a11}. \hfill \rule{6.5pt}{6.5pt}
\begin{ccc}
Let $m^{\Xi}(\lambda)$ be the solution of the RH problem 
formulated in Lemma~3.3. Then as $t \! \to \! +\infty$ and $x \! 
\to \! -\infty$ such that $\lambda_{0} \! > \! M$, for arbitrary 
$N_{o} \! \in \! \Bbb Z_{\geq 1}$,
\begin{eqnarray*}
&Q(x,t) = \sum\limits_{k={\rm odd} \atop \pm 1,\pm 3,\ldots} 
\sum\limits_{p=\vert k \vert}^{N_{o}} \sum\limits_{q=0}^{p-\vert 
k \vert} \frac{\exp \{\frac{i}{2}(k+1)(4 \lambda_{0}^{4} t - \nu
(\lambda_{0}) \ln t)\} u^{+}_{k,p,q}(\lambda_{0})(\ln t)^{q}}{t
^{p/2}} + {\cal O} \! \left(\frac{\underline{c}(\lambda_{0})(\ln 
t)^{N_{o}}}{(\lambda_{0}^{2} t)^{(N_{o}+1)/2}} \right) \!,&
\end{eqnarray*}
where, for $k \! = \! \pm 1,\pm 3,\ldots$, $\vert k \vert \! \leq 
\! p \! \leq \! N_{o}$, and $q \! > \! p \! - \! \vert k \vert$, 
$u^{+}_{k,p,q}(\lambda_{0}) \! \equiv \! 0$.
\end{ccc}

{\em Proof.\/}
{}From Eq.~(9) and Lemma~3.3, one shows that, as $t \! \to \! 
+\infty$ and $x \! \to \! -\infty$ such that $\lambda_{0} \! 
> \! M$, for arbitrary $N_{o} \! \in \! \Bbb Z_{\geq 1}$, $Q
(x,t) \! = \! -i([\sigma_{3},\int_{\Xi}((\underline{{\bf Id}}
^{\Xi} \! - \! C_{w^{\Xi}}(\xi;\! \lambda_{0}))^{-1}{\rm I})
(w_{\Xi}^{-}(\xi) \! + \! w_{\Xi}^{+}(\xi)) \frac{d \xi}{2 \pi 
i}])_{12} \! + \! {\cal O}(\underline{c}(\lambda_{0})(\lambda
_{0}^{2} t)^{-l})$, $l \! \in \! \Bbb Z_{\geq 1}$ and arbitrary. 
Since ({}from the proof of Proposition~3.5) $w_{\Xi}^{-}(\lambda) 
\! = \! 0$, it follows {}from (the analogue of) Theorem~2.1 that 
$\mu^{\Xi}(\lambda;\! \lambda_{0}) \! := \! (\underline{{\bf 
Id}}^{\Xi} \! - \! C_{w^{\Xi}}(\lambda;\! \lambda_{0}))^{-1} 
{\rm I} \! = \! m^{\Xi}_{-}(\lambda)$; hence, {}from (the analogue 
of) Theorem~2.1, namely, $\mu^{\Xi}(\lambda;\! \lambda_{0}) \! 
:= \! \widehat{\mu}^{\Xi}(\lambda) \! = \! m^{\Xi}_{+}(\lambda)
({\rm I} \! + \! w_{\Xi}^{+}(\lambda))^{-1} \! = \! m^{\Xi}_{+}
(\lambda)(v^{\Xi}_{+}(\lambda))^{-1}$, and Cauchy's theorem, one 
shows that $Q(x,t) \! = \! - 2 i (\int_{\Xi} \widehat{\mu}^{\Xi}
(\xi) w_{\Xi}^{+}(\xi) \frac{d \xi}{2 \pi i})_{12} + {\cal O}(
\underline{c}(\lambda_{0})(\lambda_{0}^{2} t)^{-l}) \! = \! - 2
i(\int_{\Xi} \widehat{\mu}^{\Xi}(\xi)(v^{\Xi}_{+}(\xi) \! - \! 
{\rm I}) \frac{d \xi}{2 \pi i})_{12} + {\cal O}(\underline{c}(
\lambda_{0})(\lambda_{0}^{2} t)^{-l}) \! = \! - 2 i (\int_{
\Xi}(m^{\Xi}_{+}(\xi) \! - \! \widehat{\mu}^{\Xi}(\xi)) 
\frac{d \xi}{2 \pi i})_{12} + {\cal O}(\underline{c}(
\lambda_{0})(\lambda_{0}^{2} t)^{-l}) \! = \! (\int_{\Xi} \mu^{
\Xi}(\xi;\! \lambda_{0}) \frac{d \xi}{\pi})_{12} + {\cal O}(
\underline{c}(\lambda_{0})(\lambda_{0}^{2} t)^{-l})$. Now, using 
the expression for $\mu^{\Xi}(\lambda;\! \lambda_{0})$ given in 
Proposition~3.5, and the corresponding error estimation for $E^{
\Xi}(\lambda;\! \lambda_{0})$, one shows that, for $l \! \geq \! 
(N_{o} \! + \! 1)/2$, $Q(x,t) \! = \! \sum_{j=1}^{N_{o}}(\int_{
\Xi}(\sum_{X \in \{{\cal B},{\cal A},{\cal C}\}} \! X^{\Xi}(\xi;
\! \lambda_{0}))^{j} \frac{d \xi}{\pi})_{12} \! + \! {\cal O}(
\underline{c}(\lambda_{0})(\ln t)^{N_{o}}(\lambda_{0}^{2} t)^{
-(N_{o}+1)/2})$: the result now follows {}from Propositions 3.4 
and 3.5, and the definition of $m^{\sharp}_{i}(\lambda)$. 
\hfill \rule{6.5pt}{6.5pt}
\begin{bbb}
For $\alpha \! = \! \pm 1,\pm 3,\ldots$, $\beta \! \geq \! \vert
\alpha \vert$, and $0 \! \leq \! \gamma^{\prime} \! \leq \! \beta 
\! - \! \vert \alpha \vert$, set $u^{+}_{\alpha,\beta}(\lambda_{
0};t) \! := \! u^{+}_{\alpha,\beta} \! := \! \sum_{\gamma^{\prime}
=0}^{\beta-\vert \alpha \vert} \! u^{+}_{\alpha,\beta,\gamma^{
\prime}}(\lambda_{0})(\ln t)^{\gamma^{\prime}}$, and, for $\gamma
^{\prime} \! > \! \beta \! - \! \vert \alpha \vert$, $u^{+}_{
\alpha,\beta,\gamma^{\prime}}(\lambda_{0}) \! \equiv \! 0$. The 
coefficients of the asymptotic expansion for $Q(x,t)$ given 
in Lemma~3.4 are determined by the following linear system,
\begin{eqnarray}
&\sum\limits_{k_{1} \, = \, {\rm odd} \atop \pm 1,\pm 3,\ldots} 
\sum\limits_{p_{1} \geq \vert k_{1} \vert} \frac{e^{i\{\frac{(k_{
1}+1)\tau^{+}}{2}\}}(k_{1}+1)}{t^{(p_{1}+2)/2}} \{2 \lambda
_{0}^{4} t u^{+}_{k_{1},p_{1}} + \frac{\nu u^{+}_{k_{1},p_{1}}}{
2} - \frac{\lambda_{0} \nu^{\prime} u^{+}_{k_{1},p_{1}} \ln t}{
4}\}& \nonumber \\
&+\sum\limits_{k_{1} \, = \, {\rm odd} \atop \pm 1,\pm 3,\ldots} 
\sum\limits_{p_{1} \geq \vert k_{1} \vert} \frac{e^{i\{\frac{(k_{
1}+1)\tau^{+}}{2}\}}}{t^{(p_{1}+2)/2}} \{it \dot{u}^{+}_{k_{
1},p_{1}}-\frac{ip_{1}u^{+}_{k_{1},p_{1}}}{2}-\frac{i\lambda_{0} 
u^{+ \prime}_{k_{1},p_{1}}}{2}\}& \nonumber \\
&+\sum\limits_{k_{1} \, = \, {\rm odd} \atop \pm 1,\pm 3,\ldots} 
\sum\limits_{p_{1} \geq \vert k_{1} \vert} \frac{e^{i\{\frac{(k_{
1}+1)\tau^{+}}{2}\}}i(k_{1}+1)}{t^{(p_{1}+2)/2}} \{\frac{\lambda
_{0} u^{+\prime}_{k_{1},p_{1}}}{4} \! + \! \frac{u^{+}_{k_{1},p_{
1}}}{4} \! - \! \frac{\nu^{\prime}u^{+ \prime}_{k_{1},p_{1}} \ln 
t}{(8 \lambda_{0})^{2} t} \! - \! \frac{\nu^{\prime \prime} u^{+}
_{k_{1},p_{1}} \ln t}{2(8 \lambda_{0})^{2} t} \! + \! \frac{\nu^{
\prime} u^{+}_{k_{1},p_{1}} \ln t}{2(8)^{2} \lambda_{0}^{3}t}\}& 
\nonumber \\
&+\sum\limits_{k_{1} \, = \, {\rm odd} \atop \pm 1,\pm 3,\ldots} 
\sum\limits_{p_{1} \geq \vert k_{1} \vert} \frac{e^{i\{\frac{(k_{
1}+1)\tau^{+}}{2}\}}(k_{1}+1)^{2}}{t^{(p_{1}+2)/2}} \{-\lambda_{
0}^{4}t u^{+}_{k_{1},p_{1}} + \frac{\lambda_{0}\nu^{\prime} u^{+}
_{k_{1},p_{1}} \ln t}{8}-\frac{(\nu^{\prime})^{2}u^{+}_{k_{1},
p_{1}} (\ln t)^{2}}{(16 \lambda_{0})^{2} t}\}& \nonumber \\
&+\sum\limits_{k_{1} \, = \, {\rm odd} \atop \pm 1,\pm 3,\ldots} 
\sum\limits_{p_{1} \geq \vert k_{1} \vert} \frac{e^{i\{\frac{(k_{
1}+1)\tau^{+}}{2}\}}}{t^{(p_{1}+2)/2}} \{\frac{u^{+ \prime 
\prime}_{k_{1},p_{1}}}{(8 \lambda_{0})^{2} t} - \frac{u^{+ \prime}
_{k_{1},p_{1}}}{(8)^{2} \lambda_{0}^{3} t}\}& \nonumber \\
&-\underbrace{\sum \sum \sum}_{k_{i} \, = \, 
{\rm odd} \atop {\pm 1,\pm 3,\ldots \atop 1 \leq i \leq 3}} 
\underbrace{\sum \sum \sum}_{p_{i} 
\geq \vert k_{i} \vert \atop 1 \leq i \leq 3} \frac{e^{i\{\frac{
(k_{2}+k_{3}-k_{1}+1)\tau^{+}}{2}\}}(k_{1}+1) \lambda_{0}^{2} 
\overline{u^{+}_{k_{1},p_{1}}} u^{+}_{k_{2},p_{2}} u^{+}_{k_{3},
p_{3}}}{t^{(p_{1}+p_{2}+p_{3})/2}}& \nonumber \\
&+\underbrace{\sum \sum \sum}_{k_{i} \, = \, {\rm odd} \atop {\pm 
1,\pm 3,\ldots \atop 1 \leq i \leq 3}} \underbrace{\sum \sum \sum}
_{p_{i} \geq \vert k_{i} \vert \atop 1 \leq i \leq 3} \frac{e^{i\{
\frac{(k_{2}+k_{3}-k_{1}+1)\tau^{+}}{2}\}}(k_{1}+1)\nu^{\prime} 
\overline{u^{+}_{k_{1},p_{1}}}u^{+}_{k_{2},p_{2}}u^{+}_{k_{3},p_{3}
} \ln t}{16 \lambda_{0} t^{(p_{1}+p_{2}+p_{3}+2)/2}}& \nonumber \\
&+\underbrace{\sum \sum \sum \sum \sum}_{k_{i} \, = \, {\rm odd} 
\atop {\pm 1,\pm 3,\ldots \atop 1 \leq i \leq 5}} \underbrace{\sum 
\sum \sum \sum \sum}_{p_{i} \geq \vert k_{i} \vert \atop 1 \leq i 
\leq 5} \frac{e^{i\{\frac{(k_{1}-k_{2}+k_{3}-k_{4}+k_{5}+1)\tau^{
+}}{2}\}} u^{+}_{k_{1},p_{1}} \overline{u^{+}_{k_{2},p_{2}}} u^{
+}_{k_{3},p_{3}} \overline{u^{+}_{k_{4},p_{4}}} u^{+}_{k_{5},p_{5}
}}{2t^{(p_{1}+p_{2}+p_{3}+p_{4}+p_{5})/2}}& \nonumber \\
&-\underbrace{\sum \sum \sum}_{k_{i} \, = \, {\rm odd} \atop {\pm 
1,\pm 3,\ldots \atop 1 \leq i \leq 3}} \underbrace{\sum \sum \sum}
_{p_{i} \geq \vert k_{i} \vert \atop 1 \leq i \leq 3} \frac{e^{i\{
\frac{(k_{2}+k_{3}-k_{1}+1)\tau^{+}}{2}\}}i(\overline{u^{+}_{k_{
1},p_{1}}})^{\prime} u^{+}_{k_{2},p_{2}}u^{+}_{k_{3},p_{3}}}{8 
\lambda_{0} t^{(p_{1}+p_{2}+p_{3}+2)/2}}=0,&
\end{eqnarray}
where $\tau^{+} \! := \! 4 \lambda_{0}^{4} t \! - \! \nu \ln t$, 
$\nu \! := \! \nu(\lambda_{0})$, $f^{\prime} \! := \! \partial
_{\lambda_{0}}f(\lambda_{0};t) \vert_{t \, = \, {\rm fixed}}$, and 
$\dot{f} \! := \! \partial_{t}f(\lambda_{0};t) \vert_{\lambda_{0} 
\, = \, {\rm fixed}}$.
\end{bbb}

{\em Proof.\/} Substituting the asymptotic expansion for $Q(x,t)$ 
given in Lemma~3.4 into Eq.~(2), one obtains, after differentiation, 
the result stated in the Proposition. \hfill \rule{6.5pt}{6.5pt}
\begin{fff}
To ${\cal O}(t^{-7/2})$, the explicit recurrence formulae for 
the coefficients of the asymptotic expansion for $Q(x,t)$ given 
in Lemma~3.4 are given in the Appendix.
\end{fff}

{\em Proof.\/} Follows {}from Proposition~3.7, system~(78), upon 
equating coefficients of powers of like terms on both left- and 
right-hand sides. \hfill \rule{6.5pt}{6.5pt}

Now, recalling {}from Proposition~3.7 that, for $\alpha \! = \! 
\pm 1,\pm 3,\ldots$, $\beta \! \geq \! \vert \alpha \vert$, and 
$0 \! \leq \! \gamma^{\prime} \! \leq \! \beta \! - \! \vert 
\alpha \vert$, $u^{+}_{\alpha,\beta} \! := \! \sum_{\gamma^{
\prime}=0}^{\beta - \vert \alpha \vert} \! u^{+}_{\alpha,\beta,
\gamma^{\prime}}(\lambda_{0})(\ln t)^{\gamma^{\prime}}$, and, for 
$\gamma^{\prime} \! > \! \beta \! - \! \vert \alpha \vert$, $u^{
+}_{\alpha,\beta,\gamma^{\prime}}(\lambda_{0}) \! \equiv \! 0$, 
by substituting these expressions into Eqs.~(A.1)--(A.30) given 
in the Appendix, one deduces, after an onerous---but otherwise 
straightforward---analysis that, in order to solve the resulting 
recurrence relations for $u^{+}_{\alpha,\beta,\gamma^{\prime}}(
\lambda_{0})$, explicit, {\em a priori} knowledge of $u^{+}_{1,
1,0}(\lambda_{0})$ and $u^{+}_{-1,p,q}(\lambda_{0})$, $p \! \geq 
\! 1$, $0 \! \leq \! q \! \leq \! p \! - \! 1$, is necessary: 
the expression for $u^{+}_{1,1,0}(\lambda_{0})$ is given in 
Theorem~2.2, Eq.~(14). Hence, there remains the problem of 
determining $u^{+}_{-1,p,q}(\lambda_{0})$, $p \! \geq \! 1$, 
$0 \! \leq \! q \! \leq \! p \! - \! 1$: up to $p \! = \! 6$, 
this is the programme of study of the following section.
\section{Explicit Representation of $u^{+}_{-1,p,q}(\lambda_{0})$, 
$1 \! \leq \! p \! \leq \! 6$, $0 \! \leq \! q \! \leq \! p \! - 
\! 1$}
In this section, the RH factorisation problem for $m^{\varsigma}
_{{\cal C}}(x,t;\! \lambda) \! := \! m^{\varsigma}_{{\cal C}}(
\lambda)$, which is associated with the first-order stationary 
phase point at the origin, is solved asymptotically as $t \! \to 
\! +\infty$ and $x \! \to \! -\infty$ such that $\lambda_{0} \! 
> \! M$ and $(x,t) \! \in \! \Bbb R^{2} \! \setminus \! \Omega_{
n}$, for those $\gamma_{n} \! \in \! (\frac{\pi}{2},\! \pi)$, up 
to ${\cal O}(\underline{c}(\lambda_{0}) (\ln t)^{2} t^{-7/2})$: 
{}from this asymptotic expansion for $m^{\varsigma}_{{\cal C}}
(\lambda)$ and the resulting expression for $2 i \lim_{\lambda 
\to \infty \atop \lambda \, \in \, \Bbb C \setminus \varsigma_{
{\cal C}}}(\lambda m^{\varsigma}_{{\cal C}}(\lambda))_{12}$, 
explicit (integral) representations for $u^{+}_{-1,p,q}(\lambda
_{0})$, $1 \! \leq \! p \! \leq \! 6$, $0 \! \leq \! q \! \leq 
\! p \! - \! 1$, are deduced.
\begin{figure}[bht]
\begin{center}
\unitlength=1cm
\begin{picture}(2,2)(0,0)
\thicklines
\put(1,1){\makebox(0,0){$\scriptstyle{}\bullet$}}
\put(1,0.75){\makebox(0,0){$\scriptstyle{}0$}}
\put(1,1){\vector(1,1){0.65}}
\put(2,2){\line(-1,-1){0.5}}
\put(1,1){\line(-1,1){0.5}}
\put(0,2){\vector(1,-1){0.565}}
\put(1,1){\vector(-1,-1){0.65}}
\put(0,0){\line(1,1){0.5}}
\put(1,1){\line(1,-1){0.5}}
\put(2,0){\vector(-1,1){0.565}}
\put(2.4,1.95){\makebox(0,0){$\scriptstyle{}\varsigma_{{\cal 
C}}^{(1)}$}}
\put(2.4,0){\makebox(0,0){$\scriptstyle{}\varsigma_{{\cal C}}
^{(4)}$}}
\put(-0.35,1.95){\makebox(0,0){$\scriptstyle{}\varsigma_{{\cal 
C}}^{(3)}$}}
\put(-0.35,0){\makebox(0,0){$\scriptstyle{}\varsigma_{{\cal 
C}}^{(2)}$}}
\end{picture}
\end{center}
\vspace{-0.50cm}
\caption{}
\end{figure}
\begin{ccc}[{\rm \cite{a2,a10}}]
Set $\varsigma_{{\cal C}} \! := \! \cup_{k=1}^{4} \varsigma_{{
\cal C}}^{(k)}$ (Fig.~5). As $t \! \to \! + \infty$ and $x \! 
\to \! - \infty$ such that $\lambda_{0} \! > \! M$ and $(x,t) \! 
\in \! \Bbb R^{2} \setminus \Omega_{n}$, for those $\gamma_{n} 
\! \in \! (\frac{\pi}{2},\! \pi)$, there exists a unique 
function $m^{\varsigma}_{{\cal C}}(\lambda) \colon \Bbb C 
\setminus \varsigma_{{\cal C}} \! \to \! {\rm SL}(2,\! \Bbb C)$ 
which solves the following RH problem:
\begin{enumerate}
\item[(1)] $m^{\varsigma}_{{\cal C}}(\lambda)$ is piecewise 
holomorphic $\forall \, \lambda \! \in \! \Bbb C \! \setminus 
\! \varsigma_{{\cal C}};$
\item[(2)] $m^{\varsigma}_{{\cal C}}(\lambda)$ satisfies the 
following jump conditions,
\begin{eqnarray*}
&m^{\varsigma}_{{\cal C} \, +}(\lambda) = m^{\varsigma}_{{\cal 
C} \, -}(\lambda) v^{\varsigma}_{{\cal C}}(\lambda), \, \, \, 
\, \, \, \, \, \lambda \in \varsigma_{{\cal C}},&
\end{eqnarray*}
where
\begin{eqnarray*}
&v^{\varsigma}_{{\cal C}}(\lambda) \vert_{\cup_{k=1}^{2} 
\varsigma_{{\cal C}}^{(k)}} = {\rm I}+{\cal R}_{{\cal C}}
(\lambda) {\cal P}_{+}(\lambda) (\delta^{+}(\lambda;\! 
\lambda_{0}))^{2} e^{-2 i t \rho(\lambda;\lambda_{0})} 
\sigma_{+},& \\
&v^{\varsigma}_{{\cal C}}(\lambda) \vert_{\cup_{k=3}^{4} 
\varsigma_{{\cal C}}^{(k)}} = {\rm I} + ({\cal R}_{{\cal C}}
(\lambda))^{\ast} ({\cal P}_{+}(\lambda))^{-1} (\delta^{+}(
\lambda;\! \lambda_{0}))^{-2} e^{2 i t \rho(\lambda;\lambda
_{0})} \sigma_{-},& \\
&\delta^{+}(\lambda;\! \lambda_{0}) = \exp \! \left\{ \int_{0}^{
\lambda_{0}} \! \frac{\xi \ln(1-\vert r(\xi) \vert^{2})}{(\xi^{
2} - \lambda^{2})} \frac{d\xi}{\pi i} \! - \! \int_{0}^{\infty} 
\! \frac{\xi \ln(1 + \vert r(i \xi) \vert^{2})}{(\xi^{2}+\lambda
^{2})} \frac{d \xi}{\pi i} \right\} \! := e^{({\rm I}_{1} + {\rm 
I}_{2})(\lambda;\lambda_{0})},&
\end{eqnarray*}
${\cal P}_{+}(\lambda) \! := \! \prod_{l=n+1}^{N} \! \left(\! 
\frac{(\lambda - \overline{\lambda_{l}})(\lambda + \overline{
\lambda_{l}})}{(\lambda - \lambda_{l})(\lambda + \lambda_{l})} \! 
\right)^{2}$, $\rho(\lambda;\! \lambda_{0}) \! = \! 2 \lambda^{2}
(\lambda^{2} \! - \! 2 \lambda_{0}^{2})$, $({\cal R}_{{\cal C}}
(\lambda))^{\ast}$ denotes the same function as ${\cal R}_{
{\cal C}}(\lambda)$ except with the complex conjugated 
coefficients, ${\cal R}_{{\cal C}}(- \lambda) \! = \! - 
{\cal R}_{{\cal C}}(\lambda)$, ${\cal R}_{{\cal C}}(\lambda)$ 
is a piecewise-rational function which decays like ${\cal O}(
\lambda^{-(k+7)})$, $k \! \in \! \Bbb Z_{\geq 1}$, as $\lambda 
\! \to \! \infty$, $\lambda \! \in \! \varsigma_{{\cal C}} \! 
\setminus \! \{0\}$, and has, for $\lambda \! \in \! (\cup_{
l=1}^{2} \varsigma_{{\cal C}}^{(l)}) \! \cap \! \{\mathstrut 
\lambda^{\prime}; \, \vert \lambda^{\prime} \vert \! < \! 
\varepsilon\}$, where $\varepsilon$ is an arbitrarily fixed, 
sufficiently small positive real number, the following Taylor 
series expansion about $\lambda \! = \! 0$,
\begin{eqnarray*}
&{\cal R}_{{\cal C}}(\lambda) {\cal P}_{+}(\lambda) := 
R_{+}^{\prime}(0) \lambda + \frac{1}{3!} R_{+}^{\prime \prime 
\prime}(0) \lambda^{3} + \frac{1}{5!} R_{+}^{V}(0) \lambda^{5} 
+ {\cal O}(\lambda^{7}),&
\end{eqnarray*}
with
\begin{eqnarray*}
&R_{+}^{\prime}(0) = r_{+}^{\prime}(0) - b_{1}^{+},& \\
&\frac{1}{3!} R_{+}^{\prime \prime \prime}(0)=(r_{+}^{\prime}
(0) a_{1}^{+} + \frac{1}{3!} r_{+}^{\prime \prime \prime}(0)) 
- (\frac{1}{3!} b_{2}^{+} - b_{1}^{+} c_{1}^{+}),& \\
&\frac{1}{5!} R_{+}^{V}(0) = r_{+}^{\prime}(0)(a_{2}^{+} + 
(a_{1}^{+})^{2}) - b_{1}^{+}((c_{1}^{+})^{2} - c_{2}^{+})& \\
&\, \, \, \, \, \, \, \, \, \, \, \, \, \, \, \, \, \, \, \, 
\, \, \, \, \, \, \, \, + \, \frac{1}{3!} r_{+}^{\prime 
\prime \prime}(0) a_{1}^{+} + \frac{1}{3!}b_{2}^{+}c_{1}^{+} 
+ \frac{1}{5!} r_{+}^{V}(0) - \frac{1}{5!} b_{3}^{+},&
\end{eqnarray*}
where
\begin{eqnarray*}
&r_{+}^{\prime}(0) := r^{\prime}(0) s_{+}, \, \, \, \, \, \, b
_{1}^{+} := r^{\prime}(i0) s_{+}, \, \, \, \, \, \, a_{1}^{+} 
:= \vert r^{\prime}(0) \vert^{2},& \\
&\frac{1}{3!} r_{+}^{\prime \prime \prime}(0) := (\frac{1}{2!} 
r^{\prime}(0)(-8 i) \sum_{+} \! \frac{\sin \gamma_{l}}{\Delta_{
l}^{2}} + \frac{1}{3!} r^{\prime \prime \prime}(0)) s_{+},& \\
&c_{1}^{+} := \vert r^{\prime}(i0) \vert^{2}, \, \, \, \, \, 
\frac{1}{3!} b_{2}^{+} := (\frac{1}{2!} r^{\prime}(i0)(8i) \sum
_{+} \! \frac{\sin \gamma_{l}}{\Delta_{l}^{2}} + \frac{1}{3!} 
r^{\prime \prime \prime}(i0)) s_{+},& \\
&a_{2}^{+} := \frac{1}{3!}(r^{\prime}(0) \overline{r^{\prime 
\prime \prime}(0)} + \overline{r^{\prime}(0)} r^{\prime \prime 
\prime}(0)),& \\
&c_{2}^{+} := \frac{1}{3!}(r^{\prime}(i0) \overline{r^{\prime 
\prime \prime}(i0)} + \overline{r^{\prime}(i0)} r^{\prime \prime 
\prime}(i0)),& \\
&\frac{1}{5!} r_{+}^{V}(0) := \frac{1}{4!} r^{\prime}(0) (-192 
(\sum_{+} \! \frac{\sin \gamma_{l}}{\Delta_{l}^{2}})^{2} + 48i 
\sum_{+} \! \frac{\sin 2 \gamma_{l}}{\Delta_{l}^{4}}) s_{+}& \\
&\, \, \, \, \, \, \, \, \, \, \, + \, \, (\frac{1}{2! 3!} r^{
\prime \prime \prime}(0)(-8i) \sum_{+} \! \frac{\sin \gamma_{l}
}{\Delta_{l}^{2}} + \frac{1}{5!} r^{V}(0)) s_{+},& \\
&\frac{1}{5!} b_{3}^{+} := \frac{1}{4!} r^{\prime}(i0) (-192 
(\sum_{+} \! \frac{\sin \gamma_{l}}{\Delta_{l}^{2}})^{2} + 48i 
\sum_{+} \! \frac{\sin 2 \gamma_{l}}{\Delta_{l}^{4}}) s_{+}& \\
&\, + \, \, (\frac{1}{2! 3!} r^{\prime \prime \prime}(i0)(8i) 
\sum_{+} \! \frac{\sin \gamma_{l}}{\Delta_{l}^{2}} + \frac{1}{
5!} r^{V}(i0)) s_{+},&
\end{eqnarray*}
and
\begin{eqnarray*}
&r^{\prime}(0) := (\frac{d r(\lambda)}{d \lambda} \vert_{\lambda 
\in \Bbb R}) \vert_{\lambda=0}, \, \, \, \, \, \, \, \, \, \, r^{
\prime}(i0) := (\frac{d r(\lambda)}{d \lambda} \vert_{\lambda 
\in i \Bbb R}) \vert_{\lambda=0},& \\
&r^{\prime \prime \prime}(0) := (\frac{d^{3}r(\lambda)}{d\lambda
^{3}} \vert_{\lambda \in \Bbb R}) \vert_{\lambda=0}, \, \, \, \, 
\, \, \, \, \, \, r^{\prime \prime \prime}(i0) := (\frac{d^{3} r
(\lambda)}{d \lambda^{3}} \vert_{\lambda \in i \Bbb R}) \vert_{
\lambda=0},& \\
&r^{V}(0) := (\frac{d^{5} r(\lambda)}{d \lambda^{5}} \vert_{
\lambda \in \Bbb R}) \vert_{\lambda=0}, \, \, \, \, \, \, \, 
\, \, \, r^{V}(i0) := (\frac{d^{5} r(\lambda)}{d \lambda^{5}} 
\vert_{\lambda \in i \Bbb R}) \vert_{\lambda=0},&
\end{eqnarray*}
with $s_{+} \! := \! \exp \{4 i \sum_{+} \gamma_{l}\}$, and 
$\sum_{+} \! := \! \sum_{l=n+1}^{N};$
\item[(3)] as $\lambda \! \to \! \infty$, $\lambda \! \in \! 
\Bbb C \! \setminus \! \varsigma_{{\cal C}}$,
\begin{eqnarray*}
&m^{\varsigma}_{{\cal C}}(\lambda)={\rm I}+{\cal O}(\lambda^{
-1}).&
\end{eqnarray*}
\end{enumerate}
Moreover, $m^{\varsigma}_{{\cal C}}(\lambda)$ satisfies the 
following symmetry reductions, $m^{\varsigma}_{{\cal C}}(
\lambda) \! = \! \sigma_{3} m^{\varsigma}_{{\cal C}}(-\lambda)
\sigma_{3}$ and $m^{\varsigma}_{{\cal C}}(\lambda) \! = \! 
\sigma_{1} \overline{m^{\varsigma}_{{\cal C}}(\overline{\lambda}
)} \sigma_{1}$.
\end{ccc}
\begin{bbb}
For $\lambda \! \in \! (\cup_{l=1}^{2} \varsigma_{{\cal C}}^{(
l)}) \cap \{\mathstrut \lambda^{\prime}; \, \vert \lambda^{
\prime} \vert \! < \! \varepsilon\}$, where $\varepsilon$ is 
an arbitrarily fixed, sufficiently small positive real number,
\begin{eqnarray*}
&({\rm I}_{1} \! + \! {\rm I}_{2})(\lambda;\! \lambda_{0}) \! = \! 
i \phi_{+}(\lambda_{0}) \! + \! \widehat{E}_{1} \lambda^{2} \ln \! 
\vert \lambda \vert \! + \! \widetilde{E}_{1} \lambda^{2} \! + \! 
\widehat{E}_{2} \lambda^{4} \ln \! \vert \lambda \vert \! + \! 
\widetilde{E}_{2} \lambda^{4} \! + \! E(\lambda;\! \lambda_{0}),&
\end{eqnarray*}
where
\begin{eqnarray*}
&\phi_{+}(\lambda_{0}) := - \int_{0}^{\lambda_{0}} \frac{\ln(1 
- \vert r(\xi) \vert^{2})}{\xi} \frac{d \xi}{\pi} + \int_{0}^{
\infty} \frac{\ln(1 + \vert r(i\xi) \vert^{2})}{\xi} \frac{d 
\xi}{\pi},& \\
&\widehat{E}_{1} := - \frac{1}{2 \pi i} (\lim\limits_{\lambda 
\to 0} [D_{\lambda} \ln(1 \! - \! \vert r(\lambda) \vert^{2})] 
+ \lim\limits_{\lambda \to 0} [D_{\lambda} \ln(1 \! + \! \vert 
r(i \lambda) \vert^{2})]),& \\
&\widetilde{E}_{1} \! := \! \frac{1}{2 \pi i (2)} (\lim\limits
_{\lambda \to 0} [D_{\lambda} \ln(1 \! - \! \vert r(\lambda) 
\vert^{2})] \! + \! \lim\limits_{\lambda \to 0} [D_{\lambda} 
\ln (1 \! + \! \vert r(i \lambda) \vert^{2})]) \! - \! \int_{
\lambda_{0}}^{\infty} \frac{\ln(1-\vert r(\xi) \vert^{2})}{
\xi^{3}} \frac{d \xi}{\pi i},& \\
&\widehat{E}_{2} := - \frac{1}{2 \pi i (2)^{2}} (\lim\limits_{
\lambda \to 0} [D_{\lambda}^{2} \ln (1 \! - \! \vert r(\lambda) 
\vert^{2})] - \lim\limits_{\lambda \to 0} [D_{\lambda}^{2} \ln 
(1 \! + \! \vert r(i \lambda) \vert^{2})]),& \\
&\widetilde{E}_{2} \! := \! \frac{(3/2)}{2 \pi i (2)^{3}} 
(\lim\limits_{\lambda \to 0} [D_{\lambda}^{2} \ln (1 \! - 
\! \vert r(\lambda) \vert^{2})] \! - \! \lim\limits_{\lambda 
\to 0} [D_{\lambda}^{2} \ln (1 \! + \! \vert r(i \lambda) 
\vert^{2})]) \! - \! \int_{\lambda_{0}}^{\infty} \frac{\ln(
1-\vert r(\xi) \vert^{2})}{\xi^{5}} \frac{d \xi}{\pi i},&
\end{eqnarray*}
$D_{\lambda}^{l} \! := \! (\frac{1}{\lambda}\frac{d}{d\lambda}
)^{l}$, $l \! \in \! \Bbb Z_{\geq 1}$, $\lim\limits_{\lambda 
\to 0}[D_{\lambda} \ln (1 \! - \! \vert r(\lambda) \vert^{2})] 
\! = \! - 2 a_{1}^{+}$, $\lim\limits_{\lambda \to 0} [D_{
\lambda} \ln (1 \! + \! \vert r(i \lambda) \vert^{2})] \! = \! 
2 c_{1}^{+}$, $\lim\limits_{\lambda \to 0} [D_{\lambda}^{2} \ln 
(1 \! - \! \vert r(\lambda) \vert^{2})] \! = \! -8(\frac{1}{2}
(a_{1}^{+})^{2} \! + \! a_{2}^{+})$, $\lim\limits_{\lambda \to 
0} [D_{\lambda}^{2} \ln (1 \! + \! \vert r(i \lambda) \vert^{2}
)] \! = \! -8(\frac{1}{2}(c_{1}^{+})^{2} \! - \! c_{2}^{+})$, 
and
\begin{eqnarray*}
&E(\lambda;\! \lambda_{0}) \! := \! \sum\limits_{l \in \Bbb Z_{
\geq 3}} (\widehat{E}_{l} \ln \! \vert \lambda \vert \! + \! 
\widetilde{E}_{l}) \lambda^{2l},&
\end{eqnarray*}
with
\begin{eqnarray*}
&\widehat{E}_{l} = \frac{2^{-2l+2}}{2 \pi i}(\widehat{K}_{l}^{
0} \lim\limits_{\lambda \to 0} [D_{\lambda}^{l} \ln (1 \! - \! 
\vert r(\lambda) \vert^{2})] + \widehat{K}_{l}^{1} \lim\limits_{
\lambda \to 0} [D_{\lambda}^{l} \ln (1 \! + \! \vert r(i \lambda) 
\vert^{2})]),& \\
&\widetilde{E}_{l} \! = \! \frac{2^{-2l+1}}{2 \pi i}(\widetilde{
K}_{l}^{0} \lim\limits_{\lambda \to 0} [D_{\lambda}^{l} \ln (1 \! 
- \! \vert r(\lambda) \vert^{2})] \! + \! \widetilde{K}_{l}^{1} 
\lim\limits_{\lambda \to 0} [D_{\lambda}^{l} \ln (1 \! + \! \vert 
r(i \lambda) \vert^{2})]) \! - \! \int_{\lambda_{0}}^{\infty} 
\frac{\ln(1 - \vert r(\xi) \vert^{2})}{\xi^{2l+1}} \frac{d \xi}{
\pi i},&
\end{eqnarray*}
where, for $l \! \in \! \Bbb Z_{\geq 3}$, $\widehat{K}_{l}^{0}$, 
$\widehat{K}_{l}^{1}$, $\widetilde{K}_{l}^{0}$, and $\widetilde{
K}_{l}^{1} \! \in \! \Bbb R \! \setminus \! \{0\}$.
\end{bbb}

{\em Proof.\/} Using the fact that $r(\lambda) \! \in \! {\cal 
S}(\widehat{\Gamma};\! \Bbb C)$ and $\vert \vert r \vert \vert
_{{\cal L}^{\infty}(\widehat{\Gamma};\Bbb C)} \! < \! 1$, the 
results stated in the Proposition follow {}from the definition of 
$({\rm I}_{1} \! + \! {\rm I}_{2})(\lambda;\! \lambda_{0})$ (and 
the expression for $\delta^{+}(\lambda;\! \lambda_{0}))$ given in 
Lemma~4.1 via an integration by parts argument and the expansion 
$\ln (1 \! - \! x) \! = \! - \sum_{l \in \Bbb Z_{\geq 1}} \! x^{
l} / l$, $\vert x \vert \! < \! 1$. \hfill \rule{6.5pt}{6.5pt}
\begin{bbb}
Set $\lambda \! = \! \frac{w}{\sqrt{t}}$. Then,
\begin{eqnarray*}
&v^{\varsigma}_{{\cal C}}(\frac{w}{\sqrt{t}}) \vert_{\sqrt{t} 
\, \cup_{l=1}^{2} \varsigma_{{\cal C}}^{(l)}}={\rm I} + {\cal 
R}_{{\cal C}}(\frac{w}{\sqrt{t}}) {\cal P}_{+}(\frac{
w}{\sqrt{t}}) e^{2({\rm I}_{1} + {\rm I}_{2})(\frac{w}{\sqrt{
t}};\lambda_{0})} e^{-2 i t \rho(\frac{w}{\sqrt{t}};\lambda_{
0})} \sigma_{+},& \\
&v^{\varsigma}_{{\cal C}}(\frac{w}{\sqrt{t}}) \vert_{\sqrt{t} 
\, \cup_{l=3}^{4} \varsigma_{{\cal C}}^{(l)}}={\rm I}+({\cal 
R}_{{\cal C}}(\frac{w}{\sqrt{t}}))^{\ast} ({\cal P}_{
+}(\frac{w}{\sqrt{t}}))^{-1} e^{-2({\rm I}_{1} + {\rm I}_{2})
(\frac{w}{\sqrt{t}};\lambda_{0})} e^{2 i t \rho(\frac{w}{
\sqrt{t}};\lambda_{0})} \sigma_{-},&
\end{eqnarray*}
where
\begin{eqnarray*}
&{\cal R}_{{\cal C}}(\frac{w}{\sqrt{t}}) {\cal P}_{+}(
\frac{w}{\sqrt{t}}) e^{2({\rm I}_{1} + {\rm I}_{2})(\frac{w}{
\sqrt{t}};\lambda_{0})} e^{-2 i t \rho(\frac{w}{\sqrt{t}};
\lambda_{0})} \! := \! \sum\limits_{p=0}^{2} \sum\limits_{q=0}
^{p} \frac{{\cal J}^{\alpha}_{p,q}(w;\lambda_{0})(\ln t)^{q}}
{t^{p+1/2}}& \\
&\, \, \, \, \, \, \, \, \, \, \, \, \, \, \, \, \, \, \, \, 
\, \, \, \, \, \, \, \, \, \, \, \, \, \, \, \, \, \, \, \,
\, \, \, \, \, \, \, \, \, \, \, \, \, \, \, \, \, \, \, \, 
\, \, \, \, \, \, \, \, \, \, \, \, \, \, \, \, \, \, \, \,
\, \, \, \, \, \, \, \, \, \, \, \, \, 
+ \, {\cal O} \! \left(\frac{{\cal F}^{\alpha}(w;\lambda_{0}) 
\exp\{8 i \lambda_{0}^{2} w^{2}\}(\ln t)^{3}}{t^{7/2}} \right) 
\!,& \\
&({\cal R}_{{\cal C}}(\frac{w}{\sqrt{t}}))^{\ast}({\cal 
P}_{+}(\frac{w}{\sqrt{t}}))^{-1} e^{-2({\rm I}_{1} + {\rm I}_{
2})(\frac{w}{\sqrt{t}};\lambda_{0})} e^{2 i t \rho(\frac{w}{
\sqrt{t}};\lambda_{0})} \! := \! \sum\limits_{p=0}^{2} 
\sum\limits_{q=0}^{p} \frac{{\cal J}^{\beta}_{p,q}(w;\lambda
_{0})(\ln t)^{q}}{t^{p+1/2}}& \\
&\, \, \, \, \, \, \, \, \, \, \, \, \, \, \, \, \, \, \, \, 
\, \, \, \, \, \, \, \, \, \, \, \, \, \, \, \, \, \, \, \,
\, \, \, \, \, \, \, \, \, \, \, \, \, \, \, \, \, \, \, \, 
\, \, \, \, \, \, \, \, \, \, \, \, \, \, \, \, \, \, \, \,
\, \, \, \, \, \, \, \, \, \, \, \, \, \, \, \, \, \, \, \, 
\, \, \, \, \, \, 
+ \, {\cal O} \! \left(\frac{{\cal F}^{\beta}(w;\lambda_{0}) 
\exp\{-8 i \lambda_{0}^{2} w^{2}\}(\ln t)^{3}}{t^{7/2}} \right) 
\!,&
\end{eqnarray*}
with
\begin{eqnarray*}
&{\cal J}^{\alpha}_{0,0}(w;\! \lambda_{0}) \! = \! w R_{+}^{
\prime}(0) e^{2i\phi_{+}(\lambda_{0})} e^{8i\lambda_{0}^{2} 
w^{2}},& \\
&{\cal J}^{\alpha}_{1,0}(w;\! \lambda_{0}) \! = \! {\cal V}^{{
\cal C}}_{1,0}(w;\! \lambda_{0}) w R_{+}^{\prime}(0) e^{8 i 
\lambda_{0}^{2} w^{2}} \! - \! 4 i w^{5} R_{+}^{\prime}(0) e^{
2 i \phi_{+}(\lambda_{0})} e^{8 i \lambda_{0}^{2} w^{2}}& \\
&\! \! \! \! \! \! \! \! \! \! \! \! \! \! \! \! \! \! \! \! 
\! \! \! \! \! \! \! \! \! \! \! \! \! \! \! \! \! \! \! \! 
\! \! \! \, \, + \, w^{3} \frac{1}{3!} R_{+}^{\prime \prime 
\prime}(0) e^{2 i \phi_{+}(\lambda_{0})} e^{8 i \lambda_{0}^{
2} w^{2}},& \\
&{\cal J}^{\alpha}_{1,1}(w;\! \lambda_{0}) \! = \! {\cal V}^{{
\cal C}}_{1,1}(w;\! \lambda_{0}) w R_{+}^{\prime}(0) e^{8 i 
\lambda_{0}^{2} w^{2}},& \\
&{\cal J}^{\alpha}_{2,0}(w;\! \lambda_{0}) \! = \! {\cal V}^{{
\cal C}}_{2,0}(w;\! \lambda_{0}) w R_{+}^{\prime} (0) e^{8 i 
\lambda_{0}^{2} w^{2}} \! - \! 4 i w^{5} {\cal V}^{{\cal C}}_{
1,0}(w;\! \lambda_{0}) R_{+}^{\prime}(0) e^{8 i \lambda_{0}^{2} 
w^{2}}& \\
&\, \, \, \, \, \, \, \, \, \, \, \, \, \, \, \, \, \,
\, \, \, \, \, \, \, \, \, \, - \, 
8 w^{9} R_{+}^{\prime}(0) e^{2 i \phi_{+}(\lambda_{0})} e^{8 
i \lambda_{0}^{2} w^{2}} \! + \! {\cal V}^{{\cal C}}_{1,0} 
(w;\! \lambda_{0}) w^{3} \frac{1}{3!} R_{+}^{\prime \prime 
\prime}(0) e^{8 i \lambda_{0}^{2} w^{2}}& \\
&\, \, \, \, \, \, \, \, \, \, \, \, \, \, \, \, \, \, 
\, \, \, \, \, \, \, \, \, \, \, \, \, \, - \, 
4 i w^{7} \frac{1}{3!} R_{+}^{\prime \prime \prime}(0) e^{2i 
\phi_{+}(\lambda_{0})} e^{8 i \lambda_{0}^{2} w^{2}} \! + \! 
w^{5} \frac{1}{5!} R_{+}^{V}(0) e^{2 i \phi_{+}(\lambda_{0})} 
e^{8 i \lambda_{0}^{2} w^{2}},& \\
&{\cal J}^{\alpha}_{2,1}(w;\! \lambda_{0}) \! = \! {\cal V}^{
{\cal C}}_{2,1}(w;\! \lambda_{0}) w R_{+}^{\prime}(0) e^{8 i 
\lambda_{0}^{2} w^{2}} \! - \! 4 i w^{5} {\cal V}^{{\cal C}}_{
1,1}(w;\! \lambda_{0}) R_{+}^{\prime}(0) e^{8 i \lambda_{0}^{2} 
w^{2}}& \\
&\! \! \! \! \! \! \! \! \! \! \! \! \! \! \! \! \! \! \! \! 
\! \! \! \! \! \! \! \! \! \! \! \! \! \! \! \! \! \! \! \! 
+ \, {\cal V}^{{\cal C}}_{1,1}(w;\! \lambda_{0}) w^{3} 
\frac{1}{3!} R_{+}^{\prime \prime \prime}(0) e^{8 i \lambda
_{0}^{2} w^{2}},& \\
&{\cal J}^{\alpha}_{2,2}(w;\! \lambda_{0}) \! = \! {\cal V}^{{
\cal C}}_{2,2}(w;\! \lambda_{0}) w R_{+}^{\prime}(0) e^{8 i 
\lambda_{0}^{2} w^{2}},& \\
&{\cal V}^{{\cal C}}_{1,0}(w;\! \lambda_{0}) \! := \! 2 (
\widehat{E}_{1} w^{2} \ln \! \vert w \vert \! + \! \widetilde{
E}_{1} w^{2}) e^{2 i \phi_{+}(\lambda_{0})},& \\
&{\cal V}^{{\cal C}}_{1,1}(w;\! \lambda_{0}) \! := \! -\widehat{
E}_{1} w^{2} e^{2 i \phi_{+}(\lambda_{0})},& \\
&{\cal V}^{{\cal C}}_{2,0}(w;\! \lambda_{0}) \! := \! 2 ((
\widehat{E}_{1} w^{2} \ln \! \vert w \vert \! + \! \widetilde{
E}_{1} w^{2})^{2} \! + \! \widehat{E}_{2} w^{4} \ln \! \vert w 
\vert \! + \! \widetilde{E}_{2} w^{4}) e^{2 i \phi_{+}(\lambda
_{0})},& \\
&{\cal V}^{{\cal C}}_{2,1}(w;\! \lambda_{0}) \! := \! (- 2 
\widehat{E}_{1} w^{2} (\widehat{E}_{1} w^{2} \ln \! \vert w 
\vert \! + \! \widetilde{E}_{1} w^{2}) \! - \! \widehat{E}_{2}
w^{4}) e^{2 i \phi_{+}(\lambda_{0})},& \\
&{\cal V}^{{\cal C}}_{2,2}(w;\! \lambda_{0}) \! := \! \frac{1}
{2}(\widehat{E}_{1})^{2} w^{4} e^{2 i \phi_{+}(\lambda_{0})},&
\end{eqnarray*}
and
\begin{eqnarray*}
&{\cal J}^{\beta}_{0,0}(w;\! \lambda_{0}) \! = \! w \overline{
R_{+}^{\prime}(0)} e^{-2 i \phi_{+}(\lambda_{0})} e^{-8 i 
\lambda_{0}^{2} w^{2}},& \\
&{\cal J}^{\beta}_{1,0}(w;\! \lambda_{0}) \! = \! {\cal W}^{{
\cal C}}_{1,0}(w;\! \lambda_{0}) w \overline{R_{+}^{\prime}(0)} 
e^{-8 i \lambda_{0}^{2} w^{2}} \! + \! 4 i w^{5} \overline{R_{
+}^{\prime}(0)} e^{-2 i \phi_{+}(\lambda_{0})} e^{-8 i \lambda
_{0}^{2} w^{2}}& \\
&\! \! \! \! \! \! \! \! \! \! \! \! \! \! \! \! \! \! \! \! 
\! \! \! \! \! \! \! \! \! \! \! \! \! \! \! \! \! \! \! \! 
\! \! \! \! \! \! \! \! 
\, \, + \, w^{3} \frac{1}{3!} \overline{R_{+}^{\prime 
\prime \prime}(0)} e^{-2 i \phi_{+}(\lambda_{0})} e^{-8 i 
\lambda_{0}^{2} w^{2}},& \\
&{\cal J}^{\beta}_{1,1}(w;\! \lambda_{0}) \! = \! {\cal W}^{{
\cal C}}_{1,1}(w;\! \lambda_{0}) w \overline{R_{+}^{\prime}(0)} 
e^{-8 i \lambda_{0}^{2} w^{2}},& \\
&{\cal J}^{\beta}_{2,0}(w;\! \lambda_{0}) \! = \! {\cal W}^{{
\cal C}}_{2,0}(w;\! \lambda_{0}) w \overline{R_{+}^{\prime}(0)} 
e^{-8 i \lambda_{0}^{2} w^{2}} \! + \! 4 i w^{5} {\cal W}^{{\cal 
C}}_{1,0}(w;\! \lambda_{0}) \overline{R_{+}^{\prime}(0)} e^{-8 i 
\lambda_{0}^{2} w^{2}}& \\
&\, \, \, \, \, \, \, \, \, \, \, \, \, \, \, \, \, \,
\, \, \, \, \, \, \, \, \, \, \, \, 
- \, 8 w^{9} \overline{R_{+}^{\prime}(0)} e^{-2 i \phi_{+}(
\lambda_{0})} e^{-8 i \lambda_{0}^{2} w^{2}} \! + \! {\cal W}
^{{\cal C}}_{1,0}(w;\! \lambda_{0}) w^{3} \frac{1}{3!} 
\overline{R_{+}^{\prime \prime \prime}(0)} e^{-8 i \lambda_{0}
^{2} w^{2}}& \\
&\, \, \, \, \, \, \, \, \, \, \, \, \, \, \, \, \, \, 
\, \, \, \, \, \, \, \, \, \, \, \, \, \, \, \, \, \, 
+ \, 4 i w^{7} \frac{1}{3!} \overline{R_{+}^{\prime \prime 
\prime}(0)} e^{-2 i \phi_{+}(\lambda_{0})} e^{-8 i \lambda_{0}
^{2} w^{2}} \! + \! w^{5} \frac{1}{5!} \overline{R_{+}^{V}(0)} 
e^{-2 i \phi_{+}(\lambda_{0})} e^{-8 i \lambda_{0}^{2} w^{2}},& 
\\
&{\cal J}^{\beta}_{2,1}(w;\! \lambda_{0}) \! = \! {\cal W}^{{
\cal C}}_{2,1}(w;\! \lambda_{0}) w \overline{R_{+}^{\prime}(0)} 
e^{-8 i \lambda_{0}^{2} w^{2}} \! + \! 4 i w^{5} {\cal W}^{{
\cal C}}_{1,1}(w;\! \lambda_{0}) \overline{R_{+}^{\prime}(0)} 
e^{-8 i \lambda_{0}^{2} w^{2}}& \\
&\! \! \! \! \! \! \! \! \! \! \! \! \! \! \! \! \! \! \! \! 
\! \! \! \! \! \! \! \! \! \! \! \! \! \! \! \! \! \! \! \!
\! \! \! \! \! \! + \, 
{\cal W}^{{\cal C}}_{1,1}(w;\! \lambda_{0}) w^{3} \frac{
1}{3!} \overline{R_{+}^{\prime \prime \prime}(0)} e^{-8 i 
\lambda_{0}^{2} w^{2}},& \\
&{\cal J}^{\beta}_{2,2}(w;\! \lambda_{0}) \! = \! {\cal W}^{{
\cal C}}_{2,2}(w;\! \lambda_{0}) w \overline{R_{+}^{\prime}(0)} 
e^{-8 i \lambda_{0}^{2} w^{2}},& \\
&{\cal W}^{{\cal C}}_{1,0}(w;\! \lambda_{0}) \! := \! - 2 (
\widehat{E}_{1} w^{2} \ln \! \vert w \vert \! + \! \widetilde{
E}_{1} w^{2}) e^{-2 i \phi_{+}(\lambda_{0})},& \\
&{\cal W}^{{\cal C}}_{1,1}(w;\! \lambda_{0}) \! := \! \widehat{
E}_{1} w^{2} e^{-2 i \phi_{+}(\lambda_{0})},& \\
&{\cal W}^{{\cal C}}_{2,0}(w;\! \lambda_{0}) \! := \! 2 ((
\widehat{E}_{1} w^{2} \ln \! \vert w \vert \! + \! \widetilde{
E}_{1} w^{2})^{2} \! - \! \widehat{E}_{2} w^{4} \ln \! \vert w 
\vert \! - \! \widetilde{E}_{2} w^{4}) e^{-2 i \phi_{+}(
\lambda_{0})},& \\
&{\cal W}^{{\cal C}}_{2,1}(w;\! \lambda_{0}) \! := \! (-2 
\widehat{E}_{1} w^{2} (\widehat{E}_{1} w^{2} \ln \! \vert w 
\vert \! + \! \widetilde{E}_{1} w^{2}) \! + \! \widehat{E}_{2} 
w^{4}) e^{-2 i \phi_{+}(\lambda_{0})},& \\
&{\cal W}^{{\cal C}}_{2,2}(w;\! \lambda_{0}) \! := \! \frac{
1}{2}(\widehat{E}_{1})^{2} w^{4} e^{-2 i \phi_{+}(\lambda_{
0})},&
\end{eqnarray*}
where, for $k \! \in \! \{1,2,\infty\}$,
\begin{eqnarray*}
&\vert \vert {\cal F}^{\alpha}(\cdot;\! \lambda_{0}) \exp \{
8 i \lambda_{0}^{2} (\cdot)^{2}\} \vert \vert_{{\cal L}^{k}
(\sqrt{t} \, \cup_{l=1}^{2} \varsigma_{{\cal C}}^{(l)};\Bbb 
C)} = \underline{c}(\lambda_{0}),& \\
&\vert \vert {\cal F}^{\beta}(\cdot;\! \lambda_{0}) \exp \{-
8 i \lambda_{0}^{2}(\cdot)^{2}\} \vert \vert_{{\cal L}^{k}(
\sqrt{t} \, \cup_{l=3}^{4} \varsigma_{{\cal C}}^{(l)};\Bbb 
C)} = \underline{c}(\lambda_{0}).&
\end{eqnarray*}
\end{bbb}

{\em Proof.\/} Follows {}from the expressions for the jumps, $v^{
\varsigma}_{{\cal C}}(\lambda) \vert_{\varsigma_{{\cal C}}^{(l)} 
\cup \varsigma_{{\cal C}}^{(l+1)}}$, $l \! \in \! \{1,3\}$, given 
in Lemma~4.1, Proposition~4.1, and the change-of-variable rule 
given in the Proposition upon expanding in reciprocal powers of 
$\sqrt{t}$. \hfill \rule{6.5pt}{6.5pt}
\begin{bbb}
The solution of the RH problem for $m^{\varsigma}_{{\cal C}}
(\lambda) \colon \Bbb C \setminus \varsigma_{{\cal C}} \! \to 
\! {\rm SL}(2,\! \Bbb C)$ formulated in Lemma~4.1 has the 
following integral representation,
\begin{eqnarray*}
m^{\varsigma}_{{\cal C}}(\lambda) = {\rm I} + \sum\limits_{k
=1}^{4} \int_{\sqrt{t} \, \varsigma_{{\cal C}}^{(k)}} \! \frac{
\mu^{\varsigma_{{\cal C}}}(\frac{w}{\sqrt{t}})(\sum_{l \in \{
\pm\}} \! \widetilde{w}_{l}^{\varsigma_{{\cal C}}^{(k)}} \! 
(\frac{w}{\sqrt{t}}))}{(w - \lambda \sqrt{t} \,)} \frac{d w}{
2 \pi i}, \, \, \, \, \, \, \, \, \lambda \in \Bbb C \setminus 
\varsigma_{{\cal C}},
\end{eqnarray*}
where, for $j \! \in \! \{1,2\}$,
\begin{eqnarray*}
&\widetilde{w}_{-}^{\varsigma_{{\cal C}}^{(j)}} \! (\cdot) = 
\widetilde{w}_{+}^{\varsigma_{{\cal C}}^{(j+2)}} \! (\cdot) = 
0,& \\
&\widetilde{w}_{+}^{\varsigma_{{\cal C}}^{(j)}} \! \! (\frac{w}
{\sqrt{t}}) \! = \! \left. {\cal R}_{{\cal C}}(\lambda) 
{\cal P}_{+}(\lambda) e^{2({\rm I}_{1} + {\rm I}_{2})(\lambda;
\lambda_{0})} e^{-2 i t \rho(\lambda;\lambda_{0})} \sigma_{
+} \right\vert_{\lambda = \frac{w}{\sqrt{t}}},& \\
&\widetilde{w}_{-}^{\varsigma_{{\cal C}}^{(j+2)}} \! \! (\frac{
w}{\sqrt{t}}) \! = \! \left. ({\cal R}_{{\cal C}}(\lambda))
^{\ast} ({\cal P}_{+}(\lambda))^{-1} e^{-2({\rm I}_{1} + 
{\rm I}_{2})(\lambda;\lambda_{0})} e^{2 i t \rho(\lambda;
\lambda_{0})} \sigma_{-} \right\vert_{\lambda = \frac{w}{
\sqrt{t}}},&
\end{eqnarray*}
and, for arbitrary $N_{o} \! \in \! \Bbb Z_{\geq 1}$,
\begin{eqnarray*}
&\mu^{\varsigma_{{\cal C}}}(\frac{w}{\sqrt{t}}) = {\rm I} + 
\sum\limits_{p^{\prime}=1}^{2} \sum\limits_{p=0}^{N_{o}} 
\sum\limits_{q=0}^{p} \frac{\mu^{p^{\prime}}_{p,q}(w;\lambda_{0}
)(\ln t)^{q}}{t^{p+p^{\prime}/2}} + {\cal E}^{\varsigma_{{\cal 
C}}}(\frac{w}{\sqrt{t}};\! \lambda_{0}),&
\end{eqnarray*}
with
\begin{eqnarray*}
&\vert \vert \mu^{p^{\prime}}_{p,q}(\cdot;\! \lambda_{0}) \vert 
\vert_{\cup_{l \in \{2,\infty\}} {\cal L}^{l}(\sqrt{t} \, 
\varsigma_{{\cal C}} \setminus \{0\};M_{2}(\Bbb C)) \to {\cal 
L}^{2}(\sqrt{t} \, \varsigma_{{\cal C}} \setminus \{0\};M_{2}(
\Bbb C))} < \infty,& \\
&\vert \vert {\cal E}^{\varsigma_{{\cal C}}}(\cdot;\! \lambda_{
0}) \vert \vert_{\cup_{l \in \{2,\infty\}} {\cal L}^{l}(\sqrt{t} 
\, \varsigma_{{\cal C}} \setminus \{0\};M_{2}(\Bbb C)) \to {\cal 
L}^{2}(\sqrt{t} \, \varsigma_{{\cal C}} \setminus \{0\};M_{2}(
\Bbb C))} = {\cal O} \! \left(\frac{\underline{c}(\lambda_{0})(
\ln t)^{N_{o}+1}}{(\lambda_{0}^{2} t)^{N_{o}+3/2}} \right) \!.&
\end{eqnarray*}
\end{bbb}

{\em Proof.\/} {}From the BC \cite{a14} formulation (the paragraph 
superseding Theorem~2.1), writing the expressions for the jump 
matrices as $v^{\varsigma}_{{\cal C}}(\lambda) \vert_{\varsigma
_{{\cal C}}^{(k)}} \! := \! ({\rm I} \! - \! \widetilde{w}_{-}^{
\varsigma_{{\cal C}}^{(k)}} \! \! (\lambda))^{-1} ({\rm I} \! + 
\! \widetilde{w}_{+}^{\varsigma_{{\cal C}}^{(k)}} \! \! (\lambda
))$, one obtains the results for $\widetilde{w}_{l}^{\varsigma_{
{\cal C}}^{(k)}} \! (\frac{w}{\sqrt{t}})$, $l \! \in \! \{\pm\}$, 
$k \! \in \! \{1,2,3,4\}$, stated in the Proposition upon 
comparison with the expressions for the jump matrices given in 
Lemma~4.1 and the change-of-variable rule $\lambda \! \to \! w / 
\sqrt{t} \, (\not= \! 0)$: the integral representation is now an 
immediate consequence of Theorem~2.1. The asymptotic expansion 
for $\mu^{\varsigma_{{\cal C}}}(\frac{w}{\sqrt{t}})$ 
(resp.~estimation for ${\cal E}^{\varsigma_{{\cal C}}}(\frac{w}
{\sqrt{t}};\! \lambda_{0}))$ is a consequence of (the analogue 
of) the asymptotic expansion (resp.~estimation) for $\mu^{{\cal 
C}}(w;\! \lambda_{0})$ (resp.~$\widetilde{{\cal E}}^{{\cal C}}
_{\mu}(w;\! \lambda_{0}))$ given in Proposition~3.3 after a 
trivial change of variable. \hfill \rule{6.5pt}{6.5pt}
\begin{eee}
{\rm Near the first-order stationary phase points ${\rm sgn}(k) 
\lambda_{0}$, $k \! \in \! \{{\cal B},{\cal A}\}$, ${\rm sgn}
({\cal B}) \! = \! -{\rm sgn}({\cal A}) \! = \! 1$, the analogues 
of the asymptotic expansions for $\mu^{\varsigma_{k}}(\frac{w}{
\sqrt{t}})$ are
\begin{eqnarray*}
&\mu^{\varsigma_{k}}(\frac{w}{\sqrt{t}}) = {\cal D}_{{\rm pc}}
(w;{\rm sgn}(k) \lambda_{0}) \exp \! \left\{i \! \left(\sum
_{j \in \Bbb Z_{\geq 0}} \sum_{l \in \Bbb Z_{\geq 0}} a^{(k)}_{
j,l}(\lambda_{0}) w^{j} (\ln w)^{l} \right) \! \sigma_{3} \! 
\right\}& \\
&\! \! \! \! \! \! \! \! \! \! \! \! \! \! \! \! \! \! \! \! 
\! \! \! \! \! \! \times \left({\rm I} + \sum_{p \in \Bbb 
Z_{\geq 1}} \sum_{0 \leq q \leq p-1} \frac{\mu^{(k)}_{p,q}
(w;{\rm sgn}(k) \lambda_{0})(\ln t)^{q}}{t^{p/2}} \right) \!,&
\end{eqnarray*}
for suitably chosen $a^{(k)}_{j,l}(\lambda_{0})$, where ${\cal 
D}_{{\rm pc}}(w;{\rm sgn}(k) \lambda_{0})$ have explicit 
representations in terms of parabolic-cylinder functions (see, 
for example, the proof of Lemma~4.1 in \cite{a10}).\/}
\end{eee}
\begin{bbb}
Let $N_{o} \! \in \! \Bbb Z_{\geq 1}$. Then for $1 \! 
\leq \! p^{\prime} \! \leq \! 2$, $0 \! \leq \! p \! \leq \! 
N_{o}$, and $0 \! \leq \! q \! \leq \! p$, $\mu^{p^{\prime}}
_{p,q}(w;\! \lambda_{0})$ satisfy the following involutions, 
$\mu^{p^{\prime}}_{p,q}(w;\! \lambda_{0}) \! = \! \sigma_{3} 
\mu^{p^{\prime}}_{p,q}(-w;\! \lambda_{0}) \sigma_{3}$ and 
$\mu^{p^{\prime}}_{p,q}(w;\! \lambda_{0}) \! = \! \sigma_{1} 
\overline{\mu^{p^{\prime}}_{p,q}(\overline{w};\! \lambda_{0})} 
\sigma_{1}$.
\end{bbb}

{\em Proof.\/}
{}From the integral representation for $m^{\varsigma}_{{\cal C}}
(\lambda)$ given in Proposition~4.3, the following symmetry 
reductions for $m^{\varsigma}_{{\cal C}}(\lambda)$, $m^{\varsigma}
_{{\cal C}}(\lambda) \! = \! \sigma_{3} m^{\varsigma}_{{\cal C}}
(-\lambda) \sigma_{3}$ and $m^{\varsigma}_{{\cal C}}(\lambda) \! = 
\! \sigma_{1} \overline{m^{\varsigma}_{{\cal C}}(\overline{
\lambda})} \sigma_{1}$, the expressions for the jumps, $v^{
\varsigma}_{{\cal C}}(\lambda) \vert_{\varsigma_{{\cal C}}^{(l)} 
\cup \varsigma_{{\cal C}}^{(l+1)}}$, $l \! \in \! \{1,3\}$, given 
in Lemma~4.1, the fact that $\delta^{+}(\pm \lambda;\! \lambda_{0}
) \! = \! (\overline{\delta^{+}(\overline{\lambda};\! \lambda_{0})
})^{-1}$ (to deduce this, one takes the principal branch of the 
logarithmic function, $\ln (\lambda \! - \! {\rm sgn}(k) 
\lambda_{0}) \! := \! \ln \! \vert \lambda \! - \! {\rm sgn}(k) 
\lambda_{0} \vert \! + \! i \! \arg(\lambda \! - \! {\rm sgn}(k) 
\lambda_{0})$, $\arg(\lambda \! - \! {\rm sgn}(k) \lambda_{0}) 
\! \in \! (-\pi,\! \pi)$, $k \! \in \! \{{\cal B},{\cal A},{\cal 
C}\}$, ${\rm sgn}({\cal B}) \! = \! - {\rm sgn}({\cal A}) \! = 
\! 1$ and ${\rm sgn}({\cal C}) \! = \! 0)$, and the properties 
of ${\cal R}_{{\cal C}}(\lambda)$ given in Lemma~4.1, one shows 
that $\mu^{\varsigma_{{\cal C}}}(\frac{w}{\sqrt{t}}) \! = \! 
\sigma_{3} \mu^{\varsigma_{{\cal C}}}(-\frac{w}{\sqrt{t}}) 
\sigma_{3}$ and $\mu^{\varsigma_{{\cal C}}}(\frac{w}{\sqrt{t}
}) \! = \! \sigma_{1} \overline{\mu^{\varsigma_{{\cal C}}}(
\frac{\overline{w}}{\sqrt{t}})} \sigma_{1}$: the results stated 
in the Proposition now follow {}from the above-derived involutions 
for $\mu^{\varsigma_{{\cal C}}}(\frac{w}{\sqrt{t}})$ and its 
corresponding asymptotic expansion given in Proposition~4.3. 
\hfill \rule{6.5pt}{6.5pt}
\begin{fff}
As $\lambda \! \to \! \infty$, $\lambda \! \in \! \Bbb C 
\setminus \varsigma_{{\cal C}}$, $m^{\varsigma}_{{\cal C}}
(\lambda)$ has the following asymptotic expansion,
\begin{eqnarray*}
&m^{\varsigma}_{{\cal C}}(\lambda) = {\rm I} + \sum\limits_{
k \in \Bbb Z_{\geq 1}} \! \frac{({\cal X}^{+}_{2k} e_{11} + 
\overline{{\cal X}^{+}_{2k}} e_{22})}{\lambda^{2k}} + 
\sum\limits_{l \in \Bbb Z_{\geq 0}} \! \frac{({\cal X}^{+}_{
2l+1} \sigma_{+} + \overline{{\cal X}^{+}_{2l+1}} \sigma_{-}
)}{\lambda^{2l+1}},&
\end{eqnarray*}
where ${\cal X}^{+}_{p} \! := \! {\cal X}_{p}^{+}(x,t;\! 
\lambda_{0})$, $p \! \in \! \Bbb Z_{\geq 0}$.
\end{fff}

{\em Proof.\/}
Consequence of the involutions $m^{\varsigma}_{{\cal C}}(\lambda) 
\! = \! \sigma_{3} m^{\varsigma}_{{\cal C}}(-\lambda) \sigma_{3}$ 
and $m^{\varsigma}_{{\cal C}}(\lambda) \! = \! \sigma_{1} 
\overline{m^{\varsigma}_{{\cal C}}(\overline{\lambda})} \sigma_{
1}$. \hfill \rule{6.5pt}{6.5pt}
\begin{ccc}
For $1 \! \leq \! p \! \leq \! 6$ and $0 \! \leq \! q \! \leq 
\! p \! - \! 1$, $u^{+}_{-1,p,q}(\lambda_{0})$ are given by 
the following expressions,
\begin{eqnarray*}
&u^{+}_{-1,1,0}(\lambda_{0}) \! = \! 0, \, \, \, \, \, \, \, u^{
+}_{-1,2,0}(\lambda_{0}) \! = \! - \frac{i R^{\prime}_{+}(0) 
s_{0}^{+}}{\pi \lambda_{0}^{2} 2^{3}}, \, \, \, \, \, \, \, u^{+}
_{-1,2,1}(\lambda_{0}) \! = \! 0,& \\
&u^{+}_{-1,3,0}(\lambda_{0}) \! = \! - \frac{i R^{\prime}_{+}(0) 
s_{0}^{+}}{\pi \lambda_{0}^{2} 2^{3}} \int_{0}^{\infty} \! \mu^{
a}_{11}(\frac{\xi^{1/2} e^{\frac{i \pi}{4}}}{2^{3/2} \lambda_{0}
};\! \lambda_{0}) e^{-\xi} d \xi, \, \, \, \, \, \, \, u^{+}_{-1,
3,1}(\lambda_{0}) \! = \! u^{+}_{-1,3,2}(\lambda_{0}) \! = \! 0,& 
\\
&u^{+}_{-1,4,0}(\lambda_{0}) \! = \! \frac{\widehat{E}_{1} R^{
\prime}_{+}(0) s_{0}^{+}}{\pi \lambda_{0}^{4} 2^{5}}(\frac{(1-
\gamma)}{2} - \ln (2^{3/2} \lambda_{0})) + \frac{\widetilde{E}
_{1} R^{\prime}_{+}(0) s_{0}^{+}}{\pi \lambda_{0}^{4} 2^{5}} + 
\frac{\frac{1}{3!} R^{\prime \prime \prime}_{+}(0) s_{0}^{+}}{
\pi \lambda_{0}^{4} 2^{6}}& \\
&+ \, \frac{R^{\prime}_{+}(0) s_{0}^{+}}{\pi \lambda_{0}^{6} 2^
{6}} - \frac{i R^{\prime}_{+}(0) s_{0}^{+}}{\pi \lambda_{0}^{2} 
2^{3}} \int_{0}^{\infty} \! \mu^{b}_{11}(\frac{\xi^{1/2} e^{
\frac{i \pi}{4}}}{2^{3/2} \lambda_{0}};\! \lambda_{0}) e^{-\xi} 
d \xi,& \\
&u^{+}_{-1,4,1}(\lambda_{0}) \! = \! - \frac{\widehat{E}_{1} R^{
\prime}_{+}(0) s_{0}^{+}}{\pi \lambda_{0}^{4} 2^{6}}, \, \, \, \, 
\, \, \, u^{+}_{-1,4,2}(\lambda_{0}) \! = \! u^{+}_{-1,4,3}(
\lambda_{0}) \! = \! 0,& \\
&u^{+}_{-1,5,0}(\lambda_{0}) \! = \! (-\frac{\widehat{E}_{1} R^{
\prime}_{+}(0) s_{0}^{+} \ln (2^{3/2} \lambda_{0})}{\pi \lambda_
{0}^{4} 2^{5}} \! + \! \frac{\widetilde{E}_{1} R^{\prime}_{+}(0) 
s_{0}^{+}}{\pi \lambda_{0}^{4} 2^{5}} \! + \! \frac{\frac{1}{3!} 
R^{\prime \prime \prime}_{+}(0) s_{0}^{+}}{\pi \lambda_{0}^{4} 
2^{6}}) \! \int_{0}^{\infty} \! \mu^{a}_{11}(\frac{\xi^{1/2} e^{
\frac{i \pi}{4}}}{2^{3/2} \lambda_{0}};\! \lambda_{0}) \xi e^{
-\xi} d \xi& \\
&\, \, \, \, \, \, \, \, \, \, \, \, \, \, \, \, \, \, \, \, 
+ \, \frac{\widehat{E}_{1} R^{\prime}_{+}(0) s_{0}^{+}}{\pi 
\lambda_{0}^{4} 2^{6}} \! \int_{0}^{\infty} \! \mu^{a}_{11}(\frac{
\xi^{1/2} e^{\frac{i \pi}{4}}}{2^{3/2} \lambda_{0}};\! \lambda_{0}
) \xi e^{- \xi} \ln \! \xi d \xi \! + \! \frac{R^{\prime}_{+}(0) 
s_{0}^{+}}{\pi \lambda_{0}^{6} 2^{7}} \! \int_{0}^{\infty} \! \mu
^{a}_{11}(\frac{\xi^{1/2} e^{\frac{i \pi}{4}}}{2^{3/2} \lambda_{0}
};\! \lambda_{0}) \xi^{2} e^{-\xi} d \xi& \\
&\! \! \! \! \! \! \! \! \! \! \! \! \! \! \! \! \! \! \! \! \! 
\! \! \! \! \! \! \! \! \! \! \! \! \! \! \! \! \! \! \! \! \! 
\! \! \! \! \! \! \! \! \! \! \! \! \! \! \! \! \! \! \! \! \! 
\! \! \! \! - \, 
\frac{i R^{\prime}_{+}(0) s_{0}^{+}}{\pi \lambda_{0}^{2} 2^{
3}} \int_{0}^{\infty} \mu^{c}_{11}(\frac{\xi^{1/2} e^{\frac{i 
\pi}{4}}}{2^{3/2} \lambda_{0}};\! \lambda_{0}) e^{-\xi} d \xi,& 
\\
&u^{+}_{-1,5,1}(\lambda_{0}) \! = \! - \frac{\widehat{E}_{1} R^{
\prime}_{+}(0) s_{0}^{+}}{\pi \lambda_{0}^{4} 2^{6}} \! \int_{0}
^{\infty} \! \mu^{a}_{11}(\frac{\xi^{1/2} e^{\frac{i \pi}{4}}}{2
^{3/2} \lambda_{0}};\! \lambda_{0}) \xi e^{-\xi} d \xi \! - \! 
\frac{iR^{\prime}_{+}(0) s_{0}^{+}}{\pi \lambda_{0}^{2} 2^{3}} \! 
\int_{0}^{\infty} \! \mu^{d}_{11}(\frac{\xi^{1/2} e^{\frac{i \pi}
{4}}}{2^{3/2} \lambda_{0}};\! \lambda_{0}) e^{-\xi} d \xi,& \\
&u^{+}_{-1,5,2}(\lambda_{0}) \! = \! u^{+}_{-1,5,3}(\lambda_{0}) 
\! = \! u^{+}_{-1,5,4}(\lambda_{0}) \! = \! 0,& \\
&u^{+}_{-1,6,0}(\lambda_{0}) \! = \! \frac{i R^{\prime}_{+}(0) 
(\widehat{E}_{1})^{2} s_{0}^{+}}{\pi \lambda_{0}^{6}2^{7}}(
\frac{(3/2-\gamma)^{2}+\zeta(2,3)}{2^{2}} \! - \! (\frac{3}
{2} \! - \! \gamma) \ln (2^{3/2} \lambda_{0}) \! + \! (\ln (2^{
3/2} \lambda_{0}))^{2})& \\
&\, \, \, \, \, \, \, \, \, \, \, \, \, \, \, \, \, + 
\, \frac{i R^{\prime}_{+}(0) s_{0}^{+} (2 \widehat{E}_{1} 
\widetilde{E}_{1} + \widehat{E}_{2})}{\pi \lambda_{0}^{6} 2^{7}} 
(\frac{3/2 - \gamma}{2} \! - \! \ln (2^{3/2} \lambda_{
0})) \! + \! \frac{i R^{\prime}_{+}(0) s_{0}^{+} ((\widetilde{E}
_{1})^{2} + \widetilde{E}_{2})}{\pi \lambda_{0}^{6} 2^{7}}& \\
&\! \! \! \! \! \! \! \! \! \! \! \! \! \! \! \! \! \! \! \! \! 
+ \, \frac{3iR^{\prime}_{+}(0) s_{0}^{+}}{\pi \lambda_{0}^{10} 
2^{9}} \! + \! \frac{3 i \frac{1}{3!} R^{\prime \prime \prime}
_{+}(0) s_{0}^{+}}{\pi \lambda_{0}^{8} 2^{9}} \! + \! \frac{3iR
^{\prime}_{+}(0) \widehat{E}_{1} s_{0}^{+}(11/6-\gamma)}{\pi 
\lambda_{0}^{8} 2^{9}}& \\
&\, \, \, \, \, \, \, \, \, \, \, \, \, \, \, \, \, \, \, \, \, 
\, \, \, \, \, \, \, \, \, \, \, \, \, \, \, \, \, \, \, \, + 
\, \frac{3 i R^{\prime}_{+}(0) \widetilde{E}_{1} s_{0}^{+}}{\pi 
\lambda_{0}^{8} 2^{8}} \! - \! \frac{3 i R^{\prime}_{+}(0) 
\widehat{E}_{1} s_{0}^{+} \ln (2^{3/2} \lambda_{0})}{\pi 
\lambda_{0}^{8} 2^{8}} \! + \! \frac{i \frac{1}{3!} R^{\prime 
\prime \prime}_{+}(0) s_{0}^{+} \widehat{E}_{1}}{\pi \lambda_{
0}^{6} 2^{7}}(\frac{3/2 - \gamma}{2} \! - \! \ln (2^{3/2} 
\lambda_{0}))& \\
&\, \, \, \, \, \, \, \, \, \, \, \, \, \, \, \, \, \, \, \, \, 
+ \, 
\frac{i \frac{1}{3!} R^{\prime \prime \prime}_{+}(0) \widetilde{
E}_{1} s_{0}^{+}}{\pi \lambda_{0}^{6} 2^{7}} \! + \! \frac{i 
\frac{1}{5!} R^{V}_{+}(0) s_{0}^{+}}{\pi \lambda_{0}^{6} 2^{8}} 
\! - \! \frac{i R^{\prime}_{+}(0) s_{0}^{+}}{\pi \lambda_{0}^{2} 
2^{3}} \int_{0}^{\infty} \mu^{e}_{11}(\frac{\xi^{1/2} e^{\frac{i 
\pi}{4}}}{2^{3/2} \lambda_{0}};\! \lambda_{0}) e^{-\xi} d \xi& 
\\
&+ \frac{R^{\prime}_{+}(0) \widehat{E}_{1} s_{0}^{+}}{\pi \lambda
_{0}^{4} 2^{6}} \! \int_{0}^{\infty} \! \mu^{b}_{11}(\frac{\xi^{
1/2} e^{\frac{i \pi}{4}}}{2^{3/2} \lambda_{0}};\! \lambda_{0}) 
\xi e^{-\xi} \ln \! \xi d \xi \! - \! \frac{R^{\prime}_{+}(0) 
\widehat{E}_{1} s_{0}^{+} \ln (2^{3/2} \lambda_{0})}{\pi \lambda
_{0}^{4} 2^{5}} \! \int_{0}^{\infty} \! \mu^{b}_{11}(\frac{\xi^{
1/2} e^{\frac{i \pi}{4}}}{2^{3/2} \lambda_{0}};\! \lambda_{0}) 
\xi e^{-\xi} d \xi& \\
&\, \, \, \, \, \, \, \, \, \, \, \, \, \, \, \, \, \, \, \, \, 
\, \, \, \, \, \, \, + \, 
\frac{R^{\prime}_{+}(0) \widetilde{E}_{1} s_{0}^{+}}{\pi 
\lambda_{0}^{4} 2^{5}} \int_{0}^{\infty} \mu^{b}_{11}(\frac{\xi
^{1/2} e^{\frac{i \pi}{4}}}{2^{3/2} \lambda_{0}};\! \lambda_{0}) 
\xi e^{-\xi} d \xi \! + \! \frac{R^{\prime}_{+}(0) s_{0}^{+}}{
\pi \lambda_{0}^{6} 2^{7}} \int_{0}^{\infty} \mu^{b}_{11}(\frac{
\xi^{1/2} e^{\frac{i \pi}{4}}}{2^{3/2} \lambda_{0}};\! \lambda_{
0}) \xi^{2} e^{-\xi} d \xi& \\
&\! \! \! \! \! \! \! \! \! \! \! \! \! \! \! \! \! \! \! \! 
\! \! \! \! \! \! \! \! \! \! \! \! \! \! \! \! \! \! \! \! \! 
+ \, \frac{\frac{1}{3!} R^{\prime \prime \prime}_{+}(0) s_{0}
^{+}}{\pi \lambda_{0}^{4} 2^{6}} \int_{0}^{\infty} \mu^{b}_{1
1}(\frac{\xi^{1/2} e^{\frac{i \pi}{4}}}{2^{3/2} \lambda_{0}};
\! \lambda_{0}) \xi e^{-\xi} d \xi,& \\
&u^{+}_{-1,6,1}(\lambda_{0}) \! = \! - \frac{3 i R^{\prime}_{+}
(0) \widehat{E}_{1} s_{0}^{+}}{\pi \lambda_{0}^{8} 2^{9}} \! - 
\! \frac{i R^{\prime}_{+}(0) (\widehat{E}_{1})^{2} s_{0}^{+}}{
\pi \lambda_{0}^{6} 2^{7}}(\frac{3/2 - \gamma}{2} \! - \! \ln 
(2^{3/2} \lambda_{0})) \! - \! \frac{i R^{\prime}_{+}(0) 
\widehat{E}_{1} \widetilde{E}_{1} s_{0}^{+}}{\pi \lambda_{0}
^{6} 2^{7}}& \\
&\, \, \, \, \, \, \, \, \, \, \, \, \, \, \, \, \, \, \, \, 
\, \, \, - \, \frac{i R^{\prime}_{+}
(0) \widehat{E}_{2} s_{0}^{+}}{\pi \lambda_{0}^{6} 2^{8}} \! 
- \! \frac{i \frac{1}{3!} R^{\prime \prime \prime}_{+}(0) 
\widehat{E}_{1} s_{0}^{+}}{\pi \lambda_{0}^{6} 2^{8}} \! - \! 
\frac{i R^{\prime}_{+}(0) s_{0}^{+}}{\pi \lambda_{0}^{2} 2^{3}} 
\int_{0}^{\infty} \mu^{f}_{11}(\frac{\xi^{1/2} e^{\frac{i \pi}{
4}}}{2^{3/2} \lambda_{0}};\! \lambda_{0}) e^{-\xi} d \xi& \\
&\! \! \! \! \! \! \! \! \! \! \! \! \! \! \! \! \! \! \! \! 
\! \! \! \! \! \! \! \! \! \! \! \! \! \! \! \! \! \! - \, 
\frac{R^{\prime}_{+}(0) \widehat{E}_{1} s_{0}^{+}}{\pi 
\lambda_{0}^{4} 2^{6}} \int_{0}^{\infty} \mu^{b}_{11}(\frac{
\xi^{1/2} e^{\frac{i \pi}{4}}}{2^{3/2} \lambda_{0}};\! \lambda
_{0}) \xi e^{-\xi} d \xi,& \\
&u^{+}_{-1,6,2}(\lambda_{0}) \! = \! \frac{i R^{\prime}_{+}(0) 
(\widehat{E}_{1})^{2} s_{0}^{+}}{\pi \lambda_{0}^{6} 2^{9}}, \, 
\, \, \, \, \, \, u^{+}_{-1,6,3}(\lambda_{0}) \! = \! u^{+}_{
-1,6,4}(\lambda_{0}) \! = \! u^{+}_{-1,6,5}(\lambda_{0}) \! = 
\! 0,&
\end{eqnarray*}
where $s_{0}^{+} \! := \! \exp \{2 i \phi_{+}(\lambda_{0})\}$, 
$R^{\prime}_{+}(0)$, $\frac{1}{3!} R^{\prime \prime \prime}_{
+}(0)$, and $\frac{1}{5!} R^{V}_{+}(0)$ are given in Lemma~4.1, 
$\phi_{+}(\lambda_{0})$, $\widehat{E}_{1}$, $\widetilde{E}_{1}$, 
$\widehat{E}_{2}$, and $\widetilde{E}_{2}$ are given in 
Proposition~4.1, $\gamma \! = \! 0.57721566490$ is Euler's 
constant, $\zeta(2,3) \! := \! \sum_{k \in \Bbb Z_{\geq 0}} \! 
(3 \! + \! k)^{-2} \! = \! \pi^{2}/6 \! - \! 5/4$, $\mu^{a}_{11}
(\cdot;\! \lambda_{0}) \! := \! (\mu^{1}_{0,0}(\cdot;\! \lambda
_{0}))_{11}$, $\mu^{b}_{11}(\cdot;\! \lambda_{0}) \! := \! (\mu
^{2}_{0,0}(\cdot;\! \lambda_{0}))_{11}$, $\mu^{c}_{11}(\cdot;\! 
\lambda_{0}) \! := \! (\mu^{1}_{1,0}(\cdot;\! \lambda_{0}))_{11}$, 
$\mu^{d}_{11}(\cdot;\! \lambda_{0}) \! := \! (\mu^{1}_{1,1}(\cdot;
\! \lambda_{0}))_{11}$, $\mu^{e}_{11}(\cdot;\! \lambda_{0}) \! := 
\! (\mu^{2}_{1,0}(\cdot;\! \lambda_{0}))_{11}$, and $\mu^{f}_{11}
(\cdot;\! \lambda_{0}) \linebreak[4] := \! (\mu^{2}_{1,1}(\cdot;\! 
\lambda_{0}))_{11}$, where $(\mu^{\alpha}_{\beta,\gamma^{\prime}}
(\cdot;\! \lambda_{0}))_{ij}$ denotes the $(i \, j)$th element of 
$\mu^{\alpha}_{\beta,\gamma^{\prime}}(\cdot;\! \lambda_{0})$.
\end{ccc}

{\em Proof.\/} {}From the integral representation for $m^{\varsigma}
_{{\cal C}} \! (\lambda)$ and the expressions for $\widetilde{w}^{
\varsigma_{{\cal C}}^{(k)}}_{l} \! (\cdot)$, $l \! \in \! \{\pm\}$, 
$k \! \in \! \{1,2,3,4\}$, given in Proposition~4.3, one derives 
the following relations for the elements of $m^{\varsigma}_{{\cal 
C}}(\lambda)$:
\begin{eqnarray*}
&(m^{\varsigma}_{{\cal C}}(\lambda))_{11} = 1 + \int_{\varepsilon 
\sqrt{t} \, e^{\frac{3 \pi i}{4}}}^{\sqrt{t} \, 0^{-}} \frac{[
(\mu^{\varsigma_{{\cal C}}}(\xi))_{12}({\cal R}_{{\cal C}}(\xi))^{
\ast} ({\cal P}_{+}(\xi))^{-1} e^{-2({\rm I}_{1} + {\rm I}_{2})(
\xi;\lambda_{0})} e^{2 i t \rho(\xi;\lambda_{0})}] \vert_{\xi = 
w/\sqrt{t}}}{(w-\lambda \sqrt{t} \,)} \frac{d w}{2 \pi 
i}& \\
&\, \, \, \, \, \, \, \, \, \, \, \, \, \, \, \, \, \, \, \, \, + 
\, \int_{\varepsilon \sqrt{t} \, e^{-\frac{i \pi}{4}}}^{\sqrt{t} 
\, 0^{+}} \frac{[(\mu^{\varsigma_{{\cal C}}}(\xi))_{12}({\cal R}
_{{\cal C}}(\xi))^{\ast} ({\cal P}_{+}(\xi))^{-1} e^{-2({\rm I}_{
1} + {\rm I}_{2})(\xi;\lambda_{0})} e^{2 i t \rho(\xi;\lambda_{0}
)}] \vert_{\xi = w/\sqrt{t}}}{(w - \lambda \sqrt{t} \,)} 
\frac{d w}{2 \pi i},& \\
&(m^{\varsigma}_{{\cal C}}(\lambda))_{12} = \int_{\sqrt{t} \, 
0^{+}}^{\varepsilon \sqrt{t} \, e^{\frac{i \pi}{4}}} \frac{
[(\mu^{\varsigma_{{\cal C}}}(\xi))_{11}{\cal R}_{{\cal C}}(\xi)
{\cal P}_{+}(\xi) e^{2({\rm I}_{1} + {\rm I}_{2})(\xi;\lambda_{
0})} e^{-2 i t \rho(\xi;\lambda_{0})}] \vert_{\xi = w/\sqrt{t}}
}{(w - \lambda \sqrt{t} \,)} \frac{d w}{2 \pi i}& \\
&\, \, \, \, \, \, \, \, \, \, \, \, \, \, \, \, \, \, \, \, 
\, \, \, \, \, \, \, \, \, \, \, \, \, \, + \, 
\int_{\sqrt{t} \, 0^{-}}^{\varepsilon \sqrt{t} \, e^{-\frac{3 
\pi i}{4}}} \frac{[(\mu^{\varsigma_{{\cal C}}}(\xi))_{11}{\cal 
R}_{{\cal C}}(\xi){\cal P}_{+}(\xi) e^{2({\rm I}_{1} + {\rm I}_{
2})(\xi;\lambda_{0})} e^{-2 i t \rho(\xi;\lambda_{0})}] \vert_{
\xi = w/\sqrt{t}}}{(w - \lambda \sqrt{t} \,)} \frac{d w}{2 \pi 
i},& \\
&(m^{\varsigma}_{{\cal C}}(\lambda))_{21} = \int_{\varepsilon 
\sqrt{t} \, e^{\frac{3 \pi i}{4}}}^{\sqrt{t} \, 0^{-}} \frac{
[(\mu^{\varsigma_{{\cal C}}}(\xi))_{22}({\cal R}_{{\cal C}}(\xi))
^{\ast} ({\cal P}_{+}(\xi))^{-1} e^{-2({\rm I}_{1} + {\rm I}_{2})
(\xi;\lambda_{0})} e^{2 i t \rho(\xi;\lambda_{0})}] \vert_{\xi 
= w/\sqrt{t}}}{(w - \lambda \sqrt{t} \,)} \frac{d w}{2 \pi i}& \\
&\, \, \, \, \, \, \, \, \, \, \, \, \, \, \, \, \, \, \, \, 
\, \, \, \, \, \, \, \, \, \, \, + \, 
\int_{\varepsilon \sqrt{t} \, e^{-\frac{i \pi}{4}}}^{\sqrt{t} 
\, 0^{+}} \frac{[(\mu^{\varsigma_{{\cal C}}}(\xi))_{22}({\cal 
R}_{{\cal C}}(\xi))^{\ast} ({\cal P}_{+}(\xi))^{-1} e^{-2({\rm I}_{
1} + {\rm I}_{2})(\xi;\lambda_{0})}e^{2 i t \rho(\xi;\lambda_{0})}] 
\vert_{\xi = w/\sqrt{t}}}{(w - \lambda \sqrt{t} \,)} 
\frac{d w}{2 \pi i},& \\
&(m^{\varsigma}_{{\cal C}}(\lambda))_{22} = 1 + \int_{\sqrt{t} \, 
0^{+}}^{\varepsilon \sqrt{t} \, e^{\frac{i \pi}{4}}} \frac{
[(\mu^{\varsigma_{{\cal C}}}(\xi))_{21}{\cal R}_{{\cal C}}(\xi){
\cal P}_{+}(\xi) e^{2({\rm I}_{1} + {\rm I}_{2})(\xi;\lambda_{0})} 
e^{-2 i t \rho(\xi;\lambda_{0})}] \vert_{\xi = w/\sqrt{t}}}{(w - 
\lambda \sqrt{t} \,)} \frac{d w}{2 \pi i}& \\
&\, \, \, \, \, \, \, \, \, \, \, \, \, \, \, \, \, \, \, \, 
\, \, \, \, \, + \, 
\int_{\sqrt{t} \, 0^{-}}^{\varepsilon \sqrt{t} \, e^{-\frac{
3 \pi i}{4}}} \frac{[(\mu^{\varsigma_{{\cal C}}}(\xi))_{21}{\cal 
R}_{{\cal C}}(\xi){\cal P}_{+}(\xi) e^{2({\rm I}_{1} + {\rm I}_{
2})(\xi;\lambda_{0})} e^{-2 i t \rho(\xi;\lambda_{0})}] \vert_{
\xi = w/\sqrt{t}}}{(w - \lambda \sqrt{t} \,)} \frac{d w}{2 \pi 
i}.& \\
\end{eqnarray*}
Since the linear contribution to the asymptotic expansion of $Q(x,
t)$ (as $t \! \to \! +\infty$ and $x \! \to \! -\infty$ such that 
$\lambda_{0} \! > \! M$ and $(x,t) \! \in \! \Bbb R^{2} \setminus 
\Omega_{n}$, for those $\gamma_{n} \! \in \! (\frac{\pi}{2},\! 
\pi))$ {}from the first-order stationary phase point at the origin 
is given by the limit $2 i \lim_{\lambda \to \infty \atop \lambda 
\, \in \, \Bbb C \setminus \varsigma_{{\cal C}}} \! (\lambda m^{
\varsigma}_{{\cal C}}(\lambda))_{12}$, the $(1 \, 2)$-element of 
$m^{\varsigma}_{{\cal C}}(\lambda)$ is analysed explicitly 
hereafter. Substituting the asymptotic expansion for ${\cal R}_{{
\cal C}}(\xi){\cal P}_{+}(\xi) \exp\{2({\rm I}_{1} \! + \! {\rm I}
_{2})(\xi;\! \lambda_{0})\} \exp\{-2it \rho(\xi;\! \lambda_{0})\} 
\vert_{\xi = \frac{w}{\sqrt{t}}}$ given in Proposition~4.2 and the 
asymptotic expansion for $\mu^{\varsigma_{{\cal C}}}(\frac{w}{
\sqrt{t}})$ given in Proposition~4.3 (with the definitions of 
$(\mu^{i^{\prime}}_{j,k}(\cdot;\! \lambda_{0}))_{11}$ given in the 
Lemma) into the above integral representation for $(m^{\varsigma}
_{{\cal C}}(\lambda))_{12}$, expanding $(w \! - \! \lambda \sqrt{
t} \,)^{-1}$ into a geometric progression in reciprocal powers of 
$\lambda \sqrt{t}$, changing variables of integration twice, 
letting the upper (resp.~lower) limit of integration of the 
resulting integrals tend to $+ \infty$ (resp.~$0)$, using the 
following integrals \cite{a21}, $\int_{0}^{\infty} \! x^{\nu_{1}-1} 
\exp\{-\mu_{1} x\} \ln \! x \, d x \! = \! \mu_{1}^{-\nu_{1}} \Gamma 
(\nu_{1})(\psi(\nu_{1}) \! - \! \ln (\mu_{1}))$ $(\Re(\mu_{1}) \! 
> \! 0$, $\Re(\nu_{1}) \! > \! 0)$ and $\int_{0}^{\infty} \! x^{\nu
_{2}-1} \exp\{-\mu_{2} x\} (\ln \! x)^{2} d x \! = \! \mu_{2}^{-\nu
_{2}} \Gamma (\nu_{2}) ((\psi (\nu_{2}) \! - \! \ln (\mu_{2}))^{
2} \! + \! \zeta(2,\nu_{2}))$ $(\Re(\mu_{2}) \! > \! 0$, $\Re(
\nu_{2}) \! > \! 0)$, where $\Gamma(\cdot)$ is the gamma function 
\cite{a21}, $\psi(\cdot)$ is the psi function $(\psi(z) \! := \! 
\frac{d \ln \Gamma(z)}{d z})$ \cite{a21}, and $\zeta(z,q) \! := 
\! \sum_{k \in \Bbb Z_{\geq 0}} (q \! + \! k)^{-z}$ $(\Re(z) \! 
> \! 1$, $q \! \not\in \! \Bbb Z \setminus \Bbb N)$ is Riemann's 
zeta function \cite{a21}, the following relation for the psi 
function \cite{a21}, $\psi(z \! + \! 1) \! = \! \psi(z) \! + \! 
\frac{1}{z}$, the fact that $\gamma \! = \! -\psi(1)$ \cite{a21}, 
where $\gamma$ is Euler's constant given in the Lemma, the 
well-known series result $\sum_{p \in \Bbb Z_{\geq 1}} \! p^{-2} \! 
= \! \pi^{2}/6$, and the involutive properties of $\mu^{p^{\prime}
}_{p,q}(\cdot;\! \lambda_{0})$ given in Proposition~4.4, after 
integration and simplification, and using the fact that $2 i \lim
_{\lambda \to \infty \atop \lambda \, \in \, \Bbb C \setminus 
\varsigma_{{\cal C}}} \! (\lambda m^{\varsigma}_{{\cal C}}(
\lambda))_{12} \! = \! \sum_{p=1}^{6} \sum_{q=0}^{p-1} \frac{u
^{+}_{-1,p,q}(\lambda_{0})(\ln t)^{q}}{t^{p/2}} \! + \! {\cal 
O}(K_{{\cal C}}^{\varsigma}(\lambda_{0}) (\ln t)^{2} t^{-7/2})$, 
one obtains the results for $u^{+}_{-1,p,q}(\lambda_{0})$, $1 \! 
\leq \! p \! \leq \! 6$, $0 \! \leq \! q \! \leq \! p \! - \! 1$, 
stated in the Lemma upon equating coefficients of like powers of 
$(\ln t)^{q} t^{-p/2}$ on both sides of the above relation, where 
$K_{{\cal C}}^{\varsigma}(\lambda_{0}) \! := \! \frac{iR^{\prime}
_{+}(0)(\widehat{E}_{1})^{2} s_{0}^{+}}{\pi \lambda_{0}^{6} 2^{
10}} \! \int_{0}^{\infty} \! \mu^{a}_{11}(\frac{\xi^{1/2} e^{
\frac{i \pi}{4}}}{2^{3/2} \lambda_{0}};\! \lambda_{0}) \xi^{2} 
e^{-\xi} d \xi \! - \! \frac{R^{\prime}_{+}(0) \widehat{E}_{1} 
s_{0}^{+}}{\pi \lambda_{0}^{4} 2^{6}} \! \int_{0}^{\infty} \! 
\mu^{d}_{11}(\frac{\xi^{1/2} e^{\frac{i \pi}{4}}}{2^{3/2} 
\lambda_{0}} \linebreak[4];\! \lambda_{0}) \xi e^{-\xi} d \xi \! 
- \! \frac{i R^{\prime}_{+}(0) s_{0}^{+}}{\pi \lambda_{0}^{2} 2
^{3}} \! \int_{0}^{\infty} \! \mu^{i}_{11}(\frac{\xi^{1/2} e^{
\frac{i \pi}{4}}}{2^{3/2} \lambda_{0}};\! \lambda_{0}) e^{-\xi} 
d \xi$, with $\mu^{i}_{11}(\cdot;\! \lambda_{0}) \! := \! (\mu
^{1}_{2,2}(\cdot;\! \lambda_{0}))_{11}$. Now, continuing the 
above-mentioned procedure to higher orders in $\lambda^{-1}$, 
one verifies that, as $\lambda \! \to \! \infty$, $\lambda \! 
\in \! \Bbb C \setminus \! \varsigma_{{\cal C}}$, 
$(m^{\varsigma}_{{\cal C}}(\lambda))_{12}$ 
satisfies---precisely---the asymptotic expansion condition 
stated in Corollary~4.1; furthermore, using the asymptotic 
expansion for $({\cal R}_{{\cal C}}(\xi))^{\ast} ({\cal P}_{
+}(\xi))^{-1} \exp\{-2({\rm I}_{1} \! + \! {\rm I}_{2})(\xi;
\! \lambda_{0})\} \exp\{2it \rho(\xi;\! \lambda_{0})\} \vert
_{\xi = \frac{w}{\sqrt{t}}}$ given in Proposition~4.2 and 
proceeding analogously as above, one shows, using the involutive 
properties of $\mu^{p^{\prime}}_{p,q}(\cdot;\! \lambda_{0})$ 
given in Proposition~4.4, that the remaining elements of $m^{
\varsigma}_{{\cal C}}(\lambda)$, namely, $(m^{\varsigma}_{{\cal 
C}}(\lambda))_{11}$, $(m^{\varsigma}_{{\cal C}}(\lambda))_{21}$ 
and $(m^{\varsigma}_{{\cal C}}(\lambda))_{22}$, satisfy, as 
$\lambda \! \to \! \infty$, $\lambda \! \in \! \Bbb C \setminus 
\varsigma_{{\cal C}}$, the asymptotic expansion conditions stated 
in Corollary~4.1: one also verifies that $m^{\varsigma}_{{\cal C}
}(\lambda)$ as constructed above satisfies the following symmetry 
reductions, $m^{\varsigma}_{{\cal C}}(\lambda) \! = \! \sigma_{3} 
m^{\varsigma}_{{\cal C}}(-\lambda) \sigma_{3}$ and $m^{\varsigma}
_{{\cal C}}(\lambda) \! = \! \sigma_{1} \overline{m^{\varsigma}_{
{\cal C}}(\overline{\lambda})} \sigma_{1}$. 
\hfill \rule{6.5pt}{6.5pt}
\begin{ccc}
As $t \! \to \! + \infty$ and $x \! \to \! - \infty$ such that 
$\lambda_{0} \! > \! M$ and $(x,t) \! \in \! \Bbb R^{2} \setminus 
\Omega_{n}$, for those $\gamma_{n} \! \in \! (\frac{\pi}{2},\pi)$, 
$Q(x,t)$ has the asymptotic expansion given in Theorem~2.2, 
Eqs.~(10)--(27).
\end{ccc}

{\em Proof.\/}
Recalling that, for $k \! = \! \pm 1, \pm 3, \ldots$, $p \! \geq 
\! \vert k \vert$, and $0 \! \leq \! q \! \leq \! p \! - \! \vert 
k \vert$, $u^{+}_{k,p} \! := \! \sum_{q=0}^{p-\vert k \vert} \! 
u^{+}_{k,p,q}(\lambda_{0}) \linebreak[4] \cdot (\ln t)^{q}$, and, 
for $q \! > \! p \! - \! \vert k \vert$, $u^{+}_{k,p} \! \equiv 
\! 0$, in accordance with the asymptotic expansion for $Q(x,\! t)$ 
given in Lemma~3.4, one must determine $u^{+}_{\pm 1,1}$, $u^{+}_{
\pm 1,2}$, $u^{+}_{\pm 1,3}$, $u^{+}_{\pm 1,4}$, $u^{+}_{\pm 3,3}$, 
and $u^{+}_{\pm 3,4}$: for this purpose, the recurrence relations 
for the coefficients of the asymptotic expansion given in the 
Appendix are necessary. The explicit representation for $u^{
+}_{-1,p}$, $1 \! \leq \! p \! \leq \! 4$, are already given in 
Lemma~4.2; therefore, it remains to determine the remaining 
coefficients listed above (throughout the proof, all explicit 
$\lambda_{0}$ dependences are suppressed, except where absolutely 
necessary). {}From Eqs.~(A.1) and (A.2), one obtains 
trivial identities. {}From Eq.~(A.3), one shows that $\nu u^{+}_{1,
1} \! + \! i t \dot{u}^{+}_{1,1} \! - \! 2 \lambda_{0}^{2} \vert 
u^{+}_{1,1} \vert^{2} u^{+}_{1,1} \! = \! 0$: using the fact that 
$u^{+}_{1,1} \! = \! u^{+}_{1,1,0}(\lambda_{0})$, and recalling 
{}from Theorem~2.2, Eq.~(14) that $u^{+}_{1,1,0} \! = \! \sqrt{\nu 
/ (2 \lambda_{0}^{2})} \exp\{i(\phi^{+}(\lambda_{0}) \! + \! 
\widehat{\Phi}^{+}(\lambda_{0}) \! + \! \pi / 2)\} \! := \! 
\sqrt{\nu / (2 \lambda_{0}^{2})} \exp\{i \varphi_{+}\}$, one shows 
that $\dot{u}^{+}_{1,1,0}(\lambda_{0}) \! = \! 0$ and that the above 
equation is satisfied. Using the fact that $u^{+}_{-1,1} \! = \! 
u^{+}_{-1,1,0}(\lambda_{0}) \! = \! 0$ (Lemma~4.2), one notes that 
Eq.~(A.4) is satisfied identically. Noting that $u^{+}_{\pm 3,3} 
\! = \! u^{+}_{\pm 3,3,0}(\lambda_{0})$, Eqs.~(A.5) and (A.6) show 
that $u^{+}_{\pm 3,3,0}(\lambda_{0}) \! = \! 0$. {}From Eq.~(A.7), 
one shows that $- \nu u^{+}_{1,2} \! + \! i t \dot{u}^{+}_{1,2} \! 
- \! \frac{i}{2} u^{+}_{1,2} \! - \! \overline{u^{+}_{1,2}} \nu 
\exp\{2i \varphi_{+}\} \! = \! 0$: substituting $u^{+}_{1,2} \! = 
\! \sum_{q=0}^{1} \! u^{+}_{1,2,q}(\lambda_{0})(\ln t)^{q}$ into 
this equation, one deduces that $-\nu u^{+}_{1,2,0}(\lambda_{0}) 
\! + \! i u^{+}_{1,2,1}(\lambda_{0}) \! - \! \frac{i}{2} u^{+}_{
1,2,0}(\lambda_{0}) \! - \! \overline{u^{+}_{1,2,0}(\lambda_{0})} 
\nu \exp\{2i \varphi_{+}\} \! = \! 0$ and $-\nu u^{+}_{1,2,1}(
\lambda_{0}) \! - \! \frac{i}{2} u^{+}_{1,2,1}(\lambda_{0}) \! - 
\! \overline{u^{+}_{1,2,1}(\lambda_{0})} \nu \exp\{2i \varphi_{+}
\} \! = \! 0$. Defining, now, $u^{+}_{1,2,1}(\lambda_{0}) \! := \! 
u^{+,{\rm r}}_{1,2,1}(\lambda_{0}) \! + \! i u^{+,{\rm i}}_{1,2,1}
(\lambda_{0})$, one shows {}from the above equation for $u^{+}_{1,2,
1}(\lambda_{0})$ that $\widetilde{{\cal M}}(\lambda_{0})(u^{+,{\rm 
r}}_{1,2,1}(\lambda_{0}),\! u^{+,{\rm i}}_{1,2,1}(\lambda_{0}))^{{
\rm T}} \! = \! (0,0)^{{\rm T}}$, where $\widetilde{{\cal M}}
(\lambda_{0}) \! := \! 
\left( \begin{array}{cc}
\! \! \! -\nu(1 \! + \! \cos 2\varphi_{+}) & \! (\frac{1}{2} \! - 
\! \nu \sin 2\varphi_{+}) \! \! \! \\
\! \! \! -(\frac{1}{2} \! + \! \nu \sin 2\varphi_{+}) & \! -\nu 
(1 \! - \! \cos 2\varphi_{+}) \! \! 
\end{array} \right)$, 
and ${\rm T}$ denotes transposition on the 
fiber bundle $\Bbb C^{2}$; however, since $\widetilde{{\cal M}}
(\lambda_{0})$ is non-degenerate $(\det (\widetilde{{\cal M}}
(\lambda_{0})) \! = \! \frac{1}{4})$, {}from a 
well-known result in linear algebra, one gets the 
trivial solution, namely, $u^{+,{\rm r}}_{1,2,1}(\lambda_{0}) \! 
= \! u^{+,{\rm i}}_{1,2,1}(\lambda_{0}) \! = \! 0$. Defining $u^{
+}_{1,2,0}(\lambda_{0}) \! := \! u^{+,{\rm r}}_{1,2,0}(\lambda_{0}
) \! + \! i u^{+,{\rm i}}_{1,2,0}(\lambda_{0})$, and reasoning 
similarly as above, one shows that $u^{+,{\rm r}}_{1,2,0}(\lambda
_{0}) \! = \! u^{+,{\rm i}}_{1,2,0}(\lambda_{0}) \! = \! 0$; hence, 
$u^{+}_{1,2} \! = \! 0$. Using the expressions for $u^{+}_{-1,2,0}
(\lambda_{0})$ and $u^{+}_{-1,2,1}(\lambda_{0})$ given in 
Lemma~4.2 and recalling that $u^{+}_{-1,2} \! = \! \sum_{q=0}^{1} 
\! u^{+}_{-1,2,q}(\lambda_{0})(\ln t)^{q}$, one shows that 
Eq.~(A.8) is satisfied identically. Using the fact that $u^{+}_{
\pm 3,4} \! = \! \sum_{q=0}^{1} \! u^{+}_{\pm 3,4,q}(\lambda_{0})
(\ln t)^{q}$, one shows, using Eqs.~(A.9) and (A.10), that $u^{+}
_{\pm 3,4} \! = \! 0$. Noting {}from the above discussion that $u^{
+}_{-1,1} \! = \! u^{+}_{1,2} \! = \! u^{+}_{\pm 3,3} \! = \! 0$, 
and substituting $u^{+}_{1,3} \! = \! \sum_{q=0}^{2} \! u^{+}_{1,
3,q}(\lambda_{0})(\ln t)^{q}$ into Eq.~(A.11), one obtains three 
linear inhomogeneous algebraic equations for the determination of 
$u^{+}_{1,3,q}(\lambda_{0})$, $0 \! \leq \! q \! \leq \! 2$: 
proceeding as above for the determination of $u^{+}_{1,2}$, one 
solves, via Cramer's rule, these linear inhomogeneous algebraic 
equations for $u^{+}_{1,3,q}(\lambda_{0})$ and derives the 
results stated in Theorem~2.2, Eqs.~(22)--(27). Recalling that 
$u^{+}_{-1,1} \! = \! u^{+}_{1,2} \! = \! u^{+}_{3,3} \! = \! 
0$, one simplifies Eq.~(A.12) to $i t \dot{u}^{+}_{-1,3} \! - \! 
\frac{3i}{2} u^{+}_{-1,3} \! - \! \frac{i \lambda_{0}}{2}(u^{
+}_{-1,3})^{\prime} \! - \! 2 \nu u^{+}_{-1,3} \! = \! 0$. Using 
the representation $u^{+}_{-1,3} \! = \! \sum_{q=0}^{2} \! u^{+}
_{-1,3,q}(\lambda_{0})(\ln t)^{q}$, and the expressions for $u^{
+}_{-1,3,q}(\lambda_{0})$, $0 \! \leq \! q \! \leq \! 2$, given 
in Lemma~4.2, namely, $u^{+}_{-1,3,2}(\lambda_{0}) \! = \! u^{+}
_{-1,3,1}(\lambda_{0}) \! = \! 0$ and $u^{+}_{-1,3,0}(\lambda_{0}
) \! = \! - \frac{i R_{+}^{\prime}(0) s_{0}^{+} {\cal I}(\lambda
_{0})}{\pi \lambda_{0}^{2} 2^{3}}$, with $R_{+}^{\prime}(0)$ given 
in Lemma~4.1, $s_{0}^{+} \! := \! \exp\{2i\phi_{+}(\lambda_{0})\}$, 
$\phi_{+}(\lambda_{0}) \! := \! - \int_{0}^{\lambda_{0}} \! \frac{
\ln (1 - \vert r(\xi) \vert^{2})}{\xi} \frac{d \xi}{\pi} \! + \! 
\int_{0}^{\infty} \! \frac{\ln (1+\vert r(i \xi) \vert^{2})}{\xi} 
\frac{d \xi}{\pi}$, and ${\cal I}(\lambda_{0}) \! := \! \int_{0}
^{\infty} \! \mu^{a}_{11}(\frac{\xi^{1/2} e^{\frac{i \pi}{4}}}{2
^{3/2} \lambda_{0}};\! \lambda_{0}) e^{-\xi} d \xi$, one reduces 
this (characteristic-like) equation to $-\frac{3i}{2} u^{+}_{-1,
3,0}(\lambda_{0}) \! - \! \frac{i \lambda_{0}}{2} \frac{\partial 
u^{+}_{-1,3,0}(\lambda_{0})}{\partial \lambda_{0}} \! - \! 2 \nu 
u^{+}_{-1,3,0}(\lambda_{0}) \! = \! 0$, which, upon recalling the 
above expression for $u^{+}_{-1,3,0}(\lambda_{0})$, is an 
integro-differential equation for $\mu^{a}_{11}(\cdot;\! \lambda
_{0})$: substituting the expression for $u^{+}_{-1,3,0}(\lambda_{
0})$ into the latter equation and recalling the definition of $
\nu$, namely, $\nu \! := \! - \frac{1}{2\pi} \ln (1 \! - \! \vert 
r(\lambda_{0}) \vert^{2})$, one shows that ${\cal I}(\lambda_{0})$ 
satisfies the following ODE, $\frac{\partial {\cal I}(\lambda_{0}
)}{\partial \lambda_{0}} \! + \! \frac{{\cal I}(\lambda_{0})}{
\lambda_{0}} \! = \! 0$; hence, ${\cal I}(\lambda_{0}) \! = \! 
\frac{{\cal K}_{1}^{+}}{\lambda_{0}}$, for some bounded, $\lambda
_{0}$-independent constant ${\cal K}_{1}^{+} \! \in \! \Bbb C$. 
Therefore, modulo the determination of ${\cal K}_{1}^{+}$, $u^{+}
_{-1,3,0}(\lambda_{0}) \! = \! - \frac{i R_{+}^{\prime}(0) s_{0}
^{+} {\cal K}_{1}^{+}}{\pi \lambda_{0}^{3} 2^{3}}$. {}From Eq.~(A.13), 
the representation 
$u^{+}_{3,5} \! = \! \sum_{q=0}^{2} \! u^{+}_{3,5,q}(\lambda_{0}
)(\ln t)^{q}$, and the results obtained above for the preceding 
coefficients, one shows that $u^{+}_{3,5,0}(\lambda_{0}) \! = \! 
u^{+}_{3,5,1}(\lambda_{0}) \! = \! u^{+}_{3,5,2}(\lambda_{0}) \! 
= \! 0$; hence, $u^{+}_{3,5} \! = \! 0$. {}From Eq.~(A.14), the 
representation $u^{+}_{-3,5} \! = \! \sum_{q=0}^{2} \! u^{+}_{-3,
5,q}(\lambda_{0})(\ln t)^{q}$, and the results obtained above for 
the preceding coefficients, one shows that $u^{+}_{-3,5,2}(\lambda
_{0}) \! = \! u^{+}_{-3,5,1}(\lambda_{0}) \! = \! 0$, and $u^{+}
_{-3,5,0}(\lambda_{0}) \! = \! - \frac{\overline{u^{+}_{1,1,0}(
\lambda_{0})}(u^{+}_{-1,2,0}(\lambda_{0}))^{2}}{4 \lambda_{0}^{
2}}$, with $u^{+}_{1,1,0}(\lambda_{0})$ given earlier in the 
proof, and $u^{+}_{-1,2,0}(\lambda_{0})$ given in Lemma~4.2. 
{}From Eqs.~(A.15) and (A.16), and the representations $u^{+}_{\pm 
5,5} \! = \! \sum_{q=0}^{0} \! u^{+}_{\pm 5,5,q}(\lambda_{0})(\ln 
t)^{q}$, one shows that $u^{+}_{\pm 5,5,0} \! = \! 0$; hence, $u
^{+}_{\pm 5,5} \! = \! 0$. {}From Eq.~(A.17) and the representation 
$u^{+}_{1,4} \! = \! \sum_{q=0}^{3} \! u^{+}_{1,4,q}(\lambda_{0})
(\ln t)^{q}$, one obtains four linear homogeneous algebraic 
equations for the determination of $u^{+}_{1,4,q}(\lambda_{0})$, 
$0 \! \leq \! q \! \leq \! 3$: decomposing $u^{+}_{1,4,q}(\lambda
_{0})$ into real and imaginary parts, $u^{+}_{1,4,q}(\lambda_{0}) 
\! := \! u^{+,{\rm r}}_{1,4,q}(\lambda_{0}) \! + \! i u^{+,{\rm i}
}_{1,4,q}(\lambda_{0})$, and noting that the coefficient matrices 
for the resulting linear homogeneous systems are non-singular, 
{}from a well-known result in linear algebra, one deduces that all 
these systems have the trivial solution, i.e., $u^{+}_{1,4,q}(
\lambda_{0}) \! = \! 0$, $0 \! \leq \! q \! \leq \! 3$; 
hence, $u^{+}_{1,4} \! = \! 0$. Substituting $u^{+}_{-1,4} \! 
= \! \sum_{q=0}^{3} \! u^{+}_{-1,4,q}(\lambda_{0})(\ln t)^{q}$ into 
Eq.~(A.18), using the expressions for $u^{+}_{-1,4,q}(\lambda_{0})$, 
$0 \! \leq \! q \! \leq \! 3$, given in Lemma~4.2, and equating 
coefficients of powers of like terms of $(\ln t)^{l}$, $0 \! \leq 
\! l \! \leq \! 2$, on both sides of the resulting equations, one 
shows that the equations resulting {}from the $(\ln t)^{2}$ and $\ln 
t$ parts are satisfied identically, and that the constant $({\cal 
O}(1))$ part gives rise to an integro-differential equation for 
$\mu^{b}_{11}(\cdot;\! \lambda_{0})$, which can be simplified to 
an integral equation of the first kind for $\mu^{b}_{11}(\cdot;\! 
\lambda_{0})$ of the following form, $\int_{0}^{\infty} \! \mu^{
b}_{11}(\frac{\xi^{1/2} e^{\frac{i \pi}{4}}}{2^{3/2} \lambda_{0}};
\! \lambda_{0}) e^{-\xi} d \xi \! = \! \widehat{{\cal G}}(\lambda
_{0})$, where $\widehat{{\cal G}}(\lambda_{0})$ is not written 
down here since it will not actually be needed (the details of 
the analysis of this integral equation for $\mu^{b}_{11}(\cdot;
\! \lambda_{0})$ will be presented elsewhere: see, also, 
Remark~4.2 below): this completes the proof. 
\hfill \rule{6.5pt}{6.5pt}
\begin{eee}
{\rm In the proof of Lemma~4.3 above, there appears a bounded, 
$\lambda_{0}$-independent complex-valued constant denoted as 
${\cal K}_{1}^{+}$. The determination of ${\cal K}_{1}^{+}$ 
requires the solution of the following integral equation of the 
first kind for $\mu^{a}_{11}(\cdot;\! \lambda_{0})$, $\int_{0}
^{\infty} \! \mu^{a}_{11}(\frac{\xi^{1/2} e^{\frac{i \pi}{4}}}{
2^{3/2} \lambda_{0}};\! \lambda_{0}) e^{-\xi} d \xi \! = \! {\cal 
K}_{1}^{+} / \lambda_{0}$, under the following conditions imposed 
on $\mu^{a}_{11}(\cdot;\! \lambda_{0})$: (1) $\mu^{a}_{11}(z;\! 
\lambda_{0})$ is an even function with respect to $z$, i.e., 
$\mu^{a}_{11}(-z;\! \lambda_{0}) \! = \! \mu^{a}_{11}(z;\! \lambda
_{0})$; and (2) $\mu^{a}_{11}(z;\! \lambda_{0}) \! \in \! {\cal L}
^{2}(\Bbb R_{\geq 0};\! \Bbb C)$, i.e., $\vert \vert \mu^{a}_{11}
(\cdot;\! \lambda_{0}) \vert \vert_{{\cal L}^{2}(\Bbb R_{\geq 0};
\Bbb C)} \! < \! \infty$. In fact, one notes that, for $\lambda_{
0} \! > \! M$ and ${\cal K}_{1}^{+}$ bounded, $\vert \vert \mu^{a}
_{11}(\frac{(\cdot) e^{\frac{i \pi}{4}}}{2^{3/2} \lambda_{0}};\! 
\lambda_{0}) \vert \vert_{{\cal L}^{k}(\Bbb R_{\geq 0};\Bbb C)} 
\! < \! \infty$, $k \! \in \! \{1,2,\infty\}$. Changing the 
variable of integration according to the rule $\xi^{1/2} / 
\lambda_{0} \! \to \! \sqrt{\tau}$, setting $s \! = \! \lambda_{
0}^{2}$ (with $\Re(s) \! = \! \Re(\lambda_{0}^{2}) \! > \! 0)$, 
and defining $({\cal K}_{1}^{+})^{-1} \mu^{a}_{11}(\frac{\sqrt{
\tau} e^{\frac{i \pi}{4}}}{2^{3/2}};\! \sqrt{s} \,) \! := \! g(
\tau;\! \sqrt{s} \,)$, one also shows that $g(\tau;\! \sqrt{s} 
\,)$ satisfies the following integral equation, $\int_{0}^{\infty} 
\! g(\tau;\! \sqrt{s} \,) e^{-s \tau} d \tau \! = \! s^{-3/2}$, 
with the standard Laplace transform-type kernel, $e^{-s \tau}$. 
Actually, one can show that, for any $f(z)$ for which $\int_{0}
^{\infty} \! \vert f(\xi) \vert^{2} d \xi \! := \! \vert \vert 
f(\cdot) \vert \vert_{{\cal L}^{2}(\Bbb R_{\geq 0};\Bbb C)}^{2} 
\! < \! \infty$ and $\int_{0}^{\infty} \! f(\xi) e^{- \xi}d \xi 
\! := \! {\rm A}(f) \! < \! \infty$, $\mu^{a}_{11}(z;\! \lambda
_{0}) \! := \! \frac{{\cal K}_{1}^{+}f(8i \lambda_{0}^{2} z^{2}
)}{\lambda_{0} {\rm A}(f)}$ solves the integral equation $\int
_{0}^{\infty} \! \mu^{a}_{11}(\frac{\xi^{1/2}e^{\frac{i \pi}{4}
}}{2^{3/2} \lambda_{0}};\! \lambda_{0}) e^{-\xi} d \xi \! = \! 
{\cal K}_{1}^{+} / \lambda_{0}$, and satisfies conditions~(1) 
and (2) stated above. The explicit determination of ${\cal K}
_{1}^{+}$ is rather involved and somewhat circuitous, and the 
details of this calculation will be presented elsewhere.\/}
\end{eee}
\section{Asymptotic Evaluation of $((\Psi^{-1}(x,t;0))_{11})^{
2}$ as $t \! \to \! +\infty$}
In this section, the phase integral, $((\Psi^{-1}(x,t;0))_{11})
^{2}$, which appears in the gauge transformation (Eq.~(3) and 
Proposition~2.3), and is also equal to $((m^{-1}(x,t;0))_{11})
^{2}$ (Lemma~2.1), is evaluated as $t \! \to \! +\infty$ $(x/t 
\! \sim \! {\cal O}(1))$ and $(x,t) \! \in \! \Bbb R^{2} 
\setminus \Omega_{n}$, for those $\gamma_{n} \! \in \! (\frac{
\pi}{2},\! \pi)$.
\begin{bbb}[{\rm \cite{a2}}]
For $Q(x,0) \! \in \! {\cal S}(\Bbb R;\! \Bbb C)$,
\begin{eqnarray*}
&\Psi(x,t;0) \! = \! m(x,t;0) \! = \! \exp\{-\frac{i \sigma_{
3}}{2} \! \int_{\infty}^{x} \vert Q(\xi,t) \vert^{2} d \xi\}.&
\end{eqnarray*}
\end{bbb}
\begin{bbb}[{\rm \cite{a2}}]
For $Q(x,0) \! \in \! {\cal S}(\Bbb R;\! \Bbb C)$,
\begin{eqnarray*}
&\vert \vert Q(\cdot,t) \vert \vert_{{\cal L}^{2}(\Bbb R;\Bbb 
C)}^{2} \! = \! \frac{2}{\pi}(\int_{0}^{\infty} \! \frac{\ln(1 + 
\vert r(i \xi) \vert^{2})}{\xi} d \xi \! - \! \int_{0}^{\infty} 
\! \frac{\ln(1 - \vert r(\xi) \vert^{2})}{\xi} d \xi).&
\end{eqnarray*}
\end{bbb}
\begin{ccc}
As $t \! \to \! +\infty$ and $x \! \to \! -\infty$ such that 
$\lambda_{0} \! > \! M$ and $(x,t) \! \in \! \Bbb R^{2} \setminus 
\Omega_{n}$, for those $\gamma_{n} \! \in \! (\frac{\pi}{2},\! 
\pi)$,
\begin{eqnarray*}
&((\Psi^{-1}(x,t;0))_{11})^{2} \! = \! \exp\{i \arg q_{+}(x,t)\} 
\! + \! {\cal O} \! \left(\frac{c^{{\cal S}}(\lambda_{0})(\ln t)
^{2}}{(\lambda_{0}^{2} t)^{3/2}} \right) \!,&
\end{eqnarray*}
where $\arg q_{+}(x,t)$ is given in Theorem~2.3, Eqs.~(46)--(51).
\end{ccc}

{\em Proof.\/} Since, {}from Eq.~(3), Proposition~2.3, Lemma~2.1, 
Lemma~2.2 (Eq.~(8)), and Proposition~5.1, $q(x,t) \! = \! Q(x,t)
((\Psi^{-1}(x,t;\! 0))_{11})^{2} \! = \! Q(x,t)((m^{-1}(x,t;\! 
0))_{11})^{2} \! = \! Q(x,t) \exp\{i \! \int_{\infty}^{x} \! 
\vert Q \linebreak[4] 
(\xi,t) \vert^{2} d \xi\}$, and $Q(x,t)$ is already given in 
Theorem~2.2, Eqs.~(10)--(27), it remains to evaluate the integral 
$\int_{\infty}^{x} \! \vert Q(\xi,t) \vert^{2} d \xi$. Writing 
$\int_{\infty}^{x} \! \vert Q(\xi,t) \vert^{2} d \xi \! = \! 
\int_{\infty}^{-\infty} \! \vert Q(\xi,t) \vert^{2} d \xi \! + 
\! \int_{-\infty}^{x} \! \vert Q(\xi,t) \vert^{2} d \xi \! = \! 
- \vert \vert Q(\cdot,t) \vert \vert_{{\cal L}^{2}(\Bbb R;\Bbb 
C)}^{2} \! + \! \int_{-\infty}^{x} \! \vert Q(\xi,t) \vert^{2} 
d \xi$, using the result for $\vert \vert Q(\cdot,t) \vert \vert
_{{\cal L}^{2}(\Bbb R;\Bbb C)}^{2}$ given in Proposition~5.2, 
the asymptotic expansion for $Q(x,t)$ given in Theorem~2.2, 
Eqs.~(10)--(27), the following inequalities, $\vert \! \exp\{
(\cdot)\} \! - \! 1 \vert \! \leq \! \vert (\cdot) \vert 
\sup_{s \in [0,1]} \vert \! \exp\{s(\cdot)\} \vert$ and $0 \! < 
\! \nu(\lambda_{0}) \! \leq \! \nu_{{\rm max}} \! := \! -\frac{
1}{2 \pi} \ln(1 \! - \! \sup_{\lambda \in \Bbb R} \vert 
r(\lambda) \vert^{2}) \! < \! \infty$, and the fact that 
$r(\lambda) \! \in \! {\cal S}(\widehat{\Gamma};\! \Bbb C)$, 
one obtains, after onerous algebraic manipulations, the result 
stated in the Lemma. \hfill \rule{6.5pt}{6.5pt}
\begin{ccc}
As $t \! \to \! + \infty$ and $x \! \to \! - \infty$ such that 
$\lambda_{0} \! > \! M$ and $(x,t) \! \in \! \Bbb R^{2} \setminus 
\Omega_{n}$, for those $\gamma_{n} \! \in \! (\frac{\pi}{2},\! 
\pi)$, $q(x,t)$ has the asymptotic expansion given in Theorem~2.3, 
Eqs.~(45)--(51).
\end{ccc}

{\em Proof.\/} Consequence of Proposition~2.3, Lemma~4.3, and 
Lemma~5.1. \hfill \rule{6.5pt}{6.5pt}
\begin{ccc}
As $t \! \to \! + \infty$ and $x \! \to \! + \infty$ such that 
$\widehat{\lambda}_{0} \! := \! \sqrt{\frac{1}{2}(\frac{x}{t} 
\! - \! \frac{1}{s})} \! > \! M$, $\frac{x}{t} \! > \! \frac{
1}{s}$, $s \! \in \! \Bbb R_{>0}$, and $(x,t) \! \in \! \Bbb 
R^{2} \setminus \widetilde{\Omega}_{n}$, for those $\gamma_{n} 
\! \in \! (\frac{\pi}{2},\! \pi)$, $u(x,t)$ has the asymptotic 
expansion given in Theorem~2.4, Eqs.~(56)--(66).
\end{ccc}

{\em Proof.\/} Consequence of Proposition~2.4 and Lemma~5.2. 
\hfill \rule{6.5pt}{6.5pt}
\section{Asymptotics as $t \! \to \! - \infty$}
In this section, the asymptotic paradigm presented in Secs.~3--5 
is succinctly presented for the case when $t \! \to \! - \infty$: 
since the proofs of all obtained results are analogous (to the $t 
\! \to \! +\infty$ case), they will be omitted. This short section 
is divided into three parts: (1) in Subsection~6.1, the model RH 
problem is formulated for the case when $t \! \to \! -\infty$ and 
$x \! \to \! +\infty$ such that $\lambda_{0} \! := \! \frac{1}{2} 
\sqrt{-\frac{x}{t}} \! > \! M$ and $(x,t) \! \in \! \Bbb R^{2} 
\setminus \Omega_{n}$ (for those $\gamma_{n} \! \in \! (\frac{\pi}
{2},\! \pi))$, the asymptotic expansion for $Q(x,t)$ is given, and 
the necessity associated with the {\em a priori} determination of 
a certain subset of the coefficients of the asymptotic expansion 
for $Q(x,t)$ is stated; (2) in Subsection~6.2, the model RH problem 
associated with the determination of the subset of the asymptotic 
series expansion coefficients for $Q(x,t)$ mentioned in (1) above 
is formulated, and solved asymptotically to ${\cal O}((-t)^{-7/2})$ 
as $t \! \to \! -\infty$; and (3) in Subsection~6.3, the phase 
integral, $((\Psi^{-1}(x,t;0))_{11})^{2}$, is evaluated asymptotically 
as $t \! \to \! -\infty$. Once again, all explicit $x,t$ dependences 
are suppressed, except where absolutely necessary.
\subsection{Model RH Problem and Asymptotic Expansion of $Q(x,t)$ 
as $t \! \to \! -\infty$}
\begin{figure}[bht]
\begin{center}
\unitlength=1cm
\begin{picture}(12,4)(0,2.5)
\thicklines
\put(10,5){\makebox(0,0){$\scriptstyle{}\bullet$}}
\put(2,5){\makebox(0,0){$\scriptstyle{}\bullet$}}
\put(6,5){\makebox(0,0){$\scriptstyle{}\bullet$}}
\put(10.05,4.6){\makebox(0,0){$\scriptstyle{}+\lambda_{0}$}}
\put(2.05,4.6){\makebox(0,0){$\scriptstyle{}-\lambda_{0}$}}
\put(6,4.6){\makebox(0,0){$\scriptstyle{}0$}}
\put(10,5){\line(1,1){0.5}}
\put(11,6){\vector(-1,-1){0.6}}
\put(10,5){\vector(1,-1){0.7}}
\put(11,4){\line(-1,1){0.5}}
\put(10,5){\vector(-1,1){0.65}}
\put(9,6){\line(1,-1){0.5}}
\put(10,5){\line(-1,-1){0.5}}
\put(9,4){\vector(1,1){0.55}}
\put(6,5){\line(1,1){0.5}}
\put(7,6){\vector(-1,-1){0.6}}
\put(6,5){\vector(1,-1){0.7}}
\put(7,4){\line(-1,1){0.5}}
\put(6,5){\vector(-1,1){0.65}}
\put(5,6){\line(1,-1){0.5}}
\put(6,5){\line(-1,-1){0.5}}
\put(5,4){\vector(1,1){0.55}}
\put(2,5){\line(1,1){0.5}}
\put(3,6){\vector(-1,-1){0.6}}
\put(2,5){\vector(1,-1){0.7}}
\put(3,4){\line(-1,1){0.5}}
\put(2,5){\vector(-1,1){0.65}}
\put(1,6){\line(1,-1){0.5}}
\put(2,5){\line(-1,-1){0.5}}
\put(1,4){\vector(1,1){0.55}}
\put(11.5,6.25){\makebox(0,0)[r]{$\scriptstyle{}\widehat{
\varsigma}_{{\cal B}}^{(1)}$}}
\put(11.5,3.75){\makebox(0,0)[r]{$\scriptstyle{}\widehat{
\varsigma}_{{\cal B}}^{(4)}$}}
\put(8.5,6.25){\makebox(0,0)[l]{$\scriptstyle{}\widehat{
\varsigma}_{{\cal B}}^{(3)}$}}
\put(8.5,3.75){\makebox(0,0)[l]{$\scriptstyle{}\widehat{
\varsigma}_{{\cal B}}^{(2)}$}}
\put(7.5,6.25){\makebox(0,0)[r]{$\scriptstyle{}\widehat{
\varsigma}_{{\cal C}}^{(1)}$}}
\put(7.5,3.75){\makebox(0,0)[r]{$\scriptstyle{}\widehat{
\varsigma}_{{\cal C}}^{(4)}$}}
\put(4.5,6.25){\makebox(0,0)[l]{$\scriptstyle{}\widehat{
\varsigma}_{{\cal C}}^{(3)}$}}
\put(4.5,3.75){\makebox(0,0)[l]{$\scriptstyle{}\widehat{
\varsigma}_{{\cal C}}^{(2)}$}}
\put(3.5,6.25){\makebox(0,0)[r]{$\scriptstyle{}\widehat{
\varsigma}_{{\cal A}}^{(1)}$}}
\put(3.5,3.75){\makebox(0,0)[r]{$\scriptstyle{}\widehat{
\varsigma}_{{\cal A}}^{(4)}$}}
\put(0.5,6.25){\makebox(0,0)[l]{$\scriptstyle{}\widehat{
\varsigma}_{{\cal A}}^{(3)}$}}
\put(0.5,3.75){\makebox(0,0)[l]{$\scriptstyle{}\widehat{
\varsigma}_{{\cal A}}^{(2)}$}}
\put(10,2.75){\makebox(0,0){$\scriptstyle{}\widehat{
\varsigma}_{{\cal B}} := \bigcup\limits_{k=1}^{4} \widehat{
\varsigma}_{{\cal B}}^{(k)}$}}
\put(6,2.75){\makebox(0,0){$\scriptstyle{}\widehat{\varsigma}
_{{\cal C}} := \bigcup\limits_{k=1}^{4} \widehat{\varsigma}_{
{\cal C}}^{(k)}$}}
\put(2,2.75){\makebox(0,0){$\scriptstyle{}\widehat{\varsigma}
_{{\cal A}} := \bigcup\limits_{k=1}^{4} \widehat{\varsigma}_{
{\cal A}}^{(k)}$}}
\end{picture}
\end{center}
\vspace{-0.30cm}
\caption{}
\end{figure}
\begin{cccc}[{\rm \cite{a2,a10}}]
Set $\widehat{\varsigma} \! := \! \cup_{l \in \{{\cal B},{\cal 
A},{\cal C}\}} \widehat{\varsigma}_{l}$ (Fig.~6). As $t \! \to 
\! -\infty$ and $x \! \to \! +\infty$ such that $\lambda_{0} \! 
> \! M$ and $(x,t) \! \in \! \Bbb R^{2} \setminus \Omega_{n}$, 
for those $\gamma_{n} \! \in \! (\frac{\pi}{2},\! \pi)$, there 
exists a unique function $\widehat{m}^{\widehat{\varsigma}}
(\lambda) \colon \Bbb C \setminus \widehat{\varsigma} \! \to \! 
{\rm SL}(2,\! \Bbb C)$ which solves the following RH problem:
\begin{enumerate}
\item[(1)] $\widehat{m}^{\widehat{\varsigma}}(\lambda)$ is 
piecewise holomorphic $\forall \, \lambda \! \in \! \Bbb C \! 
\setminus \! \widehat{\varsigma};$
\item[(2)] $\widehat{m}^{\widehat{\varsigma}}(\lambda)$ satisfies 
the following jump conditions,
\begin{eqnarray*}
&\widehat{m}^{\widehat{\varsigma}}_{+}(\lambda) = \widehat{m}^{
\widehat{\varsigma}}_{-}(\lambda) \widehat{v}^{\widehat{\varsigma}
}(\lambda), \, \, \, \, \, \, \, \, \, \lambda \in \widehat{
\varsigma},&
\end{eqnarray*}
where
\begin{eqnarray*}
&\widehat{v}^{\widehat{\varsigma}}(\lambda) \vert_{\widehat{
\varsigma}_{j}^{(1)} \cup \widehat{\varsigma}_{j}^{(2)}}={\rm 
I} + \widehat{{\cal R}}_{j}^{(1,2)}(\lambda;\! \lambda_{0}){
\cal P}_{-}(\lambda)(\delta^{-}(\lambda;\! \lambda_{0}))^{2} 
e^{-2 i t \rho(\lambda;\lambda_{0})} \sigma_{+},& \\
&\widehat{v}^{\widehat{\varsigma}}(\lambda) \vert_{\widehat{
\varsigma}_{j}^{(3)} \cup \widehat{\varsigma}_{j}^{(4)}}={\rm 
I} + \widehat{{\cal R}}_{j}^{(3,4)}(\lambda;\! \lambda_{0})
({\cal P}_{-}(\lambda))^{-1}(\delta^{-}(\lambda;\! \lambda_{0}
))^{-2} e^{2 i t \rho(\lambda;\lambda_{0})} \sigma_{-},& 
\\
&\widehat{v}^{\widehat{\varsigma}}(\lambda) \vert_{\widehat{
\varsigma}_{{\cal C}}^{(1)} \cup \widehat{\varsigma}_{{\cal 
C}}^{(2)}} = {\rm I} + \widehat{{\cal R}}_{{\cal C}}^{(1,2)}
(\lambda;\! \lambda_{0})({\cal P}_{-}(\lambda))^{-1}(\delta^{
-}(\lambda;\! \lambda_{0}))^{-2} e^{2it\rho(\lambda;\lambda_{
0})} \sigma_{-},& \\
&\widehat{v}^{\widehat{\varsigma}}(\lambda) \vert_{\widehat{
\varsigma}_{{\cal C}}^{(3)} \cup \widehat{\varsigma}_{{\cal C}
}^{(4)}} = {\rm I} + \widehat{{\cal R}}_{{\cal C}}^{(3,4)}(
\lambda;\! \lambda_{0}) {\cal P}_{-}(\lambda) (\delta^{-}(
\lambda;\! \lambda_{0}))^{2} e^{-2it\rho(\lambda;\lambda_{0})} 
\sigma_{+},&
\end{eqnarray*}
$j \! \in \! \{{\cal B},{\cal A}\}$, ${\cal P}_{-}(\lambda) \! 
:= \! \prod_{l=1}^{n-1} \! \left(\! \frac{(\lambda-\overline{
\lambda_{l}})(\lambda+\overline{\lambda_{l}})}{(\lambda-\lambda
_{l})(\lambda+\lambda_{l})} \! \right)^{2}$, $\rho(\lambda;\! 
\lambda_{0}) \! = \! 2 \lambda^{2}(\lambda^{2} \! - \! 2\lambda
_{0}^{2})$, $\{0,\pm \lambda_{0}\} \! = \! \{\mathstrut \lambda
^{\prime}; \, \partial_{\lambda} \rho \linebreak[4] (\lambda;\! 
\lambda_{0}) \vert_{\lambda=\lambda^{\prime}} \! = \! 0\}$ are 
the first-order stationary phase points,
\begin{eqnarray*}
&\delta^{-}(\lambda;\! \lambda_{0}) = \exp \! \left\{\int_{
\lambda_{0}}^{\infty} \frac{\ln(1-\vert r(\xi) \vert^{2})}{
(\xi - \lambda)} \frac{d\xi}{2 \pi i} + \int_{-\lambda_{0}}
^{-\infty} \frac{\ln(1-\vert r(\xi) \vert^{2})}{(\xi-\lambda)} 
\frac{d\xi}{2 \pi i} \right\} \!,&
\end{eqnarray*}
and $\{\widehat{{\cal R}}_{{\cal B}}^{(l,l+1)}(\lambda;\! \lambda
_{0}),\widehat{{\cal R}}_{{\cal A}}^{(l,l+1)}(\lambda;\! \lambda_
{0}),\widehat{{\cal R}}_{{\cal C}}^{(l,l+1)}(\lambda;\! \lambda_{
0})\}_{l \in \{1,3\}}$ are some rational functions which decay to 
zero as $\lambda \! \to \! \infty$, $\lambda \! \in \! \widehat{
\varsigma} \setminus \cup_{k \in \{{\cal B},{\cal A},{\cal C}\}} 
\{{\rm sgn}(k) \lambda_{0}\}$, where ${\rm sgn}({\cal B}) \! = \! 
- {\rm sgn}({\cal A}) \! = \! 1$ and ${\rm sgn}({\cal C}) \! = \! 
0$, and have, respectively, Taylor series expansions in 
$\{\mathstrut \lambda^{\prime}; \, \vert \lambda^{\prime} \! - \! 
{\rm sgn}(k) \lambda_{0} \vert \linebreak[4] < \! \varepsilon\} 
\cap \widehat{\varsigma}_{k}$, $k \! \in \! \{{\cal B},{\cal A},
{\cal C}\}$, where $\varepsilon$ is an arbitrarily fixed, 
sufficiently small positive real number;
\item[(3)] as $\lambda \! \to \! \infty$, $\lambda \! \in \! 
\Bbb C \! \setminus \! \widehat{\varsigma}$,
\begin{eqnarray*}
&\widehat{m}^{\widehat{\varsigma}}(\lambda) = {\rm I} + {\cal O}
(\lambda^{-1}).&
\end{eqnarray*}
\end{enumerate}
Moreover, for arbitrary $l^{\prime} \! \in \! \Bbb Z_{\geq 1}$,
\begin{eqnarray*}
&Q(x,t) = 2 i \lim\limits_{\lambda \to \infty \atop \lambda \, 
\in \, \Bbb C \setminus \widehat{\varsigma}}(\lambda \widehat{
m}^{\widehat{\varsigma}}(x,t;\! \lambda))_{12} + {\cal O} \! 
\left(\frac{\underline{c}(\lambda_{0})}{(\lambda_{0}^{2} t)^{l^{
\prime}}} \right) \!,&
\end{eqnarray*}
with $Q(x,\! 0) \! \in \! {\cal S}(\Bbb R;\! \Bbb C)$, satisfies 
Eq.~(2), and $\widehat{m}^{\widehat{\varsigma}}(\lambda)$ 
satisfies the following symmetry reductions, $\widehat{m}^{
\widehat{\varsigma}}(\lambda) \! = \! \sigma_{3} \widehat{m}
^{\widehat{\varsigma}}(-\lambda) \sigma_{3}$ and $\widehat{m}
^{\widehat{\varsigma}}(\lambda) \! = \! \sigma_{1} \overline{
\widehat{m}^{\widehat{\varsigma}}(\overline{\lambda})} \sigma
_{1}$.
\end{cccc}

Analysing the signature graph of $\Re(it\rho(\lambda;\! \lambda
_{0}))$ as $t \! \to \! -\infty$ $(x/t \! \sim \! {\cal O}(1))$ 
and proceeding analogously as in Sec.~3 (Higher Order Deift-Zhou 
Theory), one deduces the following.
\begin{cccc}
Let $\widehat{m}^{\widehat{\varsigma}}(\lambda)$ be the solution 
of the RH problem formulated in Lemma~6.1.1 with the condition 
$\vert \vert r \vert \vert_{{\cal L}^{\infty}(\Bbb R;\Bbb C)} \! 
< \! 1$. Then, for arbitrary $N_{o} \! \in \! \Bbb Z_{\geq 1}$, 
as $t \! \to \! - \infty$ and $x \! \to \! + \infty$ such that 
$\lambda_{0} \! > \! M$ and $(x,t) \! \in \! \Bbb R^{2} \setminus 
\Omega_{n}$, for those $\gamma_{n} \! \in \! (\frac{\pi}{2},\! 
\pi)$,
\begin{eqnarray*}
&Q(x,t) = \sum\limits_{k={\rm odd} \atop \pm 1,\pm 3,\ldots} 
\sum\limits_{p=\vert k \vert}^{N_{o}} \sum\limits_{q=0}^{p-\vert 
k \vert} \frac{\exp \{\frac{i}{2}(k+1)(4 \lambda_{0}^{4} t + \nu
(\lambda_{0}) \ln \vert t \vert)\}u^{-}_{k,p,q}(\lambda_{0})(\ln 
\vert t \vert)^{q}}{(-t)^{p/2}} + {\cal O} \! \left(\frac{
\underline{c}(\lambda_{0})(\ln \vert t \vert)^{N_{o}}}{(-\lambda
_{0}^{2} t)^{(N_{o}+1)/2}} \right) \!,&
\end{eqnarray*}
where, for $k \! = \! \pm 1,\pm 3,\ldots$, $\vert k \vert \! \leq 
\! p \! \leq \! N_{o}$, and $q \! > \! p \! - \! \vert k \vert$, 
$u^{-}_{k,p,q}(\lambda_{0}) \! \equiv \! 0$.
\end{cccc}
\begin{bbbb}
For $\alpha \! = \! \pm 1,\pm 3,\ldots$, $\beta \! \geq \! \vert 
\alpha \vert$, and $0 \! \leq \! \gamma^{\prime} \! \leq \! \beta 
\! - \! \vert \alpha \vert$, set $u^{-}_{\alpha,\beta}(\lambda_{
0};t) \! := \! u^{-}_{\alpha,\beta} \! := \! \sum_{\gamma^{\prime}
=0}^{\beta-\vert \alpha \vert} \! u^{-}_{\alpha,\beta,\gamma^{
\prime}}(\lambda_{0})(\ln \vert t \vert)^{\gamma^{\prime}}$, and, 
for $\gamma^{\prime} \! > \! \beta \! - \! \vert \alpha \vert$, 
$u^{-}_{\alpha,\beta,\gamma^{\prime}}(\lambda_{0}) \! \equiv \! 
0$. The coefficients of the asymptotic expansion for $Q(x,t)$ 
given in Lemma~6.1.2 are determined by the following linear 
system,
\begin{eqnarray}
&\sum\limits_{k_{1} \, = \, {\rm odd} \atop \pm 1,\pm 3,\ldots} 
\sum\limits_{p_{1} \geq \vert k_{1} \vert} \frac{e^{i\{\frac{(k_{
1}+1)\tau^{-}}{2}\}}(k_{1}+1)}{(-t)^{(p_{1}+2)/2}} \{-2 \lambda
_{0}^{4} t u^{-}_{k_{1},p_{1}} + \frac{\nu u^{-}_{k_{1},p_{1}}}{
2} - \frac{\lambda_{0} \nu^{\prime} u^{-}_{k_{1},p_{1}} \ln 
\vert t \vert}{4}\}& \nonumber \\
&+\sum\limits_{k_{1} \, = \, {\rm odd} \atop \pm 1,\pm 3,\ldots} 
\sum\limits_{p_{1} \geq \vert k_{1} \vert} \frac{e^{i\{\frac{(k_{
1}+1)\tau^{-}}{2}\}}}{(-t)^{(p_{1}+2)/2}} \{-it \dot{u}^{-}_{k_{
1},p_{1}}+\frac{ip_{1}u^{-}_{k_{1},p_{1}}}{2}+\frac{i\lambda_{0} 
u^{- \prime}_{k_{1},p_{1}}}{2}\}& \nonumber \\
&\! \! \! \! \! \! - \! 
\sum\limits_{k_{1} \, = \, {\rm odd} \atop \pm 1,\pm 3,
\ldots} \! \sum\limits_{p_{1} \geq \vert k_{1} \vert} \frac{e^{i
\{\frac{(k_{1}+1)\tau^{-}}{2}\}}i(k_{1}+1)}{(-t)^{(p_{1}+2)/2}}\{
\frac{\lambda_{0} u^{-\prime}_{k_{1},p_{1}}}{4} \! + \! \frac{u^{
-}_{k_{1},p_{1}}}{4} \! + \! \frac{\nu^{\prime}u^{-\prime}_{k_{1},
p_{1}} \ln \vert t \vert}{(8 \lambda_{0})^{2} t} \! + \! \frac{\nu
^{\prime \prime} u^{-}_{k_{1},p_{1}} \ln \vert t \vert}{2(8\lambda
_{0})^{2} t} \! - \! \frac{\nu^{\prime} u^{-}_{k_{1},p_{1}} \ln 
\vert t \vert}{2(8)^{2} \lambda_{0}^{3}t}\}& \nonumber \\
&+\sum\limits_{k_{1} \, = \, {\rm odd} \atop \pm 1,\pm 3,\ldots} 
\sum\limits_{p_{1} \geq \vert k_{1} \vert} \frac{e^{i\{\frac{(k_{
1}+1)\tau^{-}}{2}\}}(k_{1}+1)^{2}}{(-t)^{(p_{1}+2)/2}} \{\lambda_{
0}^{4}t u^{-}_{k_{1},p_{1}} + \frac{\lambda_{0}\nu^{\prime} u^{-}
_{k_{1},p_{1}} \ln \vert t \vert}{8}+\frac{(\nu^{\prime})^{2}u^{-}
_{k_{1},p_{1}}(\ln \vert t \vert)^{2}}{(16 \lambda_{0})^{2} t}\}& 
\nonumber \\
&+\sum\limits_{k_{1} \, = \, {\rm odd} \atop \pm 1,\pm 3,\ldots} 
\sum\limits_{p_{1} \geq \vert k_{1} \vert} \frac{e^{i\{\frac{(k_{
1}+1)\tau^{-}}{2}\}}}{(-t)^{(p_{1}+2)/2}} \{-\frac{u^{-\prime 
\prime}_{k_{1},p_{1}}}{(8 \lambda_{0})^{2} t} + \frac{u^{-\prime}
_{k_{1},p_{1}}}{(8)^{2} \lambda_{0}^{3} t}\}& \nonumber \\
&-\underbrace{\sum \sum \sum}_{k_{i} \, = \, 
{\rm odd} \atop {\pm 1,\pm 3,\ldots \atop 1 \leq i \leq 3}} 
\underbrace{\sum \sum \sum}_{p_{i} 
\geq \vert k_{i} \vert \atop 1 \leq i \leq 3} \frac{e^{i\{\frac{
(k_{2}+k_{3}-k_{1}+1)\tau^{-}}{2}\}}(k_{1}+1) \lambda_{0}^{2} 
\overline{u^{-}_{k_{1},p_{1}}} u^{-}_{k_{2},p_{2}} u^{-}_{k_{3},
p_{3}}}{(-t)^{(p_{1}+p_{2}+p_{3})/2}}& \nonumber \\
&+\underbrace{\sum \sum \sum}_{k_{i} \, = \, {\rm odd} \atop {\pm 
1,\pm 3,\ldots \atop 1 \leq i \leq 3}} \underbrace{\sum \sum \sum}
_{p_{i} \geq \vert k_{i} \vert \atop 1 \leq i \leq 3} \frac{e^{i\{
\frac{(k_{2}+k_{3}-k_{1}+1)\tau^{-}}{2}\}}(k_{1}+1)\nu^{\prime} 
\overline{u^{-}_{k_{1},p_{1}}}u^{-}_{k_{2},p_{2}}u^{-}_{k_{3},p_{
3}} \ln \vert t \vert}{16 \lambda_{0} (-t)^{(p_{1}+p_{2}+p_{3}+2)
/2}}& \nonumber \\
&+\underbrace{\sum \sum \sum \sum \sum}_{k_{i} \, = \, {\rm odd} 
\atop {\pm 1,\pm 3,\ldots \atop 1 \leq i \leq 5}} \underbrace{\sum 
\sum \sum \sum \sum}_{p_{i} \geq \vert k_{i} \vert \atop 1 \leq i 
\leq 5} \frac{e^{i\{\frac{(k_{1}-k_{2}+k_{3}-k_{4}+k_{5}+1)\tau^{
-}}{2}\}} u^{-}_{k_{1},p_{1}} \overline{u^{-}_{k_{2},p_{2}}} u^{-}
_{k_{3},p_{3}} \overline{u^{-}_{k_{4},p_{4}}} u^{-}_{k_{5},p_{5}
}}{2(-t)^{(p_{1}+p_{2}+p_{3}+p_{4}+p_{5})/2}}& \nonumber \\
&+ \underbrace{\sum \sum \sum}_{k_{i} \, = \, {\rm odd} \atop {\pm 
1,\pm 3,\ldots \atop 1 \leq i \leq 3}} \underbrace{\sum \sum \sum}
_{p_{i} \geq \vert k_{i} \vert \atop 1 \leq i \leq 3} \frac{e^{i\{
\frac{(k_{2}+k_{3}-k_{1}+1)\tau^{-}}{2}\}}i(\overline{u^{-}_{k_{
1},p_{1}}})^{\prime} u^{-}_{k_{2},p_{2}} u^{-}_{k_{3},p_{3}}}{8 
\lambda_{0} (-t)^{(p_{1}+p_{2}+p_{3}+2)/2}}=0,&
\end{eqnarray}
where $\tau^{-} \! := \! 4 \lambda_{0}^{4} t \! + \! \nu \ln 
\vert t \vert$, $\nu \! := \! \nu(\lambda_{0})$, $f^{\prime} 
\! := \! \partial_{\lambda_{0}}f(\lambda_{0};t) \vert_{t \, = 
\, {\rm fixed}}$, and $\dot{f} \! := \! \partial_{t}f(\lambda
_{0};t) \vert_{\lambda_{0} \, = \, {\rm fixed}}$.
\end{bbbb}
\begin{ffff}
To ${\cal O}((-t)^{-7/2})$, the explicit recurrence formulae 
for the coefficients of the asymptotic expansion for $Q(x,t)$ 
given in Lemma~6.1.2 are given in the Appendix.
\end{ffff}

Now, analogously as at the end of Sec.~3, one notes that, in 
order to solve the recurrence relations for $u^{-}_{\alpha,
\beta,\gamma^{\prime}}(\lambda_{0})$, explicit, {\em a priori} 
knowledge of $u^{-}_{1,1,0}(\lambda_{0})$ and $u^{-}_{-1,p,q}
(\lambda_{0})$, $p \! \geq \! 1$, $0 \! \leq \! q \! \leq \! p 
\! - \! 1$, is required: the expression for $u^{-}_{1,1,0}(
\lambda_{0})$ is given in Theorem~2.2, Eq.~(14). Hence, it 
remains to determine $u^{-}_{-1,p,q}(\lambda_{0})$, $p \! \geq 
\! 1$, $0 \! \leq \! q \! \leq \! p \! - \! 1$: up to $p \! = 
\! 6$, this is the programme of the following subsection.
\subsection{Explicit Representation of $u^{-}_{-1,p,q}(\lambda
_{0})$, $1 \! \leq \! p \! \leq \! 6$, $0 \! \leq \! q \! \leq 
\! p \! - \! 1$}
The RH factorisation problem associated with the first-order 
stationary phase point at the origin, denoted as $\widehat{m}
_{{\cal C}}^{\widehat{\varsigma}}(x,t;\! \lambda) \! := \! 
\widehat{m}_{{\cal C}}^{\widehat{\varsigma}}(\lambda)$, is 
solved asymptotically as $t \! \to \! -\infty$ up to ${\cal O}
((-t)^{-7/2})$: {}from this asymptotic expansion for $\widehat{m}
_{{\cal C}}^{\widehat{\varsigma}}(\lambda)$ and the resulting 
expression for $2 i \lim_{\lambda \to \infty \atop \lambda \, 
\in \, \Bbb C \setminus \widehat{\varsigma}_{{\cal C}}} \! (
\lambda \widehat{m}^{\widehat{\varsigma}}_{{\cal C}}(\lambda))
_{12}$, explicit (integral) representations for $u^{-}_{-1,p,q}
(\lambda_{0})$, $1 \! \leq \! p \! \leq \! 6$, $0 \! \leq \! q 
\! \leq \! p \! - \! 1$, are deduced.
\begin{figure}[bht]
\begin{center}
\unitlength=1cm
\begin{picture}(2,2)(0,0)
\thicklines
\put(1,1){\makebox(0,0){$\scriptstyle{}\bullet$}}
\put(1,0.75){\makebox(0,0){$\scriptstyle{}0$}}
\put(1,1){\line(1,1){0.5}}
\put(2,2){\vector(-1,-1){0.6}}
\put(1,1){\vector(-1,1){0.65}}
\put(0,2){\line(1,-1){0.5}}
\put(1,1){\line(-1,-1){0.5}}
\put(0,0){\vector(1,1){0.55}}
\put(1,1){\vector(1,-1){0.7}}
\put(2,0){\line(-1,1){0.5}}
\put(2.4,1.95){\makebox(0,0){$\scriptstyle{}\widehat{\varsigma}
_{{\cal C}}^{(1)}$}}
\put(2.4,0){\makebox(0,0){$\scriptstyle{}\widehat{\varsigma}_{{
\cal C}}^{(4)}$}}
\put(-0.35,1.95){\makebox(0,0){$\scriptstyle{}\widehat{\varsigma}
_{{\cal C}}^{(3)}$}}
\put(-0.35,0){\makebox(0,0){$\scriptstyle{}\widehat{\varsigma}_{
{\cal C}}^{(2)}$}}
\end{picture}
\end{center}
\vspace{-0.50cm}
\caption{}
\end{figure}
\begin{cccc}[{\rm \cite{a2,a10}}]
Set $\widehat{\varsigma}_{{\cal C}} \! := \! \cup_{k=1}^{4} 
\widehat{\varsigma}_{{\cal C}}^{(k)}$ (Fig.~7). As $t \! \to \! 
-\infty$ and $x \! \to \! +\infty$ such that $\lambda_{0} \! > 
\! M$ and $(x,t) \! \in \! \Bbb R^{2} \setminus \Omega_{n}$, for 
those $\gamma_{n} \! \in \! (\frac{\pi}{2},\! \pi)$, there exists 
a unique function $\widehat{m}^{\widehat{\varsigma}}_{{\cal C}}(
\lambda) \colon \Bbb C \setminus \widehat{\varsigma}_{{\cal C}} 
\! \to \! {\rm SL}(2,\! \Bbb C)$ which solves the following RH 
problem:
\begin{enumerate}
\item[(1)] $\widehat{m}^{\widehat{\varsigma}}_{{\cal C}}(\lambda)$ 
is piecewise holomorphic $\forall \, \lambda \! \in \! \Bbb C \! 
\setminus \! \widehat{\varsigma}_{{\cal C}};$
\item[(2)] $\widehat{m}^{\widehat{\varsigma}}_{{\cal C}}(\lambda)$ 
satisfies the following jump conditions,
\begin{eqnarray*}
&\widehat{m}^{\widehat{\varsigma}}_{{\cal C} \, +}(\lambda) = 
\widehat{m}^{\widehat{\varsigma}}_{{\cal C} \, -}(\lambda) 
\widehat{v}^{\widehat{\varsigma}}_{{\cal C}}(\lambda), \, \, \, \, 
\, \, \, \, \lambda \in \widehat{\varsigma}_{{\cal C}},&
\end{eqnarray*}
where
\begin{eqnarray*}
&\widehat{v}^{\widehat{\varsigma}}_{{\cal C}}(\lambda) \vert_{
\cup_{k=1}^{2} \widehat{\varsigma}_{{\cal C}}^{(k)}} = {\rm I} 
- (\widehat{{\cal R}}_{{\cal C}}(\lambda))^{\ast}({\cal P}_{-}
(\lambda))^{-1}(\delta^{-}(\lambda;\! \lambda_{0}))^{-2} e^{2it 
\rho(\lambda;\lambda_{0})} \sigma_{-},& \\
&\widehat{v}^{\widehat{\varsigma}}_{{\cal C}}(\lambda) \vert_{
\cup_{k=3}^{4} \widehat{\varsigma}_{{\cal C}}^{(k)}} = {\rm I} 
- \widehat{{\cal R}}_{{\cal C}}(\lambda){\cal P}_{-}(\lambda)(
\delta^{-}(\lambda;\! \lambda_{0}))^{2} e^{-2it \rho(\lambda;
\lambda_{0})} \sigma_{+},& \\
&\delta^{-}(\lambda;\! \lambda_{0}) = \exp \! \left\{\int_{
\lambda_{0}}^{\infty} \frac{\xi \ln(1-\vert r(\xi) \vert^{2})}
{(\xi^{2} - \lambda^{2})} \frac{d\xi}{\pi i} \right\} \! := 
\exp\{({\rm I}_{\alpha} \! + \! {\rm I}_{\beta})(\lambda;\! 
\lambda_{0})\},&
\end{eqnarray*}
${\cal P}_{-}(\lambda) \! := \! \prod_{l=1}^{n-1} \! \left(\! 
\frac{(\lambda - \overline{\lambda_{l}})(\lambda + \overline{
\lambda_{l}})}{(\lambda - \lambda_{l})(\lambda + \lambda_{l})} \! 
\right)^{2}$, $\rho(\lambda;\! \lambda_{0}) \! = \! 2 \lambda^{2}
(\lambda^{2} \! - \! 2 \lambda_{0}^{2})$, $(\widehat{{\cal R}}_{{
\cal C}}(\lambda))^{\ast}$ denotes the same function as $\widehat{
{\cal R}}_{{\cal C}}(\lambda)$ except with the complex conjugated 
coefficients, $\widehat{{\cal R}}_{{\cal C}}(- \lambda) \! = \! - 
\widehat{{\cal R}}_{{\cal C}}(\lambda)$, $\widehat{{\cal R}}_{{
\cal C}}(\lambda)$ is a piecewise-rational function which decays 
like ${\cal O}(\lambda^{-(k+7)})$, $k \! \in \! \Bbb Z_{\geq 
1}$, as $\lambda \! \to \! \infty$, $\lambda \! \in \! \widehat{
\varsigma}_{{\cal C}} \! \setminus \! \{0\}$, and has, for 
$\lambda \! \in \! (\cup_{l=3}^{4} \widehat{\varsigma}_{{\cal 
C}}^{(l)}) \! \cap \! \{\mathstrut \lambda^{\prime}; \, \vert 
\lambda^{\prime} \vert \! < \! \varepsilon\}$, where $\varepsilon$ 
is an arbitrarily fixed, sufficiently small positive real number, 
the following Taylor series expansion about $\lambda \! = \! 0$,
\begin{eqnarray*}
&- \widehat{{\cal R}}_{{\cal C}}(\lambda) {\cal P}_{-}(\lambda) 
:= R_{-}^{\prime}(0) \lambda + \frac{1}{3!} R_{-}^{\prime \prime 
\prime}(0) \lambda^{3} + \frac{1}{5!} R_{-}^{V}(0) \lambda^{5} + 
{\cal O}(\lambda^{7}),&
\end{eqnarray*}
with
\begin{eqnarray*}
&R_{-}^{\prime}(0) := r^{\prime}(0) g(0) - r^{\prime}(i0) 
\widehat{g}(i0),& \\
&\frac{1}{3!} R_{-}^{\prime \prime \prime}(0) := \frac{1}{2!} 
r^{\prime}(0) g^{\prime \prime}(0) + \frac{1}{3!} r^{\prime 
\prime \prime}(0) g(0) - \frac{1}{2!} r^{\prime}(i0) \widehat{g}
^{\prime \prime}(i0) - \frac{1}{3!} r^{\prime \prime \prime}(i0) 
\widehat{g}(i0),& \\
&\frac{1}{5!} R_{-}^{V}(0) := \frac{1}{4!} r^{\prime}(0) g^{IV}
(0) + \frac{1}{2! 3!} r^{\prime \prime \prime}(0) g^{\prime 
\prime}(0) + \frac{1}{5!} r^{V}(0) g(0)& \\
&\, \, \, \, \, \, \, \, \, \, \, \, \, \, \, \, \, \, \, \, \,
\, \, \, \, \, \, \, \, \, \, \, \, \, \, \, \, - \, 
\frac{1}{4!} r^{\prime}(i0) \widehat{g}^{IV}(i0) - \frac{1}{2! 
3!} r^{\prime \prime \prime}(i0) \widehat{g}^{\prime \prime}(
i0) - \frac{1}{5!} r^{V}(i0) \widehat{g}(i0),&
\end{eqnarray*}
where
\begin{eqnarray*}
&r^{\prime}(0) := (\frac{d r(\lambda)}{d \lambda} \vert_{\lambda 
\in \Bbb R}) \vert_{\lambda=0}, \, \, \, \, \, \, \, \, \, \, 
r^{\prime}(i0) := (\frac{d r(\lambda)}{d \lambda} \vert_{\lambda 
\in i \Bbb R}) \vert_{\lambda=0},& \\
&r^{\prime \prime \prime}(0) := (\frac{d^{3}r(\lambda)}{d\lambda
^{3}} \vert_{\lambda \in \Bbb R}) \vert_{\lambda=0}, \, \, \, \, 
\, \, \, \, \, \, r^{\prime \prime \prime}(i0) := (\frac{d^{3} 
r(\lambda)}{d \lambda^{3}} \vert_{\lambda \in i \Bbb R}) \vert_{
\lambda=0},& \\
&r^{V}(0) := (\frac{d^{5} r(\lambda)}{d \lambda^{5}} \vert_{
\lambda \in \Bbb R}) \vert_{\lambda=0}, \, \, \, \, \, \, \, \, 
\, \, r^{V}(i0) := (\frac{d^{5} r(\lambda)}{d \lambda^{5}} \vert
_{\lambda \in i \Bbb R}) \vert_{\lambda=0},&
\end{eqnarray*}
and
\begin{eqnarray*}
&g(0) = \widehat{g}(i0) := s_{-}, \, \, \, \, \, \, \, g^{\prime 
\prime}(0) = - \widehat{g}^{\prime \prime}(i0) := - 8 i s_{-} 
\sum_{-} \! \frac{\sin \gamma_{l}}{\Delta_{l}^{2}},& \\
&g^{IV}(0) = \widehat{g}^{IV}(i0) := -192 s_{-}(\sum_{-} \! 
\frac{\sin \gamma_{l}}{\Delta_{l}^{2}})^{2} \! + \! 48 i s_{-} 
\sum_{-} \! \frac{\sin 2 \gamma_{l}}{\Delta_{l}^{4}},&
\end{eqnarray*}
with $s_{-} \! := \! \exp \{4 i \sum_{-} \gamma_{l}\}$, and 
$\sum_{-} \! := \! \sum_{l=1}^{n-1};$
\item[(3)] as $\lambda \! \to \! \infty$, $\lambda \! \in \! 
\Bbb C \! \setminus \! \widehat{\varsigma}_{{\cal C}}$,
\begin{eqnarray*}
&\widehat{m}^{\widehat{\varsigma}}_{{\cal C}}(\lambda) = {\rm 
I} + {\cal O}(\lambda^{-1}).&
\end{eqnarray*}
\end{enumerate}
Moreover, $\widehat{m}^{\widehat{\varsigma}}_{{\cal C}}(\lambda)$ 
satisfies the following symmetry reductions, $\widehat{m}^{
\widehat{\varsigma}}_{{\cal C}}(\lambda) \! = \! \sigma_{3} 
\widehat{m}^{\widehat{\varsigma}}_{{\cal C}}(-\lambda) \sigma_{3}$ 
and $\widehat{m}^{\widehat{\varsigma}}_{{\cal C}}(\lambda) \! = \! 
\sigma_{1} \overline{\widehat{m}^{\widehat{\varsigma}}_{{\cal C}}
(\overline{\lambda})} \sigma_{1}$.
\end{cccc}
\begin{bbbb}
The solution of the RH problem for $\widehat{m}^{\widehat{
\varsigma}}_{{\cal C}}(\lambda) \colon \Bbb C \setminus \widehat{
\varsigma}_{{\cal C}} \! \to \! {\rm SL}(2,\! \Bbb C)$ formulated 
in Lemma~6.2.1 has the following integral representation,
\begin{eqnarray*}
\widehat{m}^{\widehat{\varsigma}}_{{\cal C}}(\lambda) = {\rm I} 
+ \sum\limits_{k=1}^{4} \int_{\sqrt{-t} \, \, \widehat{\varsigma}
_{{\cal C}}^{(k)}} \frac{\widehat{\mu}^{\widehat{\varsigma}_{{
\cal C}}}(\frac{w}{\sqrt{-t}})(\sum_{l \in \{\pm\}} \! \widehat{
w}_{l}^{\widehat{\varsigma}_{{\cal C}}^{(k)}} \! \! (\frac{w}{
\sqrt{-t}}))}{(w-\lambda \sqrt{-t} \,)} \frac{d w}{2 \pi i}, \, 
\, \, \, \, \, \, \, \lambda \in \Bbb C \setminus \widehat{
\varsigma}_{{\cal C}},
\end{eqnarray*}
where, for $j \! \in \! \{1,2\}$,
\begin{eqnarray*}
&\widehat{w}_{+}^{\widehat{\varsigma}_{{\cal C}}^{(j)}} \! 
(\cdot) = \widehat{w}_{-}^{\widehat{\varsigma}_{{\cal C}}^{
(j+2)}} \! (\cdot) = 0,& \\
&\widehat{w}_{-}^{\widehat{\varsigma}_{{\cal C}}^{(j)}} \! \! (
\frac{w}{\sqrt{-t}}) \! = \! - \left. (\widehat{{\cal R}}_{{\cal 
C}}(\lambda))^{\ast} ({\cal P}_{-}(\lambda))^{-1} e^{-2({\rm I}_{
\alpha} + {\rm I}_{\beta})(\lambda;\lambda_{0})} e^{2 i t \rho(
\lambda;\lambda_{0})} \sigma_{-} \right\vert_{\lambda = \frac{w}{
\sqrt{-t}}},& \\
&\widehat{w}_{+}^{\widehat{\varsigma}_{{\cal C}}^{(j+2)}} \! \! 
(\frac{w}{\sqrt{-t}}) \! = \! - \left. \widehat{{\cal R}}_{{\cal 
C}}(\lambda) {\cal P}_{-}(\lambda) e^{2({\rm I}_{\alpha} + {\rm 
I}_{\beta})(\lambda;\lambda_{0})} e^{-2 i t \rho(\lambda;\lambda
_{0})} \sigma_{+} \right\vert_{\lambda = \frac{w}{\sqrt{-t}}},&
\end{eqnarray*}
with
\begin{eqnarray*}
&- \left. (\widehat{{\cal R}}_{{\cal C}}(\lambda))^{\ast} ({\cal 
P}_{-}(\lambda))^{-1} e^{-2({\rm I}_{\alpha} + {\rm I}_{\beta})(
\lambda;\lambda_{0})} e^{2 i t \rho(\lambda;\lambda_{0})} 
\right\vert_{\lambda = \frac{w}{\sqrt{-t}}} \! := \! \sum\limits
_{p=0}^{2} \frac{\widehat{{\cal J}}^{\beta}_{p}(w;\lambda_{0})}{
(-t)^{p+1/2}}& \\
&\, \, \, \, \, \, \, \, \, \, \, \, \, \, \, \, \, \, \, \, 
\, \, \, \, \, \, \, \, \, \, \, \, \, \, \, \, \, \, \, \,
\, \, \, \, \, \, \, \, \, \, \, \, \, \, \, \, \, \, \, \, 
\, \, \, \, \, \, \, \, \, \, \, \, \, \, \, \, \, \, \, \,
\, \, \, \, \, \, \, \, \, \, \, \, \, \, \, \, \, \, \, \,
\, \, \, \, \, \, \, \, \, \, \, \, + 
\, {\cal O} \! \left(\frac{\widehat{{\cal F}}^{\beta}(w;
\lambda_{0}) \exp\{8 i \lambda_{0}^{2} w^{2}\}}{(-t)^{7/2}} 
\right) \!,& \\
&\widehat{{\cal J}}^{\beta}_{0}(w;\! \lambda_{0}) \! := \! w 
\overline{R_{-}^{\prime}(0)} e^{-2i \phi_{-}(\lambda_{0})} 
e^{8 i \lambda_{0}^{2} w^{2}},& \\
&\widehat{{\cal J}}^{\beta}_{1}(w;\! \lambda_{0}) \! := \! (-2 
w^{3} \overline{R_{-}^{\prime}(0)} a_{0}^{-} \! - \! 4 i w^{5} 
\overline{R_{-}^{\prime}(0)} \! + \! \frac{1}{3!} \overline{R
_{-}^{\prime \prime \prime}(0)} w^{3}) e^{-2i \phi_{-}(\lambda
_{0})} e^{8i \lambda_{0}^{2} w^{2}},& \\
&\widehat{{\cal J}}^{\beta}_{2}(w;\! \lambda_{0}) \! := \! (-2 
w^{5} \overline{R_{-}^{\prime}(0)} b_{0}^{-} \! + \! 2 w^{5} 
\overline{R_{-}^{\prime}(0)} (a_{0}^{-})^{2} \! + \! 8 i w^{7} 
\overline{R_{-}^{\prime}(0)} a_{0}^{-} \! - \! 8 w^{9} 
\overline{R_{-}^{\prime}(0)}& \\
&\, \, \, \, \, \, \, \, \, \, \, \, \, \, \, \, \, \, \, \,
\, \, \, \, \, \, \, \, - \, 
2 a_{0}^{-} w^{5} \frac{1}{3!} \overline{R_{-}^{\prime 
\prime \prime}(0)} \! - \! 4iw^{7} \frac{1}{3!} \overline{R_{
-}^{\prime \prime \prime}(0)} \! + \! \frac{1}{5!} \overline{
R_{-}^{V}(0)} w^{5}) e^{-2 i \phi_{-}(\lambda_{0})} e^{8 i 
\lambda_{0}^{2} w^{2}},& \\
&\vert \vert \widehat{{\cal F}}^{\beta}(\cdot;\! \lambda_{0}) 
\exp \{8 i \lambda_{0}^{2} (\cdot)^{2}\} \vert \vert_{{\cal L}
^{k}(\sqrt{-t} \, \cup_{l=1}^{2} \widehat{\varsigma}_{{\cal C}
}^{(l)};\Bbb C)} = \underline{c}(\lambda_{0}), \, \, \, \, \, 
k \in \{1,2,\infty\},& \\
&- \left. \widehat{{\cal R}}_{{\cal C}}(\lambda) {\cal P}_{-}(
\lambda) e^{2({\rm I}_{\alpha} + {\rm I}_{\beta})(\lambda;\lambda
_{0})} e^{-2 i t \rho(\lambda;\lambda_{0})} \right\vert_{\lambda 
= \frac{w}{\sqrt{-t}}} \! := \! \sum\limits_{p=0}^{2} \frac{
\widehat{{\cal J}}^{\alpha}_{p}(w;\lambda_{0})}{(-t)^{p+1/2}}& 
\\
&\, \, \, \, \, \, \, \, \, \, \, \, \, \, \, \, \, \, \, \, 
\, \, \, \, \, \, \, \, \, \, \, \, \, \, \, \, \, \, \, \,
\, \, \, \, \, \, \, \, \, \, \, \, \, \, \, \, \, \, \, \, 
\, \, \, \, \, \, \, \, \, \, \, \, \, \, \, \, \, \, \, \,
\, \, \, \, \, \, \, \, \, \, + \, 
{\cal O} \! \left(\frac{\widehat{{\cal F}}^{\alpha}
(w;\lambda_{0}) \exp\{-8 i \lambda_{0}^{2} w^{2}\}}{
(-t)^{7/2}} \right) \!,& \\
&\widehat{{\cal J}}^{\alpha}_{0}(w;\! \lambda_{0}) \! := \! w 
R_{-}^{\prime}(0) e^{2i \phi_{-}(\lambda_{0})} e^{-8 i \lambda
_{0}^{2} w^{2}},& \\
&\widehat{{\cal J}}^{\alpha}_{1}(w;\! \lambda_{0}) \! := \! (2 
w^{3} R_{-}^{\prime}(0) a_{0}^{-} \! + \! 4 i w^{5} R_{-}^{
\prime}(0) \! + \! \frac{1}{3!} R_{-}^{\prime \prime \prime}(0) 
w^{3}) e^{2i \phi_{-}(\lambda_{0})} e^{-8i \lambda_{0}^{2} w^{
2}},& \\
&\widehat{{\cal J}}^{\alpha}_{2}(w;\! \lambda_{0}) \! := \! (2 
w^{5} R_{-}^{\prime}(0) b_{0}^{-} \! + \! 2 w^{5} R_{-}^{\prime}
(0) (a_{0}^{-})^{2} \! + \! 8 i w^{7} R_{-}^{\prime}(0) a_{0}^{-} 
\! - \! 8 w^{9} R_{-}^{\prime}(0)& \\
&\, \, \, \, \, \, \, \, \, \, \, \, \, \, \, \, \, \, \, \,
\, \, \, \, \, \, \, \, \, \, \, \, \, + \, 
2 a_{0}^{-} w^{5} \frac{1}{3!} R_{-}^{\prime \prime \prime}(0) 
\! + \! 4 i w^{7} \frac{1}{3!} R_{-}^{\prime \prime \prime}(0) 
\! + \! \frac{1}{5!} R_{-}^{V}(0) w^{5}) e^{2i \phi_{-}(\lambda
_{0})} e^{-8i \lambda_{0}^{2} w^{2}},& \\
&\vert \vert \widehat{{\cal F}}^{\alpha}(\cdot;\! \lambda_{0}) 
\exp \{-8 i \lambda_{0}^{2} (\cdot)^{2}\} \vert \vert_{{\cal L}
^{k}(\sqrt{-t} \, \cup_{l=3}^{4} \widehat{\varsigma}_{{\cal C}}
^{(l)};\Bbb C)} = \underline{c}(\lambda_{0}), \, \, \, \, \, k 
\in \{1,2,\infty\},& \\
&\phi_{-}(\lambda_{0}) := - \int_{\lambda_{0}}^{\infty} \! \frac{
\ln (1 - \vert r(\xi) \vert^{2})}{\xi} \frac{d \xi}{\pi}, \, \, \, 
a_{0}^{-} := \int_{\lambda_{0}}^{\infty} \! \frac{\ln (1 - \vert 
r(\xi) \vert^{2})}{\xi^{3}} \frac{d \xi}{\pi i}, \, \, \, b_{0}^{
-} := \int_{\lambda_{0}}^{\infty} \! \frac{\ln (1 - \vert r(\xi) 
\vert^{2})}{\xi^{5}} \frac{d \xi}{\pi i},&
\end{eqnarray*}
and, for arbitrary $N_{o} \! \in \! \Bbb Z_{\geq 1}$,
\begin{eqnarray*}
&\widehat{\mu}^{\widehat{\varsigma}_{{\cal C}}}(\frac{w}{\sqrt{
-t}}) = {\rm I} + \sum\limits_{p^{\prime}=1}^{2} \sum\limits_{p
=0}^{N_{o}} \frac{\widehat{\mu}_{p^{\prime},p}(w;\lambda_{0})}{
(-t)^{p+p^{\prime}/2}} + \widehat{{\cal E}}^{\widehat{\varsigma}
_{{\cal C}}}(\frac{w}{\sqrt{-t}};\! \lambda_{0}),&
\end{eqnarray*}
with
\begin{eqnarray*}
&\vert \vert \widehat{\mu}_{p^{\prime},p}(\cdot;\! \lambda_{0}) 
\vert \vert_{\cup_{l \in \{2,\infty\}} {\cal L}^{l}(\sqrt{-t} \, 
\, \widehat{\varsigma}_{{\cal C}} \setminus \{0\};M_{2}(\Bbb C)) 
\to {\cal L}^{2}(\sqrt{-t} \, \, \widehat{\varsigma}_{{\cal C}} 
\setminus \{0\};M_{2}(\Bbb C))} < \infty,& \\
&\vert \vert \widehat{{\cal E}}^{\widehat{\varsigma}_{{\cal C}}}(
\cdot;\! \lambda_{0}) \vert \vert_{\cup_{l \in \{2,\infty\}}{\cal 
L}^{l}(\sqrt{-t} \, \, \widehat{\varsigma}_{{\cal C}} \setminus 
\{0\};M_{2}(\Bbb C)) \to {\cal L}^{2}(\sqrt{-t} \, \, \widehat{
\varsigma}_{{\cal C}} \setminus \{0\};M_{2}(\Bbb C))} = {\cal O} 
\! \left(\frac{\underline{c}(\lambda_{0})}{(-\lambda_{0}^{2} t)
^{N_{o}+3/2}} \right) \!.&
\end{eqnarray*}
\end{bbbb}
\begin{bbbb}
Let $N_{o} \! \in \! \Bbb Z_{\geq 1}$. Then for $1 \! \leq \! 
p^{\prime} \! \leq \! 2$ and $0 \! \leq \! p \! \leq \! N_{o}$, 
$\widehat{\mu}_{p^{\prime},p}(w;\! \lambda_{0})$ satisfy the 
following involutions, $\widehat{\mu}_{p^{\prime},p}(w;\! 
\lambda_{0}) \! = \! \sigma_{3} \widehat{\mu}_{p^{\prime},p}(
-w;\! \lambda_{0}) \sigma_{3}$ and $\widehat{\mu}_{p^{\prime},
p}(w;\! \lambda_{0}) \! = \! \sigma_{1} \overline{\widehat{
\mu}_{p^{\prime},p}(\overline{w};\! \lambda_{0})} \sigma_{1}$.
\end{bbbb}

Using the fact that the coefficients $u^{-}_{-1,p,q}(\lambda_{0}
) \! := \! u^{-}_{-1,p^{\prime}}(\lambda_{0})$, $1 \! \leq \! p
^{\prime} \! \leq \! 6$, are determined by the relation $2i \lim
_{\lambda \to \infty \atop \lambda \, \in \, \Bbb C \setminus 
\widehat{\varsigma}_{{\cal C}}} \! (\lambda \widehat{m}^{\widehat{
\varsigma}}_{{\cal C}}(\lambda))_{12} \! := \! \sum_{p^{\prime}=1}
^{6} \! \frac{u^{-}_{-1,p^{\prime}}(\lambda_{0})}{(-t)^{p^{\prime}
/2}}$, and proceeding as in the proof of Lemma~4.2, one obtains 
the following.
\begin{cccc}
As $t \! \to \! -\infty$ and $x \! \to \! +\infty$ such that 
$\lambda_{0} \! > \! M$ and $(x,t) \! \in \! \Bbb R^{2} \setminus 
\Omega_{n}$, for those $\gamma_{n} \! \in \! (\frac{\pi}{2},\! \pi)$,
\begin{eqnarray*}
&2 i \lim_{\lambda \to \infty \atop \lambda \, \in \, \Bbb C 
\setminus \widehat{\varsigma}_{{\cal C}}} \! (\lambda \widehat{m}^{
\widehat{\varsigma}}_{{\cal C}}(\lambda))_{12} \! := \! \sum\limits
_{p=1}^{6} \sum\limits_{q=0}^{p-1} \frac{u^{-}_{-1,p,q}(\lambda_{0}
)(\ln \vert t \vert)^{q}}{(-t)^{p/2}} + {\cal O}(\widehat{K}_{{\cal 
C}}^{\widehat{\varsigma}}(\lambda_{0})(-t)^{-7/2}),&
\end{eqnarray*}
where
\begin{eqnarray*}
&u^{-}_{-1,1,0}(\lambda_{0}) \! = \! 0, \, \, \, \, \, \, \, u^{-}
_{-1,2,0}(\lambda_{0}) \! = \! \frac{i R_{-}^{\prime}(0) s_{0}
^{-}}{\pi \lambda_{0}^{2} 2^{3}}, \, \, \, \, \, \, \, u^{-}_{-1,2,
1}(\lambda_{0}) \! = \! 0,& \\
&u^{-}_{-1,3,0}(\lambda_{0}) \! = \! \frac{i R_{-}^{\prime}(0) 
s_{0}^{-}}{\pi \lambda_{0}^{2} 2^{3}} \! \int_{0}^{\infty} \! 
\widehat{\mu}^{a}_{11}(\frac{\xi^{1/2} e^{-\frac{i \pi}{4}}}{2
^{3/2} \lambda_{0}};\! \lambda_{0}) e^{-\xi} d \xi, \, \, \, \, 
\, \, u^{-}_{-1,3,1}(\lambda_{0}) \! = \! u^{-}_{-1,3,2}(\lambda
_{0}) \! = \! 0,& \\
&u^{-}_{-1,4,0}(\lambda_{0}) \! = \! \frac{R_{-}^{\prime}(0)a_{
0}^{-} s_{0}^{-}}{\pi \lambda_{0}^{4} 2^{5}} \! + \! \frac{\frac{
1}{3!} R_{-}^{\prime \prime \prime}(0) s_{0}^{-}}{\pi \lambda_{0}
^{4} 2^{6}} \! + \! \frac{R_{-}^{\prime}(0) s_{0}^{-}}{\pi 
\lambda_{0}^{6} 2^{6}} \! + \! \frac{i R_{-}^{\prime}(0) s_{0}^{
-}}{\pi \lambda_{0}^{2} 2^{3}} \! \int_{0}^{\infty} \! \widehat{
\mu}^{b}_{11}(\frac{\xi^{1/2} e^{-\frac{i \pi}{4}}}{2^{3/2} 
\lambda_{0}};\! \lambda_{0}) e^{-\xi} d \xi,& \\
&u^{-}_{-1,4,1}(\lambda_{0}) \! = \! u^{-}_{-1,4,2}(\lambda_{0}) 
\! = \! u^{-}_{-1,4,3}(\lambda_{0}) \! = \! 0,& \\
&u^{-}_{-1,5,0}(\lambda_{0}) \! = \! \frac{R_{-}^{\prime}(0) a_{
0}^{-} s_{0}^{-}}{\pi \lambda_{0}^{4} 2^{5}} \! \int_{0}^{\infty} 
\! \widehat{\mu}^{a}_{11}(\frac{\xi^{1/2} e^{-\frac{i \pi}{4}}}{
2^{3/2} \lambda_{0}};\! \lambda_{0}) \xi e^{-\xi} d \xi \! + \! 
\frac{\frac{1}{3!} R_{-}^{\prime \prime \prime}(0) s_{0}^{-}}{\pi 
\lambda_{0}^{4} 2^{6}} \! \int_{0}^{\infty} \! \widehat{\mu}^{a}
_{11}(\frac{\xi^{1/2} e^{-\frac{i \pi}{4}}}{2^{3/2} \lambda_{0}};
\! \lambda_{0}) \xi e^{-\xi} d \xi& \\
&\, \, \, \, \, \, \, \, \, \, \, \, \, \, \, \, \, \, \, \, \, 
+ \, \frac{R_{-}^{\prime}(0) s_{0}^{-}}{\pi \lambda_{0}^{6}2^{
7}} \! \int_{0}^{\infty} \! \widehat{\mu}^{a}_{11}(\frac{\xi^{1/2} 
e^{-\frac{i \pi}{4}}}{2^{3/2} \lambda_{0}};\! \lambda_{0}) \xi^{2} 
e^{-\xi} d \xi \! + \! \frac{i R_{-}^{\prime}(0) s_{0}^{-}}{\pi 
\lambda_{0}^{2} 2^{3}} \! \int_{0}^{\infty} \! \widehat{\mu}^{c}_{
11}(\frac{\xi^{1/2} e^{-\frac{i \pi}{4}}}{2^{3/2} \lambda_{0}};\! 
\lambda_{0}) e^{-\xi} d \xi,& \\
&u^{-}_{-1,5,1}(\lambda_{0}) \! = \! u^{-}_{-1,5,2}(\lambda_{0}) 
\! = \! u^{-}_{-1,5,3}(\lambda_{0}) \! = \! u^{-}_{-1,5,4}(\lambda
_{0}) \! = \! 0,& \\
&u^{-}_{-1,6,0}(\lambda_{0}) \! = \! - \frac{i(2 R_{-}^{\prime}
(0) b_{0}^{-} + 2 R_{-}^{\prime}(0) (a_{0}^{-})^{2} + 2 a_{0}^{-} 
\frac{1}{3!} R_{-}^{\prime \prime \prime}(0) + \frac{1}{5!} R_{-}
^{V}(0)) s_{0}^{-}}{\pi \lambda_{0}^{6} 2^{8}} \! - \! \frac{3 i 
R_{-}^{\prime}(0) s_{0}^{-}}{\pi \lambda_{0}^{10} 2^{9}}& \\
&- \, \frac{3i(2 R_{-}^{\prime}(0) a_{0}^{-} + \frac{1}{3!} R_{-}
^{\prime \prime \prime}(0)) s_{0}^{-}}{\pi \lambda_{0}^{8} 2^{9}} 
\! + \! \frac{(2 R_{-}^{\prime}(0) a_{0}^{-} + \frac{1}{3!} R_{-}
^{\prime \prime \prime}(0)) s_{0}^{-}}{\pi \lambda_{0}^{4} 2^{6}} 
\! \int_{0}^{\infty} \! \widehat{\mu}^{b}_{11}(\frac{\xi^{1/2} 
e^{-\frac{i \pi}{4}}}{2^{3/2} \lambda_{0}};\! \lambda_{0}) \xi 
e^{-\xi} d \xi& \\
&+ \, \frac{R_{-}^{\prime}(0) s_{0}^{-}}{\pi \lambda_{0}^{6} 2^{7}} 
\! \int_{0}^{\infty} \! \widehat{\mu}^{b}_{11}(\frac{\xi^{1/2} e^{
-\frac{i \pi}{4}}}{2^{3/2} \lambda_{0}};\! \lambda_{0}) \xi^{2} 
e^{-\xi} d \xi \! + \! \frac{i R_{-}^{\prime}(0) s_{0}^{-}}{\pi 
\lambda_{0}^{2} 2^{3}} \! \int_{0}^{\infty} \! \widehat{\mu}^{d}
_{11}(\frac{\xi^{1/2} e^{-\frac{i \pi}{4}}}{2^{3/2} \lambda_{0}}
;\! \lambda_{0}) e^{-\xi} d \xi,& \\
&u^{-}_{-1,6,1}(\lambda_{0}) \! = \! u^{-}_{-1,6,2}(\lambda_{0}) 
\! = \! u^{-}_{-1,6,3}(\lambda_{0}) \! = \! u^{-}_{-1,6,4}(\lambda
_{0}) \! = \! u^{-}_{-1,6,5}(\lambda_{0}) \! = \! 0,& \\
&\widehat{K}^{\widehat{\varsigma}}_{{\cal C}}(\lambda_{0}) \! := \! 
- \frac{i(2 R_{-}^{\prime}(0) b_{0}^{-} + 2 R_{-}^{\prime}(0) (a_{
0}^{-})^{2} + 2 a_{0}^{-} \frac{1}{3!} R_{-}^{\prime \prime \prime}
(0) + \frac{1}{5!} R_{-}^{V}(0)) s_{0}^{-}}{\pi \lambda_{0}^{6} 2^{
9}} \! \int_{0}^{\infty} \! \widehat{\mu}^{a}_{11}(\frac{\xi^{1/2} 
e^{-\frac{i \pi}{4}}}{2^{3/2} \lambda_{0}};\! \lambda_{0}) \xi^{2} 
e^{-\xi} d \xi& \\
&- \, \frac{i(2 R_{-}^{\prime}(0) a_{0}^{-} + \frac{1}{3!}R_{-}
^{\prime \prime \prime}(0)) s_{0}^{-}}{\pi \lambda_{0}^{8}2^{10}
} \! \int_{0}^{\infty} \! \widehat{\mu}^{a}_{11}(\frac{\xi^{1/2}
e^{-\frac{i \pi}{4}}}{2^{3/2} \lambda_{0}};\! \lambda_{0}) \xi^{
3} e^{-\xi} d \xi \! + \! \frac{iR_{-}^{\prime}(0)s_{0}^{-}}{\pi 
\lambda_{0}^{2} 2^{3}} \! \int_{0}^{\infty} \! \widehat{\mu}^{e}
_{11}(\frac{\xi^{1/2} e^{-\frac{i \pi}{4}}}{2^{3/2} \lambda_{0}};
\! \lambda_{0})e^{-\xi} d \xi& \\
&\, \, \, \, \, \, \, \, \, \, \, \, \, \, \, \, \, - \, 
\frac{i R_{-}^{\prime}(0) s_{0}^{-}}{\pi \lambda_{0}^{10} 2^{12}} 
\! \int_{0}^{\infty} \! \widehat{\mu}^{a}_{11}(\frac{\xi^{1/2}e^{
-\frac{i \pi}{4}}}{2^{3/2} \lambda_{0}};\! \lambda_{0}) \xi^{4} 
e^{-\xi} d \xi \! + \! \frac{R_{-}^{\prime}(0) s_{0}^{-}}{\pi 
\lambda_{0}^{6} 2^{7}} \! \int_{0}^{\infty} \! \widehat{\mu}^{c}
_{11}(\frac{\xi^{1/2} e^{-\frac{i \pi}{4}}}{2^{3/2} \lambda_{0}};
\! \lambda_{0}) \xi^{2} e^{-\xi} d \xi& \\
&\! \! \! \! \! \! \! \! \! \! \! \! \! \! \! \! \! \! \! \! \! 
\! \! \! \! \! \! \! \! \! \! \! \! \! \! \! \! \! \! \! \! \! 
\! \! \! \! \! \! \! \, + \, 
\frac{(2 R_{-}^{\prime}(0) a_{0}^{-} + \frac{1}{3!} R_{-}^{
\prime \prime \prime}(0)) s_{0}^{-}}{\pi \lambda_{0}^{4} 2^{6}} 
\! \int_{0}^{\infty} \! \widehat{\mu}^{c}_{11}(\frac{\xi^{1/2} 
e^{-\frac{i \pi}{4}}}{2^{3/2} \lambda_{0}};\! \lambda_{0}) \xi 
e^{-\xi} d \xi,&
\end{eqnarray*}
$s_{0}^{-} \! := \! \exp \{2 i \phi_{-}(\lambda_{0})\}$, $R^{
\prime}_{-}(0)$, $\frac{1}{3!} R^{\prime \prime \prime}_{-}(0)$, 
and $\frac{1}{5!} R^{V}_{-}(0)$ are given in Lemma~6.2.1, $\phi
_{-}(\lambda_{0})$, $a_{0}^{-}$, and $b_{0}^{-}$ are given in 
Proposition~6.2.1, $\widehat{\mu}^{a}_{11}(\cdot;\! \lambda_{0}) 
\! := \! (\widehat{\mu}_{1,0}(\cdot;\! \lambda_{0}))_{11}$, 
$\widehat{\mu}^{b}_{11}(\cdot;\! \lambda_{0}) \! := \! (\widehat{
\mu}_{2,0}(\cdot;\! \lambda_{0}))_{11}$, $\widehat{\mu}^{c}_{11}(
\cdot;\! \lambda_{0}) \! := \! (\widehat{\mu}_{1,1}(\cdot;\! 
\lambda_{0}))_{11}$, $\widehat{\mu}^{d}_{11}(\cdot;\! \lambda_{0}
) \! := \! (\widehat{\mu}_{2,1}(\cdot;\! \lambda_{0}))_{11}$, and 
$\widehat{\mu}^{e}_{11}(\cdot;\! \lambda_{0}) \! := \! (\widehat{
\mu}_{1,2}(\cdot;\! \lambda_{0}))_{11}$, where $(\widehat{\mu}_{
\alpha,\beta}(\cdot;\! \lambda_{0}))_{ij}$ denotes the $(i \, 
j)$th element of $\widehat{\mu}_{\alpha,\beta}(\cdot;\! \lambda
_{0})$.
\end{cccc}

Finally, proceeding as in the proof of Lemma~4.3, one obtains the 
following.
\begin{cccc}
As $t \! \to \! -\infty$ and $x \! \to \! +\infty$ such that 
$\lambda_{0} \! > \! M$ and $(x,t) \! \in \! \Bbb R^{2} \setminus 
\Omega_{n}$, for those $\gamma_{n} \! \in \! (\frac{\pi}{2},\! 
\pi)$, $Q(x,t)$ has the asymptotic expansion given in Theorem~2.2, 
Eqs.~(10)--(27).
\end{cccc}
\subsection{Asymptotics of $((\Psi^{-1}(x,t;0))_{11})^{2}$ as $t 
\! \to \! -\infty$}
Using Proposition~5.1, and proceeding as in (the proof of) Lemma~5.1, 
one obtains the following.
\begin{cccc}
As $t \! \to \! -\infty$ and $x \! \to \! +\infty$ such that 
$\lambda_{0} \! > \! M$ and $(x,t) \! \in \! \Bbb R^{2} \setminus 
\Omega_{n}$, for those $\gamma_{n} \! \in \! (\frac{\pi}{2},\! 
\pi)$,
\begin{eqnarray*}
&((\Psi^{-1}(x,t;0))_{11})^{2} \! = \! \exp\{i \arg q_{-}(x,t)\} 
\! + \! {\cal O} \! \left(\frac{c^{{\cal S}}(\lambda_{0})(\ln 
\vert t \vert)^{2}}{(-\lambda_{0}^{2} t)^{3/2}} \right) \!,&
\end{eqnarray*}
where $\arg q_{-}(x,t)$ is given in Theorem~2.3, Eqs.~(46)--(51).
\end{cccc}
\begin{cccc}
As $t \! \to \! - \infty$ and $x \! \to \! + \infty$ such that 
$\lambda_{0} \! > \! M$ and $(x,t) \! \in \! \Bbb R^{2} \setminus 
\Omega_{n}$, for those $\gamma_{n} \! \in \! (\frac{\pi}{2},\! 
\pi)$, $q(x,t)$ has the asymptotic expansion given in Theorem~2.3, 
Eqs.~(45)--(51).
\end{cccc}
\begin{cccc}
As $t \! \to \! - \infty$ and $x \! \to \! - \infty$ such that 
$\widehat{\lambda}_{0} \! := \! \sqrt{\frac{1}{2}(\frac{x}{t} 
\! - \! \frac{1}{s})} \! > \! M$, $\frac{x}{t} \! > \! \frac{
1}{s}$, $s \! \in \! \Bbb R_{>0}$, and $(x,t) \! \in \! \Bbb 
R^{2} \setminus \widetilde{\Omega}_{n}$, for those $\gamma_{n} 
\! \in \! (\frac{\pi}{2},\! \pi)$, $u(x,t)$ has the asymptotic 
expansion given in Theorem~2.4, Eqs.~(56)--(66).
\end{cccc}
\vspace{1.15cm}
\begin{flushleft}
{\LARGE {\bf Acknowledgements}}
\end{flushleft}
The author would like to extend a profound debt of gratitude to 
A.~V.~Kitaev for extremely helpful and productive discussions 
pertaining to the results of this paper. The author is extremely 
grateful to L.~D.~Faddeev for the opportunity to work at POMI. 
The author is also very grateful to P.~A.~Deift for a copy of 
\cite{a22} prior to publication, K.~T.-R.~McLaughlin for a copy 
of \cite{a12}, S.~Venakides for a copy of \cite{a9}, X.~Zhou for 
copies of [24], and A.~R.~Its and V.~B.~Matveev for encouragement 
and support.
\clearpage
\begin{flushleft}
{\LARGE {\bf Appendix}}
\end{flushleft}
To ${\cal O}(t^{-7/2})$, the explicit recurrence formulae for 
the coefficients of the asymptotic expansions for $Q(x,t)$ given 
in Lemmae~3.4 and 6.1.2 are listed here (they are derived {}from 
systems~(78) and (79) upon equating coefficients of powers of 
like terms on both left- and right-hand sides). With respect 
to the recurrence formulae given here, the following must be 
borne in mind: (1) $\nu \! := \! \nu(\lambda_{0})$; (2) $\tau
^{\pm} \! := \! 4 \lambda_{0}^{4} t \! \mp \! \nu \ln \vert 
t \vert$; (3) $u^{\pm}_{\alpha,\beta} \! := \! u^{\pm}_{\alpha,
\beta}(\lambda_{0};t) \! := \! \sum_{\gamma^{\prime}=0}^{\beta 
- \vert \alpha \vert} \! u^{\pm}_{\alpha,\beta,\gamma^{\prime}
}(\lambda_{0})(\ln \vert t \vert)^{\gamma^{\prime}}$; (4) $f^{
\prime} \! := \! \partial_{\lambda_{0}}f(\lambda_{0};t) \vert_{
t \, = \, {\rm fixed}}$; (5) $\dot{f} \! := \! \partial_{t} f(
\lambda_{0};t) \vert_{\lambda_{0} \, = \, {\rm fixed}}$; and 
(6) the upper/lower signs are taken as $t \! \to \! \pm \infty$.
\setcounter{equation}{0}
\renewcommand{\theequation}{A.\arabic{equation}}
\begin{flushleft}
(1) ${\cal O}((\pm t)^{-1/2} \exp\{i \tau^{\pm}\})$,
\end{flushleft}
\begin{eqnarray}
&4\lambda_{0}^{4}u^{\pm}_{1,1}-4\lambda_{0}^{4}u^{\pm}_{1,1}=0;&
\end{eqnarray}
\begin{flushleft}
(2) ${\cal O}((\pm t)^{-1} \exp\{i \tau^{\pm}\})$,
\end{flushleft}
\begin{eqnarray}
&4\lambda_{0}^{4}u^{\pm}_{1,2}-4\lambda_{0}^{4}u^{\pm}_{1,2}=0;&
\end{eqnarray}
\begin{flushleft}
(3) ${\cal O}((\pm t)^{-3/2} \exp\{i \tau^{\pm}\})$,
\end{flushleft}
\begin{eqnarray}
&\nu u^{\pm}_{1,1} - \frac{\lambda_{0} \nu^{\prime} u^{\pm}_{1,1} 
\ln \vert t \vert}{2} + 4 \lambda_{0}^{4} u^{\pm}_{1,3} \pm i t 
\dot{u}^{\pm}_{1,1}+\frac{iu^{\pm}_{1,1}}{2}+\frac{i \lambda_{0} 
u^{\pm \prime}_{1,1}}{2} - \frac{i u^{\pm}_{1,1}}{2}& \nonumber \\
&- \frac{i \lambda_{0} u^{\pm \prime}_{1,1}}{2} + \frac{\lambda_{
0} \nu^{\prime} u^{\pm}_{1,1} \ln \vert t \vert}{2} - 4 \lambda_{
0}^{4} u^{\pm}_{1,3} - 2 \lambda_{0}^{2} \overline{u^{\pm}_{1,1}} 
u^{\pm}_{1,1} u^{\pm}_{1,1}=0;&
\end{eqnarray}
\begin{flushleft}
(4) ${\cal O}((\pm t)^{-3/2})$,
\end{flushleft}
\begin{eqnarray}
&\pm it \dot{u}^{\pm}_{-1,1} \mp \frac{i u^{\pm}_{-1,1}}{2} \mp 
\frac{i \lambda_{0} u^{\pm \prime}_{-1,1}}{2}-2 \lambda_{0}^{2} 
\overline{u^{\pm}_{1,1}} u^{\pm}_{1,1} u^{\pm}_{-1,1}-2\lambda_{
0}^{2} \overline{u^{\pm}_{1,1}}u^{\pm}_{-1,1}u^{\pm}_{1,1}=0;&
\end{eqnarray}
\begin{flushleft}
(5) ${\cal O}((\pm t)^{-3/2} \exp\{2i\tau^{\pm}\})$,
\end{flushleft}
\begin{eqnarray}
&8 \lambda_{0}^{4} u^{\pm}_{3,3} - 16 \lambda_{0}^{4} u^{\pm}_{
3,3}=0;&
\end{eqnarray}
\begin{flushleft}
(6) ${\cal O}((\pm t)^{-3/2} \exp\{-i\tau^{\pm}\})$,
\end{flushleft}
\begin{eqnarray}
&-4\lambda_{0}^{4}u^{\pm}_{-3,3}-4\lambda_{0}^{4}u^{\pm}_{-3,
3}-2 \lambda_{0}^{2} \overline{u^{\pm}_{1,1}} u^{\pm}_{-1,1} 
u^{\pm}_{-1,1}=0;&
\end{eqnarray}
\begin{flushleft}
(7) ${\cal O}((\pm t)^{-2} \exp\{i \tau^{\pm}\})$,
\end{flushleft}
\begin{eqnarray}
&\nu u^{\pm}_{1,2} - \frac{\lambda_{0} \nu^{\prime} u^{\pm}_{1,
2} \ln \vert t \vert}{2} + 4 \lambda_{0}^{4} u^{\pm}_{1,4} \pm 
i t \dot{u}^{\pm}_{1,2} \mp i u^{\pm}_{1,2} \mp \frac{i\lambda
_{0}u^{\pm \prime}_{1,2}}{2} \pm \frac{i \lambda_{0} u^{\pm 
\prime}_{1,2}}{2}+\frac{\lambda_{0} \nu^{\prime} u^{\pm}_{1,2} 
\ln \vert t \vert}{2}& \nonumber \\
&\pm \frac{i u^{\pm}_{1,2}}{2}-4 \lambda_{0}^{4} u^{\pm}_{1,4}-2 
\lambda_{0}^{2} \overline{u^{\pm}_{1,1}} u^{\pm}_{1,1} u^{\pm}_{
1,2}-2\lambda_{0}^{2} \overline{u^{\pm}_{1,1}}u^{\pm}_{1,2}u^{
\pm}_{1,1}-2 \lambda_{0}^{2} \overline{u^{\pm}_{1,2}} u^{\pm}_{1,
1} u^{\pm}_{1,1}=0;&
\end{eqnarray}
\begin{flushleft}
(8) ${\cal O}((\pm t)^{-2})$,
\end{flushleft}
\begin{eqnarray}
&\pm i t \dot{u}^{\pm}_{-1,2} \mp i u^{\pm}_{-1,2} \mp \frac{i 
\lambda_{0} u^{\pm \prime}_{-1,2}}{2} - 2 \lambda_{0}^{2} 
\overline{u^{\pm}_{1,1}}u^{\pm}_{1,1}u^{\pm}_{-1,2} - 2 \lambda
_{0}^{2} \overline{u^{\pm}_{1,1}}u^{\pm}_{-1,1}u^{\pm}_{1,2}& 
\nonumber \\
&-2 \lambda_{0}^{2} \overline{u^{\pm}_{1,1}}u^{\pm}_{1,2}u^{\pm}
_{-1,1} - 2 \lambda_{0}^{2} \overline{u^{\pm}_{1,1}} u^{\pm}_{
-1,2} u^{\pm}_{1,1} - 2 \lambda_{0}^{2} \overline{u^{\pm}_{1,2}} 
u^{\pm}_{1,1} u^{\pm}_{-1,1} - 2 \lambda_{0}^{2} \overline{u^{
\pm}_{1,2}} u^{\pm}_{-1,1} u^{\pm}_{1,1}=0;&
\end{eqnarray}
\begin{flushleft}
(9) ${\cal O}((\pm t)^{-2} \exp\{2i\tau^{\pm}\})$,
\end{flushleft}
\begin{eqnarray}
&8 \lambda_{0}^{4} u^{\pm}_{3,4}-16 \lambda_{0}^{4} u^{\pm}_{
3,4}=0;&
\end{eqnarray}
\begin{flushleft}
(10) ${\cal O}((\pm t)^{-2} \exp\{-i \tau^{\pm}\})$,
\end{flushleft}
\begin{eqnarray}
&-4 \lambda_{0}^{4} u^{\pm}_{-3,4}-4 \lambda_{0}^{4} u^{\pm}_{
-3,4} - 2 \lambda_{0}^{2} \overline{u^{\pm}_{1,1}} u^{\pm}_{-1,
1} u^{\pm}_{-1,2} - 2 \lambda_{0}^{2} \overline{u^{\pm}_{1,1}} 
u^{\pm}_{-1,2} u^{\pm}_{-1,1}& \nonumber \\
&- 2 \lambda_{0}^{2} \overline{u^{\pm}_{1,2}} u^{\pm}_{-1,1} 
u^{\pm}_{-1,1}=0;&
\end{eqnarray}
\begin{flushleft}
(11) ${\cal O}((\pm t)^{-5/2} \exp\{i \tau^{\pm}\})$,
\end{flushleft}
\begin{eqnarray}
&\nu u^{\pm}_{1,3} - \frac{\lambda_{0} \nu^{\prime} u^{\pm}_{1,
3} \ln \vert t \vert}{2} + 4 \lambda_{0}^{4} u^{\pm}_{1,5} \pm 
i t \dot{u}^{\pm}_{1,3} \mp \frac{3iu^{\pm}_{1,3}}{2} \mp \frac{
i \lambda_{0} u^{\pm \prime}_{1,3}}{2} \mp \frac{i \nu^{\prime} 
u^{\pm \prime}_{1,1} \ln \vert t \vert}{32 \lambda_{0}^{2}} \mp 
\frac{i \nu^{\prime \prime} u^{\pm}_{1,1} \ln \vert t \vert}{64 
\lambda_{0}^{2}}& \nonumber \\
&\pm \frac{i \nu^{\prime} 
u^{\pm}_{1,1} \ln \vert t \vert}{64 \lambda_{0}^{3}} \pm \frac{i 
\lambda_{0} u^{\pm \prime}_{1,3}}{2} \pm \frac{i u^{\pm}_{1,3}}{
2} - \frac{(\nu^{\prime})^{2} u^{\pm}_{1,1} (\ln \vert t 
\vert)^{2}}{64 \lambda_{0}^{2}} + \frac{\lambda_{0} \nu^{\prime} 
u^{\pm}_{1,3} \ln \vert t \vert}{2} - 4 \lambda_{0}^{4}u^{\pm}_{
1,5}& \nonumber \\
&+\frac{u^{\pm \prime \prime}_{1,1}}{64 \lambda_{0}^{2}}-\frac{
u^{\pm \prime}_{1,1}}{64 \lambda_{0}^{3}} - 2 \lambda_{0}^{2} 
\overline{u^{\pm}_{1,1}} u^{\pm}_{1,1} u^{\pm}_{1,3} - 2 
\lambda_{0}^{2} \overline{u^{\pm}_{1,1}} u^{\pm}_{-1,1} u^{\pm}
_{3,3} - 2 \lambda_{0}^{2} \overline{u^{\pm}_{1,1}} u^{\pm}_{1,
2} u^{\pm}_{1,2}& \nonumber \\
&-2 \lambda_{0}^{2} \overline{u^{\pm}_{1,1}} u^{\pm}_{1,3} u^{
\pm}_{1,1} - 2 \lambda_{0}^{2} \overline{u^{\pm}_{1,1}} u^{\pm}_{
3,3} u^{\pm}_{-1,1} - 2 \lambda_{0}^{2} \overline{u^{\pm}_{1,2}} 
u^{\pm}_{1,1} u^{\pm}_{1,2} - 2 \lambda_{0}^{2} \overline{u^{\pm}
_{1,2}} u^{\pm}_{1,2} u^{\pm}_{1,1}& \nonumber \\
&-2 \lambda_{0}^{2} \overline{u^{\pm}_{1,3}} u^{\pm}_{1,1} u^{
\pm}_{1,1} + 2 \lambda_{0}^{2} \overline{u^{\pm}_{-3,3}} u^{\pm}
_{-1,1} u^{\pm}_{-1,1} + \frac{\nu^{\prime} \overline{u^{\pm}
_{1,1}}u^{\pm}_{1,1}u^{\pm}_{1,1} \ln \vert t \vert}{8 \lambda_{
0}} \mp \frac{i(\overline{u^{\pm}_{1,1}})^{\prime} u^{\pm}_{1,1} 
u^{\pm}_{1,1}}{8 \lambda_{0}}& \nonumber \\
&\mp \frac{i(\overline{u^{\pm}_{-1,1}})^{\prime} u^{\pm}_{1,1} u^{
\pm}_{-1,1}}{8 \lambda_{0}} \mp \frac{i(\overline{u^{\pm}_{-1,1}})
^{\prime} u^{\pm}_{-1,1} u^{\pm}_{1,1}}{8 \lambda_{0}} + \frac{
u^{\pm}_{1,1} \overline{u^{\pm}_{1,1}} u^{\pm}_{1,1} \overline{
u^{\pm}_{1,1}} u^{\pm}_{1,1}}{2} + \frac{u^{\pm}_{1,1} \overline{
u^{\pm}_{1,1}} u^{\pm}_{1,1} \overline{u^{\pm}_{-1,1}} u^{\pm}_{
-1,1}}{2}& \nonumber \\
&+\frac{u^{\pm}_{1,1} \overline{u^{\pm}_{1,1}} u^{\pm}_{-1,1} 
\overline{u^{\pm}_{-1,1}} u^{\pm}_{1,1}}{2} + \frac{u^{\pm}_{1,1} 
\overline{u^{\pm}_{-1,1}} u^{\pm}_{1,1} \overline{u^{\pm}_{1,1}} 
u^{\pm}_{-1,1}}{2} + \frac{u^{\pm}_{1,1} \overline{u^{\pm}_{-1,1}} 
u^{\pm}_{-1,1} \overline{u^{\pm}_{1,1}} u^{\pm}_{1,1}}{2}& 
\nonumber \\
&+ \frac{u^{\pm}_{1,1} \overline{u^{\pm}_{-1,1}} u^{\pm}_{-1,1} 
\overline{u^{\pm}_{-1,1}} u^{\pm}_{-1,1}}{2} + \frac{u^{\pm}_{-1,
1} \overline{u^{\pm}_{1,1}} u^{\pm}_{1,1} \overline{u^{\pm}_{-1,
1}} u^{\pm}_{1,1}}{2} + \frac{u^{\pm}_{-1,1} \overline{u^{\pm}_{
-1,1}} u^{\pm}_{1,1} \overline{u^{\pm}_{1,1}} u^{\pm}_{1,1}}{2}&
\nonumber \\
&+ \frac{u^{\pm}_{-1,1} \overline{u^{\pm}_{-1,1}} u^{\pm}_{1,1} 
\overline{u^{\pm}_{-1,1}} u^{\pm}_{-1,1}}{2} + \frac{u^{\pm}_{
-1,1} \overline{u^{\pm}_{-1,1}} u^{\pm}_{-1,1} \overline{u^{\pm}
_{-1,1}} u^{\pm}_{1,1}}{2}=0;&
\end{eqnarray}
\begin{flushleft}
(12) ${\cal O}((\pm t)^{-5/2})$,
\end{flushleft}
\begin{eqnarray}
&\pm it\dot{u}^{\pm}_{-1,3} \mp \frac{3iu^{\pm}_{-1,3}}{2} \mp 
\frac{i \lambda_{0} u^{\pm \prime}_{-1,3}}{2} + \frac{u^{\pm}_{
-1,1}}{64\lambda_{0}^{2}} - \frac{u^{\pm}_{-1,1}}{64\lambda_{0}
^{3}}-2\lambda_{0}^{2} \overline{u^{\pm}_{1,1}} u^{\pm}_{1,3} 
u^{\pm}_{-1,1}-2 \lambda_{0}^{2} \overline{u^{\pm}_{1,1}} u^{
\pm}_{1,2} u^{\pm}_{-1,2}& \nonumber \\
&- 2 \lambda_{0}^{2} \overline{u^{\pm}_{1,1}} u^{\pm}_{-1,2} 
u^{\pm}_{1,2} - 2 \lambda_{0}^{2} \overline{u^{\pm}_{1,1}} 
u^{\pm}_{-1,3} u^{\pm}_{1,1} - 2 \lambda_{0}^{2} \overline{u^{
\pm}_{1,1}} u^{\pm}_{1,1} u^{\pm}_{-1,3} - 2 \lambda_{0}^{2} 
\overline{u^{\pm}_{1,1}} u^{\pm}_{-1,1} u^{\pm}_{1,3}& 
\nonumber \\ 
&- 2 \lambda_{0}^{2} \overline{u^{\pm}_{1,2}} u^{\pm}_{1,1} u^{
\pm}_{-1,2} - 2 \lambda_{0}^{2} \overline{u^{\pm}_{1,2}} u^{\pm}
_{-1,1} u^{\pm}_{1,2} - 2 \lambda_{0}^{2} \overline{u^{\pm}_{1,
2}} u^{\pm}_{1,2} u^{\pm}_{-1,1} - 2 \lambda_{0}^{2} \overline{
u^{\pm}_{1,2}} u^{\pm}_{-1,2} u^{\pm}_{1,1}& \nonumber \\
&- 4 \lambda_{0}^{2} \overline{u^{\pm}_{3,3}} u^{\pm}_{1,1} u^{
\pm}_{1,1} - 2 \lambda_{0}^{2} \overline{u^{\pm}_{1,3}} u^{\pm}
_{-1,1} u^{\pm}_{1,1} - 2 \lambda_{0}^{2} \overline{u^{\pm}_{1,
3}} u^{\pm}_{1,1} u^{\pm}_{-1,1} + \frac{\nu^{\prime} \overline{
u^{\pm}_{1,1}} u^{\pm}_{1,1} u^{\pm}_{-1,1} \ln \vert t \vert}{8 
\lambda_{0}}& \nonumber \\
&+ \frac{\nu^{\prime} \overline{u^{\pm}_{1,1}} u^{\pm}_{-1,1} 
u^{\pm}_{1,1} \ln \vert t \vert}{8 \lambda_{0}} \mp \frac{i 
(\overline{u^{\pm}_{1,1}})^{\prime} u^{\pm}_{1,1} u^{\pm}_{-1,
1}}{8 \lambda_{0}} \mp \frac{i(\overline{u^{\pm}_{1,1}})^{\prime} 
u^{\pm}_{-1,1} u^{\pm}_{1,1}}{8 \lambda_{0}} \mp \frac{i(
\overline{u^{\pm}_{-1,1}})^{\prime} u^{\pm}_{-1,1} u^{\pm}_{-1,1}
}{8 \lambda_{0}}& \nonumber \\
&+\frac{u^{\pm}_{1,1} \overline{u^{\pm}_{1,1}} u^{\pm}_{1,1} 
\overline{u^{\pm}_{1,1}} u^{\pm}_{-1,1}}{2} + \frac{u^{\pm}_{
1,1} \overline{u^{\pm}_{1,1}} u^{\pm}_{-1,1} \overline{u^{\pm}
_{1,1}} u^{\pm}_{1,1}}{2} + \frac{u^{\pm}_{1,1} \overline{u^{
\pm}_{1,1}} u^{\pm}_{-1,1} \overline{u^{\pm}_{-1,1}} u^{\pm}_{
-1,1}}{2}& \nonumber \\
&+ \frac{u^{\pm}_{1,1} \overline{u^{\pm}_{-1,1}} u^{\pm}_{-1,1} 
\overline{u^{\pm}_{1,1}} u^{\pm}_{-1,1}}{2} + \frac{u^{\pm}_{
-1,1} \overline{u^{\pm}_{1,1}} u^{\pm}_{1,1} \overline{u^{\pm}
_{1,1}} u^{\pm}_{1,1}}{2} + \frac{u^{\pm}_{-1,1} \overline{u^{
\pm}_{1,1}} u^{\pm}_{1,1} \overline{u^{\pm}_{-1,1}} u^{\pm}_{
-1,1}}{2}& \nonumber \\
&+ \frac{u^{\pm}_{-1,1} \overline{u^{\pm}_{1,1}} u^{\pm}_{-1,1} 
\overline{u^{\pm}_{-1,1}} u^{\pm}_{1,1}}{2} + \frac{u^{\pm}_{
-1,1} \overline{u^{\pm}_{-1,1}} u^{\pm}_{1,1} \overline{u^{\pm}
_{1,1}} u^{\pm}_{-1,1}}{2} + \frac{u^{\pm}_{-1,1} \overline{u^{
\pm}_{-1,1}} u^{\pm}_{-1,1} \overline{u^{\pm}_{1,1}} u^{\pm}_{
1,1}}{2}& \nonumber \\
&+ \frac{u^{\pm}_{-1,1} \overline{u^{\pm}_{-1,1}} u^{\pm}_{-1,1} 
\overline{u^{\pm}_{-1,1}} u^{\pm}_{-1,1}}{2}=0;&
\end{eqnarray}
\begin{flushleft}
(13) ${\cal O}((\pm t)^{-5/2} \exp\{2 i \tau^{\pm}\})$,
\end{flushleft}
\begin{eqnarray}
&2 \nu u^{\pm}_{3,3} - \lambda_{0} \nu^{\prime} u^{\pm}_{3,3} 
\ln \vert t \vert + 8 \lambda_{0}^{4} u^{\pm}_{3,5} \pm i t 
\dot{u}^{\pm}_{3,3} \mp \frac{3iu^{\pm}_{3,3}}{2} \mp \frac{i 
\lambda_{0} u^{\pm \prime}_{3,3}}{2}& \nonumber \\
&\pm i \lambda_{0} u^{\pm \prime}_{3,3} \pm i u^{\pm}_{3,3} 
+ 2 \lambda_{0} \nu^{\prime} u^{\pm}_{3,3} \ln \vert t \vert 
- 16 \lambda_{0}^{4} u^{\pm}_{3,5} - 2 \lambda_{0}^{2} 
\overline{u^{\pm}_{1,1}} u^{\pm}_{1,1} u^{\pm}_{3,3}& 
\nonumber \\
&-2\lambda_{0}^{2} \overline{u^{\pm}_{1,1}} u^{\pm}_{3,3} u^{
\pm}_{1,1} + 2 \lambda_{0}^{2} \overline{u^{\pm}_{-3,3}} u^{\pm}
_{-1,1} u^{\pm}_{1,1} + 2 \lambda_{0}^{2} \overline{u^{\pm}_{-3,
3}} u^{\pm}_{1,1} u^{\pm}_{-1,1} \mp \frac{i(\overline{u^{\pm}_{
-1,1}})^{\prime} u^{\pm}_{1,1} u^{\pm}_{1,1}}{8 \lambda_{0}}& 
\nonumber \\
&+ \frac{u^{\pm}_{1,1} \overline{u^{\pm}_{1,1}} u^{\pm}_{1,1} 
\overline{u^{\pm}_{-1,1}} u^{\pm}_{1,1}}{2} + \frac{u^{\pm}_{
1,1} \overline{u^{\pm}_{-1,1}} u^{\pm}_{1,1} \overline{u^{\pm}
_{1,1}} u^{\pm}_{1,1}}{2} + \frac{u^{\pm}_{1,1} \overline{u^{
\pm}_{-1,1}} u^{\pm}_{1,1} \overline{u^{\pm}_{-1,1}} u^{\pm}_{
-1,1}}{2}& \nonumber \\
&+\frac{u^{\pm}_{1,1} \overline{u^{\pm}_{-1,1}} u^{\pm}_{-1,1} 
\overline{u^{\pm}_{-1,1}} u^{\pm}_{1,1}}{2} + \frac{u^{\pm}_{
-1,1} \overline{u^{\pm}_{-1,1}} u^{\pm}_{1,1} \overline{u^{\pm}
_{-1,1}} u^{\pm}_{1,1}}{2}=0;&
\end{eqnarray}
\begin{flushleft}
(14) ${\cal O}((\pm t)^{-5/2} \exp\{-i\tau^{\pm}\})$,
\end{flushleft}
\begin{eqnarray}
&-\nu u^{\pm}_{-3,3} + \frac{\lambda_{0} \nu^{\prime} u^{\pm}_{
-3,3} \ln \vert t \vert}{2}-4\lambda_{0}^{4} u^{\pm}_{-3,5} \pm 
i t \dot{u}^{\pm}_{-3,3} \mp \frac{3iu^{\pm}_{-3,3}}{2} \mp 
\frac{i\lambda_{0} u^{\pm \prime}_{-3,3}}{2} \mp \frac{i \lambda
_{0} u^{\pm \prime}_{-3,3}}{2} \mp \frac{iu^{\pm}_{-3,3}}{2}& 
\nonumber \\
&+ \frac{\lambda_{0} \nu^{\prime} u^{\pm}_{-3,3}\ln \vert t 
\vert}{2} - 4 \lambda_{0}^{4} u^{\pm}_{-3,5} - 2 \lambda_{0}
^{2} \overline{u^{\pm}_{1,1}} u^{\pm}_{1,1} u^{\pm}_{-3,3} -2 
\lambda_{0}^{2} \overline{u^{\pm}_{1,1}} u^{\pm}_{-1,1} u^{
\pm}_{-1,3}-2 \lambda_{0}^{2} \overline{u^{\pm}_{1,1}} u^{\pm}_{
-1,2} u^{\pm}_{-1,2}& \nonumber \\
&-2 \lambda_{0}^{2} \overline{u^{\pm}_{1,1}} u^{\pm}_{-3,3} u^{
\pm}_{1,1}-2 \lambda_{0}^{2} \overline{u^{\pm}_{1,1}} u^{\pm}_{
-1,3} u^{\pm}_{-1,1}-2 \lambda_{0}^{2} \overline{u^{\pm}_{1,2}} 
u^{\pm}_{-1,1} u^{\pm}_{-1,2}-2 \lambda_{0}^{2} \overline{u^{\pm}
_{1,2}} u^{\pm}_{-1,2} u^{\pm}_{-1,1}& \nonumber \\
&-4 \lambda_{0}^{2} \overline{u^{\pm}_{3,3}} u^{\pm}_{-1,1} u^{
\pm}_{1,1}-4 \lambda_{0}^{2} \overline{u^{\pm}_{3,3}} u^{\pm}_{
1,1} u^{\pm}_{-1,1} - 2 \lambda_{0}^{2} \overline{u^{\pm}_{1,3}} 
u^{\pm}_{-1,1} u^{\pm}_{-1,1}& \nonumber \\
&+\frac{\nu^{\prime} \overline{u^{\pm}_{1,1}} u^{\pm}_{-1,1}
u^{\pm}_{-1,1} \ln \vert t \vert}{8 \lambda_{0}} \mp \frac{i(
\overline{u^{\pm}_{1,1}})^{\prime}u^{\pm}_{-1,1}u^{\pm}_{-1,1}}{
8 \lambda_{0}} + \frac{u^{\pm}_{1,1} \overline{u^{\pm}_{1,1}} u^
{\pm}_{-1,1} \overline{u^{\pm}_{1,1}} u^{\pm}_{-1,1}}{2}+\frac{u
^{\pm}_{-1,1} \overline{u^{\pm}_{1,1}} u^{\pm}_{1,1} \overline{u
^{\pm}_{1,1}} u^{\pm}_{-1,1}}{2}& \nonumber \\
&+ \frac{u^{\pm}_{-1,1} \overline{u^{\pm}_{1,1}} u^{\pm}_{-1,1} 
\overline{u^{\pm}_{1,1}} u^{\pm}_{1,1}}{2} + \frac{u^{\pm}_{-1,1} 
\overline{u^{\pm}_{1,1}} u^{\pm}_{-1,1} \overline{u^{\pm}_{-1,1}} 
u^{\pm}_{-1,1}}{2}+\frac{u^{\pm}_{-1,1} \overline{u^{\pm}_{-1,1}} 
u^{\pm}_{-1,1} \overline{u^{\pm}_{1,1}} u^{\pm}_{-1,1}}{2}=0;&
\end{eqnarray}
\begin{flushleft}
(15) ${\cal O}((\pm t)^{-5/2} \exp\{3i\tau^{\pm}\})$,
\end{flushleft}
\begin{eqnarray}
&12 \lambda_{0}^{4}u^{\pm}_{5,5}-36\lambda_{0}^{4}u^{\pm}_{5,5}+
2 \lambda_{0}^{2} \overline{u^{\pm}_{-3,3}}u^{\pm}_{1,1}u^{\pm}_{
1,1}+\frac{u^{\pm}_{1,1} \overline{u^{\pm}_{-1,1}} u^{\pm}_{1,1} 
\overline{u^{\pm}_{-1,1}} u^{\pm}_{1,1}}{2}=0;&
\end{eqnarray}
\begin{flushleft}
(16) ${\cal O}((\pm t)^{-5/2} \exp\{-2i\tau^{\pm}\})$,
\end{flushleft}
\begin{eqnarray}
&-8\lambda_{0}^{4}u^{\pm}_{-5,5}-16\lambda_{0}^{4}u^{\pm}_{-5,5}-
2 \lambda_{0}^{2} \overline{u^{\pm}_{1,1}}u^{\pm}_{-1,1}u^{\pm}_{
-3,3}-2\lambda_{0}^{2} \overline{u^{\pm}_{1,1}} u^{\pm}_{-3,3}u^{
\pm}_{-1,1}& \nonumber \\
&-4 \lambda_{0}^{2} \overline{u^{\pm}_{3,3}}u^{\pm}_{-1,1} u^{\pm}
_{-1,1}+\frac{u^{\pm}_{-1,1} \overline{u^{\pm}_{1,1}} u^{\pm}_{-1,
1} \overline{u^{\pm}_{1,1}} u^{\pm}_{-1,1}}{2}=0;&
\end{eqnarray}
\begin{flushleft}
(17) ${\cal O}((\pm t)^{-3} \exp\{i\tau^{\pm}\})$,
\end{flushleft}
\begin{eqnarray}
&\nu u^{\pm}_{1,4} - \frac{\lambda_{0} \nu^{\prime} u^{\pm}_{1,4} 
\ln \vert t \vert}{2} + 4 \lambda_{0}^{4} u^{\pm}_{1,6} \pm i t 
\dot{u}^{\pm}_{1,4} \mp 2iu^{\pm}_{1,4} \mp \frac{i \lambda_{0} 
u^{\pm \prime}_{1,4}}{2} \mp \frac{i \nu^{\prime} u^{\pm \prime}_
{1,2} \ln \vert t \vert}{32 \lambda_{0}^{2}} \mp \frac{i \nu^{
\prime \prime} u^{\pm}_{1,2} \ln \vert t \vert}{64 \lambda_{0}^{
2}}& \nonumber \\
&\pm \frac{i \nu^{\prime} u^{\pm}_{1,2} \ln \vert t \vert}{64 
\lambda_{0}^{3}} \pm \frac{i \lambda_{0} u^{\pm \prime}_{1,4}}{2}
\pm \frac{iu^{\pm}_{1,4}}{2}-\frac{(\nu^{\prime})^{2} u^{\pm}_{1,
2} (\ln \vert t \vert)^{2}}{64 \lambda_{0}^{2}}+\frac{\lambda_{0} 
\nu^{\prime} u^{\pm}_{1,4} \ln \vert t \vert}{2} - 4 \lambda_{0}
^{4} u^{\pm}_{1,6}& \nonumber \\
&+\frac{u^{\pm \prime \prime}_{1,2}}{64 \lambda_{0}^{2}} - \frac{
u^{\pm \prime}_{1,2}}{64 \lambda_{0}^{3}} - 2 \lambda_{0}^{2} 
\overline{u^{\pm}_{1,1}} u^{\pm}_{1,1} u^{\pm}_{1,4} - 2 \lambda
_{0}^{2} \overline{u^{\pm}_{1,1}} u^{\pm}_{-1,1} u^{\pm}_{3,4}
- 2 \lambda_{0}^{2} \overline{u^{\pm}_{1,1}} u^{\pm}_{1,2} u^{
\pm}_{1,3}& \nonumber \\
&- 2 \lambda_{0}^{2} \overline{u^{\pm}_{1,1}} u^{\pm}_{-1,2} u^{
\pm}_{3,3} - 2 \lambda_{0}^{2} \overline{u^{\pm}_{1,1}} u^{\pm}_{
1,3} u^{\pm}_{1,2} - 2 \lambda_{0}^{2} \overline{u^{\pm}_{1,1}} 
u^{\pm}_{3,3} u^{\pm}_{-1,2} - 2 \lambda_{0}^{2} \overline{u^{\pm}
_{1,1}} u^{\pm}_{1,4} u^{\pm}_{1,1}& \nonumber \\
&- 2 \lambda_{0}^{2} \overline{u^{\pm}_{1,1}} u^{\pm}_{3,4} u^{
\pm}_{-1,1} - 2 \lambda_{0}^{2} \overline{u^{\pm}_{1,2}} u^{\pm}_
{1,1} u^{\pm}_{1,3} - 2 \lambda_{0}^{2} \overline{u^{\pm}_{1,2}} 
u^{\pm}_{-1,1} u^{\pm}_{3,3} - 2 \lambda_{0}^{2} \overline{u^{\pm}
_{1,2}} u^{\pm}_{1,2} u^{\pm}_{1,2}& \nonumber \\
&- 2 \lambda_{0}^{2} \overline{u^{\pm}_{1,2}} u^{\pm}_{1,3} u^{
\pm}_{1,1} - 2 \lambda_{0}^{2} \overline{u^{\pm}_{1,2}} u^{\pm}_
{3,3} u^{\pm}_{-1,1} - 2 \lambda_{0}^{2} \overline{u^{\pm}_{1,3}} 
u^{\pm}_{1,1} u^{\pm}_{1,2} + 2 \lambda_{0}^{2} \overline{u^{\pm}
_{-3,3}} u^{\pm}_{-1,1} u^{\pm}_{-1,2}& \nonumber \\
&- 2 \lambda_{0}^{2} \overline{u^{\pm}_{1,3}} u^{\pm}_{1,2} u^{
\pm}_{1,1} + 2 \lambda_{0}^{2} \overline{u^{\pm}_{-3,3}} u^{\pm}_
{-1,2} u^{\pm}_{-1,1} - 2 \lambda_{0}^{2} \overline{u^{\pm}_{1,4}} 
u^{\pm}_{1,1} u^{\pm}_{1,1} + 2 \lambda_{0}^{2} \overline{u^{\pm}
_{-3,4}} u^{\pm}_{-1,1} u^{\pm}_{-1,1}& \nonumber \\
&+\frac{\nu^{\prime}\overline{u^{\pm}_{1,1}}u^{\pm}_{1,1}u^{\pm}_{
1,2} \ln \vert t \vert}{8 \lambda_{0}}+\frac{\nu^{\prime}\overline{
u^{\pm}_{1,1}}u^{\pm}_{1,2}u^{\pm}_{1,1} \ln \vert t \vert}{8 
\lambda_{0}}+\frac{\nu^{\prime}\overline{u^{\pm}_{1,2}}u^{\pm}_{1,1}
u^{\pm}_{1,1} \ln \vert t \vert}{8 \lambda_{0}} \mp \frac{i(
\overline{u^{\pm}_{1,1}})^{\prime}u^{\pm}_{1,1}u^{\pm}_{1,2}}{8 
\lambda_{0}}& \nonumber \\
&\mp \frac{i(\overline{u^{\pm}_{-1,1}})^{\prime}u^{\pm}_{1,1}u^{
\pm}_{-1,2}}{8 \lambda_{0}} \mp \frac{i(\overline{u^{\pm}_{-1,1}})
^{\prime}u^{\pm}_{-1,1}u^{\pm}_{1,2}}{8 \lambda_{0}} \mp \frac{i(
\overline{u^{\pm}_{1,1}})^{\prime}u^{\pm}_{1,2}u^{\pm}_{1,1}}{8 
\lambda_{0}} \mp \frac{i(\overline{u^{\pm}_{-1,1}})^{\prime}u^{\pm}
_{1,2}u^{\pm}_{-1,1}}{8 \lambda_{0}}& \nonumber \\
&\mp \frac{i(\overline{u^{\pm}_{-1,1}})^{\prime}u^{\pm}_{-1,2}u^{\pm}
_{1,1}}{8 \lambda_{0}} \mp \frac{i(\overline{u^{\pm}_{1,2}})^{\prime}
u^{\pm}_{1,1}u^{\pm}_{1,1}}{8 \lambda_{0}} \mp \frac{i(\overline{u^{
\pm}_{-1,2}})^{\prime}u^{\pm}_{1,1}u^{\pm}_{-1,1}}{8 \lambda_{0}} 
\mp \frac{i(\overline{u^{\pm}_{-1,2}})^{\prime}u^{\pm}_{-1,1}u^{\pm}
_{1,1}}{8 \lambda_{0}}& \nonumber \\
&+\frac{u^{\pm}_{1,1} \overline{u^{\pm}_{1,1}} u^{\pm}_{1,1} 
\overline{u^{\pm}_{1,1}} u^{\pm}_{1,2}}{2}+\frac{u^{\pm}_{1,1} 
\overline{u^{\pm}_{1,1}} u^{\pm}_{1,1} \overline{u^{\pm}_{-1,1}} 
u^{\pm}_{-1,2}}{2}+\frac{u^{\pm}_{1,1} \overline{u^{\pm}_{1,1}} 
u^{\pm}_{-1,1} \overline{u^{\pm}_{-1,1}} u^{\pm}_{1,2}}{2}& 
\nonumber \\
&+\frac{u^{\pm}_{1,1} \overline{u^{\pm}_{-1,1}} u^{\pm}_{1,1} 
\overline{u^{\pm}_{1,1}} u^{\pm}_{-1,2}}{2}+\frac{u^{\pm}_{1,1} 
\overline{u^{\pm}_{-1,1}} u^{\pm}_{-1,1} \overline{u^{\pm}_{1,1}} 
u^{\pm}_{1,2}}{2}+\frac{u^{\pm}_{1,1} \overline{u^{\pm}_{-1,1}} 
u^{\pm}_{-1,1} \overline{u^{\pm}_{-1,1}} u^{\pm}_{-1,2}}{2}& 
\nonumber \\
&+\frac{u^{\pm}_{-1,1} \overline{u^{\pm}_{1,1}} u^{\pm}_{1,1} 
\overline{u^{\pm}_{-1,1}} u^{\pm}_{1,2}}{2}+\frac{u^{\pm}_{-1,1}
\overline{u^{\pm}_{-1,1}} u^{\pm}_{1,1} \overline{u^{\pm}_{1,1}} 
u^{\pm}_{1,2}}{2}+\frac{u^{\pm}_{-1,1} \overline{u^{\pm}_{-1,1}} 
u^{\pm}_{1,1} \overline{u^{\pm}_{-1,1}} u^{\pm}_{-1,2}}{2}& 
\nonumber \\
&+\frac{u^{\pm}_{-1,1} \overline{u^{\pm}_{-1,1}} u^{\pm}_{-1,1} 
\overline{u^{\pm}_{-1,1}} u^{\pm}_{1,2}}{2}+\frac{u^{\pm}_{1,1}
\overline{u^{\pm}_{1,1}} u^{\pm}_{1,1} \overline{u^{\pm}_{1,2}} 
u^{\pm}_{1,1}}{2}+\frac{u^{\pm}_{1,1} \overline{u^{\pm}_{1,1}} 
u^{\pm}_{1,1} \overline{u^{\pm}_{-1,2}} u^{\pm}_{-1,1}}{2}& 
\nonumber \\
&+\frac{u^{\pm}_{1,1} \overline{u^{\pm}_{1,1}} u^{\pm}_{-1,1} 
\overline{u^{\pm}_{-1,2}} u^{\pm}_{1,1}}{2}+\frac{u^{\pm}_{1,1}
\overline{u^{\pm}_{-1,1}} u^{\pm}_{1,1} \overline{u^{\pm}_{1,2}} 
u^{\pm}_{-1,1}}{2}+\frac{u^{\pm}_{1,1} \overline{u^{\pm}_{-1,1}} 
u^{\pm}_{-1,1} \overline{u^{\pm}_{1,2}} u^{\pm}_{1,1}}{2}& 
\nonumber \\
&+\frac{u^{\pm}_{1,1} \overline{u^{\pm}_{-1,1}} u^{\pm}_{-1,1} 
\overline{u^{\pm}_{-1,2}} u^{\pm}_{-1,1}}{2}+\frac{u^{\pm}_{-1,1}
\overline{u^{\pm}_{1,1}} u^{\pm}_{1,1} \overline{u^{\pm}_{-1,2}} 
u^{\pm}_{1,1}}{2}+\frac{u^{\pm}_{-1,1} \overline{u^{\pm}_{-1,1}} 
u^{\pm}_{1,1} \overline{u^{\pm}_{1,2}} u^{\pm}_{1,1}}{2}& 
\nonumber \\
&+\frac{u^{\pm}_{-1,1} \overline{u^{\pm}_{-1,1}} u^{\pm}_{1,1} 
\overline{u^{\pm}_{-1,2}} u^{\pm}_{-1,1}}{2}+\frac{u^{\pm}_{-1,1}
\overline{u^{\pm}_{-1,1}} u^{\pm}_{-1,1} \overline{u^{\pm}_{-1,2}} 
u^{\pm}_{1,1}}{2}+\frac{u^{\pm}_{1,1} \overline{u^{\pm}_{1,1}} 
u^{\pm}_{1,2} \overline{u^{\pm}_{1,1}} u^{\pm}_{1,1}}{2}& 
\nonumber \\
&+\frac{u^{\pm}_{1,1} \overline{u^{\pm}_{1,1}} u^{\pm}_{1,2} 
\overline{u^{\pm}_{-1,1}} u^{\pm}_{-1,1}}{2}+\frac{u^{\pm}_{1,1}
\overline{u^{\pm}_{1,1}} u^{\pm}_{-1,2} \overline{u^{\pm}_{-1,1}} 
u^{\pm}_{1,1}}{2}+\frac{u^{\pm}_{1,1} \overline{u^{\pm}_{-1,1}} 
u^{\pm}_{1,2} \overline{u^{\pm}_{1,1}} u^{\pm}_{-1,1}}{2}& 
\nonumber \\
&+\frac{u^{\pm}_{1,1} \overline{u^{\pm}_{-1,1}} u^{\pm}_{-1,2} 
\overline{u^{\pm}_{1,1}} u^{\pm}_{1,1}}{2}+\frac{u^{\pm}_{1,1}
\overline{u^{\pm}_{-1,1}} u^{\pm}_{-1,2} \overline{u^{\pm}_{-1,1}} 
u^{\pm}_{-1,1}}{2}+\frac{u^{\pm}_{-1,1} \overline{u^{\pm}_{1,1}} 
u^{\pm}_{1,2} \overline{u^{\pm}_{-1,1}} u^{\pm}_{1,1}}{2}& 
\nonumber \\
&+\frac{u^{\pm}_{-1,1} \overline{u^{\pm}_{-1,1}} u^{\pm}_{1,2} 
\overline{u^{\pm}_{1,1}} u^{\pm}_{1,1}}{2}+\frac{u^{\pm}_{-1,1}
\overline{u^{\pm}_{-1,1}} u^{\pm}_{1,2} \overline{u^{\pm}_{-1,1}} 
u^{\pm}_{-1,1}}{2}+\frac{u^{\pm}_{-1,1} \overline{u^{\pm}_{-1,1}} 
u^{\pm}_{-1,2} \overline{u^{\pm}_{-1,1}} u^{\pm}_{1,1}}{2}& 
\nonumber \\
&+\frac{u^{\pm}_{1,1} \overline{u^{\pm}_{1,2}} u^{\pm}_{1,1} 
\overline{u^{\pm}_{1,1}} u^{\pm}_{1,1}}{2}+\frac{u^{\pm}_{1,1}
\overline{u^{\pm}_{1,2}} u^{\pm}_{1,1} \overline{u^{\pm}_{-1,1}} 
u^{\pm}_{-1,1}}{2}+\frac{u^{\pm}_{1,1} \overline{u^{\pm}_{1,2}} 
u^{\pm}_{-1,1} \overline{u^{\pm}_{-1,1}} u^{\pm}_{1,1}}{2}& 
\nonumber \\
&+\frac{u^{\pm}_{1,1} \overline{u^{\pm}_{-1,2}} u^{\pm}_{1,1} 
\overline{u^{\pm}_{1,1}} u^{\pm}_{-1,1}}{2}+\frac{u^{\pm}_{1,1}
\overline{u^{\pm}_{-1,2}} u^{\pm}_{-1,1} \overline{u^{\pm}_{1,1}} 
u^{\pm}_{1,1}}{2}+\frac{u^{\pm}_{1,1} \overline{u^{\pm}_{-1,2}} 
u^{\pm}_{-1,1} \overline{u^{\pm}_{-1,1}} u^{\pm}_{-1,1}}{2}& 
\nonumber \\
&+\frac{u^{\pm}_{-1,1} \overline{u^{\pm}_{1,2}} u^{\pm}_{1,1} 
\overline{u^{\pm}_{-1,1}} u^{\pm}_{1,1}}{2}+\frac{u^{\pm}_{-1,1}
\overline{u^{\pm}_{-1,2}} u^{\pm}_{1,1} \overline{u^{\pm}_{1,1}} 
u^{\pm}_{1,1}}{2}+\frac{u^{\pm}_{-1,1} \overline{u^{\pm}_{-1,2}} 
u^{\pm}_{1,1} \overline{u^{\pm}_{-1,1}} u^{\pm}_{-1,1}}{2}& 
\nonumber \\
&+\frac{u^{\pm}_{-1,1} \overline{u^{\pm}_{-1,2}} u^{\pm}_{-1,1} 
\overline{u^{\pm}_{-1,1}} u^{\pm}_{1,1}}{2}+\frac{u^{\pm}_{1,2}
\overline{u^{\pm}_{1,1}} u^{\pm}_{1,1} \overline{u^{\pm}_{1,1}} 
u^{\pm}_{1,1}}{2}+\frac{u^{\pm}_{1,2} \overline{u^{\pm}_{1,1}} 
u^{\pm}_{1,1} \overline{u^{\pm}_{-1,1}} u^{\pm}_{-1,1}}{2}& 
\nonumber \\
&+\frac{u^{\pm}_{1,2} \overline{u^{\pm}_{1,1}} u^{\pm}_{-1,1} 
\overline{u^{\pm}_{-1,1}} u^{\pm}_{1,1}}{2}+\frac{u^{\pm}_{1,2}
\overline{u^{\pm}_{-1,1}} u^{\pm}_{1,1} \overline{u^{\pm}_{1,1}} 
u^{\pm}_{-1,1}}{2}+\frac{u^{\pm}_{1,2} \overline{u^{\pm}_{-1,1}} 
u^{\pm}_{-1,1} \overline{u^{\pm}_{1,1}} u^{\pm}_{1,1}}{2}& 
\nonumber \\
&+\frac{u^{\pm}_{1,2} \overline{u^{\pm}_{-1,1}} u^{\pm}_{-1,1} 
\overline{u^{\pm}_{-1,1}} u^{\pm}_{-1,1}}{2}+\frac{u^{\pm}_{-1,2}
\overline{u^{\pm}_{1,1}} u^{\pm}_{1,1} \overline{u^{\pm}_{-1,1}} 
u^{\pm}_{1,1}}{2}+\frac{u^{\pm}_{-1,2} \overline{u^{\pm}_{-1,1}} 
u^{\pm}_{1,1} \overline{u^{\pm}_{1,1}} u^{\pm}_{1,1}}{2}& 
\nonumber \\
&+\frac{u^{\pm}_{-1,2} \overline{u^{\pm}_{-1,1}} u^{\pm}_{1,1} 
\overline{u^{\pm}_{-1,1}} u^{\pm}_{-1,1}}{2}+\frac{u^{\pm}_{-1,2}
\overline{u^{\pm}_{-1,1}} u^{\pm}_{-1,1} \overline{u^{\pm}_{-1,1}} 
u^{\pm}_{1,1}}{2}=0;&
\end{eqnarray}
\begin{flushleft}
(18) ${\cal O}((\pm t)^{-3})$,
\end{flushleft}
\begin{eqnarray}
&\pm i t \dot{u}^{\pm}_{-1,4} \mp 2iu^{\pm}_{-1,4} \mp \frac{i 
\lambda_{0} u^{\pm \prime}_{-1,4}}{2} + \frac{u^{\pm \prime 
\prime}_{-1,2}}{64 \lambda_{0}^{2}} - \frac{u^{\pm \prime}_{-1,
2}}{64 \lambda_{0}^{3}}-2 \lambda_{0}^{2} \overline{u^{\pm}_{
1,1}} u^{\pm}_{1,1} u^{\pm}_{-1,4} - 2 \lambda_{0}^{2} 
\overline{u^{\pm}_{1,1}} u^{\pm}_{-1,1} u^{\pm}_{1,4}& 
\nonumber \\
&\! \! \! \! \! \! \! 
-2 \lambda_{0}^{2} \overline{u^{\pm}_{1,1}} u^{\pm}_{1,2} u^{
\pm}_{-1,3} \! - \! 2 \lambda_{0}^{2} \overline{u^{\pm}_{1,1}} 
u^{\pm}_{-1,2} u^{\pm}_{1,3} \! - \! 2 \lambda_{0}^{2} \overline{
u^{\pm}_{1,1}} u^{\pm}_{-1,3} u^{\pm}_{1,2} \! - \! 2 \lambda_{0}
^{2} \overline{u^{\pm}_{1,1}} u^{\pm}_{1,3} u^{\pm}_{-1,2} \! + \! 
\frac{\nu^{\prime \prime} \vert u^{+}_{1,1} \vert^{2} \ln \vert t 
\vert}{8 \lambda_{0}}& \nonumber \\
&\! \! \! \! \! \! \! 
-2 \lambda_{0}^{2} \overline{u^{\pm}_{1,1}} u^{\pm}_{-1,4} u^{
\pm}_{1,1} \! - \! 2 \lambda_{0}^{2} \overline{u^{\pm}_{1,1}} 
u^{\pm}_{1,4} u^{\pm}_{-1,1} \! - \! 2 \lambda_{0}^{2} \overline{
u^{\pm}_{1,2}} u^{\pm}_{1,1} u^{\pm}_{-1,3} \! - \! 2 \lambda_{0}
^{2} \overline{u^{\pm}_{1,2}} u^{\pm}_{-1,1} u^{\pm}_{1,3} \! + \! 
\frac{\nu^{\prime} (\vert u^{+}_{1,1} \vert^{2})^{\prime} \ln 
\vert t \vert}{8}& \nonumber \\
&- 2 \lambda_{0}^{2} \overline{u^{\pm}_{1,2}} u^{\pm}_{1,2} u^{
\pm}_{-1,2} - 2 \lambda_{0}^{2} \overline{u^{\pm}_{1,2}}u^{\pm}_
{-1,2} u^{\pm}_{1,2} - 2 \lambda_{0}^{2} \overline{u^{\pm}_{1,2}} 
u^{\pm}_{-1,3} u^{\pm}_{1,1} - 2 \lambda_{0}^{2} \overline{u^{
\pm}_{1,2}} u^{\pm}_{1,3} u^{\pm}_{-1,1}& \nonumber \\
&-4 \lambda_{0}^{2} \overline{u^{\pm}_{3,3}} u^{\pm}_{1,1} u^{\pm}
_{1,2} - 2 \lambda_{0}^{2} \overline{u^{\pm}_{1,3}} u^{\pm}_{-1,1} 
u^{\pm}_{1,2} - 2 \lambda_{0}^{2} \overline{u^{\pm}_{1,3}} u^{\pm}
_{1,1} u^{\pm}_{-1,2} - 4 \lambda_{0}^{2} \overline{u^{\pm}_{3,3}} 
u^{\pm}_{1,2} u^{\pm}_{1,1}& \nonumber \\
&- 2 \lambda_{0}^{2} \overline{u^{\pm}_{1,3}} u^{\pm}_{-1,2} u^{
\pm}_{1,1} - 2 \lambda_{0}^{2} \overline{u^{\pm}_{1,3}} u^{\pm}_{
1,2} u^{\pm}_{-1,1} - 4 \lambda_{0}^{2} \overline{u^{\pm}_{3,4}} 
u^{\pm}_{1,1} u^{\pm}_{1,1} - 2 \lambda_{0}^{2} \overline{u^{\pm}
_{1,4}} u^{\pm}_{-1,1} u^{\pm}_{1,1}& \nonumber \\
&-2 \lambda_{0}^{2} \overline{u^{\pm}_{1,4}} u^{\pm}_{1,1} u^{\pm}
_{-1,1} + \frac{\nu^{\prime} \overline{u^{\pm}_{1,1}} u^{\pm}_{1,1} 
u^{\pm}_{-1,2} \ln \vert t \vert}{8 \lambda_{0}} + \frac{\nu^{
\prime} \overline{u^{\pm}_{1,1}} u^{\pm}_{-1,1} u^{\pm}_{1,2} \ln 
\vert t \vert}{8 \lambda_{0}} + \frac{\nu^{\prime} \overline{u^{
\pm}_{1,1}} u^{\pm}_{1,2} u^{\pm}_{-1,1} \ln \vert t \vert}{8 
\lambda_{0}}& \nonumber \\
&+ \frac{\nu^{\prime} \overline{u^{\pm}_{1,1}} u^{\pm}_{-1,2} u^{
\pm}_{1,1} \ln \vert t \vert}{8 \lambda_{0}} + \frac{\nu^{\prime} 
\overline{u^{\pm}_{1,2}} u^{\pm}_{1,1} u^{\pm}_{-1,1} \ln \vert t 
\vert}{8 \lambda_{0}} + \frac{\nu^{\prime} \overline{u^{\pm}_{1,
2}} u^{\pm}_{-1,1} u^{\pm}_{1,1} \ln \vert t \vert}{8 \lambda_{0}} 
\mp \frac{i(\overline{u^{\pm}_{1,1}})^{\prime} u^{\pm}_{1,1} u^{
\pm}_{-1,2}}{8 \lambda_{0}}& \nonumber \\
&\mp \frac{i(\overline{u^{\pm}_{1,1}})^{\prime} u^{\pm}_{-1,1} u^{
\pm}_{1,2}}{8 \lambda_{0}} \mp \frac{i(\overline{u^{\pm}_{-1,1}})^{
\prime} u^{\pm}_{-1,1} u^{\pm}_{-1,2}}{8 \lambda_{0}} \mp \frac{i(
\overline{u^{\pm}_{1,1}})^{\prime} u^{\pm}_{1,2} u^{\pm}_{-1,1}}{
8 \lambda_{0}} \mp \frac{i(\overline{u^{\pm}_{1,1}})^{\prime} u^{
\pm}_{-1,2} u^{\pm}_{1,1}}{8 \lambda_{0}}& \nonumber \\
&\mp \frac{i(\overline{u^{\pm}_{-1,1}})^{\prime} u^{\pm}_{-1,2} u^{
\pm}_{-1,1}}{8 \lambda_{0}} \mp \frac{i(\overline{u^{\pm}_{1,2}})^{
\prime} u^{\pm}_{1,1} u^{\pm}_{-1,1}}{8 \lambda_{0}} \mp \frac{i(
\overline{u^{\pm}_{1,2}})^{\prime} u^{\pm}_{-1,1} u^{\pm}_{1,1}}{
8 \lambda_{0}} \mp \frac{i(\overline{u^{\pm}_{-1,2}})^{\prime} u^{
\pm}_{-1,1} u^{\pm}_{-1,1}}{8 \lambda_{0}}& \nonumber \\
&+\frac{u^{\pm}_{1,1} \overline{u^{\pm}_{1,1}} u^{\pm}_{1,1} 
\overline{u^{\pm}_{1,1}} u^{\pm}_{-1,2}}{2}+\frac{u^{\pm}_{1,1} 
\overline{u^{\pm}_{1,1}} u^{\pm}_{-1,1} \overline{u^{\pm}_{1,1}} 
u^{\pm}_{1,2}}{2}+\frac{u^{\pm}_{1,1} \overline{u^{\pm}_{1,1}} 
u^{\pm}_{-1,1} \overline{u^{\pm}_{-1,1}} u^{\pm}_{-1,2}}{2}& 
\nonumber \\
&+\frac{u^{\pm}_{1,1} \overline{u^{\pm}_{-1,1}} u^{\pm}_{-1,1} 
\overline{u^{\pm}_{1,1}} u^{\pm}_{-1,2}}{2} + \frac{u^{\pm}_{-1,1} 
\overline{u^{\pm}_{1,1}} u^{\pm}_{1,1} \overline{u^{\pm}_{1,1}} 
u^{\pm}_{1,2}}{2} + \frac{u^{\pm}_{-1,1} \overline{u^{\pm}_{1,1}} 
u^{\pm}_{1,1} \overline{u^{\pm}_{-1,1}} u^{\pm}_{-1,2}}{2}& 
\nonumber \\
&+\frac{u^{\pm}_{-1,1} \overline{u^{\pm}_{1,1}} u^{\pm}_{-1,1} 
\overline{u^{\pm}_{-1,1}} u^{\pm}_{1,2}}{2} + \frac{u^{\pm}_{
-1,1}\overline{u^{\pm}_{-1,1}} u^{\pm}_{1,1} \overline{u^{\pm}_{
1,1}} u^{\pm}_{-1,2}}{2} + \frac{u^{\pm}_{-1,1} \overline{u^{\pm}
_{-1,1}} u^{\pm}_{-1,1} \overline{u^{\pm}_{1,1}} u^{\pm}_{1,2}}{
2}& \nonumber \\
&+ \frac{u^{\pm}_{-1,1} \overline{u^{\pm}_{-1,1}} u^{\pm}_{-1,1} 
\overline{u^{\pm}_{-1,1}} u^{\pm}_{-1,2}}{2} + \frac{u^{\pm}_{1,1} 
\overline{u^{\pm}_{1,1}} u^{\pm}_{1,1} \overline{u^{\pm}_{1,2}} 
u^{\pm}_{-1,1}}{2} + \frac{u^{\pm}_{1,1} \overline{u^{\pm}_{1,1}} 
u^{\pm}_{-1,1} \overline{u^{\pm}_{1,2}} u^{\pm}_{1,1}}{2}& 
\nonumber \\
&+ \frac{u^{\pm}_{1,1} \overline{u^{\pm}_{1,1}} u^{\pm}_{-1,1} 
\overline{u^{\pm}_{-1,2}} u^{\pm}_{-1,1}}{2} + \frac{u^{\pm}_{1,1} 
\overline{u^{\pm}_{-1,1}} u^{\pm}_{-1,1} \overline{u^{\pm}_{1,2}} 
u^{\pm}_{-1,1}}{2} + \frac{u^{\pm}_{-1,1} \overline{u^{\pm}_{1,1}} 
u^{\pm}_{1,1} \overline{u^{\pm}_{1,2}} u^{\pm}_{1,1}}{2}& 
\nonumber \\
&+ \frac{u^{\pm}_{-1,1} \overline{u^{\pm}_{1,1}} u^{\pm}_{1,1} 
\overline{u^{\pm}_{-1,2}} u^{\pm}_{-1,1}}{2} + \frac{u^{\pm}_{-1,1} 
\overline{u^{\pm}_{1,1}} u^{\pm}_{-1,1} \overline{u^{\pm}_{-1,2}} 
u^{\pm}_{1,1}}{2} + \frac{u^{\pm}_{-1,1} \overline{u^{\pm}_{-1,1}} 
u^{\pm}_{1,1} \overline{u^{\pm}_{1,2}} u^{\pm}_{-1,1}}{2}& 
\nonumber \\
&+ \frac{u^{\pm}_{-1,1} \overline{u^{\pm}_{-1,1}} u^{\pm}_{-1,1} 
\overline{u^{\pm}_{1,2}} u^{\pm}_{1,1}}{2} + \frac{u^{\pm}_{-1,1} 
\overline{u^{\pm}_{-1,1}} u^{\pm}_{-1,1} \overline{u^{\pm}_{-1,2}} 
u^{\pm}_{-1,1}}{2} +\frac{u^{\pm}_{1,1} \overline{u^{\pm}_{1,1}} 
u^{\pm}_{1,2} \overline{u^{\pm}_{1,1}} u^{\pm}_{-1,1}}{2}& 
\nonumber \\
&+ \frac{u^{\pm}_{1,1} \overline{u^{\pm}_{1,1}} u^{\pm}_{-1,2} 
\overline{u^{\pm}_{1,1}} u^{\pm}_{1,1}}{2} + \frac{u^{\pm}_{1,1} 
\overline{u^{\pm}_{1,1}} u^{\pm}_{-1,2} \overline{u^{\pm}_{-1,1}} 
u^{\pm}_{-1,1}}{2} + \frac{u^{\pm}_{1,1} \overline{u^{\pm}_{-1,1}} 
u^{\pm}_{-1,2} \overline{u^{\pm}_{1,1}} u^{\pm}_{-1,1}}{2}& 
\nonumber \\
&+\frac{u^{\pm}_{-1,1} \overline{u^{\pm}_{1,1}} u^{\pm}_{1,2} 
\overline{u^{\pm}_{1,1}} u^{\pm}_{1,1}}{2}+\frac{u^{\pm}_{-1,1}
\overline{u^{\pm}_{1,1}} u^{\pm}_{1,2} \overline{u^{\pm}_{-1,1}} 
u^{\pm}_{-1,1}}{2}+\frac{u^{\pm}_{-1,1} \overline{u^{\pm}_{1,1}} 
u^{\pm}_{-1,2} \overline{u^{\pm}_{-1,1}} u^{\pm}_{1,1}}{2}& 
\nonumber \\
&+\frac{u^{\pm}_{-1,1} \overline{u^{\pm}_{-1,1}} u^{\pm}_{1,2} 
\overline{u^{\pm}_{1,1}} u^{\pm}_{-1,1}}{2}+\frac{u^{\pm}_{-1,1}
\overline{u^{\pm}_{-1,1}} u^{\pm}_{-1,2} \overline{u^{\pm}_{1,1}} 
u^{\pm}_{1,1}}{2}+\frac{u^{\pm}_{-1,1} \overline{u^{\pm}_{-1,1}} 
u^{\pm}_{-1,2} \overline{u^{\pm}_{-1,1}} u^{\pm}_{-1,1}}{2}& 
\nonumber \\
&+\frac{u^{\pm}_{1,1} \overline{u^{\pm}_{1,2}} u^{\pm}_{1,1} 
\overline{u^{\pm}_{1,1}} u^{\pm}_{-1,1}}{2}+\frac{u^{\pm}_{1,1}
\overline{u^{\pm}_{1,2}} u^{\pm}_{-1,1} \overline{u^{\pm}_{1,1}} 
u^{\pm}_{1,1}}{2}+\frac{u^{\pm}_{1,1} \overline{u^{\pm}_{1,2}} 
u^{\pm}_{-1,1} \overline{u^{\pm}_{-1,1}} u^{\pm}_{-1,1}}{2}& 
\nonumber \\
&+\frac{u^{\pm}_{1,1} \overline{u^{\pm}_{-1,2}} u^{\pm}_{-1,1} 
\overline{u^{\pm}_{1,1}} u^{\pm}_{-1,1}}{2}+\frac{u^{\pm}_{-1,1}
\overline{u^{\pm}_{1,2}} u^{\pm}_{1,1} \overline{u^{\pm}_{1,1}} 
u^{\pm}_{1,1}}{2}+\frac{u^{\pm}_{-1,1} \overline{u^{\pm}_{1,2}} 
u^{\pm}_{1,1} \overline{u^{\pm}_{-1,1}} u^{\pm}_{-1,1}}{2}& 
\nonumber \\
&+\frac{u^{\pm}_{-1,1} \overline{u^{\pm}_{1,2}} u^{\pm}_{-1,1} 
\overline{u^{\pm}_{-1,1}} u^{\pm}_{1,1}}{2}+\frac{u^{\pm}_{-1,1}
\overline{u^{\pm}_{-1,2}} u^{\pm}_{1,1} \overline{u^{\pm}_{1,1}} 
u^{\pm}_{-1,1}}{2}+\frac{u^{\pm}_{-1,1} \overline{u^{\pm}_{-1,2}} 
u^{\pm}_{-1,1} \overline{u^{\pm}_{1,1}} u^{\pm}_{1,1}}{2}& 
\nonumber \\
&+\frac{u^{\pm}_{-1,1} \overline{u^{\pm}_{-1,2}} u^{\pm}_{-1,1} 
\overline{u^{\pm}_{-1,1}} u^{\pm}_{-1,1}}{2}+\frac{u^{\pm}_{1,2}
\overline{u^{\pm}_{1,1}} u^{\pm}_{1,1} \overline{u^{\pm}_{1,1}} 
u^{\pm}_{-1,1}}{2}+\frac{u^{\pm}_{1,2} \overline{u^{\pm}_{1,1}} 
u^{\pm}_{-1,1} \overline{u^{\pm}_{1,1}} u^{\pm}_{1,1}}{2}& 
\nonumber \\
&+\frac{u^{\pm}_{1,2} \overline{u^{\pm}_{1,1}} u^{\pm}_{-1,1} 
\overline{u^{\pm}_{-1,1}} u^{\pm}_{-1,1}}{2}+\frac{u^{\pm}_{1,2}
\overline{u^{\pm}_{-1,1}} u^{\pm}_{-1,1} \overline{u^{\pm}_{1,1}} 
u^{\pm}_{-1,1}}{2}+\frac{u^{\pm}_{-1,2} \overline{u^{\pm}_{1,1}} 
u^{\pm}_{1,1} \overline{u^{\pm}_{1,1}} u^{\pm}_{1,1}}{2}& 
\nonumber \\
&+\frac{u^{\pm}_{-1,2} \overline{u^{\pm}_{1,1}} u^{\pm}_{1,1} 
\overline{u^{\pm}_{-1,1}} u^{\pm}_{-1,1}}{2}+\frac{u^{\pm}_{-1,2}
\overline{u^{\pm}_{1,1}} u^{\pm}_{-1,1} \overline{u^{\pm}_{-1,1}} 
u^{\pm}_{1,1}}{2}+\frac{u^{\pm}_{-1,2} \overline{u^{\pm}_{-1,1}} 
u^{\pm}_{1,1} \overline{u^{\pm}_{1,1}} u^{\pm}_{-1,1}}{2}& 
\nonumber \\
&+\frac{u^{\pm}_{-1,2} \overline{u^{\pm}_{-1,1}} u^{\pm}_{-1,1} 
\overline{u^{\pm}_{1,1}} u^{\pm}_{1,1}}{2}+\frac{u^{\pm}_{-1,2}
\overline{u^{\pm}_{-1,1}} u^{\pm}_{-1,1} \overline{u^{\pm}_{-1,1}} 
u^{\pm}_{-1,1}}{2}=0;&
\end{eqnarray}
\begin{flushleft}
(19) ${\cal O}((\pm t)^{-3} \exp\{2i\tau^{\pm}\})$,
\end{flushleft}
\begin{eqnarray}
&2 \nu u^{\pm}_{3,4} - \lambda_{0} \nu^{\prime} u^{\pm}_{3,4} 
\ln \vert t \vert + 8 \lambda_{0}^{4} u^{\pm}_{3,6} \pm i t 
\dot{u}^{\pm}_{3,4} \mp 2iu^{\pm}_{3,4} \mp \frac{i \lambda_{0} 
u^{\pm \prime}_{3,4}}{2}& \nonumber \\
&\pm i \lambda_{0} u^{\pm \prime}_{3,4} \pm i u^{\pm}_{3,4} + 
2 \lambda_{0} \nu^{\prime} u^{\pm}_{3,4} \ln \vert t \vert - 16 
\lambda_{0}^{4} u^{\pm}_{3,6} - 2 \lambda_{0}^{2} \overline{u^{
\pm}_{1,1}} u^{\pm}_{1,1} u^{\pm}_{3,4}& \nonumber \\
&- 2 \lambda_{0}^{2} \overline{u^{\pm}_{1,1}} u^{\pm}_{1,2} u^{
\pm}_{3,3} - 2 \lambda_{0}^{2} \overline{u^{\pm}_{1,1}} u^{\pm}_{
3,3} u^{\pm}_{1,2} - 2 \lambda_{0}^{2} \overline{u^{\pm}_{1,1}} 
u^{\pm}_{3,4} u^{\pm}_{1,1} - 2 \lambda_{0}^{2} \overline{u^{\pm}
_{1,2}} u^{\pm}_{1,1} u^{\pm}_{3,3}& \nonumber \\
&- 2 \lambda_{0}^{2} \overline{u^{\pm}_{1,2}} u^{\pm}_{3,3} u^{
\pm}_{1,1} + 2 \lambda_{0}^{2} \overline{u^{\pm}_{-3,3}} u^{\pm}_
{-1,1} u^{\pm}_{1,2} + 2 \lambda_{0}^{2} \overline{u^{\pm}_{-3,3}
} u^{\pm}_{1,1} u^{\pm}_{-1,2} + 2 \lambda_{0}^{2} \overline{u^{
\pm}_{-3,3}} u^{\pm}_{-1,2} u^{\pm}_{1,1}& \nonumber \\
&+2 \lambda_{0}^{2} \overline{u^{\pm}_{-3,3}} u^{\pm}_{1,2} u^{
\pm}_{-1,1} + 2 \lambda_{0}^{2} \overline{u^{\pm}_{-3,4}} u^{\pm}_
{-1,1} u^{\pm}_{1,1} + 2 \lambda_{0}^{2} \overline{u^{\pm}_{-3,4}
} u^{\pm}_{1,1} u^{\pm}_{-1,1} \mp \frac{i(\overline{u^{\pm}_{-1,
1}})^{\prime}u^{\pm}_{1,1}u^{\pm}_{1,2}}{8 \lambda_{0}}& 
\nonumber \\
&\mp \frac{i(\overline{u^{\pm}_{-1,1}})^{\prime}u^{\pm}_{1,2} u^{
\pm}_{1,1}}{8 \lambda_{0}} \mp \frac{i(\overline{u^{\pm}_{-1,2}})
^{\prime}u^{\pm}_{1,1}u^{\pm}_{1,1}}{8 \lambda_{0}} + \frac{u^{
\pm}_{1,1} \overline{u^{\pm}_{1,1}} u^{\pm}_{1,1} \overline{u^{
\pm}_{-1,1}} u^{\pm}_{1,2}}{2} + \frac{u^{\pm}_{1,1} \overline{
u^{\pm}_{-1,1}} u^{\pm}_{1,1} \overline{u^{\pm}_{1,1}} u^{\pm}_{
1,2}}{2}& \nonumber \\
&+\frac{u^{\pm}_{1,1} \overline{u^{\pm}_{-1,1}} u^{\pm}_{1,1} 
\overline{u^{\pm}_{-1,1}} u^{\pm}_{-1,2}}{2} + \frac{u^{\pm}_{1,
1} \overline{u^{\pm}_{-1,1}} u^{\pm}_{-1,1} \overline{u^{\pm}_{
-1,1}} u^{\pm}_{1,2}}{2} + \frac{u^{\pm}_{-1,1} \overline{u^{\pm}
_{-1,1}} u^{\pm}_{1,1} \overline{u^{\pm}_{-1,1}} u^{\pm}_{1,2}}{
2}& \nonumber \\
&+\frac{u^{\pm}_{1,1} \overline{u^{\pm}_{1,1}} u^{\pm}_{1,1} 
\overline{u^{\pm}_{-1,2}} u^{\pm}_{1,1}}{2} + \frac{u^{\pm}_{1,1} 
\overline{u^{\pm}_{-1,1}} u^{\pm}_{1,1} \overline{u^{\pm}_{1,2}} 
u^{\pm}_{1,1}}{2} + \frac{u^{\pm}_{1,1} \overline{u^{\pm}_{-1,1}} 
u^{\pm}_{1,1} \overline{u^{\pm}_{-1,2}} u^{\pm}_{-1,1}}{2}& 
\nonumber \\
&+ \frac{u^{\pm}_{1,1} \overline{u^{\pm}_{-1,1}} u^{\pm}_{-1,1} 
\overline{u^{\pm}_{-1,2}} u^{\pm}_{1,1}}{2}+\frac{u^{\pm}_{-1,1} 
\overline{u^{\pm}_{-1,1}} u^{\pm}_{1,1} \overline{u^{\pm}_{-1,2}} 
u^{\pm}_{1,1}}{2} + \frac{u^{\pm}_{1,1} \overline{u^{\pm}_{1,1}} 
u^{\pm}_{1,2} \overline{u^{\pm}_{-1,1}} u^{\pm}_{1,1}}{2}& 
\nonumber \\ 
&+ \frac{u^{\pm}_{1,1} \overline{u^{\pm}_{-1,1}} u^{\pm}_{1,2} 
\overline{u^{\pm}_{1,1}} u^{\pm}_{1,1}}{2} + \frac{u^{\pm}_{1,1} 
\overline{u^{\pm}_{-1,1}} u^{\pm}_{1,2} \overline{u^{\pm}_{-1,1}} 
u^{\pm}_{-1,1}}{2} + \frac{u^{\pm}_{1,1} \overline{u^{\pm}_{-1,1}} 
u^{\pm}_{-1,2} \overline{u^{\pm}_{-1,1}} u^{\pm}_{1,1}}{2}& 
\nonumber \\
&+ \frac{u^{\pm}_{-1,1} \overline{u^{\pm}_{-1,1}} u^{\pm}_{1,2} 
\overline{u^{\pm}_{-1,1}} u^{\pm}_{1,1}}{2} + \frac{u^{\pm}_{1,1} 
\overline{u^{\pm}_{1,2}} u^{\pm}_{1,1} \overline{u^{\pm}_{-1,1}} 
u^{\pm}_{1,1}}{2} + \frac{u^{\pm}_{1,1} \overline{u^{\pm}_{-1,2}} 
u^{\pm}_{1,1} \overline{u^{\pm}_{1,1}} u^{\pm}_{1,1}}{2}& 
\nonumber \\
&+ \frac{u^{\pm}_{1,1} \overline{u^{\pm}_{-1,2}} u^{\pm}_{1,1} 
\overline{u^{\pm}_{-1,1}} u^{\pm}_{-1,1}}{2} + \frac{u^{\pm}_{1,1} 
\overline{u^{\pm}_{-1,2}} u^{\pm}_{-1,1} \overline{u^{\pm}_{-1,1}} 
u^{\pm}_{1,1}}{2} + \frac{u^{\pm}_{-1,1} \overline{u^{\pm}_{-1,2}} 
u^{\pm}_{1,1} \overline{u^{\pm}_{-1,1}} u^{\pm}_{1,1}}{2}& 
\nonumber \\
&+ \frac{u^{\pm}_{1,2} \overline{u^{\pm}_{1,1}} u^{\pm}_{1,1} 
\overline{u^{\pm}_{-1,1}} u^{\pm}_{1,1}}{2} + \frac{u^{\pm}_{1,2} 
\overline{u^{\pm}_{-1,1}} u^{\pm}_{1,1} \overline{u^{\pm}_{1,1}} 
u^{\pm}_{1,1}}{2} + \frac{u^{\pm}_{1,2} \overline{u^{\pm}_{-1,1}} 
u^{\pm}_{1,1} \overline{u^{\pm}_{-1,1}} u^{\pm}_{-1,1}}{2}& 
\nonumber \\
&+ \frac{u^{\pm}_{1,2} \overline{u^{\pm}_{-1,1}} u^{\pm}_{-1,1} 
\overline{u^{\pm}_{-1,1}} u^{\pm}_{1,1}}{2} + \frac{u^{\pm}_{-1,2} 
\overline{u^{\pm}_{-1,1}} u^{\pm}_{1,1} \overline{u^{\pm}_{-1,1}} 
u^{\pm}_{1,1}}{2}=0;&
\end{eqnarray}
\begin{flushleft}
(20) ${\cal O}((\pm t)^{-3} \exp\{-i\tau^{\pm}\})$,
\end{flushleft}
\begin{eqnarray}
&-\nu u^{\pm}_{-3,4} + \frac{\lambda_{0} \nu^{\prime} u^{\pm}_{
-3,4} \ln \vert t \vert}{2} - 4 \lambda_{0}^{4} u^{\pm}_{-3,6} 
\pm i t \dot{u}^{\pm}_{-3,4} \mp 2 i u^{\pm}_{-3,4} \mp \frac{i 
\lambda_{0} u^{\pm \prime}_{-3,4}}{2}& \nonumber \\
&\mp \frac{i \lambda_{0} u^{\pm \prime}_{-3,4}}{2} \mp \frac{i 
u^{\pm}_{-3,4}}{2} + \frac{\lambda_{0} \nu^{\prime} u^{\pm}_{-3,
4} \ln \vert t \vert}{2} - 4 \lambda_{0}^{4} u^{\pm}_{-3,6} - 2 
\lambda_{0}^{2} \overline{u^{\pm}_{1,1}} u^{\pm}_{1,1}u^{\pm}_{
-3,4}& \nonumber \\
&- 2 \lambda_{0}^{2} \overline{u^{\pm}_{1,1}} u^{\pm}_{-1,1} u^{
\pm}_{-1,4} - 2 \lambda_{0}^{2} \overline{u^{\pm}_{1,1}} u^{\pm}
_{1,2} u^{\pm}_{-3,3} - 2 \lambda_{0}^{2} \overline{u^{\pm}_{1,1}
} u^{\pm}_{-1,2} u^{\pm}_{-1,3} - 2 \lambda_{0}^{2} \overline{u^{
\pm}_{1,1}} u^{\pm}_{-3,3} u^{\pm}_{1,2}& \nonumber \\
&- 2 \lambda_{0}^{2} \overline{u^{\pm}_{1,1}} u^{\pm}_{-1,3} u^{
\pm}_{-1,2} - 2 \lambda_{0}^{2} \overline{u^{\pm}_{1,1}} u^{\pm}_
{-3,4} u^{\pm}_{1,1} - 2 \lambda_{0}^{2} \overline{u^{\pm}_{1,1}
} u^{\pm}_{-1,4} u^{\pm}_{-1,1} - 2 \lambda_{0}^{2} \overline{u^{
\pm}_{1,2}} u^{\pm}_{1,1} u^{\pm}_{-3,3}& \nonumber \\
&-2 \lambda_{0}^{2} \overline{u^{\pm}_{1,2}} u^{\pm}_{-1,1} u^{
\pm}_{-1,3} - 2 \lambda_{0}^{2} \overline{u^{\pm}_{1,2}} u^{\pm}_
{-1,2} u^{\pm}_{-1,2} - 2 \lambda_{0}^{2} \overline{u^{\pm}_{1,2}
} u^{\pm}_{-3,3} u^{\pm}_{1,1} - 2 \lambda_{0}^{2} \overline{u^{
\pm}_{1,2}} u^{\pm}_{-1,3} u^{\pm}_{-1,1}& \nonumber \\
&-4 \lambda_{0}^{2} \overline{u^{\pm}_{3,3}} u^{\pm}_{-1,1} u^{
\pm}_{1,2} - 4 \lambda_{0}^{2} \overline{u^{\pm}_{3,3}} u^{\pm}_
{1,1} u^{\pm}_{-1,2} - 2 \lambda_{0}^{2} \overline{u^{\pm}_{1,3}
} u^{\pm}_{-1,1} u^{\pm}_{-1,2} - 4 \lambda_{0}^{2} \overline{u^
{\pm}_{3,3}} u^{\pm}_{-1,2} u^{\pm}_{1,1}& \nonumber \\
&-4 \lambda_{0}^{2} \overline{u^{\pm}_{3,3}} u^{\pm}_{1,2} u^{
\pm}_{-1,1} - 2 \lambda_{0}^{2} \overline{u^{\pm}_{1,3}} u^{\pm}_
{-1,2} u^{\pm}_{-1,1} - 4 \lambda_{0}^{2} \overline{u^{\pm}_{3,4}} 
u^{\pm}_{-1,1} u^{\pm}_{1,1} - 4 \lambda_{0}^{2} \overline{u^{\pm}
_{3,4}} u^{\pm}_{1,1} u^{\pm}_{-1,1}& \nonumber \\
&-2 \lambda_{0}^{2} \overline{u^{\pm}_{1,4}} u^{\pm}_{-1,1} u^{
\pm}_{-1,1} + \frac{\nu^{\prime} \overline{u^{\pm}_{1,1}} u^{\pm}
_{-1,1} u^{\pm}_{-1,2} \ln \vert t \vert}{8 \lambda_{0}} + \frac{
\nu^{\prime} \overline{u^{\pm}_{1,1}} u^{\pm}_{-1,2} u^{\pm}_{-1,
1} \ln \vert t \vert}{8 \lambda_{0}} + \frac{\nu^{\prime} 
\overline{u^{\pm}_{1,2}} u^{\pm}_{-1,1} u^{\pm}_{-1,1} \ln \vert 
t \vert}{8 \lambda_{0}}& \nonumber \\
&\mp \frac{i(\overline{u^{\pm}_{1,1}})^{\prime}u^{\pm}_{-1,1}u^{
\pm}_{-1,2}}{8 \lambda_{0}} \mp \frac{i(\overline{u^{\pm}_{1,1}})
^{\prime}u^{\pm}_{-1,2} u^{\pm}_{-1,1}}{8 \lambda_{0}} \mp \frac{
i(\overline{u^{\pm}_{1,2}})^{\prime}u^{\pm}_{-1,1}u^{\pm}_{-1,1}}
{8 \lambda_{0}} + \frac{u^{\pm}_{1,1} \overline{u^{\pm}_{1,1}} 
u^{\pm}_{-1,1} \overline{u^{\pm}_{1,1}} u^{\pm}_{-1,2}}{2}& 
\nonumber \\
&+ \frac{u^{\pm}_{-1,1} \overline{u^{\pm}_{1,1}} u^{\pm}_{1,1} 
\overline{u^{\pm}_{1,1}} u^{\pm}_{-1,2}}{2} + \frac{u^{\pm}_{-1,1} 
\overline{u^{\pm}_{1,1}} u^{\pm}_{-1,1} \overline{u^{\pm}_{1,1}} 
u^{\pm}_{1,2}}{2} + \frac{u^{\pm}_{-1,1} \overline{u^{\pm}_{1,1}} 
u^{\pm}_{-1,1} \overline{u^{\pm}_{-1,1}} u^{\pm}_{-1,2}}{2}& 
\nonumber \\
&+ \frac{u^{\pm}_{-1,1} \overline{u^{\pm}_{-1,1}} u^{\pm}_{-1,1} 
\overline{u^{\pm}_{1,1}} u^{\pm}_{-1,2}}{2} + \frac{u^{\pm}_{1,1} 
\overline{u^{\pm}_{1,1}} u^{\pm}_{-1,1} \overline{u^{\pm}_{1,2}} 
u^{\pm}_{-1,1}}{2} + \frac{u^{\pm}_{-1,1} \overline{u^{\pm}_{1,1}} 
u^{\pm}_{1,1} \overline{u^{\pm}_{1,2}} u^{\pm}_{-1,1}}{2}& 
\nonumber \\
&+ \frac{u^{\pm}_{-1,1} \overline{u^{\pm}_{1,1}} u^{\pm}_{-1,1} 
\overline{u^{\pm}_{1,2}} u^{\pm}_{1,1}}{2} + \frac{u^{\pm}_{-1,1} 
\overline{u^{\pm}_{1,1}} u^{\pm}_{-1,1} \overline{u^{\pm}_{-1,2}} 
u^{\pm}_{-1,1}}{2} + \frac{u^{\pm}_{-1,1} \overline{u^{\pm}_{-1,1}} 
u^{\pm}_{-1,1} \overline{u^{\pm}_{1,2}} u^{\pm}_{-1,1}}{2}& 
\nonumber \\
&+ \frac{u^{\pm}_{1,1} \overline{u^{\pm}_{1,1}} u^{\pm}_{-1,2} 
\overline{u^{\pm}_{1,1}} u^{\pm}_{-1,1}}{2} + \frac{u^{\pm}_{-1,1} 
\overline{u^{\pm}_{1,1}} u^{\pm}_{1,2} \overline{u^{\pm}_{1,1}} 
u^{\pm}_{-1,1}}{2} + \frac{u^{\pm}_{-1,1} \overline{u^{\pm}_{1,1}} 
u^{\pm}_{-1,2} \overline{u^{\pm}_{1,1}} u^{\pm}_{1,1}}{2}& 
\nonumber \\
&+ \frac{u^{\pm}_{-1,1} \overline{u^{\pm}_{1,1}} u^{\pm}_{-1,2} 
\overline{u^{\pm}_{-1,1}} u^{\pm}_{-1,1}}{2} + \frac{u^{\pm}_{-1,1} 
\overline{u^{\pm}_{-1,1}} u^{\pm}_{-1,2} \overline{u^{\pm}_{1,1}} 
u^{\pm}_{-1,1}}{2} + \frac{u^{\pm}_{1,1} \overline{u^{\pm}_{1,2}} 
u^{\pm}_{-1,1} \overline{u^{\pm}_{1,1}} u^{\pm}_{-1,1}}{2}& 
\nonumber \\
&+ \frac{u^{\pm}_{-1,1} \overline{u^{\pm}_{1,2}} u^{\pm}_{1,1} 
\overline{u^{\pm}_{1,1}} u^{\pm}_{-1,1}}{2} + \frac{u^{\pm}_{-1,1} 
\overline{u^{\pm}_{1,2}} u^{\pm}_{-1,1} \overline{u^{\pm}_{1,1}} 
u^{\pm}_{1,1}}{2} + \frac{u^{\pm}_{-1,1} \overline{u^{\pm}_{1,2}} 
u^{\pm}_{-1,1} \overline{u^{\pm}_{-1,1}} u^{\pm}_{-1,1}}{2}& 
\nonumber \\
&+ \frac{u^{\pm}_{-1,1} \overline{u^{\pm}_{-1,2}} u^{\pm}_{-1,1} 
\overline{u^{\pm}_{1,1}} u^{\pm}_{-1,1}}{2} + \frac{u^{\pm}_{1,2} 
\overline{u^{\pm}_{1,1}} u^{\pm}_{-1,1} \overline{u^{\pm}_{1,1}} 
u^{\pm}_{-1,1}}{2} + \frac{u^{\pm}_{-1,2} \overline{u^{\pm}_{1,1}} 
u^{\pm}_{1,1} \overline{u^{\pm}_{1,1}} u^{\pm}_{-1,1}}{2}& 
\nonumber \\
&+ \frac{u^{\pm}_{-1,2} \overline{u^{\pm}_{1,1}} u^{\pm}_{-1,1} 
\overline{u^{\pm}_{1,1}} u^{\pm}_{1,1}}{2} + \frac{u^{\pm}_{-1,2} 
\overline{u^{\pm}_{1,1}} u^{\pm}_{-1,1} \overline{u^{\pm}_{-1,1}} 
u^{\pm}_{-1,1}}{2} + \frac{u^{\pm}_{-1,2} \overline{u^{\pm}_{-1,1}} 
u^{\pm}_{-1,1} \overline{u^{\pm}_{1,1}} u^{\pm}_{-1,1}}{2}=0;&
\end{eqnarray}
\begin{flushleft}
(21) ${\cal O}((\pm t)^{-3} \exp\{3 i \tau^{\pm}\})$,
\end{flushleft}
\begin{eqnarray}
&12 \lambda_{0}^{4} u^{\pm}_{5,6}-36 \lambda_{0}^{4} u^{\pm}_{5,6} 
+ 2 \lambda_{0}^{2} \overline{u^{\pm}_{-3,3}} u^{\pm}_{1,1}u^{\pm}
_{1,2} + 2 \lambda_{0}^{2} \overline{u^{\pm}_{-3,3}} u^{\pm}_{1,2} 
u^{\pm}_{1,1} + 2 \lambda_{0}^{2} \overline{u^{\pm}_{-3,4}}u^{\pm}
_{1,1} u^{\pm}_{1,1}& \nonumber \\
&+ \frac{u^{\pm}_{1,1} \overline{u^{\pm}_{-1,1}} u^{\pm}_{1,1} 
\overline{u^{\pm}_{-1,1}} u^{\pm}_{1,2}}{2} + \frac{u^{\pm}_{1,1} 
\overline{u^{\pm}_{-1,1}} u^{\pm}_{1,1} \overline{u^{\pm}_{-1,2}} 
u^{\pm}_{1,1}}{2} + \frac{u^{\pm}_{1,1} \overline{u^{\pm}_{-1,1}} 
u^{\pm}_{1,2} \overline{u^{\pm}_{-1,1}} u^{\pm}_{1,1}}{2}& 
\nonumber \\
&+ \frac{u^{\pm}_{1,1} \overline{u^{\pm}_{-1,2}} u^{\pm}_{1,1} 
\overline{u^{\pm}_{-1,1}} u^{\pm}_{1,1}}{2} + \frac{u^{\pm}_{1,2} 
\overline{u^{\pm}_{-1,1}} u^{\pm}_{1,1} \overline{u^{\pm}_{-1,1}} 
u^{\pm}_{1,1}}{2}=0;&
\end{eqnarray}
\begin{flushleft}
(22) ${\cal O}((\pm t)^{-3} \exp\{-2i\tau^{\pm}\})$,
\end{flushleft}
\begin{eqnarray}
&-8\lambda_{0}^{4}u^{\pm}_{-5,6}-16\lambda_{0}^{4}u^{\pm}_{-5,6} 
-2 \lambda_{0}^{2} \overline{u^{\pm}_{1,1}} u^{\pm}_{-1,1}u^{\pm}
_{-3,4}-2\lambda_{0}^{2} \overline{u^{\pm}_{1,1}} u^{\pm}_{-1,2} 
u^{\pm}_{-3,3}-2 \lambda_{0}^{2} \overline{u^{\pm}_{1,1}}u^{\pm}
_{-3,3} u^{\pm}_{-1,2}& \nonumber \\
&-2 \lambda_{0}^{2} \overline{u^{\pm}_{1,1}}u^{\pm}_{-3,4}u^{\pm}
_{-1,1}-2 \lambda_{0}^{2} \overline{u^{\pm}_{1,2}} u^{\pm}_{-1,1} 
u^{\pm}_{-3,3}-2 \lambda_{0}^{2} \overline{u^{\pm}_{1,2}} u^{\pm}
_{-3,3} u^{\pm}_{-1,1}-4 \lambda_{0}^{2} \overline{u^{\pm}_{3,3}} 
u^{\pm}_{-1,1} u^{\pm}_{-1,2}& \nonumber \\
&-4 \lambda_{0}^{2} \overline{u^{\pm}_{3,3}} u^{\pm}_{-1,2} u^{
\pm}_{-1,1}-4 \lambda_{0}^{2} \overline{u^{\pm}_{3,4}} u^{\pm}_{
-1,1} u^{\pm}_{-1,1} + \frac{u^{\pm}_{-1,1} \overline{u^{\pm}_{1,
1}} u^{\pm}_{-1,1} \overline{u^{\pm}_{1,1}} u^{\pm}_{-1,2}}{2} + 
\frac{u^{\pm}_{-1,1} \overline{u^{\pm}_{1,1}} u^{\pm}_{-1,1} 
\overline{u^{\pm}_{1,2}} u^{\pm}_{-1,1}}{2}& \nonumber \\
&+\frac{u^{\pm}_{-1,1} \overline{u^{\pm}_{1,1}} u^{\pm}_{-1,2} 
\overline{u^{\pm}_{1,1}} u^{\pm}_{-1,1}}{2} + \frac{u^{\pm}_{-1,1} 
\overline{u^{\pm}_{1,2}} u^{\pm}_{-1,1} \overline{u^{\pm}_{1,1}} 
u^{\pm}_{-1,1}}{2} + \frac{u^{\pm}_{-1,2} \overline{u^{\pm}_{1,1}} 
u^{\pm}_{-1,1} \overline{u^{\pm}_{1,1}} u^{\pm}_{-1,1}}{2}=0;& 
\end{eqnarray}
\begin{flushleft}
(23) ${\cal O}((\pm t)^{-7/2} \exp\{i\tau^{\pm}\})$,
\end{flushleft}

\begin{flushleft}
(25) ${\cal O}((\pm t)^{-7/2} \exp\{2i\tau^{\pm}\})$,
\end{flushleft}
\begin{eqnarray}
&2\nu u^{\pm}_{3,5} - \lambda_{0} \nu^{\prime} u^{\pm}_{3,5} 
\ln \vert t \vert + 8 \lambda_{0}^{4} u^{\pm}_{3,7} \pm i t 
\dot{u}^{\pm}_{3,5} \mp \frac{5iu^{\pm}_{3,5}}{2} \mp \frac{i 
\lambda_{0} u^{\pm \prime}_{3,5}}{2} \mp \frac{i \nu^{\prime}
u^{\pm \prime}_{3,3} \ln \vert t \vert}{16 \lambda_{0}^{2}}& 
\nonumber \\
&\mp \frac{i \nu^{\prime \prime} u^{\pm}_{3,3} \ln \vert t 
\vert}{32 \lambda_{0}^{2}} 
\pm \frac{i \nu^{\prime} u^{\pm}_{3,3} \ln \vert t \vert}{32 
\lambda_{0}^{3}} \pm i \lambda_{0} u^{\pm \prime}_{3,5} \pm i
u^{\pm}_{3,5}-\frac{(\nu^{\prime})^{2} u^{\pm}_{3,3}(\ln \vert 
t \vert)^{2}}{16 \lambda_{0}^{2}} + 2 \lambda_{0} \nu^{\prime} 
u^{\pm}_{3,5} \ln \vert t \vert& \nonumber \\
&- 16 \lambda_{0}^{4} u^{\pm}_{3,7} 
+ \frac{u^{\pm \prime \prime}_{3,3}}{64 \lambda_{0}^{2}}-\frac{
u^{\pm \prime}_{3,3}}{64 \lambda_{0}^{3}} - 2 \lambda_{0}^{2} 
\overline{u^{\pm}_{1,1}} u^{\pm}_{1,1} u^{\pm}_{3,5} - 2 \lambda
_{0}^{2} \overline{u^{\pm}_{1,1}} u^{\pm}_{-1,1} u^{\pm}_{5,5}
- 2 \lambda_{0}^{2} \overline{u^{\pm}_{1,1}} u^{\pm}_{1,2} u^{
\pm}_{3,4}& \nonumber \\
&-2 \lambda_{0}^{2} \overline{u^{\pm}_{1,1}} u^{\pm}_{1,3} u^{
\pm}_{3,3} - 2 \lambda_{0}^{2} \overline{u^{\pm}_{1,1}} u^{\pm}
_{3,3} u^{\pm}_{1,3} - 2 \lambda_{0}^{2} \overline{u^{\pm}_{1,1}
} u^{\pm}_{3,4} u^{\pm}_{1,2}-2\lambda_{0}^{2} \overline{u^{
\pm}_{1,1}} u^{\pm}_{3,5} u^{\pm}_{1,1}& \nonumber \\
&-2\lambda_{0}^{2} \overline{u^{\pm}_{1,1}} u^{\pm}_{5,5}u^{\pm}
_{-1,1}-2\lambda_{0}^{2} \overline{u^{\pm}_{1,2}} u^{\pm}_{1,1}
u^{\pm}_{3,4}-2\lambda_{0}^{2} \overline{u^{\pm}_{1,2}} u^{\pm}
_{1,2} u^{\pm}_{3,3}-2 \lambda_{0}^{2} \overline{u^{\pm}_{1,2}} 
u^{\pm}_{3,3} u^{\pm}_{1,2}& \nonumber \\
&-2 \lambda_{0}^{2} \overline{u^{\pm}_{1,2}} u^{\pm}_{3,4} u^{
\pm}_{1,1} - 2 \lambda_{0}^{2} \overline{u^{\pm}_{1,3}} u^{\pm}_
{1,1} u^{\pm}_{3,3}+2\lambda_{0}^{2} \overline{u^{\pm}_{-3,3}} 
u^{\pm}_{1,1} u^{\pm}_{-1,3}+2 \lambda_{0}^{2} \overline{u^{
\pm}_{-3,3}} u^{\pm}_{-1,1} u^{\pm}_{1,3}& \nonumber \\
&+2 \lambda_{0}^{2} \overline{u^{\pm}_{-3,3}} u^{\pm}_{-1,2}u^{
\pm}_{1,2}+2\lambda_{0}^{2} \overline{u^{\pm}_{-3,3}} u^{\pm}_
{1,2} u^{\pm}_{-1,2}-2\lambda_{0}^{2} \overline{u^{\pm}_{1,3}} 
u^{\pm}_{3,3} u^{\pm}_{1,1}+2\lambda_{0}^{2} \overline{u^{\pm}
_{-3,3}} u^{\pm}_{-1,3} u^{\pm}_{1,1}& \nonumber \\
&+2 \lambda_{0}^{2} \overline{u^{\pm}_{-3,3}} u^{\pm}_{1,3} u^{
\pm}_{-1,1}+2\lambda_{0}^{2} \overline{u^{\pm}_{-3,4}} u^{\pm}_{
-1,1} u^{\pm}_{1,2}+2\lambda_{0}^{2} \overline{u^{\pm}_{-3,4}} 
u^{\pm}_{1,1} u^{\pm}_{-1,2}+2\lambda_{0}^{2} \overline{u^{\pm}
_{-3,4}} u^{\pm}_{-1,2} u^{\pm}_{1,1}& \nonumber \\
&+2 \lambda_{0}^{2} \overline{u^{\pm}_{-3,4}} u^{\pm}_{1,2} u^{
\pm}_{-1,1}+2\lambda_{0}^{2} \overline{u^{\pm}_{-3,5}} u^{\pm}_{
-1,1} u^{\pm}_{1,1}+2\lambda_{0}^{2} \overline{u^{\pm}_{-3,5}} 
u^{\pm}_{1,1} u^{\pm}_{-1,1}+4\lambda_{0}^{2} \overline{u^{\pm}
_{-5,5}} u^{\pm}_{-1,1} u^{\pm}_{-1,1}& \nonumber \\
&+\frac{\nu^{\prime}\overline{u^{\pm}_{1,1}}u^{\pm}_{1,1}u^{\pm}
_{3,3} \ln \vert t \vert}{8 \lambda_{0}} + \frac{\nu^{\prime}
\overline{u^{\pm}_{1,1}}u^{\pm}_{3,3} u^{\pm}_{1,1} \ln \vert t 
\vert}{8 \lambda_{0}}-\frac{\nu^{\prime}\overline{u^{\pm}_{-3,
3}}u^{\pm}_{-1,1} u^{\pm}_{1,1} \ln \vert t \vert}{8 \lambda_{0}
} - \frac{\nu^{\prime}\overline{u^{\pm}_{-3,3}}u^{\pm}_{1,1}u^{
\pm}_{-1,1} \ln \vert t \vert}{8 \lambda_{0}}& \nonumber \\
&\mp \frac{i(\overline{u^{\pm}_{1,1}})^{\prime}u^{\pm}_{1,1}u^{
\pm}_{3,3}}{8 \lambda_{0}} \mp \frac{i(\overline{u^{\pm}_{-1,1}})
^{\prime}u^{\pm}_{1,1}u^{\pm}_{1,3}}{8 \lambda_{0}} \mp \frac{i(
\overline{u^{\pm}_{-1,1}})^{\prime}u^{\pm}_{-1,1}u^{\pm}_{3,3}}{8 
\lambda_{0}} \mp \frac{i(\overline{u^{\pm}_{-1,1}})^{\prime}u^{
\pm}_{1,2}u^{\pm}_{1,2}}{8 \lambda_{0}}& \nonumber \\
&\mp \frac{i(\overline{u^{\pm}_{1,1}})^{\prime}u^{\pm}_{3,3}u^{
\pm}_{1,1}}{8 \lambda_{0}} \mp \frac{i(\overline{u^{\pm}_{-1,1}})
^{\prime}u^{\pm}_{1,3} u^{\pm}_{1,1}}{8 \lambda_{0}} \mp \frac{i
(\overline{u^{\pm}_{-1,1}})^{\prime} u^{\pm}_{3,3}u^{\pm}_{-1,1}}
{8 \lambda_{0}} \mp \frac{i(\overline{u^{\pm}_{-1,2}})^{\prime}
u^{\pm}_{1,1}u^{\pm}_{1,2}}{8 \lambda_{0}}& \nonumber \\
&\mp \frac{i(\overline{u^{\pm}_{-1,2}})^{\prime}u^{\pm}_{1,2}u^{
\pm}_{1,1}}{8 \lambda_{0}} \mp \frac{i(\overline{u^{\pm}_{-3,3}}
)^{\prime}u^{\pm}_{-1,1}u^{\pm}_{1,1}}{8 \lambda_{0}} \mp \frac{
i(\overline{u^{\pm}_{-3,3}})^{\prime}u^{\pm}_{1,1}u^{\pm}_{-1,1}}
{8 \lambda_{0}} \mp \frac{i(\overline{u^{\pm}_{-1,3}})^{\prime}
u^{\pm}_{1,1}u^{\pm}_{1,1}}{8 \lambda_{0}}& \nonumber \\
&+\frac{u^{\pm}_{1,1} \overline{u^{\pm}_{1,1}} u^{\pm}_{1,1} 
\overline{u^{\pm}_{-1,2}} u^{\pm}_{1,2}}{2}+\frac{u^{\pm}_{
1,1} \overline{u^{\pm}_{-1,1}} u^{\pm}_{1,1} \overline{u^{\pm}
_{1,2}} u^{\pm}_{1,2}}{2} + \frac{u^{\pm}_{1,1} \overline{u^{
\pm}_{-1,1}} u^{\pm}_{1,1} \overline{u^{\pm}_{-1,2}} u^{\pm}_{
-1,2}}{2}& \nonumber \\
&+\frac{u^{\pm}_{1,1} \overline{u^{\pm}_{-1,1}} u^{\pm}_{-1,1} 
\overline{u^{\pm}_{-1,2}} u^{\pm}_{1,2}}{2} + \frac{u^{\pm}_{
-1,1} \overline{u^{\pm}_{-1,1}} u^{\pm}_{1,1} \overline{u^{\pm}
_{-1,2}} u^{\pm}_{1,2}}{2} + \frac{u^{\pm}_{1,1} \overline{u^
{\pm}_{1,1}} u^{\pm}_{1,2} \overline{u^{\pm}_{-1,1}} u^{\pm}_{
1,2}}{2}& \nonumber \\
&+\frac{u^{\pm}_{1,1} \overline{u^{\pm}_{-1,1}} u^{\pm}_{1,2} 
\overline{u^{\pm}_{1,1}} u^{\pm}_{1,2}}{2} + \frac{u^{\pm}_{
1,1} \overline{u^{\pm}_{-1,1}} u^{\pm}_{1,2} \overline{u^{\pm}
_{-1,1}} u^{\pm}_{-1,2}}{2} + \frac{u^{\pm}_{1,1} \overline{u
^{\pm}_{-1,1}} u^{\pm}_{-1,2} \overline{u^{\pm}_{-1,1}} u^{\pm}
_{1,2}}{2}& \nonumber \\
&+\frac{u^{\pm}_{-1,1} \overline{u^{\pm}_{-1,1}} u^{\pm}_{1,2} 
\overline{u^{\pm}_{-1,1}} u^{\pm}_{1,2}}{2} + \frac{u^{\pm}_{
1,1} \overline{u^{\pm}_{1,1}} u^{\pm}_{1,2} \overline{u^{\pm}
_{-1,2}} u^{\pm}_{1,1}}{2} + \frac{u^{\pm}_{1,1} \overline{u
^{\pm}_{-1,1}} u^{\pm}_{1,2} \overline{u^{\pm}_{1,2}} u^{\pm}
_{1,1}}{2}& \nonumber \\
&+\frac{u^{\pm}_{1,1} \overline{u^{\pm}_{-1,1}} u^{\pm}_{1,2} 
\overline{u^{\pm}_{-1,2}} u^{\pm}_{-1,1}}{2} + \frac{u^{\pm}_{
1,1} \overline{u^{\pm}_{-1,1}} u^{\pm}_{-1,2} \overline{u^{\pm}
_{-1,2}} u^{\pm}_{1,1}}{2} + \frac{u^{\pm}_{-1,1} \overline{u
^{\pm}_{-1,1}} u^{\pm}_{1,2} \overline{u^{\pm}_{-1,2}}u^{\pm}
_{1,1}}{2}& \nonumber \\
&+\frac{u^{\pm}_{1,1} \overline{u^{\pm}_{1,2}} u^{\pm}_{1,1} 
\overline{u^{\pm}_{-1,1}} u^{\pm}_{1,2}}{2} + \frac{u^{\pm}_{
1,1} \overline{u^{\pm}_{-1,2}} u^{\pm}_{1,1} \overline{u^{\pm}
_{1,1}} u^{\pm}_{1,2}}{2}+\frac{u^{\pm}_{1,1} \overline{u^{
\pm}_{-1,2}} u^{\pm}_{1,1} \overline{u^{\pm}_{-1,1}}u^{\pm}
_{-1,2}}{2}& \nonumber \\
&+\frac{u^{\pm}_{1,1} \overline{u^{\pm}_{-1,2}} u^{\pm}_{-1,1} 
\overline{u^{\pm}_{-1,1}} u^{\pm}_{1,2}}{2} + \frac{u^{\pm}_{
-1,1} \overline{u^{\pm}_{-1,2}} u^{\pm}_{1,1} \overline{u^{\pm}
_{-1,1}} u^{\pm}_{1,2}}{2}+\frac{u^{\pm}_{1,1} \overline{u
^{\pm}_{1,2}} u^{\pm}_{1,1} \overline{u^{\pm}_{-1,2}}u^{\pm}
_{1,1}}{2}& \nonumber \\
&+\frac{u^{\pm}_{1,1} \overline{u^{\pm}_{-1,2}} u^{\pm}_{1,1} 
\overline{u^{\pm}_{1,2}} u^{\pm}_{1,1}}{2} + \frac{u^{\pm}_{
1,1} \overline{u^{\pm}_{-1,2}} u^{\pm}_{1,1} \overline{u^{\pm}
_{-1,2}} u^{\pm}_{-1,1}}{2} + \frac{u^{\pm}_{1,1} \overline{u
^{\pm}_{-1,2}} u^{\pm}_{-1,1} \overline{u^{\pm}_{-1,2}}u^{\pm}
_{1,1}}{2}& \nonumber \\
&+\frac{u^{\pm}_{-1,1} \overline{u^{\pm}_{-1,2}} u^{\pm}_{1,1} 
\overline{u^{\pm}_{-1,2}} u^{\pm}_{1,1}}{2} + \frac{u^{\pm}_{
1,1} \overline{u^{\pm}_{1,2}} u^{\pm}_{1,2} \overline{u^{\pm}
_{-1,1}} u^{\pm}_{1,1}}{2} + \frac{u^{\pm}_{1,1} \overline{u
^{\pm}_{-1,2}} u^{\pm}_{1,2} \overline{u^{\pm}_{1,1}}u^{\pm}
_{1,1}}{2}& \nonumber \\
&+\frac{u^{\pm}_{1,1} \overline{u^{\pm}_{-1,2}} u^{\pm}_{1,2} 
\overline{u^{\pm}_{-1,1}} u^{\pm}_{-1,1}}{2} + \frac{u^{\pm}_{
1,1} \overline{u^{\pm}_{-1,2}} u^{\pm}_{-1,2} \overline{u^{
\pm}_{-1,1}} u^{\pm}_{1,1}}{2}+\frac{u^{\pm}_{-1,1} \overline{
u^{\pm}_{-1,2}} u^{\pm}_{1,2} \overline{u^{\pm}_{-1,1}}u^{\pm}
_{1,1}}{2}& \nonumber \\
&+\frac{u^{\pm}_{1,2} \overline{u^{\pm}_{1,1}} u^{\pm}_{1,1} 
\overline{u^{\pm}_{-1,1}} u^{\pm}_{1,2}}{2} + \frac{u^{\pm}_
{1,2} \overline{u^{\pm}_{-1,1}} u^{\pm}_{1,1} \overline{u^{
\pm}_{1,1}} u^{\pm}_{1,2}}{2}+\frac{u^{\pm}_{1,2} \overline{
u^{\pm}_{-1,1}} u^{\pm}_{1,1} \overline{u^{\pm}_{-1,1}}u^{
\pm}_{-1,2}}{2}& \nonumber \\
&+\frac{u^{\pm}_{1,2} \overline{u^{\pm}_{-1,1}} u^{\pm}_{-1,1} 
\overline{u^{\pm}_{-1,1}} u^{\pm}_{1,2}}{2} + \frac{u^{\pm}_{
-1,2} \overline{u^{\pm}_{-1,1}} u^{\pm}_{1,1} \overline{u^{
\pm}_{-1,1}} u^{\pm}_{1,2}}{2}+\frac{u^{\pm}_{1,2} \overline{
u^{\pm}_{1,1}} u^{\pm}_{1,1} \overline{u^{\pm}_{-1,2}}u^{\pm}
_{1,1}}{2}& \nonumber \\
&+\frac{u^{\pm}_{1,2} \overline{u^{\pm}_{-1,1}} u^{\pm}_{1,1} 
\overline{u^{\pm}_{1,2}} u^{\pm}_{1,1}}{2}+\frac{u^{\pm}_{1,
2} \overline{u^{\pm}_{-1,1}} u^{\pm}_{1,1} \overline{u^{\pm}
_{-1,2}} u^{\pm}_{-1,1}}{2}+\frac{u^{\pm}_{1,2} \overline{
u^{\pm}_{-1,1}} u^{\pm}_{-1,1} \overline{u^{\pm}_{-1,2}}u^{\pm}
_{1,1}}{2}& \nonumber \\
&+\frac{u^{\pm}_{-1,2} \overline{u^{\pm}_{-1,1}} u^{\pm}_{1,1} 
\overline{u^{\pm}_{-1,2}} u^{\pm}_{1,1}}{2}+\frac{u^{\pm}_{1,
2} \overline{u^{\pm}_{1,1}} u^{\pm}_{1,2} \overline{u^{\pm}
_{-1,1}} u^{\pm}_{1,1}}{2}+\frac{u^{\pm}_{1,2} \overline{
u^{\pm}_{-1,1}} u^{\pm}_{1,2} \overline{u^{\pm}_{1,1}}u^{\pm}
_{1,1}}{2}& \nonumber \\
&+\frac{u^{\pm}_{1,2} \overline{u^{\pm}_{-1,1}} u^{\pm}_{1,2} 
\overline{u^{\pm}_{-1,1}} u^{\pm}_{-1,1}}{2} + \frac{u^{\pm}_{
1,2} \overline{u^{\pm}_{-1,1}} u^{\pm}_{-1,2} \overline{u^{\pm}
_{-1,1}} u^{\pm}_{1,1}}{2}+\frac{u^{\pm}_{-1,2} \overline{u^{
\pm}_{-1,1}} u^{\pm}_{1,2} \overline{u^{\pm}_{-1,1}} u^{\pm}
_{1,1}}{2}& \nonumber \\
&+\frac{u^{\pm}_{1,2} \overline{u^{\pm}_{1,2}} u^{\pm}_{1,1} 
\overline{u^{\pm}_{-1,1}} u^{\pm}_{1,1}}{2}+\frac{u^{\pm}_{
1,2} \overline{u^{\pm}_{-1,2}} u^{\pm}_{1,1} \overline{u^{\pm}
_{1,1}} u^{\pm}_{1,1}}{2}+\frac{u^{\pm}_{1,2} \overline{u^{
\pm}_{-1,2}} u^{\pm}_{1,1} \overline{u^{\pm}_{-1,1}} u^{\pm}
_{-1,1}}{2}& \nonumber \\
&+\frac{u^{\pm}_{1,2} \overline{u^{\pm}_{-1,2}} u^{\pm}_{-1,1} 
\overline{u^{\pm}_{-1,1}} u^{\pm}_{1,1}}{2} + \frac{u^{\pm}_{
-1,2} \overline{u^{\pm}_{-1,2}} u^{\pm}_{1,1} \overline{u^{\pm}
_{-1,1}} u^{\pm}_{1,1}}{2} + \frac{u^{\pm}_{1,1} \overline{u^{
\pm}_{1,1}} u^{\pm}_{1,1} \overline{u^{\pm}_{1,1}} u^{\pm}
_{3,3}}{2}& \nonumber \\
&+\frac{u^{\pm}_{1,1} \overline{u^{\pm}_{1,1}} u^{\pm}_{1,1} 
\overline{u^{\pm}_{-1,1}} u^{\pm}_{1,3}}{2}+\frac{u^{\pm}_{
1,1} \overline{u^{\pm}_{1,1}} u^{\pm}_{-1,1} \overline{u^{\pm}
_{-1,1}} u^{\pm}_{3,3}}{2}+\frac{u^{\pm}_{1,1} \overline{u^
{\pm}_{-1,1}} u^{\pm}_{1,1} \overline{u^{\pm}_{1,1}} u^{\pm}
_{1,3}}{2}& \nonumber \\
&+\frac{u^{\pm}_{1,1} \overline{u^{\pm}_{-1,1}} u^{\pm}_{1,1} 
\overline{u^{\pm}_{-1,1}} u^{\pm}_{-1,3}}{2}+\frac{u^{\pm}_{
1,1} \overline{u^{\pm}_{-1,1}} u^{\pm}_{-1,1} \overline{u^{
\pm}_{1,1}} u^{\pm}_{3,3}}{2}+\frac{u^{\pm}_{1,1} \overline{
u^{\pm}_{-1,1}} u^{\pm}_{-1,1} \overline{u^{\pm}_{-1,1}} u^{
\pm}_{1,3}}{2}& \nonumber \\
&+\frac{u^{\pm}_{-1,1} \overline{u^{\pm}_{1,1}} u^{\pm}_{1,1} 
\overline{u^{\pm}_{-1,1}} u^{\pm}_{3,3}}{2}+\frac{u^{\pm}_{-1,
1} \overline{u^{\pm}_{-1,1}} u^{\pm}_{1,1} \overline{u^{\pm}
_{1,1}} u^{\pm}_{3,3}}{2}+\frac{u^{\pm}_{-1,1}\overline{u^{
\pm}_{-1,1}} u^{\pm}_{1,1} \overline{u^{\pm}_{-1,1}} u^{\pm}_{
1,3}}{2}& \nonumber \\
&+\frac{u^{\pm}_{-1,1} \overline{u^{\pm}_{-1,1}} u^{\pm}_{-1,1} 
\overline{u^{\pm}_{-1,1}} u^{\pm}_{3,3}}{2} + \frac{u^{\pm}_{
1,1} \overline{u^{\pm}_{1,1}} u^{\pm}_{1,1} \overline{u^{\pm}
_{-1,3}} u^{\pm}_{1,1}}{2}+\frac{u^{\pm}_{1,1} \overline{u^{
\pm}_{1,1}} u^{\pm}_{1,1} \overline{u^{\pm}_{-3,3}} u^{\pm}_{
-1,1}}{2}& \nonumber \\
&+\frac{u^{\pm}_{1,1} \overline{u^{\pm}_{1,1}} u^{\pm}_{-1,1} 
\overline{u^{\pm}_{-3,3}} u^{\pm}_{1,1}}{2} + \frac{u^{\pm}_{
1,1} \overline{u^{\pm}_{-1,1}} u^{\pm}_{1,1} \overline{u^{\pm}
_{1,3}} u^{\pm}_{1,1}}{2}+\frac{u^{\pm}_{1,1} \overline{u^{
\pm}_{-1,1}} u^{\pm}_{1,1} \overline{u^{\pm}_{-1,3}} u^{\pm}_{
-1,1}}{2}& \nonumber \\
&+\frac{u^{\pm}_{1,1} \overline{u^{\pm}_{-1,1}} u^{\pm}_{-1,1} 
\overline{u^{\pm}_{-1,3}} u^{\pm}_{1,1}}{2}+\frac{u^{\pm}_{1,1} 
\overline{u^{\pm}_{-1,1}} u^{\pm}_{-1,1} \overline{u^{\pm}_{-3,
3}} u^{\pm}_{-1,1}}{2}+\frac{u^{\pm}_{-1,1} \overline{u^{\pm}_{
1,1}} u^{\pm}_{1,1} \overline{u^{\pm}_{-3,3}} u^{\pm}_{1,1}}{
2}& \nonumber \\
&+\frac{u^{\pm}_{-1,1} \overline{u^{\pm}_{-1,1}} u^{\pm}_{1,1} 
\overline{u^{\pm}_{-1,3}} u^{\pm}_{1,1}}{2} + \frac{u^{\pm}_{
-1,1} \overline{u^{\pm}_{-1,1}} u^{\pm}_{1,1} \overline{u^{\pm}
_{-3,3}} u^{\pm}_{-1,1}}{2}+\frac{u^{\pm}_{-1,1} \overline{u^{
\pm}_{-1,1}} u^{\pm}_{-1,1} \overline{u^{\pm}_{-3,3}} u^{\pm}_
{1,1}}{2}& \nonumber \\
&+\frac{u^{\pm}_{1,1} \overline{u^{\pm}_{1,1}} u^{\pm}_{3,3} 
\overline{u^{\pm}_{1,1}} u^{\pm}_{1,1}}{2}+\frac{u^{\pm}_{
1,1} \overline{u^{\pm}_{1,1}} u^{\pm}_{1,3} \overline{u^{\pm}
_{-1,1}} u^{\pm}_{1,1}}{2}+\frac{u^{\pm}_{1,1} \overline{u^{
\pm}_{1,1}} u^{\pm}_{3,3} \overline{u^{\pm}_{-1,1}} u^{\pm}_
{-1,1}}{2}& \nonumber \\
&+\frac{u^{\pm}_{1,1} \overline{u^{\pm}_{-1,1}} u^{\pm}_{1,3} 
\overline{u^{\pm}_{1,1}} u^{\pm}_{1,1}}{2}+\frac{u^{\pm}_{1,1} 
\overline{u^{\pm}_{-1,1}} u^{\pm}_{3,3} \overline{u^{\pm}_{1,
1}} u^{\pm}_{-1,1}}{2}+\frac{u^{\pm}_{1,1} \overline{u^{\pm}_{
-1,1}} u^{\pm}_{-1,3} \overline{u^{\pm}_{-1,1}} u^{\pm}_{1,1}}
{2}& \nonumber \\
&+\frac{u^{\pm}_{1,1} \overline{u^{\pm}_{-1,1}} u^{\pm}_{1,3} 
\overline{u^{\pm}_{-1,1}} u^{\pm}_{-1,1}}{2}+\frac{u^{\pm}_{
-1,1} \overline{u^{\pm}_{1,1}} u^{\pm}_{3,3} \overline{u^{
\pm}_{-1,1}} u^{\pm}_{1,1}}{2}+\frac{u^{\pm}_{-1,1} \overline{
u^{\pm}_{-1,1}} u^{\pm}_{3,3} \overline{u^{\pm}_{1,1}} u^{
\pm}_{1,1}}{2}& \nonumber \\
&+\frac{u^{\pm}_{-1,1} \overline{u^{\pm}_{-1,1}} u^{\pm}_{1,3} 
\overline{u^{\pm}_{-1,1}} u^{\pm}_{1,1}}{2} + \frac{u^{\pm}_{
-1,1} \overline{u^{\pm}_{-1,1}} u^{\pm}_{3,3} \overline{u^{\pm}
_{-1,1}} u^{\pm}_{-1,1}}{2}+\frac{u^{\pm}_{1,1} \overline{u^{
\pm}_{-1,3}} u^{\pm}_{1,1} \overline{u^{\pm}_{1,1}} u^{\pm}_
{1,1}}{2}& \nonumber \\
&+\frac{u^{\pm}_{1,1} \overline{u^{\pm}_{-3,3}} u^{\pm}_{1,1} 
\overline{u^{\pm}_{1,1}} u^{\pm}_{-1,1}}{2}+\frac{u^{\pm}_{
1,1} \overline{u^{\pm}_{1,3}} u^{\pm}_{1,1} \overline{u^{\pm}
_{-1,1}} u^{\pm}_{1,1}}{2}+\frac{u^{\pm}_{1,1} \overline{u^{
\pm}_{-1,3}} u^{\pm}_{1,1} \overline{u^{\pm}_{-1,1}} u^{\pm}_
{-1,1}}{2}& \nonumber \\
&+\frac{u^{\pm}_{1,1} \overline{u^{\pm}_{-3,3}} u^{\pm}_{-1,1} 
\overline{u^{\pm}_{1,1}} u^{\pm}_{1,1}}{2}+\frac{u^{\pm}_{1,1} 
\overline{u^{\pm}_{-1,3}} u^{\pm}_{-1,1} \overline{u^{\pm}_{
-1,1}} u^{\pm}_{1,1}}{2}+\frac{u^{\pm}_{1,1} \overline{u^{
\pm}_{-3,3}} u^{\pm}_{-1,1} \overline{u^{\pm}_{-1,1}} u^{\pm}
_{-1,1}}{2}& \nonumber \\
&+\frac{u^{\pm}_{-1,1} \overline{u^{\pm}_{-3,3}} u^{\pm}_{1,1} 
\overline{u^{\pm}_{1,1}} u^{\pm}_{1,1}}{2} + \frac{u^{\pm}_{
-1,1} \overline{u^{\pm}_{-1,3}} u^{\pm}_{1,1} \overline{u^{\pm}
_{-1,1}} u^{\pm}_{1,1}}{2}+\frac{u^{\pm}_{-1,1} \overline{u^{
\pm}_{-3,3}} u^{\pm}_{1,1} \overline{u^{\pm}_{-1,1}} u^{\pm}_{
-1,1}}{2}& \nonumber \\
&+\frac{u^{\pm}_{-1,1} \overline{u^{\pm}_{-3,3}} u^{\pm}_{-1,1} 
\overline{u^{\pm}_{-1,1}} u^{\pm}_{1,1}}{2} + \frac{u^{\pm}_{
3,3} \overline{u^{\pm}_{1,1}} u^{\pm}_{1,1} \overline{u^{\pm}
_{1,1}} u^{\pm}_{1,1}}{2}+\frac{u^{\pm}_{1,3} \overline{u^{
\pm}_{-1,1}} u^{\pm}_{1,1} \overline{u^{\pm}_{1,1}} u^{\pm}_{
1,1}}{2}& \nonumber \\
&+\frac{u^{\pm}_{3,3} \overline{u^{\pm}_{-1,1}} u^{\pm}_{-1,1} 
\overline{u^{\pm}_{1,1}} u^{\pm}_{1,1}}{2} + \frac{u^{\pm}_{1,3} 
\overline{u^{\pm}_{1,1}} u^{\pm}_{1,1} \overline{u^{\pm}_{-1,1}} 
u^{\pm}_{1,1}}{2}+\frac{u^{\pm}_{-1,3} \overline{u^{\pm}_{-1,1}} 
u^{\pm}_{1,1} \overline{u^{\pm}_{-1,1}} u^{\pm}_{1,1}}{2}& 
\nonumber \\
&+\frac{u^{\pm}_{3,3} \overline{u^{\pm}_{1,1}} u^{\pm}_{-1,1} 
\overline{u^{\pm}_{-1,1}} u^{\pm}_{1,1}}{2} + \frac{u^{\pm}_{
1,3} \overline{u^{\pm}_{-1,1}} u^{\pm}_{-1,1} \overline{u^{\pm}
_{-1,1}} u^{\pm}_{1,1}}{2}+\frac{u^{\pm}_{3,3} \overline{u^{
\pm}_{-1,1}} u^{\pm}_{1,1} \overline{u^{\pm}_{1,1}} u^{\pm}_{
-1,1}}{2}& \nonumber \\
&+\frac{u^{\pm}_{3,3} \overline{u^{\pm}_{1,1}} u^{\pm}_{1,1} 
\overline{u^{\pm}_{-1,1}} u^{\pm}_{-1,1}}{2}+\frac{u^{\pm}_{
1,3} \overline{u^{\pm}_{-1,1}} u^{\pm}_{1,1} \overline{u^{\pm}
_{-1,1}} u^{\pm}_{-1,1}}{2}+\frac{u^{\pm}_{3,3} \overline{u^{
\pm}_{-1,1}} u^{\pm}_{-1,1} \overline{u^{\pm}_{-1,1}} u^{\pm}
_{-1,1}}{2}=0;&
\end{eqnarray}
\begin{flushleft}
(26) ${\cal O}((\pm t)^{-7/2} \exp\{-i\tau^{\pm}\})$,
\end{flushleft}
\begin{eqnarray}
&-\nu u^{\pm}_{-3,5} + \frac{\lambda_{0} \nu^{\prime} u^{\pm}_{
-3,5} \ln \vert t \vert}{2}-4\lambda_{0}^{4} u^{\pm}_{-3,7} \pm 
i t \dot{u}^{\pm}_{-3,5} \mp \frac{5iu^{\pm}_{-3,5}}{2} \mp 
\frac{i \lambda_{0} u^{\pm \prime}_{-3,5}}{2} \pm \frac{i \nu^{
\prime}u^{\pm \prime}_{-3,3} \ln \vert t \vert}{32 \lambda_{0}^
{2}}& \nonumber \\
&\pm \frac{i \nu^{\prime \prime} u^{\pm}_{-3,3} \ln \vert t 
\vert}{64 \lambda_{0}^{2}} 
\mp \frac{i \nu^{\prime} u^{\pm}_{-3,3} \ln \vert t \vert}{64 
\lambda_{0}^{3}} \mp \frac{i \lambda_{0} u^{\pm \prime}_{-3,5}}
{2} \mp \frac{iu^{\pm}_{-3,5}}{2}-\frac{(\nu^{\prime})^{2} u^{
\pm}_{-3,3}(\ln \vert t \vert)^{2}}{64 \lambda_{0}^{2}}+\frac{
\lambda_{0} \nu^{\prime} u^{\pm}_{-3,5} \ln \vert t \vert}{2}& 
\nonumber \\
&-4 \lambda_{0}^{4} u^{\pm}_{-3,7} 
+ \frac{u^{\pm \prime \prime}_{-3,3}}{64 \lambda_{0}^{2}}-\frac{
u^{\pm \prime}_{-3,3}}{64 \lambda_{0}^{3}} - 2 \lambda_{0}^{2} 
\overline{u^{\pm}_{1,1}} u^{\pm}_{1,1} u^{\pm}_{-3,5} - 2 \lambda
_{0}^{2} \overline{u^{\pm}_{1,1}} u^{\pm}_{-1,1} u^{\pm}_{-1,5}
- 2 \lambda_{0}^{2} \overline{u^{\pm}_{1,1}} u^{\pm}_{1,2} u^{
\pm}_{-3,4}& \nonumber \\
&-2 \lambda_{0}^{2} \overline{u^{\pm}_{1,1}} u^{\pm}_{-1,2} u^{
\pm}_{-1,4} - 2 \lambda_{0}^{2} \overline{u^{\pm}_{1,1}} u^{\pm}
_{1,3} u^{\pm}_{-3,3}-2 \lambda_{0}^{2} \overline{u^{\pm}_{1,1}
} u^{\pm}_{-1,3} u^{\pm}_{-1,3}-2\lambda_{0}^{2} \overline{u^{
\pm}_{1,1}} u^{\pm}_{-3,3} u^{\pm}_{1,3}& \nonumber \\
&-2\lambda_{0}^{2} \overline{u^{\pm}_{1,1}} u^{\pm}_{-3,4}u^{\pm}
_{1,2}-2\lambda_{0}^{2} \overline{u^{\pm}_{1,1}} u^{\pm}_{-1,4}
u^{\pm}_{-1,2}-2\lambda_{0}^{2} \overline{u^{\pm}_{1,1}} u^{\pm}
_{-3,5} u^{\pm}_{1,1}-2 \lambda_{0}^{2} \overline{u^{\pm}_{1,1}} 
u^{\pm}_{-1,5} u^{\pm}_{-1,1}& \nonumber \\
&-2 \lambda_{0}^{2} \overline{u^{\pm}_{1,2}} u^{\pm}_{1,1} u^{
\pm}_{-3,4}-2 \lambda_{0}^{2} \overline{u^{\pm}_{1,2}} u^{\pm}_
{-1,1} u^{\pm}_{-1,4}-2\lambda_{0}^{2} \overline{u^{\pm}_{1,2}} 
u^{\pm}_{1,2} u^{\pm}_{-3,3}-2 \lambda_{0}^{2} \overline{u^{
\pm}_{1,2}} u^{\pm}_{-1,2} u^{\pm}_{-1,3}& \nonumber \\
&-2 \lambda_{0}^{2} \overline{u^{\pm}_{1,2}} u^{\pm}_{-3,3}u^{
\pm}_{1,2}-2\lambda_{0}^{2} \overline{u^{\pm}_{1,2}} u^{\pm}_
{-1,3} u^{\pm}_{-1,2}-2\lambda_{0}^{2} \overline{u^{\pm}_{1,2}} 
u^{\pm}_{-3,4} u^{\pm}_{1,1}-2\lambda_{0}^{2} \overline{u^{\pm}
_{1,2}} u^{\pm}_{-1,4} u^{\pm}_{-1,1}& \nonumber \\
&-2 \lambda_{0}^{2} \overline{u^{\pm}_{1,3}} u^{\pm}_{1,1} u^{
\pm}_{-3,3}-4\lambda_{0}^{2} \overline{u^{\pm}_{3,3}} u^{\pm}_{
1,1} u^{\pm}_{-1,3}-2\lambda_{0}^{2} \overline{u^{\pm}_{1,3}} 
u^{\pm}_{-1,1} u^{\pm}_{-1,3}-4\lambda_{0}^{2} \overline{u^{\pm}
_{3,3}} u^{\pm}_{-1,1} u^{\pm}_{1,3}& \nonumber \\
&-4\lambda_{0}^{2} \overline{u^{\pm}_{3,3}} u^{\pm}_{-1,2} u^{
\pm}_{1,2}-4\lambda_{0}^{2} \overline{u^{\pm}_{3,3}} u^{\pm}_{
1,2} u^{\pm}_{-1,2}-2\lambda_{0}^{2} \overline{u^{\pm}_{1,3}} 
u^{\pm}_{-1,2} u^{\pm}_{-1,2}-2\lambda_{0}^{2} \overline{u^{\pm}
_{1,3}} u^{\pm}_{-3,3} u^{\pm}_{1,1}& \nonumber \\
&-4\lambda_{0}^{2} \overline{u^{\pm}_{3,3}} u^{\pm}_{-1,3} u^{
\pm}_{1,1}-2\lambda_{0}^{2} \overline{u^{\pm}_{1,3}} u^{\pm}_{
-1,3} u^{\pm}_{-1,1}-4\lambda_{0}^{2} \overline{u^{\pm}_{3,3}} 
u^{\pm}_{1,3} u^{\pm}_{-1,1}-4\lambda_{0}^{2} \overline{u^{\pm}
_{3,4}} u^{\pm}_{-1,1} u^{\pm}_{1,2}& \nonumber \\
&-4\lambda_{0}^{2} \overline{u^{\pm}_{3,4}} u^{\pm}_{1,1} u^{
\pm}_{-1,2}-2\lambda_{0}^{2} \overline{u^{\pm}_{1,4}} u^{\pm}_{
-1,1} u^{\pm}_{-1,2}-4\lambda_{0}^{2} \overline{u^{\pm}_{3,4}} 
u^{\pm}_{-1,2} u^{\pm}_{1,1}-4\lambda_{0}^{2} \overline{u^{\pm}
_{3,4}} u^{\pm}_{1,2} u^{\pm}_{-1,1}& \nonumber \\
&-2\lambda_{0}^{2} \overline{u^{\pm}_{1,4}} u^{\pm}_{-1,2} u^{
\pm}_{-1,1}-6\lambda_{0}^{2} \overline{u^{\pm}_{5,5}} u^{\pm}_{
1,1} u^{\pm}_{1,1}-4\lambda_{0}^{2} \overline{u^{\pm}_{3,5}} 
u^{\pm}_{-1,1} u^{\pm}_{1,1}-4\lambda_{0}^{2} \overline{u^{\pm}
_{3,5}} u^{\pm}_{1,1} u^{\pm}_{-1,1}& \nonumber \\
&-2\lambda_{0}^{2} \overline{u^{\pm}_{1,5}} u^{\pm}_{-1,1} u^{
\pm}_{-1,1} + \frac{\nu^{\prime}\overline{u^{\pm}_{1,1}}u^{\pm}
_{1,1}u^{\pm}_{-3,3} \ln \vert t \vert}{8 \lambda_{0}} + \frac{
\nu^{\prime}\overline{u^{\pm}_{1,1}}u^{\pm}_{-1,1} u^{\pm}_{-1,
3} \ln \vert t \vert}{8 \lambda_{0}} + \frac{\nu^{\prime}
\overline{u^{\pm}_{1,1}}u^{\pm}_{-1,2} u^{\pm}_{-1,2} \ln \vert 
t \vert}{8 \lambda_{0}}& \nonumber \\
&+\frac{\nu^{\prime}\overline{u^{\pm}_{1,1}}u^{\pm}_{-3,3}u^{
\pm}_{1,1} \ln \vert t \vert}{8 \lambda_{0}}+\frac{\nu^{\prime}
\overline{u^{\pm}_{1,1}}u^{\pm}_{-1,3}u^{\pm}_{-1,1} \ln \vert t 
\vert}{8 \lambda_{0}}+\frac{\nu^{\prime}\overline{u^{\pm}_{1,2}}
u^{\pm}_{-1,1}u^{\pm}_{-1,2} \ln \vert t \vert}{8 \lambda_{0}}
+\frac{\nu^{\prime}\overline{u^{\pm}_{1,2}}u^{\pm}_{-1,2}u^{
\pm}_{-1,1} \ln \vert t \vert}{8 \lambda_{0}}& \nonumber \\
&+\frac{\nu^{\prime}\overline{u^{\pm}_{3,3}}u^{\pm}_{-1,1}u^{
\pm}_{1,1} \ln \vert t \vert}{4 \lambda_{0}}+\frac{\nu^{\prime}
\overline{u^{\pm}_{3,3}}u^{\pm}_{1,1}u^{\pm}_{-1,1} \ln \vert t 
\vert}{4\lambda_{0}}+\frac{\nu^{\prime}\overline{u^{\pm}_{1,3}}
u^{\pm}_{-1,1}u^{\pm}_{-1,1} \ln \vert t \vert}{8\lambda_{0}} 
\mp \frac{i(\overline{u^{\pm}_{1,1}})^{\prime}u^{\pm}_{1,1}u^{
\pm}_{-3,3}}{8 \lambda_{0}}& \nonumber \\
&\mp \frac{i(\overline{u^{\pm}_{1,1}})^{\prime}u^{\pm}_{-1,1}u^{
\pm}_{-1,3}}{8 \lambda_{0}} \mp \frac{i(\overline{u^{\pm}_{-1,1}
})^{\prime}u^{\pm}_{-1,1}u^{\pm}_{-3,3}}{8 \lambda_{0}} \mp 
\frac{i(\overline{u^{\pm}_{1,1}})^{\prime}u^{\pm}_{-1,2}u^{\pm}_{
-1,2}}{8 \lambda_{0}} \mp \frac{i(\overline{u^{\pm}_{1,1}})^{
\prime}u^{\pm}_{-3,3}u^{\pm}_{1,1}}{8 \lambda_{0}}& \nonumber \\
&\mp \frac{i(\overline{u^{\pm}_{1,1}})^{\prime}u^{\pm}_{-1,3} u^{
\pm}_{-1,1}}{8 \lambda_{0}} \mp \frac{i(\overline{u^{\pm}_{-1,1}}
)^{\prime} u^{\pm}_{-3,3}u^{\pm}_{-1,1}}{8 \lambda_{0}} \mp \frac{
i(\overline{u^{\pm}_{1,2}})^{\prime}u^{\pm}_{-1,1}u^{\pm}_{-1,2}}{
8 \lambda_{0}} \mp \frac{i(\overline{u^{\pm}_{1,2}})^{\prime}u^{
\pm}_{-1,2}u^{\pm}_{-1,1}}{8 \lambda_{0}}& \nonumber \\
&\mp \frac{i(\overline{u^{\pm}_{3,3}})^{\prime}u^{\pm}_{-1,1}u^{
\pm}_{1,1}}{8 \lambda_{0}} \mp \frac{i(\overline{u^{\pm}_{3,3}})^{
\prime}u^{\pm}_{1,1}u^{\pm}_{-1,1}}{8 \lambda_{0}} \mp \frac{i(
\overline{u^{\pm}_{1,3}})^{\prime}u^{\pm}_{-1,1}u^{\pm}_{-1,1}}{8 
\lambda_{0}} + \frac{u^{\pm}_{1,1} \overline{u^{\pm}_{1,1}}u^{\pm}
_{-1,1} \overline{u^{\pm}_{1,2}} u^{\pm}_{-1,2}}{2}& \nonumber \\
&+\frac{u^{\pm}_{-1,1} \overline{u^{\pm}_{1,1}} u^{\pm}_{1,1} 
\overline{u^{\pm}_{1,2}} u^{\pm}_{-1,2}}{2} + \frac{u^{\pm}_{-1,1} 
\overline{u^{\pm}_{1,1}} u^{\pm}_{-1,1} \overline{u^{\pm}_{1,2}} 
u^{\pm}_{1,2}}{2} + \frac{u^{\pm}_{-1,1} \overline{u^{\pm}_{1,1}} 
u^{\pm}_{-1,1} \overline{u^{\pm}_{-1,2}} u^{\pm}_{-1,2}}{2}& 
\nonumber \\
&+\frac{u^{\pm}_{-1,1} \overline{u^{\pm}_{-1,1}} u^{\pm}_{-1,1} 
\overline{u^{\pm}_{1,2}} u^{\pm}_{-1,2}}{2} + \frac{u^{\pm}_{1,1} 
\overline{u^{\pm}_{1,1}} u^{\pm}_{-1,2} \overline{u^{\pm}_{1,1}} 
u^{\pm}_{-1,2}}{2} + \frac{u^{\pm}_{-1,1} \overline{u^{\pm}_{1,1}} 
u^{\pm}_{1,2} \overline{u^{\pm}_{1,1}} u^{\pm}_{-1,2}}{2}& 
\nonumber \\
&+ \frac{u^{\pm}_{-1,1} \overline{u^{\pm}_{1,1}} u^{\pm}_{-1,2} 
\overline{u^{\pm}_{1,1}} u^{\pm}_{1,2}}{2} + \frac{u^{\pm}_{-1,1} 
\overline{u^{\pm}_{1,1}} u^{\pm}_{-1,2} \overline{u^{\pm}_{-1,1}} 
u^{\pm}_{-1,2}}{2} + \frac{u^{\pm}_{-1,1} \overline{u^{\pm}_{-1,1}} 
u^{\pm}_{-1,2} \overline{u^{\pm}_{1,1}} u^{\pm}_{-1,2}}{2}& 
\nonumber \\
&+ \frac{u^{\pm}_{1,1} \overline{u^{\pm}_{1,1}} u^{\pm}_{-1,2} 
\overline{u^{\pm}_{1,2}} u^{\pm}_{-1,1}}{2} + \frac{u^{\pm}_{-1,1} 
\overline{u^{\pm}_{1,1}} u^{\pm}_{1,2} \overline{u^{\pm}_{1,2}} 
u^{\pm}_{-1,1}}{2} + \frac{u^{\pm}_{-1,1} \overline{u^{\pm}_{1,1}} 
u^{\pm}_{-1,2} \overline{u^{\pm}_{1,2}} u^{\pm}_{1,1}}{2}& 
\nonumber \\
&+ \frac{u^{\pm}_{-1,1} \overline{u^{\pm}_{1,1}} u^{\pm}_{-1,2} 
\overline{u^{\pm}_{-1,2}} u^{\pm}_{-1,1}}{2}+\frac{u^{\pm}_{-1,
1} \overline{u^{\pm}_{-1,1}} u^{\pm}_{-1,2} \overline{u^{\pm}_{
1,2}}u^{\pm}_{-1,1}}{2} + \frac{u^{\pm}_{1,1} \overline{u^{\pm}
_{1,2}} u^{\pm}_{-1,1} \overline{u^{\pm}_{1,1}} u^{\pm}_{-1,2}}{
2}& \nonumber \\
&+ \frac{u^{\pm}_{-1,1} \overline{u^{\pm}_{1,2}} u^{\pm}_{1,1} 
\overline{u^{\pm}_{1,1}} u^{\pm}_{-1,2}}{2}+\frac{u^{\pm}_{-1,1} 
\overline{u^{\pm}_{1,2}} u^{\pm}_{-1,1} \overline{u^{\pm}_{1,1}}
u^{\pm}_{1,2}}{2} + \frac{u^{\pm}_{-1,1} \overline{u^{\pm}_{1,2}} 
u^{\pm}_{-1,1} \overline{u^{\pm}_{-1,1}} u^{\pm}_{-1,2}}{2}&
\nonumber \\
&+ \frac{u^{\pm}_{-1,1} \overline{u^{\pm}_{-1,2}} u^{\pm}_{-1,1} 
\overline{u^{\pm}_{1,1}} u^{\pm}_{-1,2}}{2}+\frac{u^{\pm}_{1,1} 
\overline{u^{\pm}_{1,2}} u^{\pm}_{-1,1} \overline{u^{\pm}_{1,2}}
u^{\pm}_{-1,1}}{2} + \frac{u^{\pm}_{-1,1} \overline{u^{\pm}_{1,2}} 
u^{\pm}_{1,1} \overline{u^{\pm}_{1,2}} u^{\pm}_{-1,1}}{2}& 
\nonumber \\
&+ \frac{u^{\pm}_{-1,1} \overline{u^{\pm}_{1,2}} u^{\pm}_{-1,1} 
\overline{u^{\pm}_{1,2}} u^{\pm}_{1,1}}{2}+\frac{u^{\pm}_{-1,1} 
\overline{u^{\pm}_{1,2}} u^{\pm}_{-1,1} \overline{u^{\pm}_{-1,2}}
u^{\pm}_{-1,1}}{2} + \frac{u^{\pm}_{-1,1} \overline{u^{\pm}_{-1,
2}} u^{\pm}_{-1,1} \overline{u^{\pm}_{1,2}} u^{\pm}_{-1,1}}{2}&
\nonumber \\
&+\frac{u^{\pm}_{1,1} \overline{u^{\pm}_{1,2}} u^{\pm}_{-1,2} 
\overline{u^{\pm}_{1,1}} u^{\pm}_{-1,1}}{2} + \frac{u^{\pm}_{
-1,1} \overline{u^{\pm}_{1,2}} u^{\pm}_{1,2} \overline{u^{\pm}_{
1,1}}u^{\pm}_{-1,1}}{2} + \frac{u^{\pm}_{-1,1} \overline{u^{\pm}
_{1,2}} u^{\pm}_{-1,2} \overline{u^{\pm}_{1,1}} u^{\pm}_{1,1}}{
2}& \nonumber \\
&+ \frac{u^{\pm}_{-1,1} \overline{u^{\pm}_{1,2}} u^{\pm}_{-1,2} 
\overline{u^{\pm}_{-1,1}} u^{\pm}_{-1,1}}{2}+\frac{u^{\pm}_{-1,1} 
\overline{u^{\pm}_{-1,2}} u^{\pm}_{-1,2} \overline{u^{\pm}_{1,1}}
u^{\pm}_{-1,1}}{2} + \frac{u^{\pm}_{1,2} \overline{u^{\pm}_{1,1}} 
u^{\pm}_{-1,1} \overline{u^{\pm}_{1,1}} u^{\pm}_{-1,2}}{2}&
\nonumber \\
&+\frac{u^{\pm}_{-1,2} \overline{u^{\pm}_{1,1}} u^{\pm}_{1,1} 
\overline{u^{\pm}_{1,1}} u^{\pm}_{-1,2}}{2}+\frac{u^{\pm}_{-1,
2} \overline{u^{\pm}_{1,1}} u^{\pm}_{-1,1} \overline{u^{\pm}_{
1,1}}u^{\pm}_{1,2}}{2} + \frac{u^{\pm}_{-1,2} \overline{u^{\pm}
_{1,1}} u^{\pm}_{-1,1} \overline{u^{\pm}_{-1,1}} u^{\pm}_{-1,2}
}{2}& \nonumber \\
&+\frac{u^{\pm}_{-1,2} \overline{u^{\pm}_{-1,1}} u^{\pm}_{-1,1} 
\overline{u^{\pm}_{1,1}} u^{\pm}_{-1,2}}{2}+\frac{u^{\pm}_{1,2} 
\overline{u^{\pm}_{1,1}} u^{\pm}_{-1,1} \overline{u^{\pm}_{1,2}}
u^{\pm}_{-1,1}}{2} + \frac{u^{\pm}_{-1,2} \overline{u^{\pm}_{1,
1}} u^{\pm}_{1,1} \overline{u^{\pm}_{1,2}} u^{\pm}_{-1,1}}{2}&
\nonumber \\
&+ \frac{u^{\pm}_{-1,2} \overline{u^{\pm}_{1,1}} u^{\pm}_{-1,1} 
\overline{u^{\pm}_{1,2}} u^{\pm}_{1,1}}{2}+\frac{u^{\pm}_{-1,2} 
\overline{u^{\pm}_{1,1}} u^{\pm}_{-1,1} \overline{u^{\pm}_{-1,2}
}u^{\pm}_{-1,1}}{2}+\frac{u^{\pm}_{-1,2} \overline{u^{\pm}_{-1,
1}} u^{\pm}_{-1,1} \overline{u^{\pm}_{1,2}} u^{\pm}_{-1,1}}{2}&
\nonumber \\
&+\frac{u^{\pm}_{1,2} \overline{u^{\pm}_{1,1}} u^{\pm}_{-1,2} 
\overline{u^{\pm}_{1,1}} u^{\pm}_{-1,1}}{2}+\frac{u^{\pm}_{-1,2} 
\overline{u^{\pm}_{1,1}} u^{\pm}_{1,2} \overline{u^{\pm}_{1,1}} 
u^{\pm}_{-1,1}}{2} + \frac{u^{\pm}_{-1,2} \overline{u^{\pm}_{1,1}
} u^{\pm}_{-1,2} \overline{u^{\pm}_{1,1}} u^{\pm}_{1,1}}{2}& 
\nonumber \\
&+ \frac{u^{\pm}_{-1,2} \overline{u^{\pm}_{1,1}} u^{\pm}_{-1,2} 
\overline{u^{\pm}_{-1,1}} u^{\pm}_{-1,1}}{2} + \frac{u^{\pm}_{
-1,2} \overline{u^{\pm}_{-1,1}} u^{\pm}_{-1,2} \overline{u^{\pm}
_{1,1}} u^{\pm}_{-1,1}}{2} + \frac{u^{\pm}_{1,2} \overline{u^{
\pm}_{1,2}} u^{\pm}_{-1,1} \overline{u^{\pm}_{1,1}} u^{\pm}_{-1,
1}}{2}& \nonumber \\
&+\frac{u^{\pm}_{-1,2} \overline{u^{\pm}_{1,2}} u^{\pm}_{1,1} 
\overline{u^{\pm}_{1,1}} u^{\pm}_{-1,1}}{2}+\frac{u^{\pm}_{-1,2} 
\overline{u^{\pm}_{1,2}} u^{\pm}_{-1,1} \overline{u^{\pm}_{1,1}} 
u^{\pm}_{1,1}}{2} + \frac{u^{\pm}_{-1,2} \overline{u^{\pm}_{1,2}} 
u^{\pm}_{-1,1} \overline{u^{\pm}_{-1,1}} u^{\pm}_{-1,1}}{2}&
\nonumber \\
&+ \frac{u^{\pm}_{-1,2} \overline{u^{\pm}_{-1,2}} u^{\pm}_{-1,1} 
\overline{u^{\pm}_{1,1}} u^{\pm}_{-1,1}}{2} + \frac{u^{\pm}_{1,1} 
\overline{u^{\pm}_{1,1}} u^{\pm}_{1,1} \overline{u^{\pm}_{1,1}} 
u^{\pm}_{-3,3}}{2} + \frac{u^{\pm}_{1,1} \overline{u^{\pm}_{1,1}} 
u^{\pm}_{-1,1} \overline{u^{\pm}_{1,1}} u^{\pm}_{-1,3}}{2}& 
\nonumber \\
&+\frac{u^{\pm}_{1,1} \overline{u^{\pm}_{1,1}} u^{\pm}_{-1,1} 
\overline{u^{\pm}_{-1,1}} u^{\pm}_{-3,3}}{2}+\frac{u^{\pm}_{1,1} 
\overline{u^{\pm}_{-1,1}} u^{\pm}_{-1,1} \overline{u^{\pm}_{1,1}
} u^{\pm}_{-3,3}}{2} + \frac{u^{\pm}_{-1,1} \overline{u^{\pm}_{
1,1}} u^{\pm}_{1,1} \overline{u^{\pm}_{1,1}} u^{\pm}_{-1,3}}{2}&
\nonumber \\
&+\frac{u^{\pm}_{-1,1} \overline{u^{\pm}_{1,1}} u^{\pm}_{1,1} 
\overline{u^{\pm}_{-1,1}} u^{\pm}_{-3,3}}{2} + \frac{u^{\pm}_{
-1,1} \overline{u^{\pm}_{1,1}} u^{\pm}_{-1,1} \overline{u^{\pm}
_{1,1}} u^{\pm}_{1,3}}{2} + \frac{u^{\pm}_{-1,1} \overline{u^{
\pm}_{1,1}} u^{\pm}_{-1,1} \overline{u^{\pm}_{-1,1}} u^{\pm}_{
-1,3}}{2}& \nonumber \\
&+\frac{u^{\pm}_{-1,1} \overline{u^{\pm}_{-1,1}} u^{\pm}_{1,1} 
\overline{u^{\pm}_{1,1}} u^{\pm}_{-3,3}}{2} + \frac{u^{\pm}_{
-1,1} \overline{u^{\pm}_{-1,1}} u^{\pm}_{-1,1} \overline{u^{\pm}
_{1,1}} u^{\pm}_{-1,3}}{2} + \frac{u^{\pm}_{-1,1} \overline{u^{
\pm}_{-1,1}} u^{\pm}_{-1,1} \overline{u^{\pm}_{-1,1}} u^{\pm}_{
-3,3}}{2}& \nonumber \\
&+\frac{u^{\pm}_{1,1} \overline{u^{\pm}_{1,1}} u^{\pm}_{1,1} 
\overline{u^{\pm}_{3,3}} u^{\pm}_{-1,1}}{2} + \frac{u^{\pm}_{
1,1} \overline{u^{\pm}_{1,1}} u^{\pm}_{-1,1} \overline{u^{\pm}
_{3,3}} u^{\pm}_{1,1}}{2}+\frac{u^{\pm}_{1,1} \overline{u^{
\pm}_{1,1}} u^{\pm}_{-1,1} \overline{u^{\pm}_{1,3}} u^{\pm}_{
-1,1}}{2}& \nonumber \\
&+\frac{u^{\pm}_{1,1} \overline{u^{\pm}_{-1,1}} u^{\pm}_{-1,1} 
\overline{u^{\pm}_{3,3}} u^{\pm}_{-1,1}}{2} + \frac{u^{\pm}_{
-1,1} \overline{u^{\pm}_{1,1}} u^{\pm}_{1,1} \overline{u^{\pm}
_{3,3}} u^{\pm}_{1,1}}{2}+\frac{u^{\pm}_{-1,1} \overline{u^{
\pm}_{1,1}} u^{\pm}_{1,1} \overline{u^{\pm}_{1,3}} u^{\pm}_{
-1,1}}{2}& \nonumber \\
&+\frac{u^{\pm}_{-1,1} \overline{u^{\pm}_{1,1}} u^{\pm}_{-1,1} 
\overline{u^{\pm}_{1,3}} u^{\pm}_{1,1}}{2} + \frac{u^{\pm}_{
-1,1} \overline{u^{\pm}_{1,1}} u^{\pm}_{-1,1} \overline{u^{\pm}
_{-1,3}} u^{\pm}_{-1,1}}{2}+\frac{u^{\pm}_{-1,1} \overline{u^{
\pm}_{-1,1}} u^{\pm}_{1,1} \overline{u^{\pm}_{3,3}} u^{\pm}_{
-1,1}}{2}& \nonumber \\
&+\frac{u^{\pm}_{-1,1} \overline{u^{\pm}_{-1,1}} u^{\pm}_{-1,1} 
\overline{u^{\pm}_{3,3}} u^{\pm}_{1,1}}{2} + \frac{u^{\pm}_{
-1,1} \overline{u^{\pm}_{-1,1}} u^{\pm}_{-1,1} \overline{u^{\pm}
_{1,3}} u^{\pm}_{-1,1}}{2}+\frac{u^{\pm}_{1,1} \overline{u^{
\pm}_{1,1}} u^{\pm}_{-3,3} \overline{u^{\pm}_{1,1}} u^{\pm}_{
1,1}}{2}& \nonumber \\
&+\frac{u^{\pm}_{1,1} \overline{u^{\pm}_{1,1}} u^{\pm}_{-1,3} 
\overline{u^{\pm}_{1,1}} u^{\pm}_{-1,1}}{2}+\frac{u^{\pm}_{1,
1} \overline{u^{\pm}_{1,1}} u^{\pm}_{-3,3} \overline{u^{\pm}
_{-1,1}} u^{\pm}_{-1,1}}{2}+\frac{u^{\pm}_{1,1} \overline{u^{
\pm}_{-1,1}} u^{\pm}_{-3,3} \overline{u^{\pm}_{1,1}} u^{\pm}_{
-1,1}}{2}& \nonumber \\
&+\frac{u^{\pm}_{-1,1} \overline{u^{\pm}_{1,1}} u^{\pm}_{-1,3} 
\overline{u^{\pm}_{1,1}} u^{\pm}_{1,1}}{2}+\frac{u^{\pm}_{-1,
1} \overline{u^{\pm}_{1,1}} u^{\pm}_{1,3} \overline{u^{\pm}
_{1,1}} u^{\pm}_{-1,1}}{2}+\frac{u^{\pm}_{-1,1} \overline{u^{
\pm}_{1,1}} u^{\pm}_{-3,3} \overline{u^{\pm}_{-1,1}} u^{\pm}_{
1,1}}{2}& \nonumber \\
&+\frac{u^{\pm}_{-1,1} \overline{u^{\pm}_{1,1}} u^{\pm}_{-1,3} 
\overline{u^{\pm}_{-1,1}} u^{\pm}_{-1,1}}{2}+\frac{u^{\pm}_{-1,
1} \overline{u^{\pm}_{-1,1}} u^{\pm}_{-3,3} \overline{u^{\pm}
_{1,1}} u^{\pm}_{1,1}}{2}+\frac{u^{\pm}_{-1,1} \overline{u^{
\pm}_{-1,1}} u^{\pm}_{-1,3} \overline{u^{\pm}_{1,1}} u^{\pm}_
{-1,1}}{2}& \nonumber \\
&+\frac{u^{\pm}_{-1,1} \overline{u^{\pm}_{-1,1}} u^{\pm}_{-3,3} 
\overline{u^{\pm}_{-1,1}} u^{\pm}_{-1,1}}{2}+\frac{u^{\pm}_{1,
1} \overline{u^{\pm}_{3,3}} u^{\pm}_{1,1} \overline{u^{\pm}
_{1,1}} u^{\pm}_{-1,1}}{2}+\frac{u^{\pm}_{1,1} \overline{u^{
\pm}_{3,3}} u^{\pm}_{-1,1} \overline{u^{\pm}_{1,1}} u^{\pm}_
{1,1}}{2}& \nonumber \\
&+\frac{u^{\pm}_{1,1} \overline{u^{\pm}_{1,3}} u^{\pm}_{-1,1} 
\overline{u^{\pm}_{1,1}} u^{\pm}_{-1,1}}{2}+\frac{u^{\pm}_{1,
1} \overline{u^{\pm}_{3,3}} u^{\pm}_{-1,1} \overline{u^{\pm}
_{-1,1}} u^{\pm}_{-1,1}}{2}+\frac{u^{\pm}_{-1,1} \overline{u^{
\pm}_{3,3}} u^{\pm}_{1,1} \overline{u^{\pm}_{1,1}} u^{\pm}_{
1,1}}{2}& \nonumber \\
&+\frac{u^{\pm}_{-1,1} \overline{u^{\pm}_{1,3}} u^{\pm}_{1,1} 
\overline{u^{\pm}_{1,1}} u^{\pm}_{-1,1}}{2}+\frac{u^{\pm}_{-1,
1} \overline{u^{\pm}_{3,3}} u^{\pm}_{1,1} \overline{u^{\pm}
_{-1,1}} u^{\pm}_{-1,1}}{2}+\frac{u^{\pm}_{-1,1} \overline{u^{
\pm}_{1,3}} u^{\pm}_{-1,1} \overline{u^{\pm}_{1,1}} u^{\pm}_{
1,1}}{2}& \nonumber \\
&+\frac{u^{\pm}_{-1,1} \overline{u^{\pm}_{-1,3}} u^{\pm}_{-1,1} 
\overline{u^{\pm}_{1,1}} u^{\pm}_{-1,1}}{2}+\frac{u^{\pm}_{-1,
1} \overline{u^{\pm}_{3,3}} u^{\pm}_{-1,1} \overline{u^{\pm}
_{-1,1}} u^{\pm}_{1,1}}{2}+\frac{u^{\pm}_{-1,1} \overline{u^{
\pm}_{1,3}} u^{\pm}_{-1,1} \overline{u^{\pm}_{-1,1}} u^{\pm}_{
-1,1}}{2}& \nonumber \\
&+\frac{u^{\pm}_{-3,3} \overline{u^{\pm}_{1,1}} u^{\pm}_{1,1} 
\overline{u^{\pm}_{1,1}} u^{\pm}_{1,1}}{2}+\frac{u^{\pm}_{-1,
3} \overline{u^{\pm}_{1,1}} u^{\pm}_{-1,1} \overline{u^{\pm}
_{1,1}} u^{\pm}_{1,1}}{2}+\frac{u^{\pm}_{-3,3} \overline{u^{
\pm}_{-1,1}} u^{\pm}_{-1,1} \overline{u^{\pm}_{1,1}} u^{\pm}_{
1,1}}{2}& \nonumber \\
&+\frac{u^{\pm}_{-3,3} \overline{u^{\pm}_{1,1}} u^{\pm}_{-1,1} 
\overline{u^{\pm}_{-1,1}} u^{\pm}_{1,1}}{2}+\frac{u^{\pm}_{-1,
3} \overline{u^{\pm}_{1,1}} u^{\pm}_{1,1} \overline{u^{\pm}
_{1,1}} u^{\pm}_{-1,1}}{2}+\frac{u^{\pm}_{-3,3} \overline{u^{
\pm}_{-1,1}} u^{\pm}_{1,1} \overline{u^{\pm}_{1,1}} u^{\pm}_{
-1,1}}{2}& \nonumber \\
&+\frac{u^{\pm}_{1,3} \overline{u^{\pm}_{1,1}} u^{\pm}_{-1,1} 
\overline{u^{\pm}_{1,1}} u^{\pm}_{-1,1}}{2}+\frac{u^{\pm}_{-1,
3} \overline{u^{\pm}_{-1,1}} u^{\pm}_{-1,1} \overline{u^{\pm}
_{1,1}} u^{\pm}_{-1,1}}{2}+\frac{u^{\pm}_{-3,3} \overline{u^{
\pm}_{1,1}} u^{\pm}_{1,1} \overline{u^{\pm}_{-1,1}} u^{\pm}_{
-1,1}}{2}& \nonumber \\
&+\frac{u^{\pm}_{-1,3} \overline{u^{\pm}_{1,1}} u^{\pm}_{-1,1} 
\overline{u^{\pm}_{-1,1}} u^{\pm}_{-1,1}}{2}+\frac{u^{\pm}_{-3,
3} \overline{u^{\pm}_{-1,1}} u^{\pm}_{-1,1} \overline{u^{\pm}
_{-1,1}} u^{\pm}_{-1,1}}{2}=0;&
\end{eqnarray}
\begin{flushleft}
(27) ${\cal O}((\pm t)^{-7/2} \exp\{3i\tau^{\pm}\})$,
\end{flushleft}
\begin{eqnarray}
&3\nu u^{\pm}_{5,5}-\frac{3\lambda_{0} \nu^{\prime} u^{\pm}_{5,5} 
\ln \vert t \vert}{2} + 12 \lambda_{0}^{4} u^{\pm}_{5,7} \pm i t 
\dot{u}^{\pm}_{5,5} \mp \frac{5iu^{\pm}_{5,5}}{2} \mp \frac{i 
\lambda_{0} u^{\pm \prime}_{5,5}}{2}& \nonumber \\
&\pm \frac{3i\lambda_{0}u^{\pm \prime}_{5,5}}{2} \pm \frac{3i
u^{\pm}_{5,5}}{2}+\frac{9 \lambda_{0}\nu^{\prime}u^{\pm}_{5,5} 
\ln \vert t \vert}{2}-36 \lambda_{0}^{4} u^{\pm}_{5,7}-2\lambda_{
0}^{2} \overline{u^{\pm}_{1,1}} u^{\pm}_{1,1} u^{\pm}_{5,5}& 
\nonumber \\
&- 2 \lambda_{0}^{2} \overline{u^{\pm}_{1,1}} u^{\pm}_{3,3} u^{
\pm}_{3,3}-2 \lambda_{0}^{2} \overline{u^{\pm}_{1,1}} u^{\pm}_{
5,5} u^{\pm}_{1,1}+2\lambda_{0}^{2} \overline{u^{\pm}_{-3,3}}u^{
\pm}_{1,1} u^{\pm}_{1,3} + 2 \lambda_{0}^{2} \overline{u^{\pm}_{
-3,3}} u^{\pm}_{-1,1} u^{\pm}_{3,3}& \nonumber \\
&+2\lambda_{0}^{2} \overline{u^{\pm}_{-3,3}} u^{\pm}_{1,2}u^{\pm}
_{1,2}+2\lambda_{0}^{2} \overline{u^{\pm}_{-3,3}} u^{\pm}_{1,3}
u^{\pm}_{1,1}+2\lambda_{0}^{2}\overline{u^{\pm}_{-3,3}} u^{\pm}
_{3,3} u^{\pm}_{-1,1}+2\lambda_{0}^{2} \overline{u^{\pm}_{-3,4}} 
u^{\pm}_{1,1} u^{\pm}_{1,2}& \nonumber \\
&+2 \lambda_{0}^{2} \overline{u^{\pm}_{-3,4}} u^{\pm}_{1,2} u^{
\pm}_{1,1}+2 \lambda_{0}^{2} \overline{u^{\pm}_{-3,5}} u^{\pm}_
{1,1} u^{\pm}_{1,1}+4\lambda_{0}^{2} \overline{u^{\pm}_{-5,5}} 
u^{\pm}_{-1,1} u^{\pm}_{1,1}+4\lambda_{0}^{2} \overline{u^{
\pm}_{-5,5}} u^{\pm}_{1,1} u^{\pm}_{-1,1}& \nonumber \\
&- \frac{\nu^{\prime}\overline{u^{\pm}_{-3,3}} u^{\pm}_{1,1} 
u^{\pm}_{1,1} \ln \vert t \vert}{8 \lambda_{0}} \mp \frac{i
(\overline{u^{\pm}_{-1,1}})^{\prime}u^{\pm}_{1,1}u^{\pm}_{3,
3}}{8 \lambda_{0}} \mp \frac{i(\overline{u^{\pm}_{-1,1}})^{
\prime}u^{\pm}_{3,3}u^{\pm}_{1,1}}{8 \lambda_{0}} \mp \frac{
i(\overline{u^{\pm}_{-3,3}})^{\prime}u^{\pm}_{1,1}u^{\pm}_{
1,1}}{8 \lambda_{0}}& \nonumber \\
&+\frac{u^{\pm}_{1,1} \overline{u^{\pm}_{-1,1}} u^{\pm}_{1,1} 
\overline{u^{\pm}_{-1,2}} u^{\pm}_{1,2}}{2}+\frac{u^{\pm}_{1,1} 
\overline{u^{\pm}_{-1,1}} u^{\pm}_{1,2} \overline{u^{\pm}_{-1,1}} 
u^{\pm}_{1,2}}{2}+\frac{u^{\pm}_{1,1} \overline{u^{\pm}_{-1,1}} 
u^{\pm}_{1,2} \overline{u^{\pm}_{-1,2}} u^{\pm}_{1,1}}{2}& 
\nonumber \\
&+\frac{u^{\pm}_{1,1} \overline{u^{\pm}_{-1,2}} u^{\pm}_{1,1} 
\overline{u^{\pm}_{-1,1}} u^{\pm}_{1,2}}{2}+\frac{u^{\pm}_{1,1} 
\overline{u^{\pm}_{-1,2}} u^{\pm}_{1,1} \overline{u^{\pm}_{-1,
2}} u^{\pm}_{1,1}}{2}+\frac{u^{\pm}_{1,1} \overline{u^{\pm}_{
-1,2}} u^{\pm}_{1,2} \overline{u^{\pm}_{-1,1}} u^{\pm}_{1,1}}{
2}& \nonumber \\
&+ \frac{u^{\pm}_{1,2} \overline{u^{\pm}_{-1,1}} u^{\pm}_{1,1} 
\overline{u^{\pm}_{-1,1}} u^{\pm}_{1,2}}{2}+\frac{u^{\pm}_{1,2} 
\overline{u^{\pm}_{-1,1}} u^{\pm}_{1,1} \overline{u^{\pm}_{-1,2}} 
u^{\pm}_{1,1}}{2}+\frac{u^{\pm}_{1,2} \overline{u^{\pm}_{-1,1}} 
u^{\pm}_{1,2} \overline{u^{\pm}_{-1,1}} u^{\pm}_{1,1}}{2}& 
\nonumber \\
&+\frac{u^{\pm}_{1,2} \overline{u^{\pm}_{-1,2}} u^{\pm}_{1,1} 
\overline{u^{\pm}_{-1,1}} u^{\pm}_{1,1}}{2}+\frac{u^{\pm}_{1,1} 
\overline{u^{\pm}_{1,1}} u^{\pm}_{1,1} \overline{u^{\pm}_{-1,1}} 
u^{\pm}_{3,3}}{2}+\frac{u^{\pm}_{1,1} \overline{u^{\pm}_{-1,1}} 
u^{\pm}_{1,1} \overline{u^{\pm}_{1,1}} u^{\pm}_{3,3}}{2}& 
\nonumber \\
&+\frac{u^{\pm}_{1,1} \overline{u^{\pm}_{-1,1}} u^{\pm}_{1,1} 
\overline{u^{\pm}_{-1,1}} u^{\pm}_{1,3}}{2}+\frac{u^{\pm}_{1,1} 
\overline{u^{\pm}_{-1,1}} u^{\pm}_{-1,1} \overline{u^{\pm}_{-1,
1}}u^{\pm}_{3,3}}{2}+\frac{u^{\pm}_{-1,1} \overline{u^{\pm}_{
-1,1}} u^{\pm}_{1,1} \overline{u^{\pm}_{-1,1}} u^{\pm}_{3,3}}{
2}& \nonumber \\
&+\frac{u^{\pm}_{1,1} \overline{u^{\pm}_{1,1}} u^{\pm}_{1,1} 
\overline{u^{\pm}_{-3,3}} u^{\pm}_{1,1}}{2}+\frac{u^{\pm}_{1,1} 
\overline{u^{\pm}_{-1,1}} u^{\pm}_{1,1} \overline{u^{\pm}_{-1,3}}
u^{\pm}_{1,1}}{2}+\frac{u^{\pm}_{1,1} \overline{u^{\pm}_{-1,1}} 
u^{\pm}_{1,1} \overline{u^{\pm}_{-3,3}} u^{\pm}_{-1,1}}{2}&
\nonumber \\
&+\frac{u^{\pm}_{1,1} \overline{u^{\pm}_{-1,1}} u^{\pm}_{-1,1} 
\overline{u^{\pm}_{-3,3}} u^{\pm}_{1,1}}{2}+\frac{u^{\pm}_{-1,1} 
\overline{u^{\pm}_{-1,1}} u^{\pm}_{1,1} \overline{u^{\pm}_{-3,3}}
u^{\pm}_{1,1}}{2}+\frac{u^{\pm}_{1,1} \overline{u^{\pm}_{1,1}} 
u^{\pm}_{3,3} \overline{u^{\pm}_{-1,1}} u^{\pm}_{1,1}}{2}& 
\nonumber \\
&+ \frac{u^{\pm}_{1,1} \overline{u^{\pm}_{-1,1}} u^{\pm}_{3,3} 
\overline{u^{\pm}_{1,1}} u^{\pm}_{1,1}}{2}+\frac{u^{\pm}_{1,1} 
\overline{u^{\pm}_{-1,1}} u^{\pm}_{1,3} \overline{u^{\pm}_{-1,1}}
u^{\pm}_{1,1}}{2}+\frac{u^{\pm}_{1,1} \overline{u^{\pm}_{-1,1}} 
u^{\pm}_{3,3} \overline{u^{\pm}_{-1,1}} u^{\pm}_{-1,1}}{2}&
\nonumber \\
&+\frac{u^{\pm}_{-1,1} \overline{u^{\pm}_{-1,1}} u^{\pm}_{3,3} 
\overline{u^{\pm}_{-1,1}} u^{\pm}_{1,1}}{2}+\frac{u^{\pm}_{1,1} 
\overline{u^{\pm}_{-3,3}} u^{\pm}_{1,1} \overline{u^{\pm}_{1,1}} 
u^{\pm}_{1,1}}{2}+\frac{u^{\pm}_{1,1} \overline{u^{\pm}_{-1,3}} 
u^{\pm}_{1,1} \overline{u^{\pm}_{-1,1}} u^{\pm}_{1,1}}{2}& 
\nonumber \\
&+\frac{u^{\pm}_{1,1} \overline{u^{\pm}_{-3,3}} u^{\pm}_{1,1} 
\overline{u^{\pm}_{-1,1}} u^{\pm}_{-1,1}}{2}+\frac{u^{\pm}_{1,1} 
\overline{u^{\pm}_{-3,3}} u^{\pm}_{-1,1} \overline{u^{\pm}_{-1,1}}
u^{\pm}_{1,1}}{2}+\frac{u^{\pm}_{-1,1} \overline{u^{\pm}_{-3,3}} 
u^{\pm}_{1,1} \overline{u^{\pm}_{-1,1}} u^{\pm}_{1,1}}{2}&
\nonumber \\
&+\frac{u^{\pm}_{3,3} \overline{u^{\pm}_{-1,1}} u^{\pm}_{1,1} 
\overline{u^{\pm}_{1,1}} u^{\pm}_{1,1}}{2}+\frac{u^{\pm}_{3,3} 
\overline{u^{\pm}_{1,1}} u^{\pm}_{1,1} \overline{u^{\pm}_{-1,1}} 
u^{\pm}_{1,1}}{2}+\frac{u^{\pm}_{1,3} \overline{u^{\pm}_{-1,1}} 
u^{\pm}_{1,1} \overline{u^{\pm}_{-1,1}} u^{\pm}_{1,1}}{2}& 
\nonumber \\
&+\frac{u^{\pm}_{3,3} \overline{u^{\pm}_{-1,1}} u^{\pm}_{-1,1} 
\overline{u^{\pm}_{-1,1}}u^{\pm}_{1,1}}{2}+\frac{u^{\pm}_{3,3} 
\overline{u^{\pm}_{-1,1}}u^{\pm}_{1,1}\overline{u^{\pm}_{-1,1}}
u^{\pm}_{-1,1}}{2}=0;&
\end{eqnarray}
\begin{flushleft}
(28) ${\cal O}((\pm t)^{-7/2} \exp\{-2i\tau^{\pm}\})$,
\end{flushleft}
\begin{eqnarray}
&-2\nu u^{\pm}_{-5,5}+\lambda_{0} \nu^{\prime} u^{\pm}_{-5,5} 
\ln \vert t \vert - 8 \lambda_{0}^{4} u^{\pm}_{-5,7} \pm i t 
\dot{u}^{\pm}_{-5,5} \mp \frac{5iu^{\pm}_{-5,5}}{2} \mp \frac{
i \lambda_{0} u^{\pm \prime}_{-5,5}}{2}& \nonumber \\
&\mp i \lambda_{0}u^{\pm \prime}_{-5,5} \mp i u^{\pm}_{-5,5}
+2\lambda_{0}\nu^{\prime}u^{\pm}_{-5,5} \ln \vert t \vert 
- 16 \lambda_{0}^{4} u^{\pm}_{-5,7}-2 \lambda_{0}^{2} 
\overline{u^{\pm}_{1,1}} u^{\pm}_{1,1} u^{\pm}_{-5,5}& 
\nonumber \\
&-2 \lambda_{0}^{2} \overline{u^{\pm}_{1,1}} u^{\pm}_{-1,1} u^{
\pm}_{-3,5}-2\lambda_{0}^{2} \overline{u^{\pm}_{1,1}} u^{\pm}_{
-1,2} u^{\pm}_{-3,4}-2\lambda_{0}^{2}\overline{u^{\pm}_{1,1}}u^{
\pm}_{-1,3} u^{\pm}_{-3,3}-2 \lambda_{0}^{2} \overline{u^{\pm}_{
1,1}} u^{\pm}_{-3,3} u^{\pm}_{-1,3}& \nonumber \\
&-2\lambda_{0}^{2} \overline{u^{\pm}_{1,1}} u^{\pm}_{-3,4}u^{\pm}
_{-1,2}-2\lambda_{0}^{2} \overline{u^{\pm}_{1,1}} u^{\pm}_{-5,5}
u^{\pm}_{1,1}-2\lambda_{0}^{2}\overline{u^{\pm}_{1,1}} u^{\pm}
_{-3,5} u^{\pm}_{-1,1}-2\lambda_{0}^{2} \overline{u^{\pm}_{1,2}} 
u^{\pm}_{-1,1} u^{\pm}_{-3,4}& \nonumber \\
&-2\lambda_{0}^{2} \overline{u^{\pm}_{1,2}} u^{\pm}_{-1,2} u^{
\pm}_{-3,3}-2\lambda_{0}^{2} \overline{u^{\pm}_{1,2}} u^{\pm}_
{-3,3} u^{\pm}_{-1,2}-2\lambda_{0}^{2} \overline{u^{\pm}_{1,2}} 
u^{\pm}_{-3,4} u^{\pm}_{-1,1}-4\lambda_{0}^{2} \overline{u^{
\pm}_{3,3}} u^{\pm}_{1,1} u^{\pm}_{-3,3}& \nonumber \\
&-2\lambda_{0}^{2} \overline{u^{\pm}_{1,3}} u^{\pm}_{-1,1} u^{
\pm}_{-3,3}-4\lambda_{0}^{2} \overline{u^{\pm}_{3,3}} u^{\pm}_
{-1,1} u^{\pm}_{-1,3}-4\lambda_{0}^{2} \overline{u^{\pm}_{3,3}} 
u^{\pm}_{-1,2} u^{\pm}_{-1,2}-4\lambda_{0}^{2} \overline{u^{
\pm}_{3,3}} u^{\pm}_{-3,3} u^{\pm}_{1,1}& \nonumber \\
&-2\lambda_{0}^{2} \overline{u^{\pm}_{1,3}} u^{\pm}_{-3,3} u^{
\pm}_{-1,1}-4\lambda_{0}^{2} \overline{u^{\pm}_{3,3}} u^{\pm}_
{-1,3} u^{\pm}_{-1,1}-4\lambda_{0}^{2} \overline{u^{\pm}_{3,4}} 
u^{\pm}_{-1,1} u^{\pm}_{-1,2}-4\lambda_{0}^{2} \overline{u^{
\pm}_{3,4}} u^{\pm}_{-1,2} u^{\pm}_{-1,1}& \nonumber \\
&-6\lambda_{0}^{2} \overline{u^{\pm}_{5,5}} u^{\pm}_{-1,1} u^{
\pm}_{1,1}-6\lambda_{0}^{2} \overline{u^{\pm}_{5,5}} u^{\pm}_
{1,1} u^{\pm}_{-1,1}-4\lambda_{0}^{2} \overline{u^{\pm}_{3,5}} 
u^{\pm}_{-1,1} u^{\pm}_{-1,1} + \frac{\nu^{\prime}\overline{
u^{\pm}_{1,1}} u^{\pm}_{-1,1} u^{\pm}_{-3,3} \ln \vert t 
\vert}{8 \lambda_{0}}& \nonumber \\
&+ \frac{\nu^{\prime}\overline{u^{\pm}_{1,1}} u^{\pm}_{-3,3} 
u^{\pm}_{-1,1} \ln \vert t \vert}{8 \lambda_{0}}+\frac{\nu^{
\prime}\overline{u^{\pm}_{3,3}} u^{\pm}_{-1,1} u^{\pm}_{-1,1} 
\ln \vert t \vert}{4 \lambda_{0}} \mp \frac{i(\overline{u^{
\pm}_{1,1}})^{\prime}u^{\pm}_{-1,1}u^{\pm}_{-3,3}}{8 \lambda_{
0}} \mp \frac{i(\overline{u^{\pm}_{1,1}})^{\prime}u^{\pm}_{-3,3}
u^{\pm}_{-1,1}}{8 \lambda_{0}}& \nonumber \\
&\mp \frac{i(\overline{u^{\pm}_{3,3}})^{\prime}u^{\pm}_{-1,1} 
u^{\pm}_{-1,1}}{8 \lambda_{0}} + \frac{u^{\pm}_{-1,1} \overline{
u^{\pm}_{1,1}} u^{\pm}_{-1,1} \overline{u^{\pm}_{1,2}} u^{\pm}_{
-1,2}}{2}+\frac{u^{\pm}_{-1,1} \overline{u^{\pm}_{1,1}}u^{\pm}_{
-1,2} \overline{u^{\pm}_{1,1}}u^{\pm}_{-1,2}}{2}+\frac{u^{\pm}_{
-1,1} \overline{u^{\pm}_{1,1}} u^{\pm}_{-1,2}\overline{u^{\pm}_{
1,2}} u^{\pm}_{-1,1}}{2}& \nonumber \\
&+\frac{u^{\pm}_{-1,1} \overline{u^{\pm}_{1,2}} u^{\pm}_{-1,1} 
\overline{u^{\pm}_{1,1}} u^{\pm}_{-1,2}}{2}+\frac{u^{\pm}_{-1,1} 
\overline{u^{\pm}_{1,2}} u^{\pm}_{-1,1} \overline{u^{\pm}_{1,2}} 
u^{\pm}_{-1,1}}{2}+\frac{u^{\pm}_{-1,1} \overline{u^{\pm}_{1,2}} 
u^{\pm}_{-1,2} \overline{u^{\pm}_{1,1}} u^{\pm}_{-1,1}}{2}& 
\nonumber \\
&+ \frac{u^{\pm}_{-1,2} \overline{u^{\pm}_{1,1}} u^{\pm}_{-1,1} 
\overline{u^{\pm}_{1,1}} u^{\pm}_{-1,2}}{2}+\frac{u^{\pm}_{-1,2} 
\overline{u^{\pm}_{1,1}} u^{\pm}_{-1,1} \overline{u^{\pm}_{1,2}} 
u^{\pm}_{-1,1}}{2}+\frac{u^{\pm}_{-1,2} \overline{u^{\pm}_{1,1}} 
u^{\pm}_{-1,2} \overline{u^{\pm}_{1,1}} u^{\pm}_{-1,1}}{2}& 
\nonumber \\
&+\frac{u^{\pm}_{-1,2} \overline{u^{\pm}_{1,2}} u^{\pm}_{-1,1} 
\overline{u^{\pm}_{1,1}} u^{\pm}_{-1,1}}{2}+\frac{u^{\pm}_{1,1} 
\overline{u^{\pm}_{1,1}} u^{\pm}_{-1,1} \overline{u^{\pm}_{1,1}} 
u^{\pm}_{-3,3}}{2}+\frac{u^{\pm}_{-1,1} \overline{u^{\pm}_{1,1}} 
u^{\pm}_{1,1} \overline{u^{\pm}_{1,1}} u^{\pm}_{-3,3}}{2}& 
\nonumber \\
&+\frac{u^{\pm}_{-1,1} \overline{u^{\pm}_{1,1}} u^{\pm}_{-1,1} 
\overline{u^{\pm}_{1,1}} u^{\pm}_{-1,3}}{2}+\frac{u^{\pm}_{-1,1} 
\overline{u^{\pm}_{1,1}} u^{\pm}_{-1,1} \overline{u^{\pm}_{-1,
1}}u^{\pm}_{-3,3}}{2}+\frac{u^{\pm}_{-1,1} \overline{u^{\pm}_{
-1,1}} u^{\pm}_{-1,1} \overline{u^{\pm}_{1,1}} u^{\pm}_{-3,3}}{
2}& \nonumber \\
&+\frac{u^{\pm}_{1,1} \overline{u^{\pm}_{1,1}} u^{\pm}_{-1,1} 
\overline{u^{\pm}_{3,3}} u^{\pm}_{-1,1}}{2}+\frac{u^{\pm}_{-1,1} 
\overline{u^{\pm}_{1,1}} u^{\pm}_{1,1} \overline{u^{\pm}_{3,3}}
u^{\pm}_{-1,1}}{2}+\frac{u^{\pm}_{-1,1} \overline{u^{\pm}_{1,1}} 
u^{\pm}_{-1,1} \overline{u^{\pm}_{3,3}} u^{\pm}_{1,1}}{2}&
\nonumber \\
&+\frac{u^{\pm}_{-1,1} \overline{u^{\pm}_{1,1}} u^{\pm}_{-1,1} 
\overline{u^{\pm}_{1,3}} u^{\pm}_{-1,1}}{2}+\frac{u^{\pm}_{-1,1} 
\overline{u^{\pm}_{-1,1}} u^{\pm}_{-1,1} \overline{u^{\pm}_{3,3}}
u^{\pm}_{-1,1}}{2}+\frac{u^{\pm}_{1,1} \overline{u^{\pm}_{1,1}} 
u^{\pm}_{-3,3} \overline{u^{\pm}_{1,1}} u^{\pm}_{-1,1}}{2}& 
\nonumber \\
&+ \frac{u^{\pm}_{-1,1} \overline{u^{\pm}_{1,1}} u^{\pm}_{-3,3} 
\overline{u^{\pm}_{1,1}} u^{\pm}_{1,1}}{2}+\frac{u^{\pm}_{-1,1} 
\overline{u^{\pm}_{1,1}} u^{\pm}_{-1,3} \overline{u^{\pm}_{1,1}}
u^{\pm}_{-1,1}}{2}+\frac{u^{\pm}_{-1,1} \overline{u^{\pm}_{1,1}} 
u^{\pm}_{-3,3} \overline{u^{\pm}_{-1,1}} u^{\pm}_{-1,1}}{2}&
\nonumber \\
&+\frac{u^{\pm}_{-1,1} \overline{u^{\pm}_{-1,1}} u^{\pm}_{-3,3} 
\overline{u^{\pm}_{1,1}} u^{\pm}_{-1,1}}{2}+\frac{u^{\pm}_{1,1} 
\overline{u^{\pm}_{3,3}} u^{\pm}_{-1,1} \overline{u^{\pm}_{1,1}} 
u^{\pm}_{-1,1}}{2}+\frac{u^{\pm}_{-1,1} \overline{u^{\pm}_{3,3}} 
u^{\pm}_{1,1} \overline{u^{\pm}_{1,1}} u^{\pm}_{-1,1}}{2}& 
\nonumber \\
&+\frac{u^{\pm}_{-1,1} \overline{u^{\pm}_{3,3}} u^{\pm}_{-1,1} 
\overline{u^{\pm}_{1,1}} u^{\pm}_{1,1}}{2}+\frac{u^{\pm}_{-1,1} 
\overline{u^{\pm}_{1,3}} u^{\pm}_{-1,1} \overline{u^{\pm}_{1,1}}
u^{\pm}_{-1,1}}{2}+\frac{u^{\pm}_{-1,1} \overline{u^{\pm}_{3,3}} 
u^{\pm}_{-1,1} \overline{u^{\pm}_{-1,1}} u^{\pm}_{-1,1}}{2}&
\nonumber \\
&+\frac{u^{\pm}_{-3,3} \overline{u^{\pm}_{1,1}} u^{\pm}_{-1,1} 
\overline{u^{\pm}_{1,1}} u^{\pm}_{1,1}}{2}+\frac{u^{\pm}_{-3,3} 
\overline{u^{\pm}_{1,1}} u^{\pm}_{1,1} \overline{u^{\pm}_{1,1}} 
u^{\pm}_{-1,1}}{2}+\frac{u^{\pm}_{-1,3} \overline{u^{\pm}_{1,1}} 
u^{\pm}_{-1,1} \overline{u^{\pm}_{1,1}} u^{\pm}_{-1,1}}{2}& 
\nonumber \\
&+\frac{u^{\pm}_{-3,3} \overline{u^{\pm}_{-1,1}} u^{\pm}_{-1,1} 
\overline{u^{\pm}_{1,1}}u^{\pm}_{-1,1}}{2}+\frac{u^{\pm}_{-3,3} 
\overline{u^{\pm}_{1,1}}u^{\pm}_{-1,1}\overline{u^{\pm}_{-1,1}}
u^{\pm}_{-1,1}}{2}=0;&
\end{eqnarray}
\begin{flushleft}
(29) ${\cal O}((\pm t)^{-7/2} \exp\{4i\tau^{\pm}\})$,
\end{flushleft}
\begin{eqnarray}
&+2\lambda_{0}^{2} \overline{u^{\pm}_{-3,3}} u^{\pm}_{1,1} u^{
\pm}_{3,3}+2\lambda_{0}^{2} \overline{u^{\pm}_{-3,3}} u^{\pm}_{
3,3} u^{\pm}_{1,1}+4\lambda_{0}^{2} \overline{u^{\pm}_{-5,5}} 
u^{\pm}_{1,1} u^{\pm}_{1,1}& \nonumber \\
&+\frac{u^{\pm}_{1,1} \overline{u^{\pm}_{-1,1}} u^{\pm}_{1,1} 
\overline{u^{\pm}_{-1,1}} u^{\pm}_{3,3}}{2}+\frac{u^{\pm}_{1,1} 
\overline{u^{\pm}_{-1,1}} u^{\pm}_{1,1} \overline{u^{\pm}_{-3,3}} 
u^{\pm}_{1,1}}{2}+\frac{u^{\pm}_{1,1} \overline{u^{\pm}_{-1,1}} 
u^{\pm}_{3,3} \overline{u^{\pm}_{-1,1}} u^{\pm}_{1,1}}{2}& 
\nonumber \\
&+\frac{u^{\pm}_{1,1} \overline{u^{\pm}_{-3,3}} u^{\pm}_{1,1} 
\overline{u^{\pm}_{-1,1}}u^{\pm}_{1,1}}{2}+\frac{u^{\pm}_{3,3} 
\overline{u^{\pm}_{-1,1}}u^{\pm}_{1,1}\overline{u^{\pm}_{-1,1}}
u^{\pm}_{1,1}}{2}=0;&
\end{eqnarray}
\begin{flushleft}
(30) ${\cal O}((\pm t)^{-7/2} \exp\{-3i\tau^{\pm}\})$,
\end{flushleft}
\begin{eqnarray}
&-2\lambda_{0}^{2} \overline{u^{\pm}_{1,1}} u^{\pm}_{-1,1} u^{
\pm}_{-5,5}-2\lambda_{0}^{2} \overline{u^{\pm}_{1,1}} u^{\pm}_{
-3,3} u^{\pm}_{-3,3}-2\lambda_{0}^{2} \overline{u^{\pm}_{1,1}} 
u^{\pm}_{-5,5} u^{\pm}_{-1,1}-4\lambda_{0}^{2} \overline{u^{
\pm}_{3,3}} u^{\pm}_{-1,1} u^{\pm}_{-3,3}& \nonumber \\
&-4\lambda_{0}^{2} \overline{u^{\pm}_{3,3}} u^{\pm}_{-3,3} u^{
\pm}_{-1,1}-6\lambda_{0}^{2}\overline{u^{\pm}_{5,5}}u^{\pm}_{
-1,1} u^{\pm}_{-1,1}+\frac{u^{\pm}_{-1,1} \overline{u^{\pm}_{1,1}} 
u^{\pm}_{-1,1} \overline{u^{\pm}_{1,1}} u^{\pm}_{-3,3}}{2}+\frac{
u^{\pm}_{-1,1} \overline{u^{\pm}_{1,1}} u^{\pm}_{-1,1} \overline{
u^{\pm}_{3,3}} u^{\pm}_{-1,1}}{2}& \nonumber \\
&+\frac{u^{\pm}_{-1,1} \overline{u^{\pm}_{1,1}} u^{\pm}_{-3,3} 
\overline{u^{\pm}_{1,1}}u^{\pm}_{-1,1}}{2}+\frac{u^{\pm}_{-1,1} 
\overline{u^{\pm}_{3,3}}u^{\pm}_{-1,1}\overline{u^{\pm}_{1,1}}
u^{\pm}_{-1,1}}{2}+\frac{u^{\pm}_{-3,3} \overline{u^{\pm}_{1,1}} 
u^{\pm}_{-1,1} \overline{u^{\pm}_{1,1}} u^{\pm}_{-1,1}}{2}=0.&
\end{eqnarray}

\end{document}